%% file: spt_profs.tex
\documentclass[usenatbib]{mnras}
\pdfoutput=1

\usepackage[utf8]{inputenc}
\usepackage[pdftex]{graphicx}
\usepackage{mathptmx}
\usepackage{fixltx2e}
\usepackage{color}
\usepackage{amssymb}
\usepackage{amsmath}

\input{defn}

\begin{document}

\title
[Evolution and cores of SPT-cluster profiles]
{Hydrostatic \emph{Chandra} X-ray analysis of SPT-selected galaxy clusters - I. Evolution of profiles and core properties}
\author
[J.~S. Sanders et al.]
{
  \begin{minipage}[b]{\linewidth}
    \begin{flushleft}
      J.~S.~Sanders,$^1$\thanks{E-mail: jsanders@mpe.mpg.de}
      A.~C.~Fabian,$^2$
      H.~R.~Russell$^2$ and
      S.~A.~Walker$^3$
    \end{flushleft}
  \end{minipage}
  \\
  $^1$ Max-Planck-Institut f\"ur extraterrestrische Physik,
  Giessenbachstrasse 1, D-85748 Garching, Germany\\
  $^2$ Institute of Astronomy, Madingley Road, Cambridge, CB3 0HA\\
  $^3$ Astrophysics Science Division, X-ray Astrophysics Laboratory, Code 662, NASA / Goddard Space Flight Center, Greenbelt, MD 20771, USA
}

\maketitle

\begin{abstract}
We analyse \emph{Chandra X-ray Observatory} observations of a set of galaxy clusters selected by the South Pole Telescope using a new publicly-available forward-modelling projection code, \textsc{mbproj2}, assuming hydrostatic equilibrium.
By fitting a powerlaw plus constant entropy model we find no evidence for a central entropy floor in the lowest-entropy systems.
A model of the underlying central entropy distribution shows a narrow peak close to zero entropy which accounts for 60 per cent of the systems, and a second broader peak around $130 \keV\cm^{2}$.
We look for evolution over the 0.28 to 1.2 redshift range of the sample in density, pressure, entropy and cooling time at $0.015 R_{500}$ and at 10~kpc radius.
By modelling the evolution of the central quantities with a simple model, we find no evidence for a non-zero slope with redshift.
In addition, a non-parametric sliding median shows no significant change.
The fraction of cool-core clusters with central cooling times below 2~Gyr is consistent above and below $z=0.6$ ($\sim 30-40$ per cent).
Both by comparing the median thermodynamic profiles, centrally biased towards cool cores, in two redshift bins, and by modelling the evolution of the unbiased average profile as a function of redshift, we find no significant evolution beyond self-similar scaling in any of our examined quantities.
Our average modelled radial density, entropy and cooling-time profiles appear as powerlaws with breaks around $0.2 R_{500}$.
The dispersion in these quantities rises inwards of this radius to around $0.4$ dex, although some of this scatter can be fit by a bimodal model.
\end{abstract}


\section{Introduction}
Within the dark-matter-dominated potential well of galaxy clusters lies the intracluster medium (ICM), a hot atmosphere primarily seen by its emission in the X-ray waveband.
By examining the morphology and temperature of the ICM and assuming that it is in hydrostatic equilibrium, the mass, profile and shape of the underlying dark-matter halo can be inferred \citep[e.g.][]{Allen01A2390}.
The ICM is also sensitive to baryonic physics, such as the input of energy by AGN in galactic nuclei within clusters \citep[e.g.][]{BohringerPer93}.

In many nearby clusters the central cooling times of the ICM are short.
In the absence of heating a cooling flow would develop \citep{Fabian94}, where material would rapidly cool out of the X-ray waveband at rates of $10-1000\Msunpyr$.
Such high rates of cooling are not observed \citep[e.g.][]{PetersonFabian06} and so there must be a mechanism by which the rapid cooling is prevented.
AGN in cluster cores are observed to put energy into their surroundings by the inflation of bubbles of radio-emitting plasma, seen as cavities in X-ray images of the ICM \citep[e.g.][]{McNamaraNulsen07}.
The balance between the energy lost by X-ray emission and the cavity heating rates estimated from observations imply AGN feedback is the mechanism for how cooling flows are prevented \citep[e.g.][]{Fabian12}.
An important question is how AGN maintain the close heating-cooling balance in nearby clusters and whether this balance is maintained in earlier epochs.
This is not only of relevance to the cores of galaxy clusters and their central galaxies, but has widespread relevance to understanding galaxy formation.

One of the strongest indicators of non-gravitational heating in clusters is the entropy \citep{Voit02}.
The specific value can be written as $K_\mathrm{e} = kT \: n_\mathrm{e}^{-2/3}$, if $n_\mathrm{e}$ is the electron density and $T$ is the temperature.
In the absence of conduction, convection ensures that low-entropy material moves to the centre, while high-entropy material goes to the outskirts.
Non-gravitational processes show as deviations from the entropy distribution expected in pure-gravitational distributions.

Previous studies have found different behaviour of the entropy in cluster cores.
Several groups have found evidence for entropy flattening in the cores of clusters and groups \citep[e.g.][]{David96,Cavagnolo09,McDonald13SPT}.
In a volume-limited sample of local clusters \citep{Panagoulia14} instead found that the entropy profiles were consistent with being powerlaws in the central regions.
The measurement of central cluster properties is difficult because of projection effects, substructure, multiphase material and metallicity gradients.
\cite{Hogan17} also found no evidence for a floor in a small sample of clusters with deep X-ray observations.
Resolving these disagreements over the central entropy is important for understanding the heating and cooling processes taking place in the centres of these objects.

The X-ray emission observed from galaxy clusters is projected along the line of sight.
To extract the three-dimensional information some assumptions of the geometry have to be made, such as spherical symmetry.
Various methods have been previously used to extract the three-dimensional thermodynamical properties, including deprojection of the X-ray surface-brightness profile \citep{Fabian81}, projecting a spectral model in shells to fit projected spectra, as implemented as the PROJCT model in \textsc{xspec} \citep{ArnaudXspec}, correcting projected quantities \citep{Ettori02}, deprojection of the X-ray spectra \citep{SandersPer07,Russell08}, forward-fitting of a mass and temperature model to spectra extracted from shells \citep{Mahdavi08,Nulsen10} and fitting a model to the X-ray event dataset \citep{Olamaie15}.

In \cite{Sanders14PKS0745} we introduced a new forward-fitting code, \textsc{mbproj}, which fits surface-brightness profiles in multiple energy bands.
It fits a model density profile and either a mass (assuming hydrostatic equilibrium) or a temperature model.
An MCMC (Markov Chain Monte Carlo) analysis is used to generate profiles of physical quantities.
The advantage a surface-brightness profile modelling code has over spectral fitting is that it is easier to visually inspect how the goodness of the fit changes as a function of radius and to check that the background modelling is correct.
It is also easier to adapt the size of the radial bins as the modelling requires and to connect the obtained profiles to images of the cluster.

Selection of galaxy clusters using the Sunyaev-Zel'dovich (SZ; \citealt{Sunyaev72}) effect has some advantages compared to other methods, including a uniform mass selection to higher redshifts and not being sensitive to cool cores \citep[e.g.][]{Birkinshaw99}.
Nevertheless, it may be the case that there could be a large population of contaminating AGN at higher redshifts which could affect the SZ signal or selection \citep[e.g.][]{Bufanda17}.

In this paper we analyse \emph{Chandra} observations of an SZ-selected sample of galaxy clusters obtained by the SPT telescope \citep{Carlstrom11} with the aid of a new version of the multi-band X-ray projection algorithm, \textsc{mbproj2}.
In this paper we focus on the thermodynamic profiles from our hydrostatic analysis, to examine the evolution of the profiles and core properties, while we leave the comparison of our obtained hydrostatic masses with other mass measurements to a future work.

This analysis differs in several respects from the analyses of almost-same samples by \cite{McDonald13SPT} [hereafter referred to as MD13] and \cite{McDonald14SPT} [hereafter MD14].
Firstly, we did a self-consistent modelling of the X-ray profiles to obtain the profiles of physical quantities, using both parametric and binned density profiles.
In MD13, the surface-brightness profiles were fit by a parametric model and the projected temperature profiles obtained by fitting spectra in wide spatial bins.
A hydrostatic model was fit to both to obtain deprojected quantities.
Our modelling is more sensitive to variations in temperature as we do not use these wide spectral bins.
In addition, the published density profiles of MD13 can have central densities a factor of a few away from updated values due to an error in the fitting procedure (see Section \ref{sect:MD13compar}).

In MD14 a joint analysis was done to all the clusters in different subsamples to obtain average physical profiles.
In their analysis they assumed that the clusters in a subsample shared a common modelled temperature profile.
The projected spectra in radial bins for the clusters were fit jointly with this temperature model, allowing the normalisation profiles to be different.
To deproject the temperature profile a parametric model was fitted to the previous projected profile.
The deprojected density profiles were taken from \cite{McDonald13SPT}, but the mean density profile of a subsample was computed by weighting the individual density profiles by the number of counts in each radial bin.
In comparison, our modelling assumed hydrostatic equilibrium but did not assume the same temperature profile for each system.
We also examine some median cluster quantities, which are independent on the relative data quality of different clusters.

Our analysis also differs from MD13 and MD14 by the choice of cluster centres.
We use the X-ray peak defined using a small (50 count) aperture as a cluster centre, whereas they use the centroid of a 250-500 kpc annulus for the main part of their analysis.
When they compare to results using the X-ray peak, the aperture used to find the peak position is substantially larger than ours.
This difference cluster centre is important for the differences in results we obtain in the cores of these systems from MD13 and MD14.

We assume a cosmology where $H_0 = 70 \kmpspMpc$, $\Omega_m=0.3$ and $\Omega_\Lambda=0.7$.

\section{Data analysis}
Our sample of systems (listed in Appendix \ref{append:sample}) includes the sample of SPT clusters of \cite{Bleem15} marked as having X-ray data, excluding SPT-CLJ0037-5047 which has low signal to noise.
We also include two further systems in that paper which also have X-ray data and were examined in MD13, SPT-CLJ0236-4938 and SPT-CLJ0310-4647.
We do not include SPT-CLJ0446-5849 which has low signal to noise.
We also exclude SPT-CLJ0330-5228 and SPT-CLJ0551-5709 which are contaminated (MD14).
As MD13 describes, their sample consists of strongly-detected clusters by SPT, with SPT detection significances between 5.7 and 43.
The mass range of $M_{500}$ is between $2\times 10^{14}$ and $2\times10^{15}\Msun$, while the redshift range is between 0.3 and 1.2.
At the median redshift, the sample should be around 50 per cent complete at $M_{500} =4\times 10^{14}$, increasing to 100 per cent at $6\times 10^{14} \Msun$.
We make use of any new public observations from the \emph{Chandra} archive, where possible.

As detailed below, the cluster surface-brightness profiles are fit in multiple X-ray bands using the \textsc{mbproj2} code (described in Appendix \ref{append:mbproj}), which can compute profiles assuming hydrostatic equilibrium or in its absence.
A surface-brightness profile in a single band would only be sensitive to variations in gas density.
However, by using profiles from multiple bands simultaneously, we are sensitive to temperature variations due to the change in the spectral shape.
With low numbers of bands, this is similar to using X-ray colours to measure temperature \cite[e.g.][]{Allen97}.
As the number of bands increases this becomes equivalent to spectral fitting, which would also allow the metallicity to be fit given sufficient data quality.

When using fine radial bins it is difficult to obtain the gas temperature due to the lack of counts.
However, by introducing the assumption of hydrostatic equilibrium with some underlying dark-matter potential, we can compute the pressure, and given the densities, the temperature.
\textsc{mbproj2} computes the projected surface-brightness profiles in multiple bands for a given gas-density profile and dark-matter profile.
The uncertainties on the fits are explored using the observed profiles and MCMC.

\subsection{Initial data preparation}
We downloaded the data for each cluster from the \emph{Chandra} archive.
The datasets were reprocessed using \textsc{ciao} \citep{Fruscione06} \textsc{acis\_process\_events}, applying very-faint event grading where possible.
We excluded bad time periods by iteratively $\sigma$-clipping the lightcurve to remove periods $2.5\sigma$ away from the median value, where $\sigma$ is the Poisson error on the median number of counts in a 200s time bin.
For observations using ACIS-S the lightcurve was constructed in the 2.5 to 7 keV band, otherwise the 0.5 to 12 keV band was used.
For each cluster, we reprojected observations to a common coordinate system.

Standard blank-sky background datasets were obtained for each CCD of each observation.
We used the bad-pixel table from an observation to remove bad pixels from the respective background observations.
For a particular observation, the exposure times of the background event files for each CCDs were adjusted so that each had the same 9 to 12 keV band count rate as the respective cluster data (this spectral range is dominated by the particle background).
These background event files for an observation were then adjusted to have the same exposure time as the lowest exposure file by randomly discarding events.
When multiple datasets were used for a cluster, we similarly adjusted the exposure times of the background datasets to have the same ratio of their exposure to the total background exposure, as their respective cluster observation to the total cluster exposure.
The background event files for each cluster were reprojected to match the coordinate system of the respective cluster observation and then the common coordinate system for the cluster.

Total images were created using detector pixel binning in the 0.5 to 7 keV band.
We detected point sources in these images using \textsc{wavdetect}, using scales of 2, 2.828, 4, 5.657, 8, 11.314 and 16 pixels, with a maximum of 5 iterations.
The resulting point sources were manually verified, removing obvious false detections and adding missed sources, as appropriate.
Some of the observations were contaminated by other extended sources and structures.
We identified these in smoothed images and excluded them in our analysis.
In a couple of cases the systems were too close to be separated and so they remain in the analysis.

\subsection{Surface-brightness profiles}
\label{sect:sbprofs}
To identify the centre of each cluster, we initially found the brightest pixel in an adaptively-smoothed map, smoothed to have a minimum signal to noise of 10 in a top hat kernel.
This position was then refined iteratively by repeatedly finding the centroid of a circle with a radius chosen to contain 50 counts (although its minimum radius was four 0.492 arcsec pixels).
These peak positions are given in Appendix~\ref{append:posns}.
Also shown in this table for each object is a second position computed from the centroid of an annulus between radii of 250 and 500 kpc, which is the same technique as used by MD13 to define their cluster centres.
As the centroid of this annulus does not always converge to a single point, we used the mean position of 100 iterations, after discarding an initial 100 iterations.
The offset between the two positions in arcsec and kpc on the sky is also shown in the table.
Although we used the same technique as MD13 for the annulus centroids, some of our positions show large differences from MD13.
These include the positions for SPT-CLJ0217-5245 and SPT-CLJ0252-4824, which show differences by 47 and 40 arcsec, respectively.
It is unclear why the positions differ given we use the same method.

We extracted total cluster and background images in ten energy bands between
neighbouring energies of 0.5, 0.75, 1, 1.25, 1.5, 2, 3, 4, 5, 6 and 7 keV.
These bands were chosen to capture most of the spectral information without overly increasing the processing time and storage used.
We also created total exposure maps assuming monochromatic energies in the centre of each band.
Radial profiles of cluster counts, background counts, average exposure and sky area were extracted around the cluster centre.
These profiles were created with single-pixel (0.492 arcsec) radial binning, masking out excluded regions and not splitting pixels between bins.
The profiles were truncated at radii where the cluster was not distinguishable from the background, found by manually examining the adaptively-smoothed maps and profiles (listed in in Appendix \ref{append:sample}).
We did this truncation to improve the robustness of the radial binning procedure and to greatly reduce the time to model the cluster profiles.
We manually chose a maximum radius for each cluster as automated procedures were insufficiently robust.

\subsection{Radial binning}
For the standard binned analysis (Section \ref{sect:nfw}), we binned the projected profiles to have root-mean-square uncertainties on the inferred deprojected emissivities below a threshold (typically 20 per cent -- see Appendix~\ref{append:sample}), which we refer to as `wide binning'.
This was done by optimizing the edges of the annuli to minimize the total-squared uncertainty on the emissivities.
To compute the uncertainties on the deprojected emissivities we propagated the covariance matrix of the uncertainties in the surface brightness in the projected bins.
This optimization procedure works well in most cases, although it can occasionally produce bins which do not meet the requirement.

We also created surface-brightness profiles used for the parametric fits, referred to as `fine binning'.
In this case we split each of the wide-binned annuli into five annuli with approximately-equal radial size, rounding to integer pixel radii and splitting into fewer bins if it was not possible to split into five.
For SPT-CLJ0658-5556 and SPT-CLJ0102-4915 we split by three instead to reduce the number of bins and increase the analysis speed.
\textsc{mbproj2} does not require the input profiles to be binned for parametric fits, but this substantially decreases the computing time required to analyse the profiles.

\subsection{Background modelling}
\label{sect:background}
As our background model for each of the bands, we used background surface-brightness profiles extracted from the background event files using the same binning as the cluster observations.
For each band these background profiles were rescaled to match the surface brightness of the cluster observation at large radius beyond the cluster emission, unless the cluster emission fills the entire field of view.
The scaling was to account for cluster-to-cluster variation in the astrophysical background from the blank-sky backgrounds.
At low energies, the scaling accounts for variation in Galactic and extragalactic emission, while at high energies the scaling accounts for particle background changes.
In the softest bands the clusters are scaled by factors with a standard deviation of around 10 percent, which declines to 4 per cent at the highest energies.
The softest 0.5 to 0.75 keV band is scaled up on average by around 6 per cent relative to the standard blank-sky data, while the other bands are consistent with no scaling on average.

The background was treated as an additional component added to the total cluster model.
The exposure times of the background datasets are typically 10 times greater than the cluster observations (although in the case of SPT-CLJ0102-4915 this ratio decreases to 4).
We can therefore ignore the Poisson uncertainty on the background as the statistical uncertainty on the surface brightness is dominated by the cluster emission, providing there are sufficient counts per radial bin.
For the wide-binned profiles, the total cluster signal is on average 9 per cent above the background in the outermost bins.
This decreases to around 2 per cent for the finely-binned profiles.
The median number of total counts in the outermost bins of the wide-binned profiles is around 2400 in the data and 39000 in the background.

Although we match the backgrounds to the observed profiles for those clusters which do not fill the field of view, there may be additional unresolved substructures and point sources in the source.
Larger scale fluctuations will be removed by the background scaling, but smaller scale features can remain.
We quantified this by measuring the fluctuations in the cluster surface-brightness profiles in the radial range used for background matching, relative to the background model, assuming Gaussian fluctuations and taking account of the Poisson noise.
We examined the variation on scales of $16$ arcsec between 0.5 and 7 keV.
The typical variation is 3 per cent, with 90 per cent lying between 1 and 6 percent and a tail up to 10 per cent.
We therefore conservatively add an additional free parameter allowing scaling of the background profiles using a Gaussian prior with $\sigma=10$ per cent.

\begin{table*}
\caption{
Models fitted to the surface-brightness profiles.
Listed are the model names, the section in this paper which describes a model, whether wide- or finely-binned data are fitted, the mass models used and the parametrizations in the analysis.
Other parameters included in the model are a background scaling parameter and the log outer pressure (if a mass model is used).
}
\begin{tabular}{lllll}
Name & Section & Binning & Mass model & Parametrization \\ \hline
BIN-NFW & \ref{sect:nfw} & Wide & NFW & Density in bins; mass model \\
BIN-GNFW & \ref{sect:gnfw} & Wide & GNFW & Density in bins; mass model \\
BIN-NONHYDRO & \ref{sect:nonhydro} & Wide & None & Density and temperatures in bins \\
INT-NFW & \ref{sect:interpol} & Fine & NFW & Density at centres of wide bins with interpolation; mass model \\
MBETA-NFW & \ref{sect:mbeta} & Fine & NFW & Modified-$\beta$ density profile; mass model \\
STEP-NFW & \ref{sect:mod_step} & Fine & NFW & Constant densities between fixed radii; mass model \\
GRAD-NFW & \ref{sect:mod_gstep} & Fine & NFW & Constant gradients between fixed radii; mass model\\
KPLAW-NFW & \ref{sect:plawentropy} & Fine & NFW & Powerlaw entropies inside and outside 300 kpc plus constant; mass model \\ \hline
\end{tabular}
\label{tab:models}
\end{table*}

\subsection{Profile modelling}
\label{sect:proffits}
The profiles were fit using the \textsc{mbproj2} multiband projection code (described in Appendix \ref{append:mbproj}).
Given the predicted model (including background) and observed profiles for a cluster a total Poisson likelihood can be computed.
An affine-invariant MCMC sampler \citep{Goodman10} as implemented in \textsc{emcee} \citep{ForemanMackey12} was used to sample the parameters for the model, starting from around the maximum-likelihood position.
This sampler has the advantage of not requiring a proposal distribution and being affine-invariant, handles well covariance between the parameters.
In our analyses we used a chain length of 2000 steps, a burn-in period of 1000 steps and 800 walkers.
Profiles of various physical properties were then calculated by iterating through the resulting chain in jumps of 10 steps and computing the profile given the model parameters for each position.
We then obtained the median profiles and uncertainties enclosing 68.3 percent of the produced profiles.

The metallicity was assumed to be $0.3\Zsun$ using the solar relative abundances of \cite{AndersGrevesse89}.
Equivalent hydrogen absorbing column densities were obtained from the LAB survey \citep{Kalberla05} and redshifts from \cite{Bleem15}, both given in Table \ref{tab:sample}.

Our main results come from fitting an NFW mass model to the binned profiles (model BIN-NFW).
In some cases we use a different modelling to check sensitivity to the assumed model, to look at fixed radius or when modelling the entropy profiles.
The models are detailed below and listed in Table \ref{tab:models}.
In Appendix \ref{append:goodness} we calculate the goodness of the fits for a subset of the models.

\subsubsection{Binned NFW fits (model BIN-NFW)}
\label{sect:nfw}
Our most simple analysis is to use widely-binned surface-brightness profiles, chosen to have constant fractional uncertainties in derived emissivities (Section \ref{sect:sbprofs}).
One logarithmic parameter with a flat prior was used to parametrize the electron-density value in each radial bin.
The dark matter was modelled with an NFW profile \citep{NFW96}, using parameters which are the logarithms of $R_{200,\mathrm{DM}}$ (the radius of an average overdensity of dark matter of 200 times the critical value) and $c$ (the concentration).
The density of dark matter at a radius $r$ is given by
\begin{equation}
  \rho(r) = \frac{ \delta_\mathrm{c} \rho_\mathrm{c} }
  { (r/r_\mathrm{s}) (1 + r/r_\mathrm{s})^{2} },
\end{equation}
where the scale radius $r_\mathrm{s} = R_{200,\mathrm{DM}} / c$, the critical density of the universe at redshift $z$ is $\rho_\mathrm{c} = (3 H^2(z)) / (8 \pi G)$, $H(z)$ is the Hubble constant at the redshift and $G$ is the gravitational constant.
The characteristic overdensity of the halo is
\begin{equation}
  \delta_\mathrm{c} = \frac{200}{3}
  \frac{c^3} { \log (1+c) - c/(1+c) }.
\end{equation}
Note that $R_{200,\mathrm{DM}}$ is not the usual $R_{200}$ for a galaxy cluster, as it is the radius where dark matter has an average overdensity of 200, rather than the total matter in the cluster having this overdensity.

As the data quality are often limited in the outskirts of the profiles, $R_{200,\mathrm{DM}}$ and $c$ can be strongly correlated parameters.
We therefore used a flat prior of $0 \le \log_{10} c \le 1$, to give a range similar to that found in high-mass systems in simulations \citep[e.g.][]{Duffy08} and observed using gravitational lensing \citep[e.g.][]{Merten15}.
A flat prior on the log cluster radius $-1 \le (\log_{10} R_{200,\mathrm{DM}} / \mathrm{Mpc}) \le 1$ was used.
We used wide, flat logarithmic priors on the outer pressure of the cluster.

\subsubsection{Binned GNFW fits (model BIN-GNFW)}
\label{sect:gnfw}
To check the sensitivity of our results on the assumed form of the dark-matter profile, we used the generalized NFW (GNFW) profile \citep{Zhao96, Wyithe01}, applying it to the widely-binned surface-brightness profiles.
The GNFW profile has a parametrized inner density slope, $\alpha$, which if $\alpha=1$ is the NFW profile.
The mass density follows the form
\begin{equation}
  \rho(r) = \frac{\rho_0}{(1/r_\mathrm{s})^\alpha (1+r/r_\mathrm{s})^{3-\alpha}},
\end{equation}
where $\rho_0$ and $r_\mathrm{s}$ are the central density and scale radius, respectively,  calculated from $c$ and $r_{200}$.
In the analysis $\alpha$ was allowed to vary between 0 and 2.5, assuming a flat prior.
$c$ and $r_{200}$ were allowed to vary in the same ranges as for the NFW fits.
We do not quantitatively examine the results of this model, but plot the resulting profiles.

\subsubsection{Binned non-hydrostatic fits (model BIN-NONHYDRO)}
\label{sect:nonhydro}
To check the assumption of hydrostatic equilibrium we also fitted a non-hydrostatic model.
Like BIN-NFW, the density was modelled by a logarithmic density parameter in each radial bin.
The temperature was parametrized in every third bin, assuming a flat log prior between 0.1 and 50 keV.
Values in intermediate bins were calculated using interpolation of log temperature in the radial bin index.

\subsubsection{Interpolated NFW profiles (model INT-NFW)}
\label{sect:interpol}
To examine the effect of binning, we also fitted the finely-binned surface-brightness profiles (Section \ref{sect:sbprofs}) with an interpolated density model.
The model parametrizes the density at the particular radii with the density at intermediate radii calculated by log interpolation in log radius.
We set the parametrized radii to be the centres of each of the wide bins, giving the same number of free parameters as BIN-NFW.
In these fits we assume hydrostatic equilibrium using the NFW model, using the same flat priors as BIN-NFW.

\subsubsection{Modified-$\beta$ model (model MBETA-NFW)}
\label{sect:mbeta}
The MD13 paper assumes a parametric model for the gas density, given by
\begin{equation}
  n_\mathrm{e}^2 = n_0^2
  \frac{(r/r_\mathrm{c})^{-\alpha}}{(1+r^2 / r_\mathrm{c}^2)^{3\beta-\alpha/2}}
  \frac{1}{(1 + r^\gamma / r_\mathrm{s}^\gamma)^{\epsilon / \gamma}},
\label{eqn:modbeta}
\end{equation}
which we refer to as a modified-$\beta$ profile.
This form is based on the profile described by \cite{Vikhlinin06}, not including their second $\beta$ component.
We set $\gamma$ to be 3 (following MD13) and apply the following flat priors: $0 \le \beta \le 4$, $-1 \le \log_{10} (r_\mathrm{c} / \mathrm{kpc}) \le 3.7$,  $1 \le \log_{10} (r_\mathrm{s} / \mathrm{kpc}) \le 3.7$, $0\le \epsilon \le 5$ and $0 \le \alpha \le 4$.
We note that $r_\mathrm{s}$ and $\alpha$ are not the same parameters as used in the GNFW or NFW models.
The model was fit to the data assuming hydrostatic equilibrium (with the same NFW dark-matter profile and priors as BIN-NFW).

We note that this functional form is not capable of fitting all possible X-ray surface-brightness profiles.
For example, we tested fitting the profiles of Abell~1795, a relatively nearby cluster with cool low-entropy gas in its core.
The model was unable to fit the steep central X-ray peak inside 10 kpc radius.
The parameters of parametric models are driven by the brightest radial regions and may not produce good results outside these regions.

Forcing the parameter $\alpha$ to be positive (following MD13) ensures the model central profiles are either flat or inwards-rising.
We examine the effect of allowing $\alpha$ to be negative in Section \ref{sect:MD13compar}.

\subsubsection{Stepped-density models (model STEP-NFW)}
\label{sect:mod_step}
To check how well we can measure the central density, we fitted a model where the density is constant within annuli with edges of fixed radii of 20, 40, 80, 160, 320, 640 and 1280~kpc.
We assume hydrostatic equilibrium with an NFW dark-matter mass component and the usual priors.

\subsubsection{Gradient model (model GRAD-NFW)}
\label{sect:mod_gstep}
To investigate the density gradient (log density in log radius), we parametrized it at radii of $<50$, $50$--$100$, $100$--$200$, $200$--$400$, $400$--$800$ and $>800$~kpc.
We used flat priors on the gradients, to better be able to examine their distributions.
The model was normalised by the density at $200$~kpc.
The surface brightness profiles were fitted with fine binning assuming hydrostatic equilibrium with an NFW dark-matter mass component and the priors used previously.

\subsubsection{Powerlaw entropy (model KPLAW-NFW)}
\label{sect:plawentropy}
The previous models parametrize the density profile of the cluster.
\textsc{mbproj2} also allows a parametrization of the entropy profile of the cluster (Appendix \ref{append:mbproj}).
To examine whether there is an entropy floor we assumed the form
\begin{equation}
K(r) = K_{0} + K_{300} \left( \frac{r}{\mathrm{300 \: kpc}} \right)^{\alpha_{k}(r)},
\label{eqn:entropy}
\end{equation}
where $\alpha_{K}(r)$ is $\alpha_{K,\mathrm{inner}}$ at $r<300\kpc$ and $\alpha_{K,\mathrm{outer}}$ otherwise.
We assumed a flat prior on $K_0$ between $10^{-5}$ and $1000\keV\cm^{2}$.
$K_{300}$ was given a flat prior in log space between 1 and $10^4 \keV \cm^{2}$.
We gave $\alpha_{K,\mathrm{inner}}$ and $\alpha_{K,\mathrm{outer}}$ Jeffreys priors between values of 0 and 4.
A Jeffreys prior is an uninformative prior invariant under monotonic transformations.
In the case of a gradient, gradient values which increase linearly give profiles which are ever more closely separated.
A flat prior would therefore be weighted towards steep slopes, as every value is assumed equally likely.
The Jeffreys prior removes this bias towards steep slopes.

We used the NFW gravitational potential with the same priors as the BIN-NFW model.
The entropy model was fitted to the finely-binned surface-brightness profiles to better-resolve the core region.
As described in Appendix \ref{append:mbproj}, under the entropy parametrization the model surface-brightness profiles were predicted from density and temperature profiles which were themselves calculated from the entropy and gravitational profiles assuming hydrostatic equilibrium.
Densities and temperatures were calculated at the centres of each radial bin.

\subsubsection{Central galaxy}
The mass models we fitted in this paper do not include a special mass component for the central galaxy, which could have some effect on the central properties of the cluster.
Whether this is important depends on how well resolved the centre is and whether the centre used actually lies on a central galaxy (or whether the cluster is relaxed).
To check this, we fitted NFW models adding a point source to the potential to account for the central galaxy.
The typical effect of this was to reduce temperatures, entropies and pressures by around 5 to 10 per cent.
In most clusters there is no significant point source mass component, but in 10 clusters the inclusion of a central component led to largely unconstrained central temperatures and unphysical central masses.
Most of these systems were disturbed and unlikely to have a real central galaxy and so this component appears unphysical.

If we compare the BIN-GNFW fits, where the central slope of the mass profile is a free parameter, with BIN-NFW, the inner temperature is increased by $5 \pm 16$ per cent and the density by $0 \pm 1$ per cent ($1\sigma$ percentile ranges).
This also suggests that the bias caused by the exclusion of the central galaxy is small.
However, forcing the addition of a King mass model with a velocity dispersion of $300\kmps$ and a core radius of $10$~kpc into the BIN-NFW model increases inner temperatures by $12_{-5}^{+10}$ per cent and densities by $-1_{-2}^{+1}$ per cent ($1\sigma$ percentile ranges).
\cite{Zappacosta06} note pure-NFW profiles are good fits to relaxed cluster mass profiles, in particular Abell~2589 and Abell~2029 \citep{Lewis03}; in Abell~2589 a central galaxy component degrades the fit (although this is based on one data point).
This result suggests that a central galaxy component should not be forced into the mass model.
In the remainder of this paper we do not include a central galaxy mass component.

\subsubsection{Effect of priors}
In the hydrostatic analyses, there are multiple priors assumed on the fitting parameters: the density/entropy profile priors, the mass model priors, the outer pressure prior and the background rescaling factor prior.
These model parameters except density and entropy are not interesting for the purposes of this paper.
The priors on these parameters are folded implicitly into the derived thermodynamic profiles.
We found that the effects of the priors on the derived profiles are small.

For most of the parameters we have assumed wide, physically non-informative priors.
In the binned and interpolated analysis, BIN-NFW and INT-NFW, we do not assume a strong parametric form for the density.
We assume a flat logarithmic prior on the density parameters, which are very well constrained by the data and so any prior is unimportant.
Likewise, we assume a flat logarithmic prior on the outer pressure, which does not influence the results significantly.

In the mass model, we use a flat logarithmic prior on the concentration between values of $c=1$ and $10$, which is a large range given existing simulations and datasets.
The main effect of this prior is to constrain the dark matter mass at large radius, which is not examined in this paper.
We tested increasing the range to $c=0.5-20$, finding there was no systematic change in average temperature, density, entropy or pressure values, while the uncertainties increased by around 6 per cent for temperature, entropy and pressure.
The effect on the profiles was typically much smaller than the size of the error bars given.
The prior on the background scaling factor affects whether the density is well constrained in around 14 systems in the outermost bins of the profiles.
Inside the outermost bin the profiles are unaffected if the allowed range is increased or decreased by a factor of 2.

In our analysis we assume a fixed metallicity of $0.3\Zsun$.
As the clusters are relatively hot and massive, the effect of this assumption is weak.
For example, if we vary the metallicity assumed to $0.2$ or $0.4\Zsun$, in SPT-CLJ0000-5748, the temperatures are changed on average by around 2.5 per cent, densities by 1.5 per cent and entropies by 1.5 per cent.

\begin{figure}
  \includegraphics[width=\columnwidth]{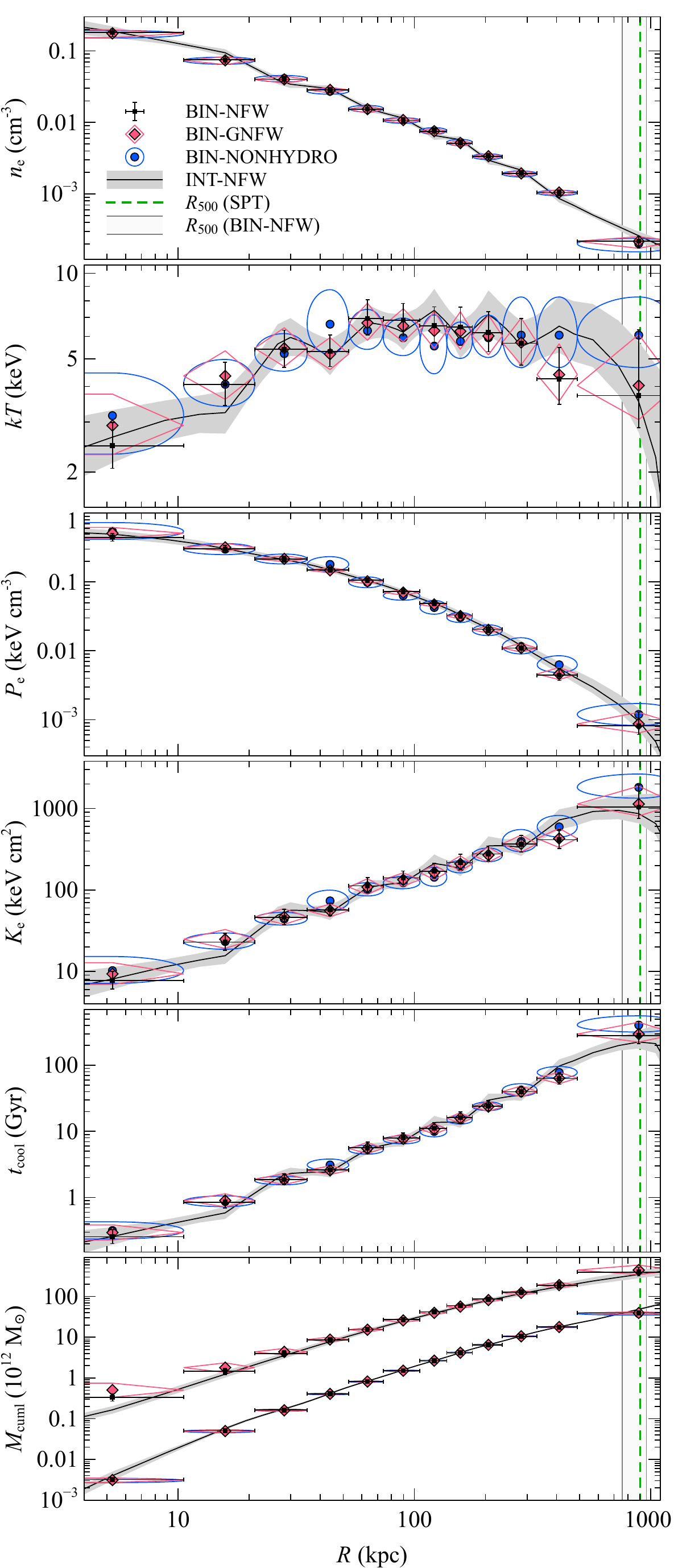}
  \caption{
Radial profiles of density, temperature, pressure, entropy, cooling time, cumulative-gas mass and cumulative-total mass for SPT~CLJ0000-5748.
The binned results are for the BIN-NFW and BIN-GNFW hydrostatic cluster models and the BIN-NONHYDRO non-hydrostatic model.
The shaded region is for the INT-NFW interpolated model.
The vertical line is the SPT value of $R_{500}$, while the vertical bounded region is $R_{500}$ calculated from the BIN-NFW fitting.
}
\label{fig:example}
\end{figure}

\subsection{Example cluster}
As an example we consider SPT~CLJ0000-5748, the first cluster in the sample, which is at a redshift of 0.70.
The \emph{Chandra} data have $\sim 1800$ counts in the 0.5 to 7 keV band after subtracting background, which is not atypical in our sample.
The system appears relaxed with a bright central peak.
Fig.~\ref{fig:example} shows our profiles of the physical quantities using different models and binning.
The quantities shown include the electron density ($n_\mathrm{e}$), temperature ($kT$), electron pressure ($P_\mathrm{e}$), electron entropy ($K_\mathrm{e}$), radiative cooling time ($t_\mathrm{cool}$), cumulative-gas mass ($M_\mathrm{gas}$) and cumulative-total mass ($M$).
Cooling time here is defined as the ratio between the enthalpy per unit volume  of the intracluster medium ($5nkT/2$, where $n$ is the total particle density) and its emissivity.
The results plotted include those from the binned data assuming an NFW potential (BIN-NFW), a GNFW profile (BIN-GNFW) and without assuming hydrostatic equilibrium (BIN-NONHYDRO).
We also show as a shaded region the interpolated density profile fit to the finely-binned data (INT-NFW).
In the plot is marked the BIN-NFW $1\sigma$ range of $R_{500}$, obtained by scaling the dark-matter mass profile assuming a baryon fraction of 0.175, and the SPT value of $R_{500}$ \citep{Bleem15}.

The results show good agreement between the different modelling techniques.
The main differences occur in the outer bin, where the background is a large component of the observed surface brightness.
The non-hydrostatic results agree well with the hydrostatic models, indicating that hydrostatic equilibrium is a reasonable assumption for this system.
BIN-GNFW shows a flatter mass profile in the core, becoming steeper in the outskirts, with $\alpha=1.31^{+0.20}_{-0.31}$.
In the central region of the cluster the density is high, there is a cool core, the cooling time drops to around $0.3\Gyr$ and the entropy to $10 \keV\cm^{2}$.
The entropy profile shows no evidence for a central floor.

We show similar individual profiles for each of the clusters in Appendix \ref{append:individual}.

\section{Central thermodynamic quantities}
Here we examine the central thermodynamical properties of the clusters to look for the existence of an entropy or cooling time floor.

\subsection{Central quantities as a function of radius}
\label{sect:dataquality}
\begin{figure}
\includegraphics[width=\columnwidth]{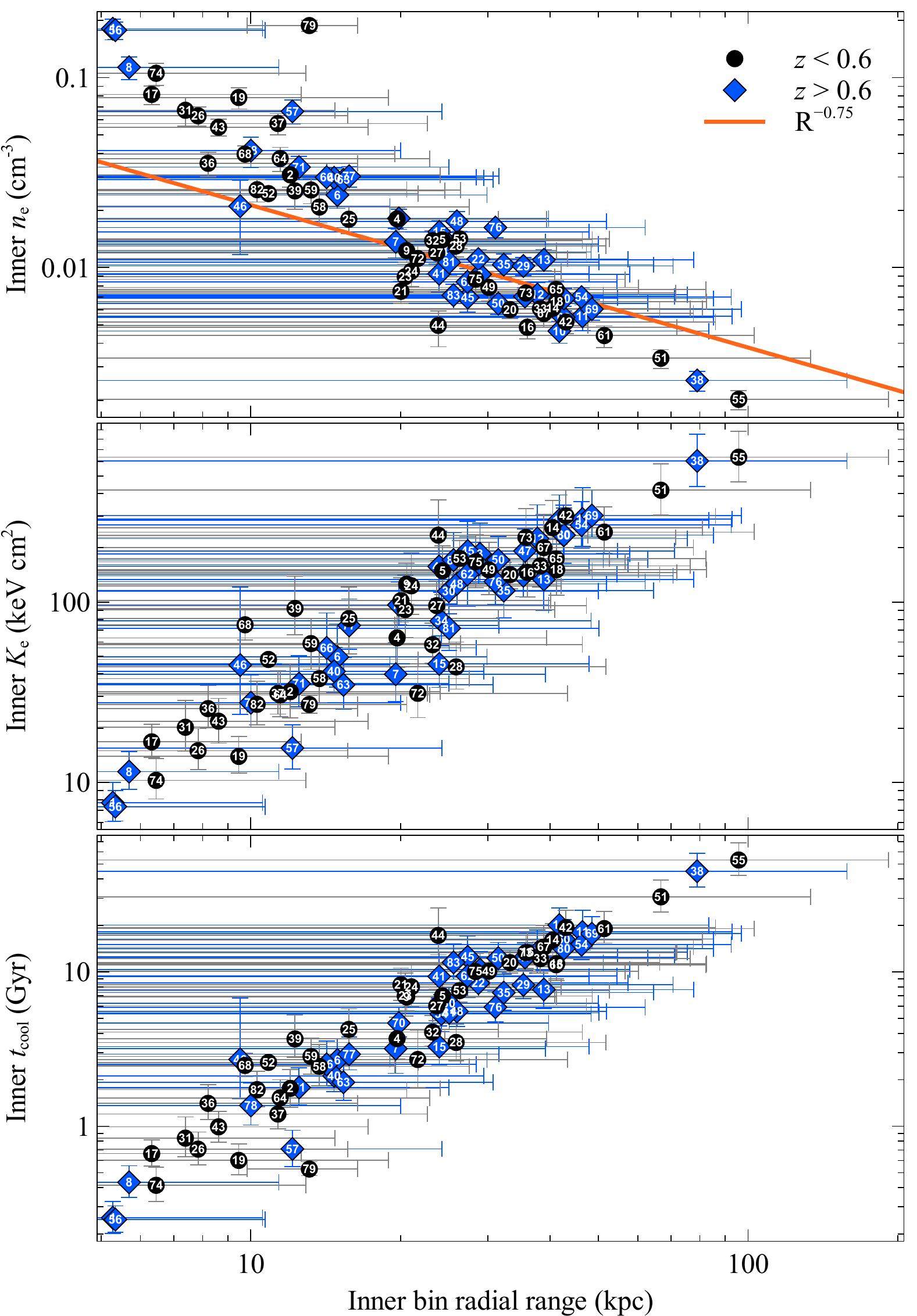}
\caption{
Inner-bin BIN-NFW values of the density, entropy and cooling time plotted against the inner-bin radial range for the sample of clusters.
The majority of clusters were binned to have a 20 per cent emissivity uncertainty in each spatial bin.
}
\label{fig:cen_radius}
\end{figure}

Fig.~\ref{fig:cen_radius} shows the BIN-NFW model central values of the density, entropy and cooling time profiles plotted against the radial range of the inner bin.
The results clearly show that the larger the size of the inner bin, the lower the density, the longer the cooling time and the higher the entropy.
These bins were chosen to give the same 10 per cent uncertainty (for most objects) on the density.
As found by \cite{Panagoulia14}, our ability to resolve these inner values is limited by the quality of the data.
Therefore these central measured density values (the average in the annulus) are lower limits to the central density and the entropy and cooling times are upper limits.
The values roughly scale (or inversely scale) with the size of the central bin to the power 1.5.
Despite the density being correlated to bin radius (and therefore data quality), our ability to resolve spatial regions in a cluster also depends on the density of the ICM as more counts are emitted from denser regions.
There are typically 100 to 200 counts in total in the central bin of the clusters with 20 per cent emissivity uncertainties.
In order to measure a central entropy to a reasonable accuracy requires this number of counts, given the hydrostatic model.

\begin{figure}
  \includegraphics[width=\columnwidth]{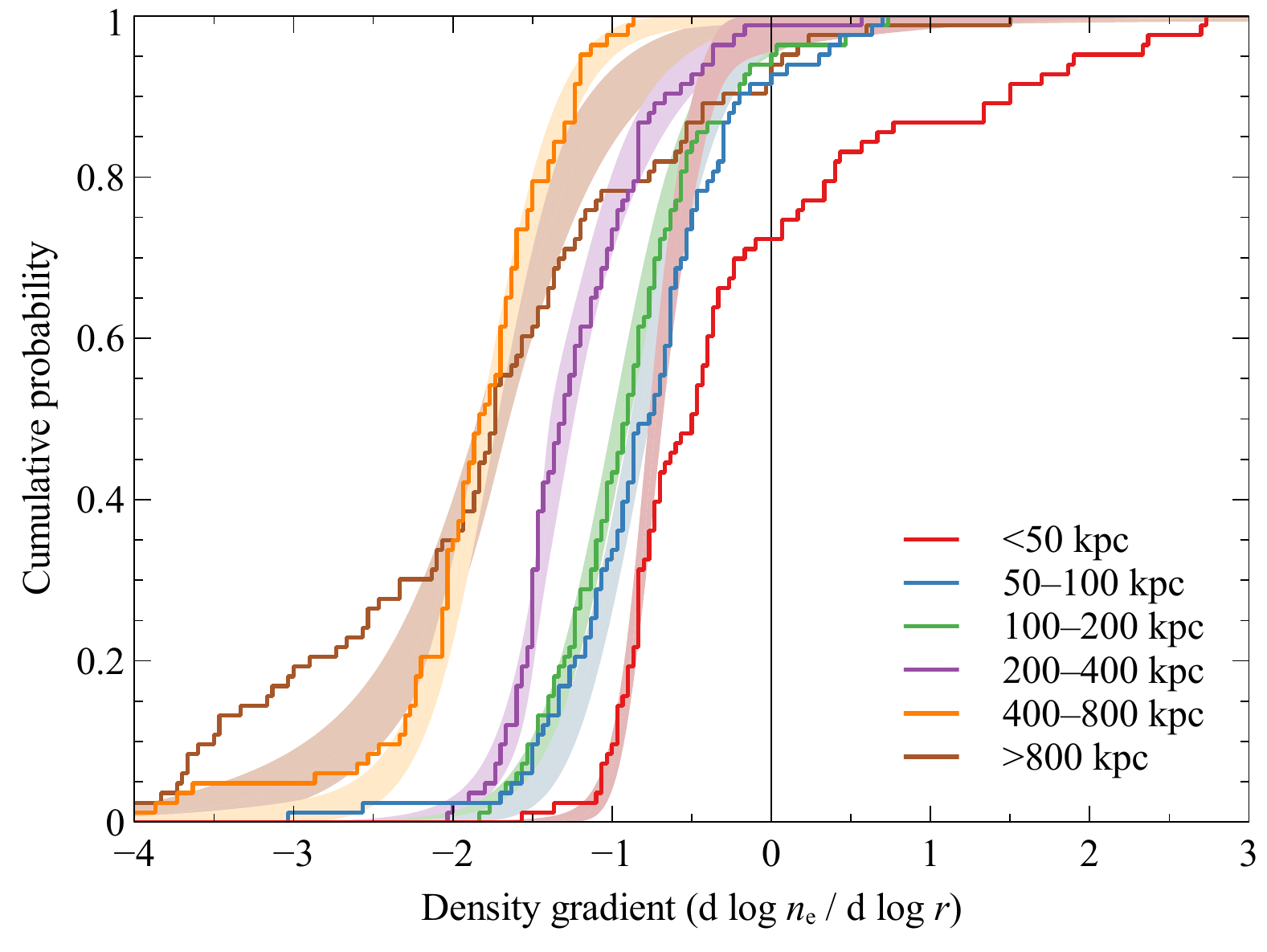}
  \includegraphics[width=\columnwidth]{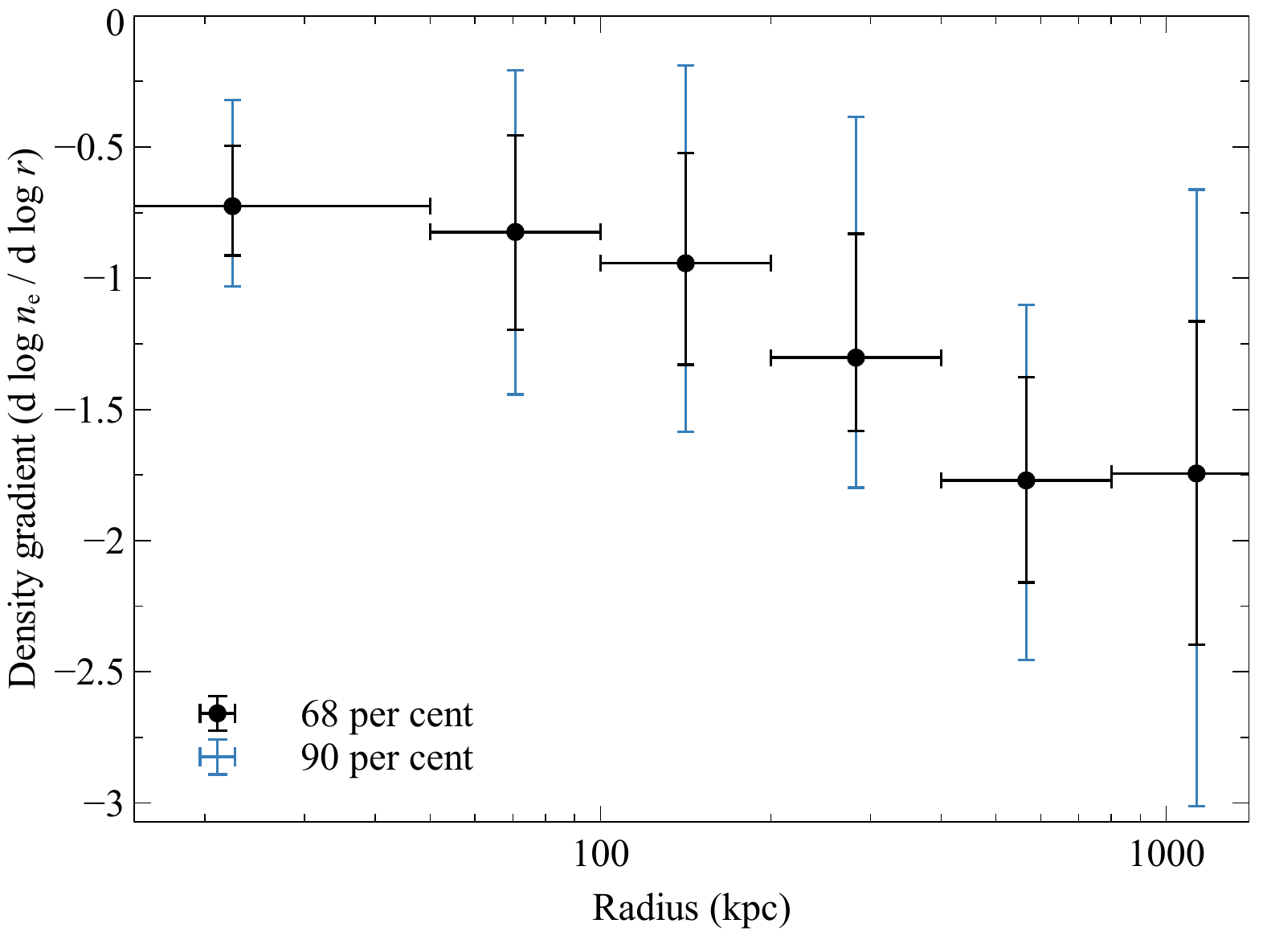}
  \caption{
    (Top panel) Measured (solid line) and modelled (shaded region) cumulative distributions of the density gradient in radial regions of the sample, obtained using the GRAD-NFW fits.
For the measured distribution we plot the distribution of median values from the MCMC chains.
For the modelled distribution we show the results from a two-component Gaussian model fitted to the gradient posterior probability distributions for the sample, showing the $1\sigma$ range.
Note that the large difference between the data and model in the inner region is due to large uncertainties for a subset of clusters with best-fitting flat cores.
    (Bottom panel) 68 and 90 per cent widths of the model density gradient distributions in each radial bin.
    }
    \label{fig:grad_rad}
\end{figure}

As clusters with flat cores will have lower surface brightnesses than those with steep cores, it is not clear what fraction of the trend in Fig.~\ref{fig:cen_radius} is due to data quality or cluster morphology.
Using the results from the GRAD-NFW model we can look at the distribution of density gradients as a function of radius in the sample.
Fig.~\ref{fig:grad_rad} (top panel) plots the cumulative distribution of measured density gradients in the sample (taking median values from each chain) in radial regions and models fitted to the gradient posterior probability distributions for the sample in those regions.
This modelling accounts for the uncertainties on the measurements, in particular in the central region, to obtain the intrinsic distribution.
In this case we assume the gradient distributions can be modelled by a two-Gaussian-component model, but the results are very similar assuming a skewed-normal distribution instead. To see the radial distribution we plot the median model slope as a function of radius (bottom panel), showing the width of the distribution.

The results show that the average cluster is consistent with a central density gradient of around $-0.75$, with very few systems with completely flat cores.
This average inner gradient is consistent with our later modelling of the thermodynamic profiles of the clusters (Section \ref{sect:modelevo}) and the results for a representative sample of nearby clusters \citep{Croston08}.
Therefore around half the $\sim -1.5$ power density trend seen in Fig.~\ref{fig:cen_radius} is attributable to data quality and half due to the bias towards steeply-peaked surface brightness profiles.

\subsection{Density comparison}
\label{sect:MD13compar}

\begin{figure}
  \includegraphics[width=\columnwidth]{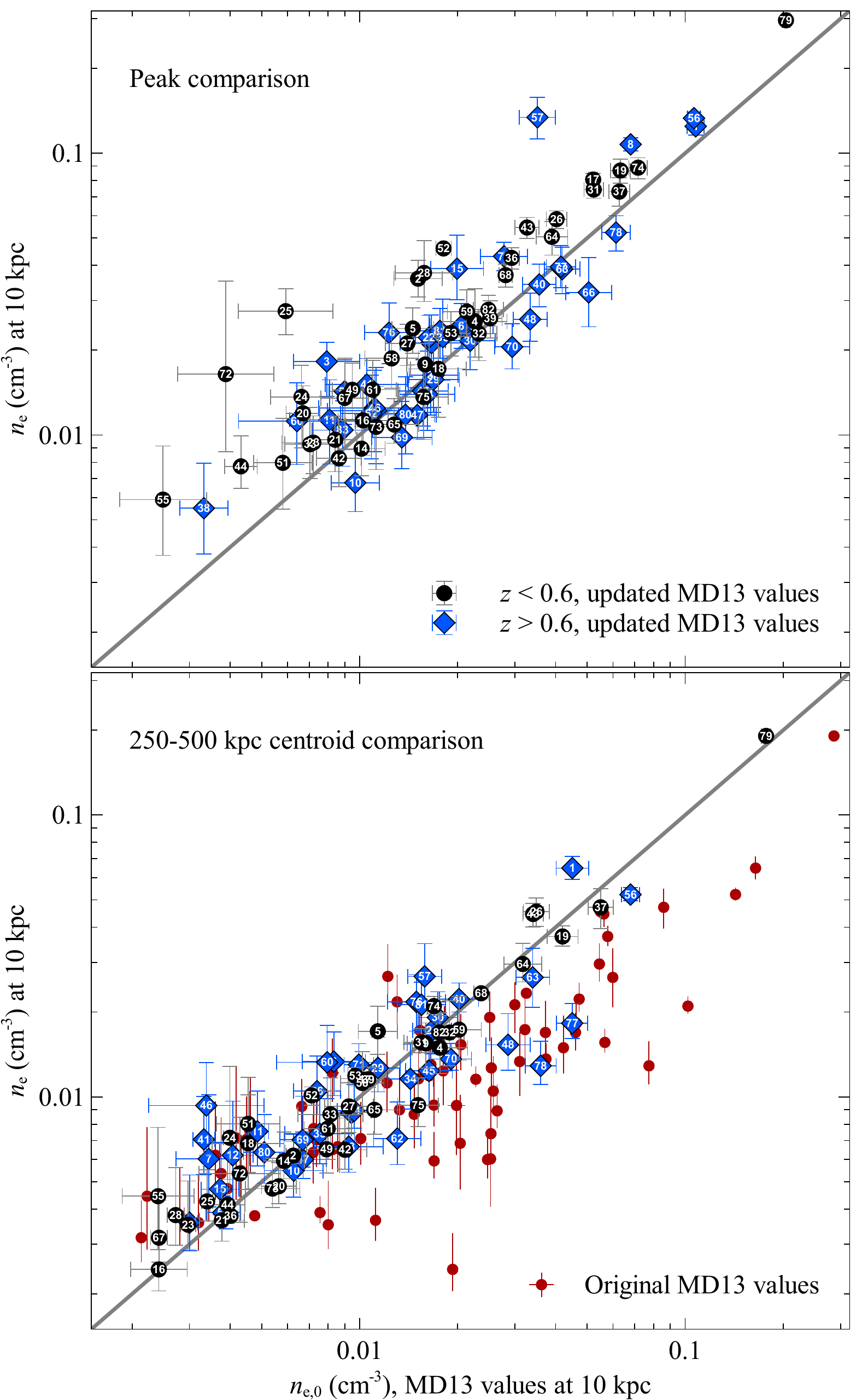}
  \caption{
Comparison of our MBETA-NFW densities at 10 kpc radius with the updated values from MD13 using the same functional form.
The top panel shows the densities if the cluster centred on the X-ray peak, while the bottom panel shows the values centred on a 250--500 kpc centroid.
Also shown in the centroid panel are the originally-published results from MD13.
}
\label{fig:mcd_ne_compare}
\end{figure}

It is important to check that our densities are accurate as the density profile is an important contribution to the other profiles, due to it being a parametrized profile in the majority of our analysis.
We tested our \textsc{mbproj2} binned density profiles by comparing with those produced by the PROJCT spectral model in \textsc{xspec}, using the same radial bins and assuming isothermality.
We reproduced the density profiles well in these cases, subject to small factors due to the lack of temperature variation.
In addition, the hydrostatic assumption does not appear to bias the densities, with excellent agreement between the non-hydrostatic BIN-NONHYDRO and the hydrostatic BIN-NFW central densities.

MD13 assumes the same functional form for the density as we use for the MBETA-NFW model.
We compared our results at fixed 10 kpc radius against MD13, initially finding poor agreement.
There were problems in the method used to obtain the published MD13 values,
{
but the updated values in \cite{McDonaldSPT17}
}
matched our results much better (Fig.~\ref{fig:mcd_ne_compare}).
The top panel compares the results for the profiles centred on the X-ray peak, while the lower panel shows the results using a 250--500 kpc annulus centroid (the main method used by MD13).
Note that both these sets of cluster centres were independently obtained by us and MD13.
For the annulus centroid we also show the originally-published MD13 results for comparison.

For the peak densities, there is a reasonable agreement between the two sets of results, with our densities being on average 30 per cent larger than the updated results of MD13.
Using the 250--500 kpc annulus centroid, our densities are higher than the updated values of MD13 by 7 per cent.
We have also checked the profiles for a few systems against those obtained by McDonald (private communication), finding reasonable agreement.
The differences in the peak densities are due to the differences in how the centre is chosen.
We optimize the position using a circle containing at least 50 counts, while MD13 considers a larger region.

\begin{figure}
\includegraphics[width=\columnwidth]{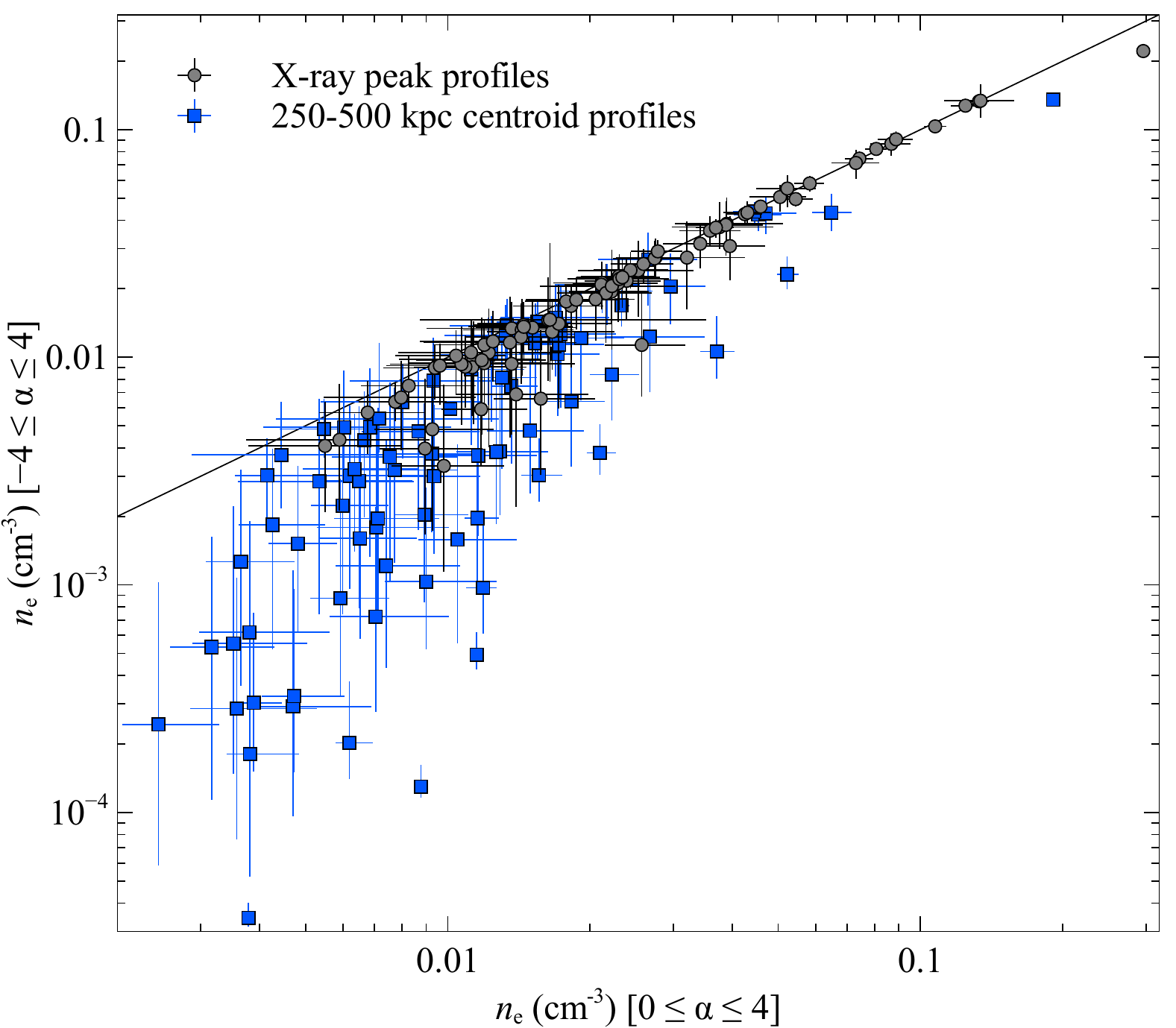}
\caption{
The effect of allowing centrally-declining profiles in the MBETA-NFW fits.
Plotted are the median densities at 10 kpc obtained assuming $-4 \le \alpha \le 4$ against those for $0 \le \alpha \le 4$ (our standard results).
The results are shown for fitting X-ray profiles centred on the X-ray peaks and for the 250-500 kpc centroid positions.
}
\label{fig:vikh_alpha}
\end{figure}

However, when fitting profiles centred on the X-ray annulus the measured inner densities are strongly affected by the modified-$\beta$ model assumptions.
Fig.~\ref{fig:vikh_alpha} compares our standard densities at 10 kpc radius (assuming centrally flat or rising-inward profiles following MD13; $0 \le \alpha \le 4$) against those obtained allowing declining central profiles ($-4 \le \alpha \le 4$).
For the profiles using the X-ray peak position, as might be expected there is little difference between the two results, with a trend allowing slightly lower central densities for systems with low central densities.
However, if the annulus centre is used, the assumption of flat or inward-rising profiles has a large effect on the obtained densities.
If the centre is not on the X-ray peak then it is physically possible to have a declining central profile.
Assuming positive slopes will bias the central densities upwards in unrelaxed systems.

\begin{figure}
\includegraphics[width=\columnwidth]{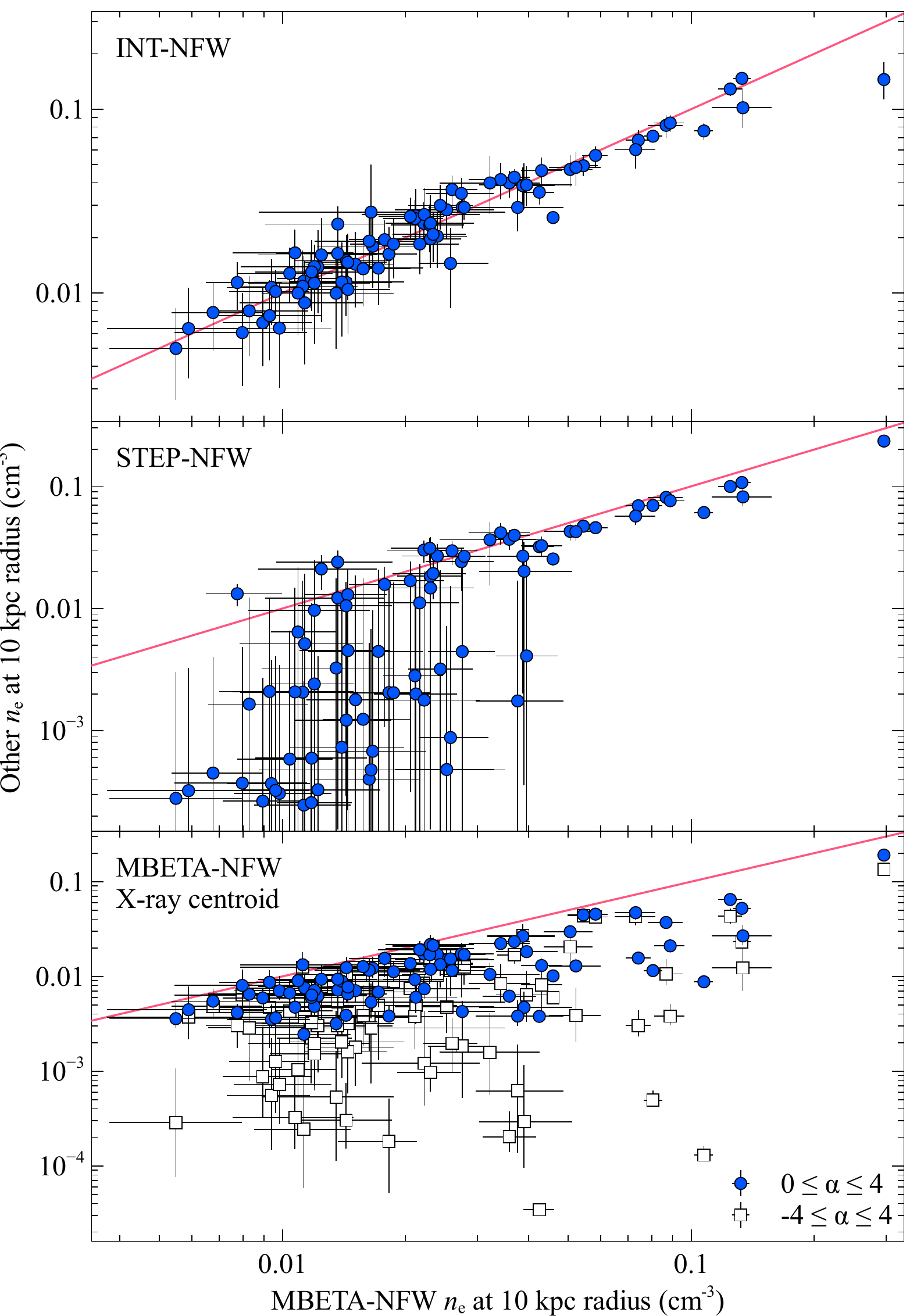}
\caption{
Comparison with the X-ray peak MBETA-NFW densities at 10 kpc radius.
Compared are the INT-NFW densities (top panel), the STEP-NFW densities (centre panel) and the MBETA-NEW densities from profiles centred on the 250--500 kpc centroid, for the cases where $\alpha$ is forced to be positive or allowed to be negative (lower panel).
}
\label{fig:ne_cent}
\end{figure}

The density at 10~kpc radius is poorly constrained by the data (Fig.~\ref{fig:cen_radius}), so the assumed functional form can have a large impact on the resulting values.
The STEP-NFW density model assumes a constant density inside a radius of 20~kpc.
Even with this relatively large region, we can only constrain the density within this region to better than an order of magnitude in around 40 per cent of systems (Fig.~\ref{fig:ne_cent} centre panel).
Despite the difficulty in directly measuring densities at 10 kpc radius, if we compare the INT-NFW model, which interpolates in density between the wide bin centres, with MBETA-NFW, we obtain good agreement (Fig.~\ref{fig:ne_cent} top panel).
The INT-NFW densities are $10 \pm 4$ per cent greater than the MBETA-NFW densities (examining median differences in log space), implying the central slopes are similar between the two models.

While the choice of density model has some effect on the central densities, the choice of centre also strongly influences the obtained values.
Fig.~\ref{fig:ne_cent} (bottom panel) compares the MBETA-NFW densities for the standard X-ray peak profile centre and for profiles using the 250--500 kpc radius centroid (following MD13).
Some of the points are an order of magnitude lower than the peak densities.
We note that these results assume a positive $\alpha$ parameter in the modified-$\beta$ model fits.
If this is allowed to be negative (as in Fig.~\ref{fig:vikh_alpha}), the densities are moved to lower values, with some three orders of magnitude below the peak-centre profile values.

In conclusion, although our density values agree with MD13 given the same cluster centre and density model, the choice of cluster centre and model strongly influences the obtained central densities and other derived quantities such as entropy.
Due to the variable quality of data (Section \ref{sect:dataquality} and Fig.~\ref{fig:ne_cent} centre panel) extrapolation has to be used to obtain the cluster properties at the 10 kpc radius used by MD13.
The density profiles obtained by MD13 when using the annulus as cluster centre, are strongly biased upwards by the choice to force the inner density profiles to be flat or inwards-rising in the functional fit.
However, this is partially compensated for by choosing a cluster centre based on the larger scale emission and missing the central peak.
Our use of the X-ray peak as cluster centre is more robust against the choice of density model.

We note that using the X-ray peak as the centre of our NFW mass model may be inconsistent with simulations which use a mass centroid, possibly impacting the derived deprojected quantities.
However, the NFW model is being used to fit a smooth pressure profile and not the cluster masses here, and so as long as the pressure profile is consistent with the data this should not be a problem.
Indeed, if we compare the BIN-GNFW profiles, which have freedom in the inner slope of the pressure profile, with the BIN-NFW profiles (Appendix \ref{append:individual}) we see good consistency between the two, indicating that the NFW model and assumed priors is sufficient to fit the data.

\subsection{Inner entropy values}
\label{sect:innerentropy}
\cite{Cavagnolo09} fitted entropy profiles from a large sample of clusters with the functional form
\begin{equation}
K(r) = K_{0} + K_{100} \left( \frac{r}{\mathrm{100 \: kpc}} \right)^{\alpha},
\label{eqn:cav09}
\end{equation}
finding evidence for a bimodal distribution of values with peaks of $K_0$ at $\sim 15$ and $\sim 150 \keV\cm^{2}$.

We examine our cluster entropy profiles using the KPLAW-NFW model, which uses a similar functional form for the entropy (equation \ref{eqn:entropy}), to see whether there is evidence for a floor.
It parametrizes the slope of the entropy profile inside and outside a radius of 300~kpc separately to avoid the central profile fits being biased by the outskirts (300~kpc was a typical outer radius of a profile analysed by \citealt{Cavagnolo09}).
The entropy at 300 kpc, $K_{300}$, is parametrized instead of $K_{100}$, to avoid covariance with the $K_0$ parameter.
We choose to fit the cluster X-ray profiles directly with the entropy model rather than fitting the posterior entropy profiles from the MCMC chain, to better take account of the covariance between the radial bins, which might otherwise bias the fit parameters \citep{Lakhchaura16}, particularly if the posterior distributions are non-Gaussian.
It should be noted, however, that a few clusters are fit poorly with this simple model (Appendix \ref{append:goodness}), which could lead to biased results for those objects.

\begin{figure}
\includegraphics[width=\columnwidth]{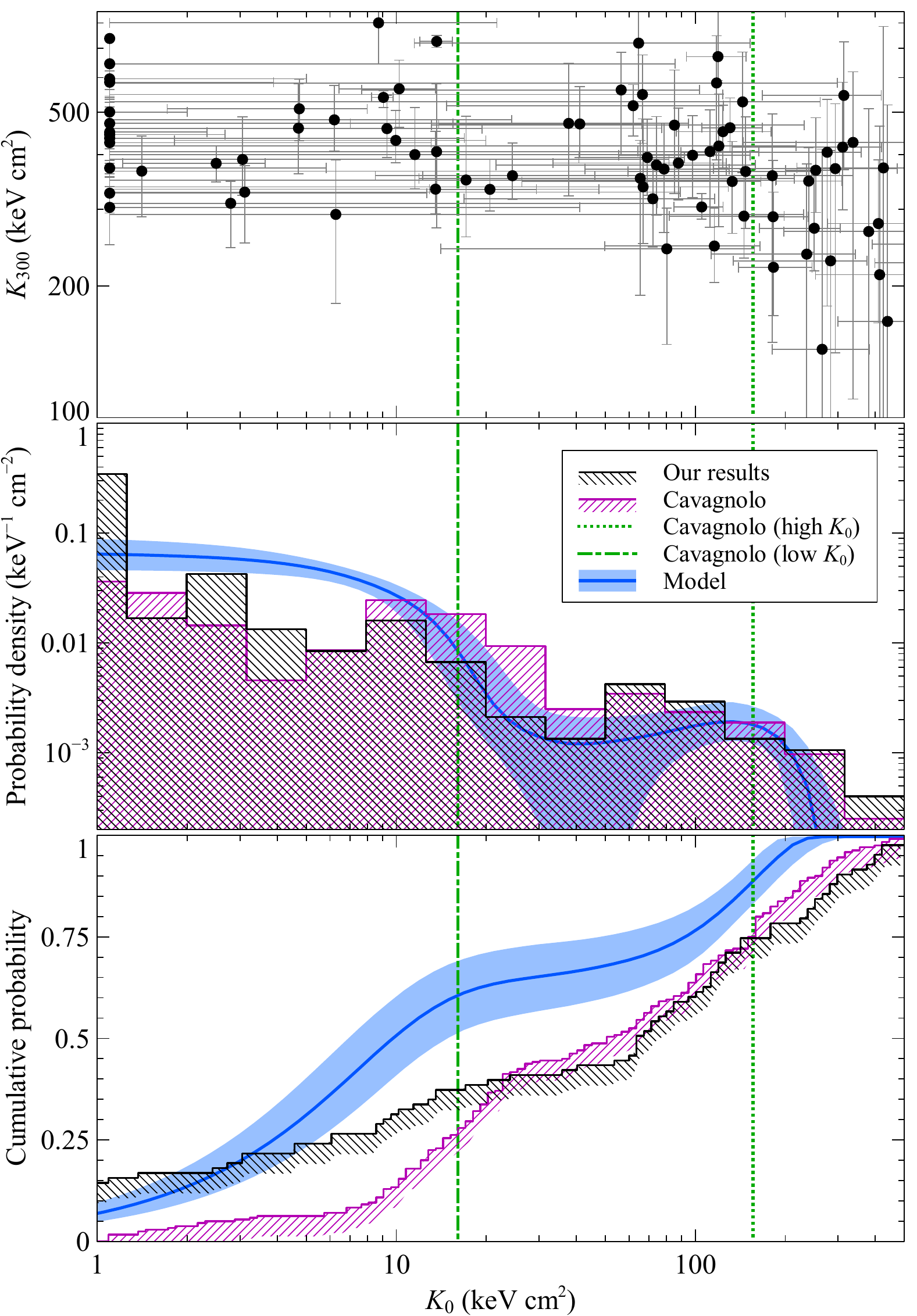}
\caption{
Inner entropy ($K_0$) values and distributions.
(Top panel) $K_0$ plotted against the entropy increase at a radius of 300~kpc ($K_\mathrm{300}$).
Vertical lines show the \protect\cite{Cavagnolo09} bimodal distribution peaks at $K_0 \sim 15$ and $\sim 150 \keV\cm^{2}$.
$K_0$ values below $1.1\keV\cm2$ are shown at this value and included in the histograms there.
(Centre panel) Histograms showing the probability density of the median $K_0$ values and those of \protect\cite{Cavagnolo09}, with logarithmic bin widths.
A two-component model for the underlying $K_0$ distribution of our data before measurement errors is also plotted with its $1\sigma$ range.
(Bottom panel)
Cumulative probability histograms of our median $K_0$ values, the values of \protect\cite{Cavagnolo09} and the model distribution.
}
\label{fig:innerK}
\end{figure}

Fig.~\ref{fig:innerK} (top panel) plots the $K_0$ and $K_{300}$ parameters for each system.
The points plotted are the most-likely values in the chain, to avoid biasing the $K_0$ values upwards due to the lower bound on the parameter.
In the centre panel is a probability density histogram of the median $K_0$ values.
We also plot the histogram of the entropies obtained by \cite{Cavagnolo09} and indicate the peaks of their bimodal distribution in the centre panel.
The cumulative distribution of the two sets of $K_0$ values is shown in the bottom panel.
Many of the clusters with entropies around zero are at lower entropies than the lower peak of values obtained by \cite{Cavagnolo09}.
The strong zero peak is consistent with the findings of \cite{Panagoulia14} and \cite{Hogan17}, who found no evidence for a common floor in entropy in their cluster samples.

We obtained a similar distribution of points using a break radius of 100~kpc instead of 300~kpc, or when we forced the parameter $\alpha_{K,inner}$ to be $0.64$, the average value obtained by \cite{Panagoulia14}.
The peak around zero also remains if we fit the BIN-NFW posterior entropy profiles for each cluster, rather than reanalyse the surface-brightness profiles.

We modelled the underlying distribution of $K_0$ values before measurement errors with a probability-density function (PDF) made of two skew-normal components
\footnote{The Python code and input data used to model the distribution can be found at \url{https://github.com/jeremysanders/K0dist/} and included with this paper as online-only material.}.
We used MCMC to sample this distribution given the marginalised $K_0$ PDFs for each cluster.
The median model PDF and $1\sigma$ range is shown in the centre panel, while its cumulative distribution is shown in the bottom panel.
In this analysis we excluded SPT-CLJ0658-5556 due to it having a very tightly constrained $K_0$ value given the high quality of data, but very poor fit quality.

In detail, to calculate the likelihood for a given a model PDF, for each cluster we multiplied the model PDF by the $K_0$ posterior PDF and integrated to compute the per-cluster likelihood.
The log likelihoods for the individual clusters were summed with the prior to calculate a total log likelihood.
We assumed the two components had centres between $0$ and $1000\keV\cm^{2}$ and widths between $1$ and $500\keV\cm^{2}$, with flat priors.
The model PDFs were forced to have no likelihood for $K_0<0$, adjusting to have a total integrated probability of 1.
The skew parameters had a normal priors with a width of 20 in the analysis.
The relative strength of the two components was modelled with a parameter $b$, where the fraction of probability give for the first component was the sigmoid function $1/(1+e^{-b})$, with a normal prior on $b$ of width 20.

The modelling appears to favour two components, although the second peak could be consistent with broad wings on the first peak to higher entropy values.
Around 60 per cent of the integrated probability is in the lower entropy component.
The second component is centred around $130 \keV\cm^{2}$.
Adding a third component does not produce an identifiable peak, but increases the width of the tails in the probability distribution.
We note that we obtained similar model distributions if the $K_0$ model parameter was allowed to go negative, to avoid having a hard limit at 0, or if we did not bin the surface-brightness profiles within the central annulus.
If we excluded those clusters with the worst fits for the KPLAW-NFW model (a goodness of fit greater than 2; Appendix \ref{append:goodness}), we obtained a consistent model.

\section{Evolution with redshift}

\begin{figure*}
\centering
\includegraphics[width=\columnwidth]{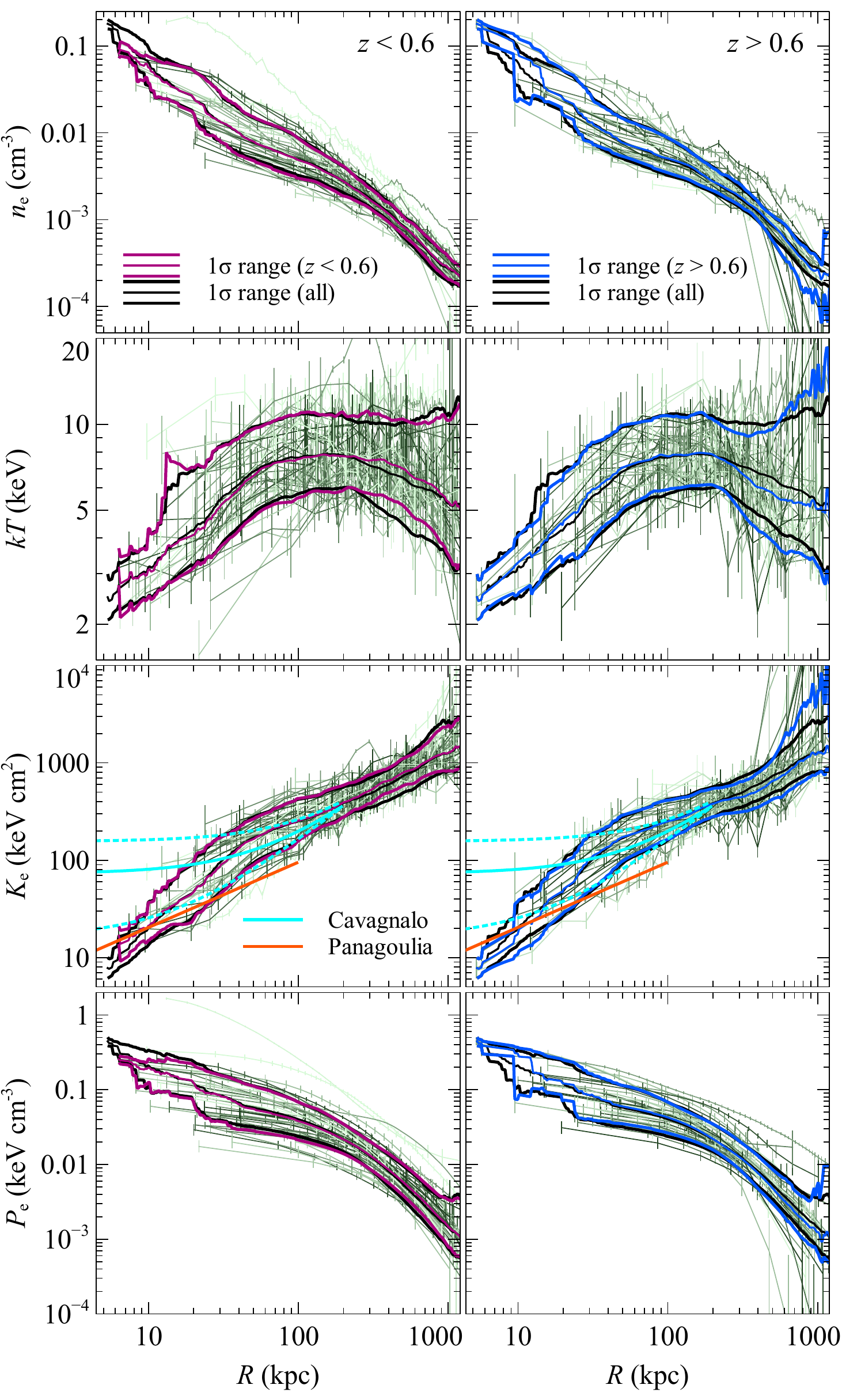} \hspace{5mm}
\includegraphics[width=\columnwidth]{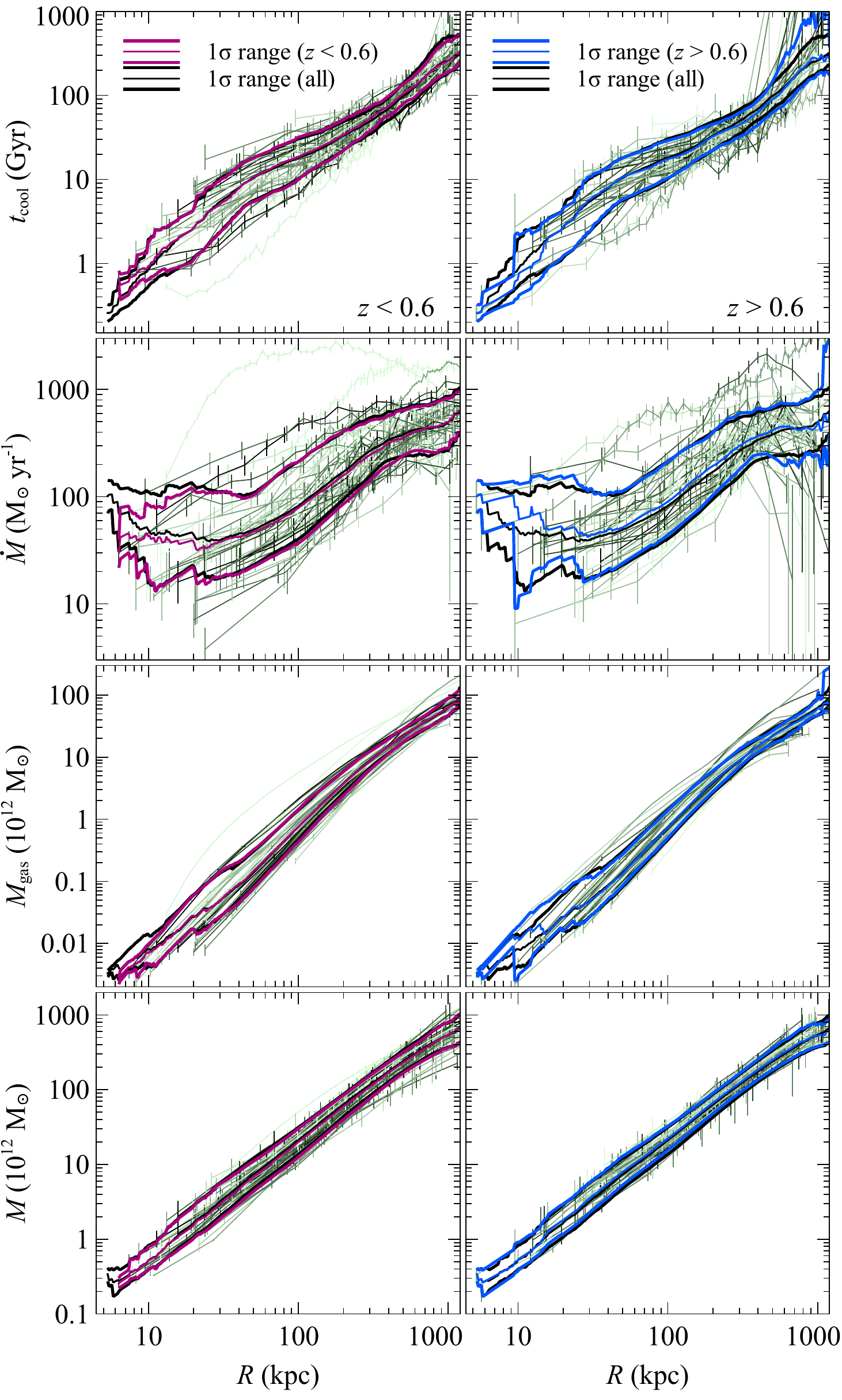}
\caption{
Individual profiles (from model BIN-NFW) of electron density, temperature, entropy and pressure (left side), and radiative cooling time, mass deposition rate and cumulative gas and total mass (right side) as a function of physical radius.
The clusters have been split into low (left panels) and high redshift (right panels) subsamples.
The thick lines show the median and 1$\sigma$ ranges for all clusters and the particular redshift subsample.
In the entropy plot are also plotted the powerlaw inner profile of \protect\cite{Panagoulia14} and the cored entropy profiles of \protect\cite{Cavagnolo09} (for their whole, low and high central entropy subsets).
}
\label{fig:phys_profiles}
\end{figure*}

\subsection{Median profiles, with central cool-core bias}
\label{sect:profs}
To look for evolution of the cluster properties in a model-independent fashion we examined the median profiles in two subsamples, $z<0.6$ and $z>0.6$, giving approximately equal numbers of systems (44 and 39, respectively).
This approach has the advantage of being non-parametric, but we also examine the evolution using Gaussian and two-component modelling of the distribution of thermodynamic quantities in Sections \ref{sect:modelevo} and \ref{sect:bimodal}, respectively.

Examining the unscaled BIN-NFW profiles in physical units, Fig.~\ref{fig:phys_profiles} shows the density, temperature, entropy, pressure, cooling time, mass deposition rate, cumulative gas and total mass profiles for the two subsamples.
Plotted on each are the median and 68 per cent range of the profiles (calculated from percentiles) for the whole sample and redshift subset.
To compute the medians and percentiles at a particular radius, we took all clusters where this radius was between the central radii of their inner and outer bins.
For each cluster we constructed a sample of profiles using the MCMC chains, interpolating in radial log space between the bin centres.
The median and range was computed from the combined set of samples from each cluster, weighting equally.
Note that only including clusters which have valid profiles in the radial range examined, as we do here, is correct if data quality is the primary reason for poor spatial resolution (as indicated by Fig.~\ref{fig:cen_radius}), but could introduce biases if the cluster properties also affect the data quality (see Section \ref{sect:bimodal}).
As cool-core systems have brighter cores and therefore smaller central annuli, this will bias the median profile towards cool-core systems (see Section \ref{sect:dataquality}).
Assuming the density profiles are flat inwards when computing the median would produce the opposite bias.
This bias is not present when we model the core (Section \ref{sect:modelcore}) or profile (Section \ref{sect:modelevo}) evolution.

\begin{figure*}
\centering
\includegraphics[width=\columnwidth]{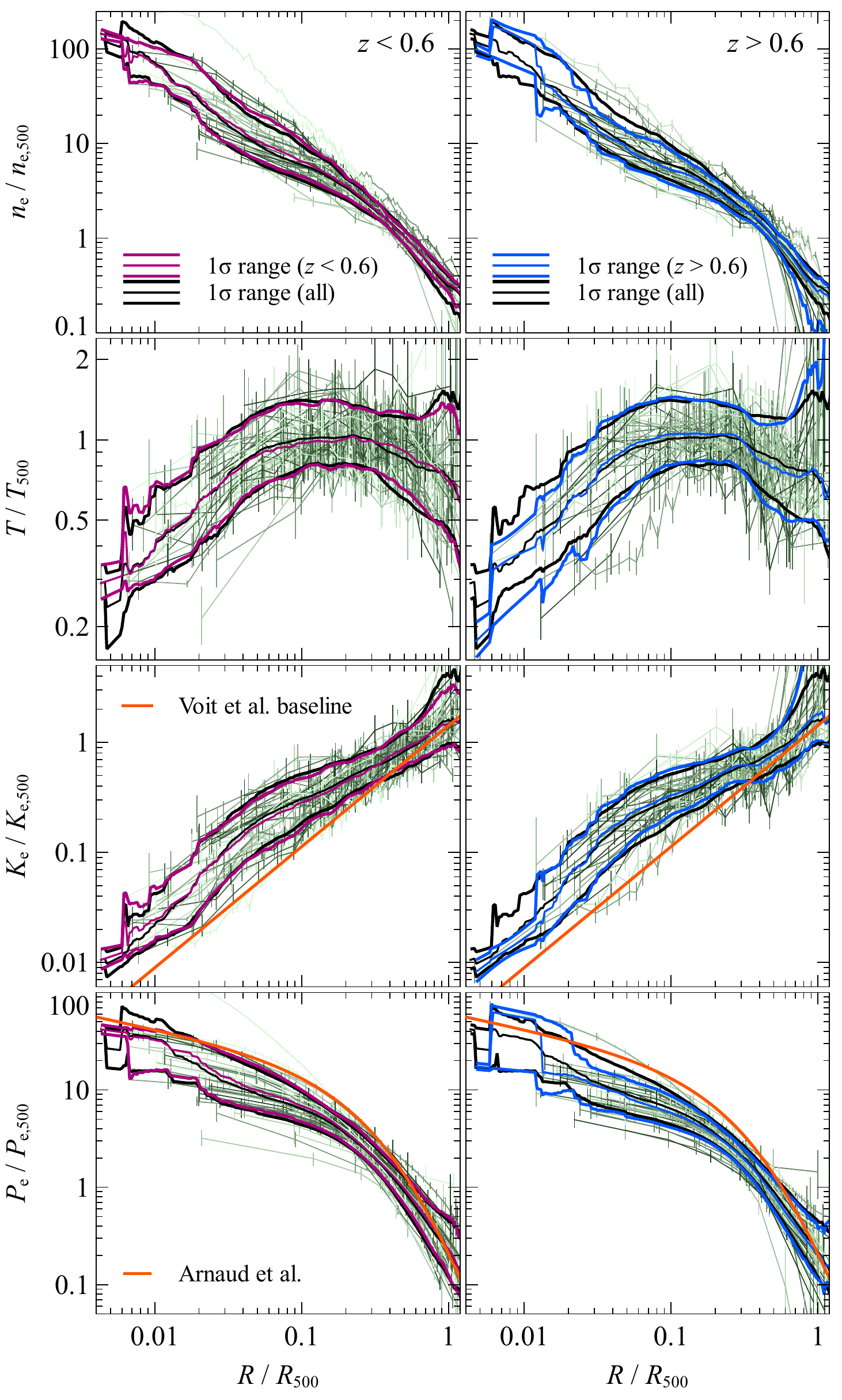} \hspace{5mm}
\includegraphics[width=\columnwidth]{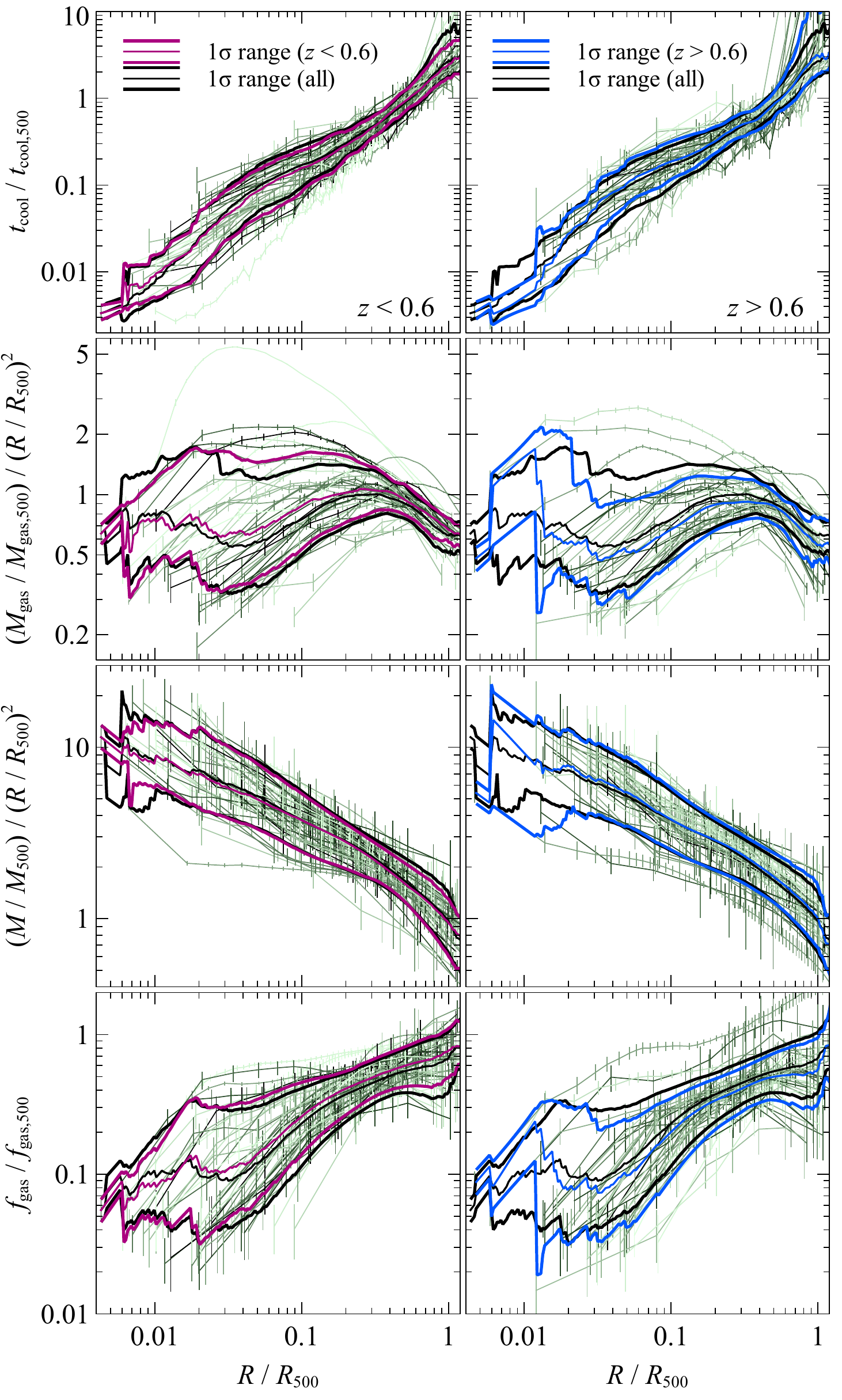}
\caption{
Self-similar-scaled profiles (from model BIN-NFW) of electron density, temperature, entropy and pressure (left side) and radiative cooling time, cumulative-gas and -total mass and gas-mass fraction (right side) as a function of scaled radius.
The scaled radius is the physical radius divided by the SPT $R_{500}$.
The profiles were scaled by the self-similar values at $R_{500}$.
The gas-mass and total-mass profiles have also been divided by $(R/R_{500})^2$ to highlight the differences between profiles.
The clusters have been split into low- (left panels) and high-redshift (right panels) bins.
The thick lines show the median and 1$\sigma$ ranges for all clusters and each redshift subsample.
The baseline entropy profile of \protect\cite{Voit05}, scaled to $R_{500}$ by \protect\cite{Pratt10}, is also shown.
We plot the ``universal'' pressure profile of \protect\cite{Arnaud10}, for the median mass of each subset, scaled due to calibration differences following MD14.
}
\label{fig:ss_profiles}
\end{figure*}

Following \cite{Arnaud10}, we also scaled our physical profiles to the characteristic values of the isothermal self-similar model at a radius of $R_{500}$.
When scaling we assumed $R_{500}$ and $M_{500}$ were the SPT-derived values (the hydrostatic values have larger and non-uniform uncertainties) and we did not account for the uncertainties in these quantities.
We scaled the physical chains of quantities by the self-similar values and repeated the same analysis as for the median physical profiles.
Fig.~\ref{fig:ss_profiles} show the profiles after scaling, plotted as a function of scaled radius.
Similarly to the physical profiles we also plot the median and range of the scaled profiles.

\begin{figure}
  \includegraphics[width=\columnwidth]{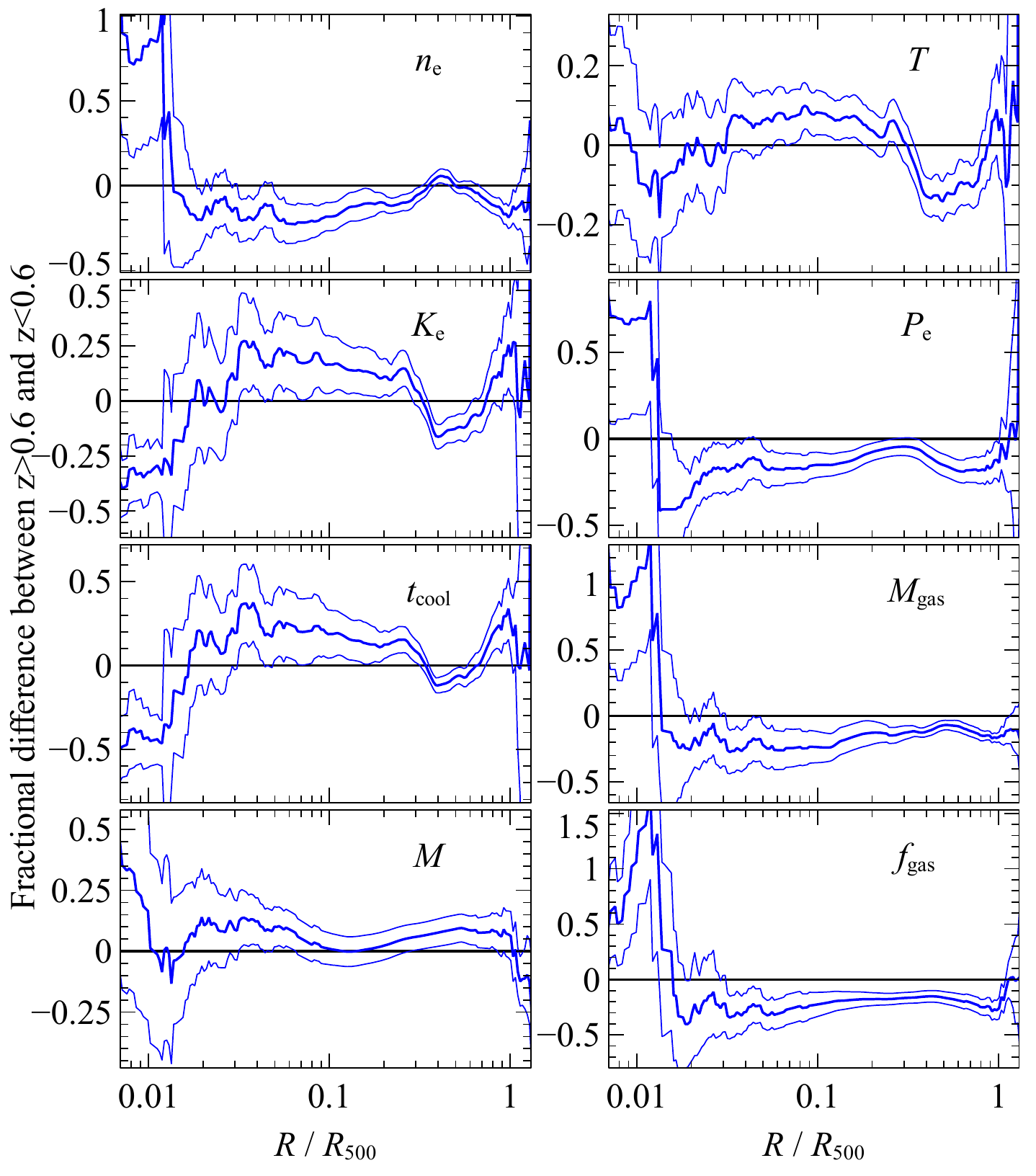}
  \caption{
Fractional difference between the self-similar-scaled median profiles for the high ($z>0.6$) and low ($z<0.6$) redshift subsamples (Fig.~\ref{fig:ss_profiles}).
The thick central line is the fractional difference, while the other two lines mark its 68 percent uncertainties.
}
\label{fig:ss_ratio}
\end{figure}

To compare the median self-similar profiles of the subsamples, we computed the fractional difference between the two median profiles for each physical quantity as a function of scaled radius (Fig.~\ref{fig:ss_ratio}).
The uncertainties were obtained using bootstrap resampling, where for each quantity the uncertainty was taken as the standard deviation in the median profile from 1000 new random cluster subsamples created from the original cluster sample with replacement.
The uncertainty on the fractional difference was calculated using the standard error propagation formulae.

Examining each of the physical quantities, there is no significant difference ($>2\sigma$) between the two subsamples at any scaled radius.
The differences between the two samples are, if present, a small fraction of the dispersion within the sample (comparing the profiles in Fig.~\ref{fig:ss_profiles}).

The median cluster increases in density inwards to around $0.02 \pcmcu$ at a radius of 20 kpc (Fig.~\ref{fig:phys_profiles}).
The low-redshift clusters have core densities which are 20 per cent greater than at higher redshift, although the difference is insignificant.
The gas-mass profiles also show this similar trend.
The median temperature profiles decline inwards to around 4 keV, or 1/2 of $T_{500}$ at this radius.
There is little evidence for difference in the two samples, except at the largest radii where the background is more important.

The core entropies for the best-resolved clusters lie below the floor of \cite{Cavagnolo09} (Fig.~\ref{fig:phys_profiles}) and look similar to the central entropy powerlaw of \cite{Panagoulia14} (note that their profile was fitted to radii of below 20 kpc which are not resolved in many of our systems).
The central cooling times fall inwards to $\sim 1$~Gyr or around 1 per cent of $t_\mathrm{cool,500}$ (Fig.~\ref{fig:ss_profiles}).
Comparing the low- to the high-redshift samples shows mild indications for the entropy and cooling times being 20 per cent lower at low redshifts.
If we compare our scaled entropy profiles to the baseline profile from the \cite{Voit05} hydrodynamical simulations (Fig.~\ref{fig:ss_profiles}), we see at $R_{500}$ that our results are consistent.
At lower radii, we see entropy enhancement above this baseline, similar to that seen by \cite{Pratt10}, consistent with a picture of additional centrally-concentrated entropy increase not associated with the shocks in the outskirts of clusters.
MD14 found that their entropy profiles flattened above $0.5 R_{500}$ for their $z>0.6$ clusters, which we do not see in our median profile for this redshift bin.

In Fig.~\ref{fig:ss_profiles} we also plot the ``universal'' pressure profile of \cite{Arnaud10} on our scaled pressure profiles, assuming median SPT masses for the two redshift subsamples.
The profile was scaled following MD14, to account for the relatively cooler temperatures measured by \emph{XMM-Newton} compared to \emph{Chandra} \citep{Schellenberger15}, by 10 per cent in pressure and 3 per cent in radius.
However, it is not completely clear whether this scaling is valid in our case as our profiles were created assuming the SPT values of $R_{500}$, but the amount scaled is small.
Our profiles match the \cite{Arnaud10} values at $R_{500}$, as was also found by MD14 in their analysis of the same data, without assuming hydrostatic equilibrium.
If there was some significant non-thermal pressure component, then it might be expected to appear as a difference at large radius between the hydrostatic and non-hydrostatic pressure profiles, which appears not to be the case.
Scaling the pressure profiles by 20-30 per cent moves them significantly at large radius from the ``universal'' profile.
At low radius, there is increasingly poor agreement between the \cite{Arnaud10} profiles and ours, with ours around a factor of two lower at $0.03R_{500}$.
MD14 also found that their pressure profiles were significantly lower than the ``universal'' profile at smaller radii.
We note that the there is increasing scatter in the pressure profiles of \cite{Arnaud10} below $0.2 R_{500}$, with roughly an order of magnitude variation at $0.1 R_{500}$, so the disagreement between our results and the universal profile is unlikely to be significant.
If we compute the median pressure profile from the non-hydrostatic fits, it agrees well with the hydrostatic profiles at radii above $0.06R_{500}$, but increases at lower radii as the temperatures become more uncertain.

\subsection{Modelling the core evolution}
\label{sect:modelcore}

\begin{figure*}
\includegraphics[width=0.47\textwidth]{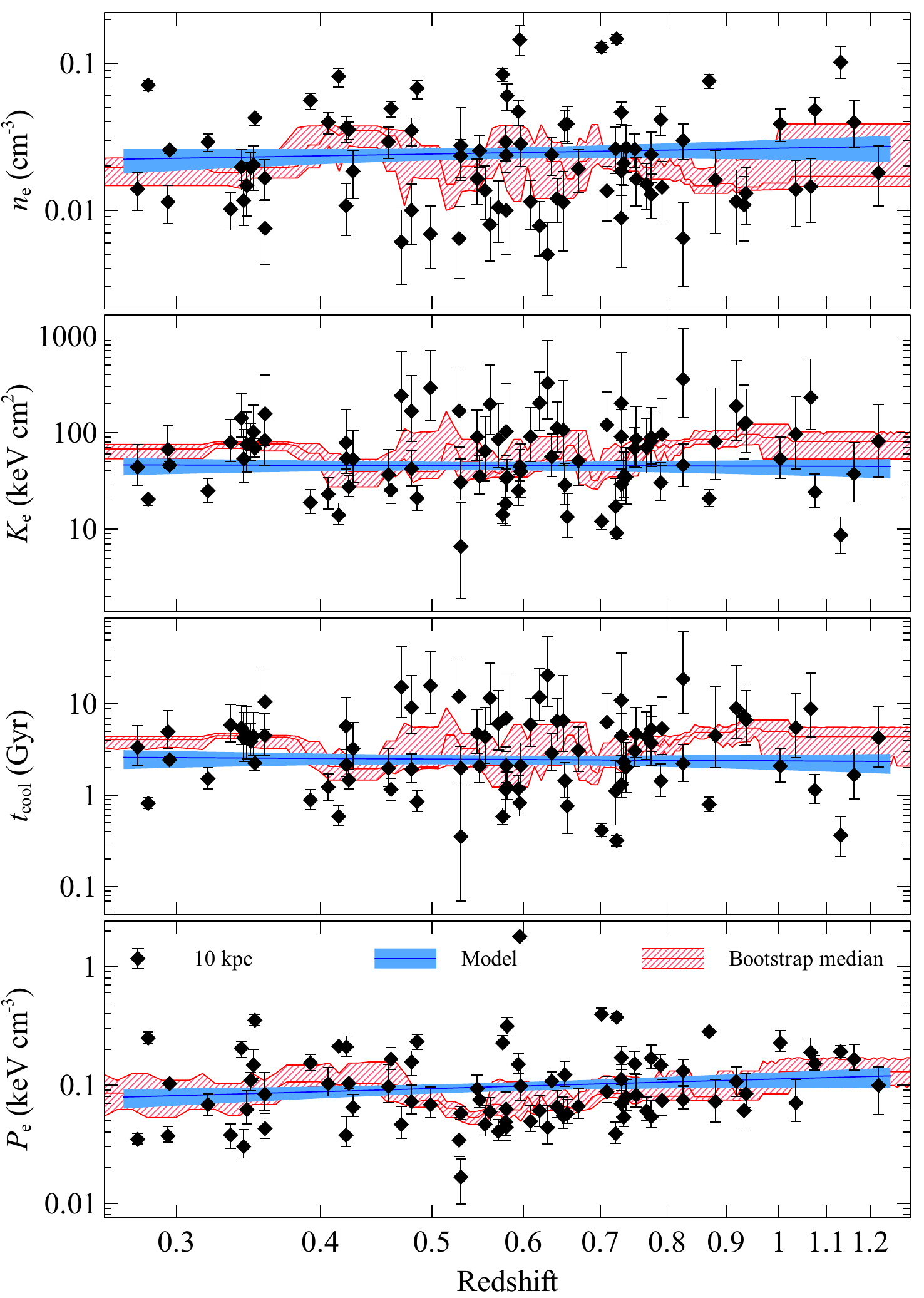} \hspace{5mm}
\includegraphics[width=0.47\textwidth]{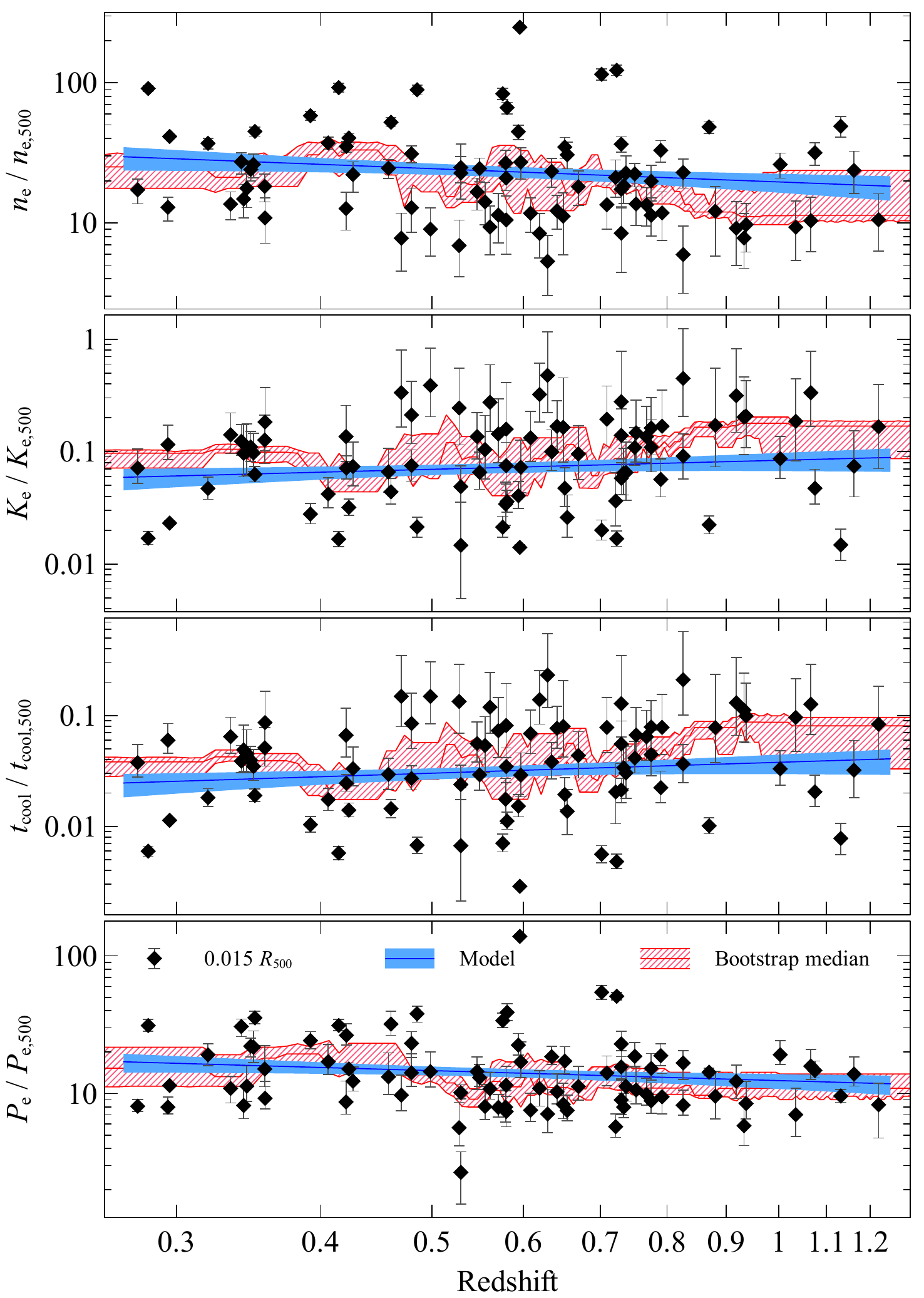}
\caption{
Cluster near-central properties as a function of redshift for the INT-NFW fits, showing the physical quantities at 10 kpc radius (left panel) and scaled quantities at radii of $0.015 R_{500}$ relative to the self-similar model values at $R_{500}$ (right panel).
For each set of radial data we show the evolution in redshift using a non-parametric method, the median and bootstrap errors of the 12 clusters with redshifts nearest the value plotted, and a logarithmic fit to the data and its uncertainties.
Plotted are profiles of the the electron density, entropy, radiative cooling time and pressure.
}
\label{fig:evo_centre}
\end{figure*}

We looked for evolution in the ICM properties near the centre of the cluster.
The binned profiles cannot be evaluated at a particular radius, so we used the interpolated density fits (the INT-NFW model).
The results are very similar if the MBETA-NFW model is used instead, although we decided to use INT-NFW due to the MBETA-NFW fits being poor in some disturbed systems.
Note that as we showed in Section \ref{sect:MD13compar}, the values at this radius are model-dependent, although here the MBETA-NFW and INT-NFW models agree.
We examine the gas properties at 10 kpc radius, although there is model uncertainty here.
This choice is to compare against MD13 and MD14 who used this radius.

In Section \ref{sect:modelevo} and Section \ref{sect:bimodal} we examine the evolution of the whole profile, finding a consistent picture.
Fig.~\ref{fig:evo_centre} (left panel) shows the density, entropy, cooling time and pressure at 10 kpc radius for each cluster as a function of redshift.
The right panel shows the quantities divided by the value at $R_{500}$ in the self-similar model at a radius of $0.015 R_{500}$.

Shown in the plot is a sliding median, showing the median (with bootstrap uncertainties) of the 12 clusters nearest the redshift shown.

\begin{table}
\centering
\caption{
Parameters from model fits to the inner quantities as a function of redshift (Fig.~\ref{fig:evo_centre}).
The values given are the median and $1\sigma$ uncertainties from the marginalised distributions.
}
\begin{tabular}{llccc}
Radius & Value & $c$ & $m$ & $w$ \\ \hline
\input{tab2.tex}
\end{tabular}
\label{tab:evo_params}
\end{table}

We also fit the evolution of the quantities with a simple parametric model, included in the figure.
Taking the $\log_{10}$ data values (physical or scaled to the self-similar value), we assume they can be fit as a function of log redshift by the relation $c+m\log_{10}(z/0.6)$, where $c$ is the value at a redshift of 0.6 and $m$ is the gradient in log redshift.
We assumed the distribution of the points in log space has constant Gaussian width, $w$.
MCMC was used to sample the model parameters, assuming a Jeffreys prior for $m$ (see Section \ref{sect:plawentropy}) and flat priors for $c$ and $w$.
The numerical values of the parameters are given in Table \ref{tab:evo_params}.
If the width, $w$, is allowed to vary as a function of redshift this does not affect the other parameters, and its slope is consistent with zero.

Examining the physical values at fixed 10 kpc radius, the non-parametric sliding median (Fig.~\ref{fig:evo_centre} left panel) shows no significant evolution.
In addition, the slopes from the model fits (Table \ref{tab:evo_params}) are consistent with zero at the $1.5\sigma$ level.

We also examine the quantities measured relative to the self-similar model (Fig.~\ref{fig:evo_centre} right panel).
Again, examining both the non-parametric sliding median and the model fit, there is no evidence for any redshift evolution.
If there is no evolution in the unscaled physical quantities, by introducing the self-similar scaling we should see evolution.
However, the changes in $m$ caused by self-similar scaling are too small to make significant changes to the fits.
From a redshift of 1.2 to 0.3, the density at $R_{500}$ should scale by a factor of $\sim 0.35$ for a $5\times10^{14} \Msun$ cluster. Entropy should scale by a smaller factor of $\sim 1.4$.

The Gaussian model assumed for the distribution of values may not be a good representation of the real distribution, particularly given the bimodal entropy distribution for $K_0$ (Section \ref{sect:innerentropy}).
We also tested other distributions, including bimodal Gaussian, skewed normal and student-t distributions.
The best-fitting evolution parameters were very similar for the different models.
The only parameter affected by the choice of distribution was the pressure parameter, which shows less-significant evolution using a bimodal or skewed distribution by about $1\sigma$ (see Section \ref{sect:bimodal}).
The choice of a Gaussian model gives an indication of the average values.

The cool-core fraction is the fraction of clusters with a cooling time less than some threshold value.
If we examine the median value of the INT-NFW cooling time at 10 kpc radius from the MCMC chains and use a strong-cool-core threshold of 2~Gyr, then the cool-core fraction is $39 \pm 9$ per cent (17 objects) below $z=0.6$ and $28 \pm 9$ per cent (11 objects) above this redshift.
These fractions are similar to the fraction of 34 per cent obtained in a lower-redshift $z=0.14-0.4$ sample \citep{Bauer05}.
Using a threshold of 4~Gyr also produces fractions consistent in the two redshift bins ($59 \pm 12$ and $49 \pm 11$ per cent, respectively).
Therefore, there is no evidence for any evolution in the cool-core fraction.

\subsection{Modelling the average cluster profile and its evolution}
\label{sect:modelevo}

\begin{figure*}
\includegraphics[width=\textwidth]{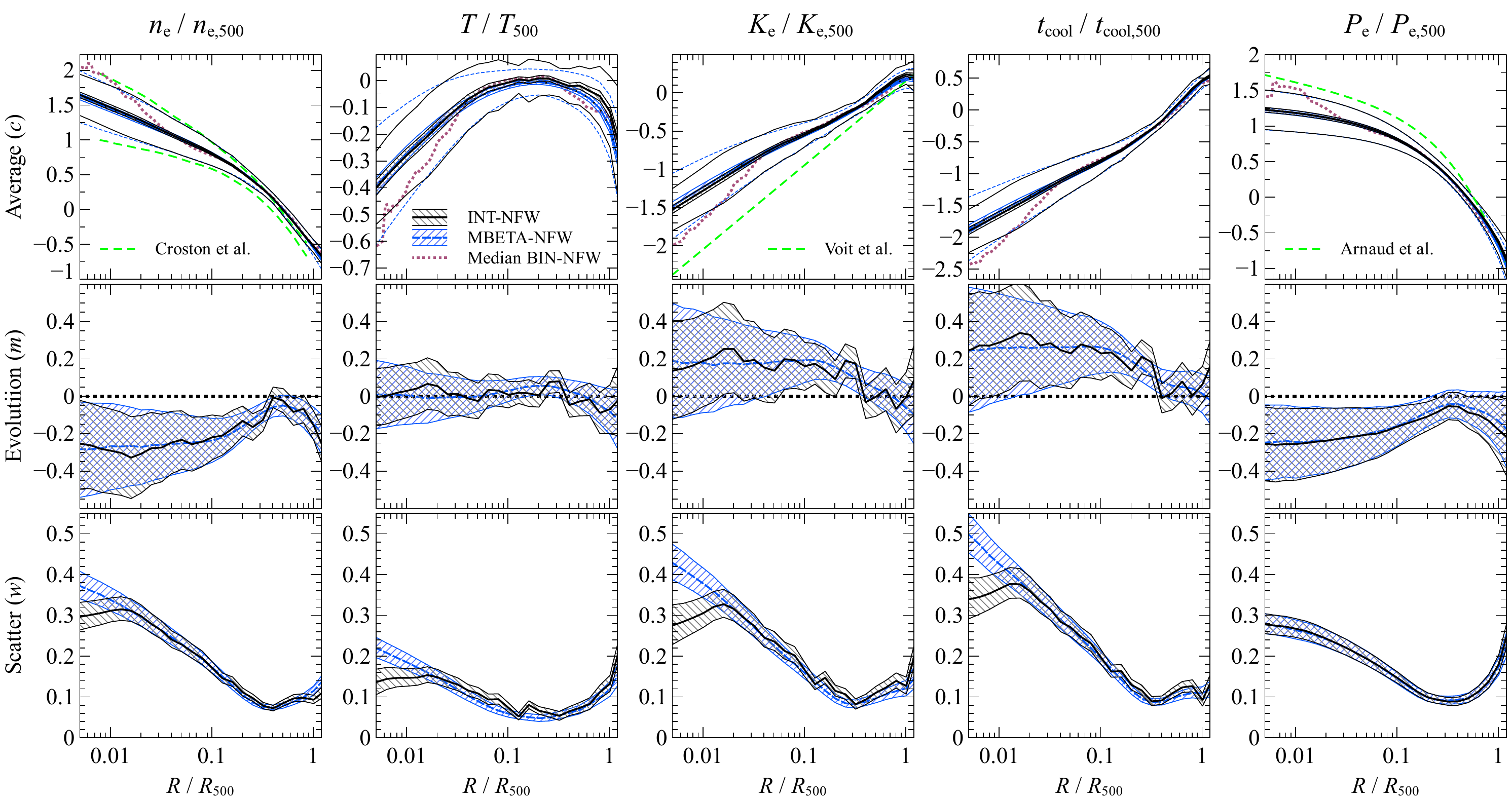}
\caption{
Results of a model fit to the self-similar-scaled thermodynamic profiles, showing the average, evolution and distribution width.
Columns show different thermodynamic quantities, as labelled.
The rows show $c$ ($\log_{10}$ distribution average value at $z=0.6$), $m$ (evolution slope) and $w$ ($\log_{10}$ distribution width).
The shaded region shows the median and $1\sigma$ range as a function of scaled radius.
In the average, $c$, results we also show the width, $w$, as the lines above and below the shaded region.
We plot the results for two different sets of profiles, INT-NFW and MBETA-NFW.
Also plotted are the median profiles from Fig.~\ref{fig:ss_profiles}, the $1\sigma$ range in density from \protect\cite{Croston08},
the universal pressure profile of \protect\cite{Arnaud10} and the baseline entropy profile of \protect\cite{Voit05}.
}
\label{fig:model_evo}
\end{figure*}

\begin{figure*}
\includegraphics[width=\textwidth]{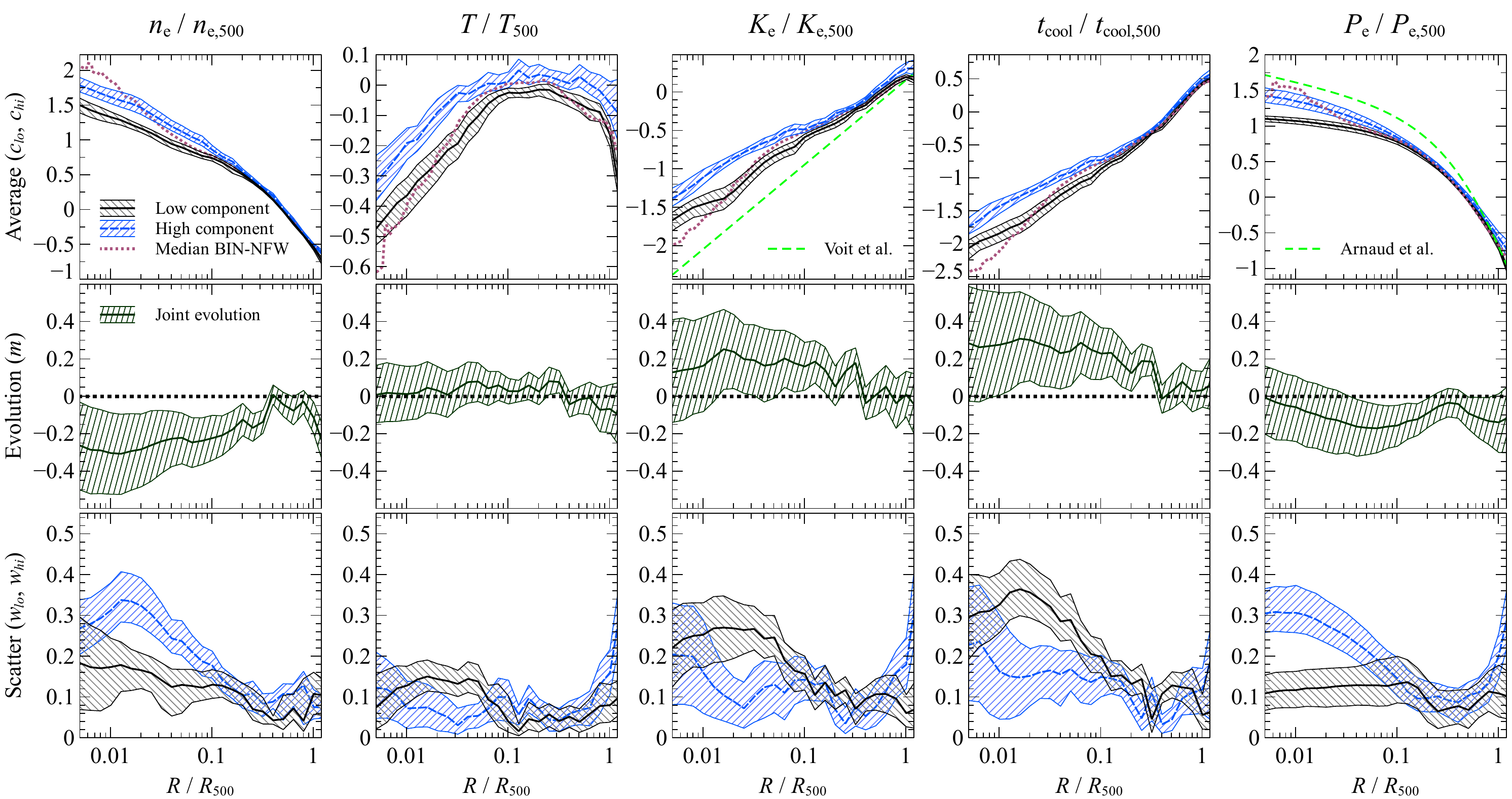}
\caption{
Parameters for the two-component modelling of the self-similar-scaled INT-NFW profiles.
Shown are the Gaussian centres at $z=0.6$ ($c_\mathrm{lo}$ and $c_\mathrm{hi}$), their joint evolution with redshift ($m$) and their width ($w_\mathrm{lo}$ and $w_\mathrm{hi}$).
Also plotted are the median profiles from Fig.~\ref{fig:ss_profiles}.
The poorly-constrained relative normalisations are not shown.
Note that the each quantity was modelled separately, so the upper density component is not the upper entropy component.
}
\label{fig:model_evo_bimodal}
\end{figure*}

In Section \ref{sect:modelcore} we modelled the evolution with redshift of the central scaled thermodynamic quantities at a fixed scaled radius, finding the logarithmic average ($c$), scatter ($w$) and slope with redshift ($m$).
Here we extend this approach, applying this model to the data as a function of scaled radius, to find the average cluster profile for each quantity, its scatter with radius and how it evolves with redshift.

This is similar to the median profile analysis in Section \ref{sect:profs}, but we assume here a parametric redshift model instead of directly comparing two redshift bins.
We also use the INT-NFW or MBETA-NFW narrow-binned cluster profiles, rather than the wide-binned BIN-NFW profiles, giving better spatial resolution in the core.
Another difference here is that we do not exclude the parts of the individual cluster profiles inside or outside the inner or outer radial regions, which biased our previous results towards cool cores.
Finally, the clusters are not equally-weighted as in the median analysis, but are effectively weighted by the uncertainties on the quantities being modelled.
This analysis assumes a Gaussian distribution of quantities at each radius.
{
By taking account of the error bars, more statistical weight will be put on the clusters with cool cores, but this will not bias our results, providing the distribution assumed is reasonable.
}
In Section \ref{sect:bimodal} we investigate the results using a two-Gaussian-component distribution model, finding similar results.

Fig.~\ref{fig:model_evo} shows the results, plotting the parameters for each of the self-similar-scaled quantities as a function of self-similar-scaled radius.
We examined both the INT-NFW and MBETA-NFW profiles, although the results are very similar except for the scatter-width parameter, $w$, in the very centre.
In the top panels are the average, $c$, profiles at $z=0.6$, also showing the width of the distribution, $w$.
These profiles look very similar to the median profiles (Section \ref{sect:profs}), except in the very centre.
The density profile keeps rising inwards, with no evidence for a flat core.
The density profile and dispersion are very similar to that obtained by \cite{Croston08} for a representative sample of nearby clusters (taking values from their fig. 13).
Likewise, the entropy and cooling time profiles are powerlaws in the central regions.
However, there are breaks in density, entropy and cooling time between $0.1$ and $0.2 R_{500}$.

Fig.~\ref{fig:model_evo} (centre row) shows the radial variation of the evolution slope parameter, $m$.
Over the redshift range examined, this can be multiplied by $\sim 1.2$ to give the total change in the self-similar-scaled physical value in dex.
As in Section \ref{sect:profs}, we see no evidence for non-zero evolution parameters.
Mildly-decreasing central density and pressure with redshift, and mildly-increasing entropy and cooling time would be consistent with the data, however.

The radial profiles of the scatter, $w$, are shown in Fig.~\ref{fig:model_evo} (bottom row).
The entropy, density and cooling time profiles show a similar pattern, where there is relatively little scatter ($\sim 0.1$ dex or 26 per cent) outside $0.3 R_{500}$.
Going into the centre, the scatter increases to around 0.3~dex (around $100$ per cent) for each of these quantities using the INT-NFW model.
The central scatter is higher for the MBETA-NFW model, but this model fails to fit the surface brightness of some highly-disturbed clusters.
The difference shows inside $0.015R_{500}$ the results become more model-dependent.
The temperature scatter profile shows a flatter distribution, although there is a minimum of 0.06 dex (15 per cent) around $0.2 R_{500}$, increasing to around 0.15 dex (40 per cent) at small and large radii.
The pressure profiles show a similar picture, with 0.27 dex (90 per cent) scatter in the core and 0.2 dex (60 per cent) around $R_{500}$.

As a check on the scaling of profiles we repeated the analysis using masses from a $Y_\mathrm{X}$ scaling relation, finding the resulting average, evolution and scatter profiles were very similar to the ones shown here.
Minor changes were that the peak temperatures were slightly higher ($\sim0.025$ dex) and the pressure profiles showed less scatter ($0.05$~dex) at 0.5~$R_{500}$, but increased scatter inside 0.3~$R_{500}$.

\subsection{Two-component modelling of the average profile and its evolution}
\label{sect:bimodal}
As the distributions of thermodynamic quantities at each radius may not be Gaussian, possibly biasing the results, we extended the analysis from the previous section with a more complex distribution.
We choose a two-Gaussian-component model, although the underlying shape may also be consistent with other distributions, such as a skewed normal.
The bimodal distribution of $K_0$ values gives some justification for the use of a bimodal fit.

For each thermodynamic quantity, we modelled the distribution of log-space values at each scaled radius with two Gaussian components, applying it to the INT-NFW results using MCMC.
The two components were parametrized by their centres at $z=0.6$, $c_1$ and $c_2$, their widths, $w_1$ and $w_2$, their common evolution with redshift, $m$, and their relative normalisation.
We allowed the normalisation of the first component to lie between a fraction of 0.25 and 0.75 of the total, assuming a flat prior.
By forcing similarly-sized components we exclude weak, noisy contributions which obscure the interpretation of the parameter values.
If the underlying distribution is Gaussian, forcing comparable normalisations aligns the component centres to the same value.
We allow evolution with redshift of the data points, assuming both components evolve with the same gradient in logarithmic redshift ($m$) and that there is no evolution in the relative normalizations of the two components.

Fig.~\ref{fig:model_evo_bimodal} shows the results from our analysis.
As there are two components at every radius, for each entry in the MCMC chain we choose the one with the lowest-value centre as the lower component and the other as higher component.
We refer to the centres of the lower and higher $c$ components as $c_\mathrm{lo}$ and $c_\mathrm{hi}$, respectively, with their widths $w_\mathrm{lo}$ and $w_\mathrm{hi}$.
The top panels show $c_\mathrm{lo}$ and $c_\mathrm{hi}$ as a function of scaled radius, calculated from their medians and $1\sigma$ percentiles.
At radii beyond $0.2 R_{500}$ the components have consistent centres (i.e. the data do not show evidence for bimodality).
This radius is where a break is seen in the density, entropy and cooling time profiles, and close to where the temperature declines towards the centre.
Towards the centres of the clusters, the $c$ values significantly diverge, although in the temperature profiles the evidence is weaker.
The difference between the two components is similar to the width of the distribution $w$ in the single component fits (Fig.~\ref{fig:model_evo}).
In the inner parts of the cluster, the highest density component is similar to the median BIN-NFW profile.
This is due to the median BIN-NFW profiles excluding at a particular radius clusters where this radius is not between the centres of their inner and outer bins.
This makes the median profiles more sensitive to clusters with higher central densities.

The evolution parameters as a function of scaled radius ($m$; centre row) are consistent with the results assuming the Gaussian distribution.
In the outer regions the width parameters ($w_\mathrm{lo}$ and $w_\mathrm{hi}$; bottom row) are also consistent with the results of the single-component modelling.
In the centres, the widths of the two components diverge in the density, entropy, cooling time and pressure plots.
The higher density, lower entropy, lower cooling time and higher pressure components have a larger spread on their distribution.
The other components have widths which are roughly consistent with scaled radius.

Comparing the median profiles with the results from this analysis, we see as we go inwards in radius that the median profile moves from the low density, high entropy, long cooling time edge to the high density, low entropy, short cooling time component.
This shows there is likely some central bias in the median profiles, caused by the data quality being systematically lower in those clusters with less-dense cores.

The bimodality we find from our modelling is rather weak overall.
In the centres of the clusters there is only an offset between 0.25 and 0.5 dex between the centres of the two components.
The distribution width around these centres can increase to 0.4 dex however, for density, pressure and cooling time.

It may be the case that cool-core and non-cool-core clusters could evolve differently with redshift.
We examined a model with two separate evolution parameters for the lower and higher $c$ components.
We did not see evidence for evolution in either case, although the uncertainties on these parameters are increased over the analysis above with a single evolution parameter.

\section{Discussion}

\subsection{Central entropies and cooling times}
By modelling the obtained central entropy ($K_0$) values for the sample (Section \ref{sect:innerentropy}), we find the distribution has a narrow tail upwards from zero entropy and a second, broader peak around $130 \keV\cmsq$.
Our model implies around 60 per cent of systems are part of the narrow low-entropy peak.
\cite{Cavagnolo09} examined a large sample of archival cluster data, finding a bimodal distribution with peaks at $K_0=16.1 \keV\cmsq$ and $150 \keV\cmsq$, with clusters split roughly equally between the two.
The significant positive centre of their lower entropy peak is inconsistent with our finding that there is no evidence for a floor of this level, for our lowest entropy systems.
\cite{Panagoulia14} found powerlaw entropy profiles for a volume-limited sample of local clusters and groups observed using \emph{Chandra}.
They explain the presence of a floor in \cite{Cavagnolo09} as being due to their use of wide-temperature bins, projected temperature measurements and the inclusion of lower quality datasets.
\cite{Pratt10} examined $K_0$ in a representative sample of clusters, finding entropy peaks at $3$ and $75 \keV\cmsq$, close to our results.
They also claim the difference with \cite{Cavagnolo09} is due to the use of projected temperatures by \cite{Cavagnolo09}.
\cite{Hogan17} recently examined a small sample of clusters with deep \emph{Chandra} observations also finding no evidence for an entropy floor.
\cite{Lakhchaura16} studied the inner entropy for a set of \emph{Chandra}-observed clusters finding floors in their powerlaw profiles.
However, the cores they found were mostly within the inner-bin probed and so it is unlikely that they have the spatial resolution to actually measure these floors.
In our analysis the \textsc{mbproj2} code allows us to trace the temperature and density on small spatial scales.
Like in \cite{Panagoulia14} we find that the size of the central region probed (Fig.~\ref{fig:cen_radius}) is strongly correlated with entropy and so care must be taken in comparing data of different quality.

MD13 found a bimodal distribution of their parameter $K_0$ for almost the same sample of clusters as us.
However, the parameter MD13 called $K_0$ is not the parameter used by \cite{Cavagnolo09}, but is the value of the entropy at a radius of 10 kpc, and so it is difficult to compare directly.
We used the X-ray peak as the cluster centre, whereas MD13 used the centroid of a 250-500 kpc annulus.
The peak densities are much larger than the centroid densities (Fig.~\ref{fig:ne_cent} lower panel).
MD13 also fitted a modified-$\beta$ model forcing the inner slopes to be flat or inward-rising, which biases densities upwards if the X-ray peak is not the cluster centre (Fig.~\ref{fig:vikh_alpha}).
These two effects somewhat cancel each other out, as seen in Fig.~\ref{fig:ne_cent} (lower panel), giving densities up to 10 times lower than the peak densities (or entropies up to 20 times higher assuming isothermality).

The differences between our results and MD13 are mainly due to the choice of a cluster centre and the priors used by MD13 on the $\alpha$ parameter in the modified-$\beta$ model fitting.
We assert the X-ray brightest peak of the cluster is the appropriate location, as it is where the lowest entropy gas is located.
However, in highly disturbed systems, such as the Bullet cluster (SPT-CLJ0658-5556) undergoing a major merger, or the Coma cluster which has two X-ray nuclei, the question arises as whether we should treat these bright peaks in the same way as a relaxed cool-core cluster.
We believe the X-ray peak is a better definition, as this is likely to be where any AGN is likely to be located, which may not know about the wider morphology of the galaxy cluster.

Our separate analyses modelling the thermodynamic profiles as a function of radius (Sections \ref{sect:modelevo} and \ref{sect:bimodal}) obtained average density, entropy and cooling time profiles for the clusters at $z=0.6$.
These profiles extend inwards as powerlaws with no evidence for a central floor.

\subsection{Cluster evolution}
We looked for evidence for evolution in cluster properties, relative to the self-similar scaling model, between clusters above and below a redshift of $0.6$ (Section \ref{sect:profs}).
By computing the ratio of the median profiles relative to self-similar values, each with errors calculated using bootstrap resampling, we find very mild evidence for the low-redshift clusters being slightly more gas dense, cooler, having lower entropy and higher pressure.
We note, however, that the median profile analysis is likely biased towards cool cores.
The total-scaled mass profiles look similar overall, but the gas-mass fraction appears higher in the low-mass systems by 20 to 30 per cent.
However, the bootstrap-uncertainty error bars suggest that each of these findings has rather low significance (equivalent to 1 to 2 $\sigma$).

We also did more sophisticated modelling of the profiles as a function of redshift, both using single- (Section \ref{sect:modelevo}) and two-component modelling of the distribution of values in clusters as a function of scaled radius (Section \ref{sect:bimodal}).
This analysis method is not biased towards cool cores.
The profiles were assumed to have scatter with a modelled width and have evolving averages with redshift.
These analyses differ from the median analysis by not treating clusters equally, by having more spatial resolution in the centre and by not excluding the central regions of clusters where the profiles are poorly defined.
With these analyses we also found no evidence for evolution in any of the examined thermodynamic quantities as a function of radius.

MD14 conducted joint fits on sets of clusters, divided into subsamples based on central density and redshift (using the same redshift bins as here).
They found that the inner pressure values in their high-$z$ subsample were lower in the cores than in their low-$z$ clusters.
The high-$z$ clusters had pressures 3 to 6 times less than those at intermediate or low $z$.
Neither our median profiles (Fig.~\ref{fig:ss_ratio}) nor our modelling with radius (Fig.~\ref{fig:model_evo} and Fig.~\ref{fig:model_evo_bimodal}) shows evidence for redshift evolution.
In the central regions, the difference with MD14 is likely due to the difference in choice in cluster centres, as seen by the sensitivity of the central density to the cluster centre.

MD14 did not find entropy evolution, which agrees with our very-mild difference between the two redshift bins (Fig.~\ref{fig:ss_ratio} and Fig.~\ref{fig:evo_centre}).
MD13 found no significant evolution in the central entropy, cooling time and mass deposition rate.
Similarly, we find little evidence for any evolution in these quantities.
MD14, however, claimed that the entropy at radii beyond $0.5 R_{500}$ in $z>0.6$ clusters flattened relative to low-$z$ systems, which we see no significant evidence for in our scaled profiles.
It is not clear what the difference might be due to.
We assume hydrostatic equilibrium, which may not apply in the cluster outskirts, while MD14 assume a common temperature profile for their set of systems.

One of their strongest claims is that there is strong evolution in central density (increasing by an order of magnitude between $z\sim1$ and 0).
We find only mild evidence for a density increase from high to low redshifts, with our median profiles (calculated from the binned analysis) increasing by around 25 per cent in the central regions (Fig.~\ref{fig:ss_ratio}), while our analysis using the INT-NFW fits shows an increase by $\sim 50$ per cent at $0.015 R_{500}$ (Fig.~\ref{fig:evo_centre} right panel).
However, both of these changes have significances of just over $1\sigma$.
We also find that the shape of the median density profiles in our analysis appears roughly consistent in our two redshift bins and there is no significant change in the shape of our modelled profiles.
We reiterate that our densities agree very well between the hydrostatic and non-hydrostatic analyses, so the assumption of hydrostatic equilibrium is not relevant.

In Section \ref{sect:MD13compar} we show that the central densities are highly sensitive to the choice of cluster centre and, if the cluster centre is not on the X-ray peak, to the assumed form for density profile.
These effects could introduce a redshift dependence if the data quality changes with redshift or the choice of centre is redshift-dependent.
We also note that the published results in MD13 do not use the peak of the cluster to compute the densities.
It may be the case that the evolution is induced by the way the centre is chosen as a function of redshift.
In addition our modelled evolution of the profiles (Fig.~\ref{fig:model_evo}) shows no evidence for density evolution at larger radii.
We used the SPT SZ masses to do the scaling of the clusters, while MD14 used $Y_\mathrm{X}$ masses.
This did not cause of the differences between our conclusions, as we found good agreement between our own $Y_\mathrm{X}$ scaling-relation-derived masses and the published SPT values, and our reanalysis in Section \ref{sect:modelevo} gave almost identical results with the two scalings.
The evidence for density evolution is dependent on whether it is more reasonable to choose the X-ray peak as the cluster core, as we do, or a larger-scale centroid and a functional form for the density profile which is forced to not centrally decline, as is the case for the MD13 and MD14 analyses.

In conclusion, the lack of evolution in the overall shape of the profiles for each thermodynamic quantity shows that the population of clusters at fixed mass appears remarkably stable over the 0.28 to 1.2 redshift range we examine.
We also see that cool cores have existed in clusters over the last 8~Gyr, with similar frequencies to the local universe and with consistent profiles and core entropies.
Cool cores are an extremely stable phenomena.
The physical processes which balance the growth of cool cores in the centres of clusters must have been put in place at even higher redshifts.

\section{Conclusions}
We analyse X-ray data from a set of galaxy clusters selected by the SPT telescope and observed by the \emph{Chandra X-ray Observatory}.
The surface-brightness profiles in multiple X-ray bands were fit using a new hydrostatic modelling code, \textsc{mbproj2}, to obtain dark-matter and thermodynamic profiles.
Our conclusions are as follows:
\begin{enumerate}
\item
By fitting an entropy profile to the entropy profiles consisting of a constant $K_0$ plus a powerlaw, we find many systems are consistent with no entropy floor (zero $K_0$)
By modelling the underlying distribution we find a narrow peak close to zero entropy containing 60 per cent of the systems and a second broader peak at $\sim 130 \keV\cm^{2}$.

\item
The central density, entropy and cooling time values in the clusters are strongly affected by the size of the region used to probe them.

\item
We compute median thermodynamic profiles and scaled profiles, biased centrally towards cool-core clusters, scaled by the self-similar quantities at $R_{500}$, both for the combined sample and in redshift bins around $z=0.6$.
All of the computed quantities, including the temperature, density, pressure, entropy, cooling time, gas mass, total mass and gas mass fraction, show no significant evolution at the $2\sigma$ level.
We model the evolution of the average thermodynamic profiles as a function of scaled radius, without the cool-core bias, also finding no evidence for evolution.
Although there should be evolution in either or both the unscaled physical or scaled profiles, the data are insufficient to see evolution in either case.

\item
The modelled average profiles at $z=0.6$ are centrally powerlaw profiles with no floor.
The density, entropy and cooling time profiles show a break at around $0.2 R_{500}$ and are approximately powerlaws inside and outside that radius.

\item
We look for evolution in the core properties with redshift, finding no evidence for evolution in both the physical and scaled profiles of density, entropy, cooling time or pressure.

\item
There is no significant difference in the cool-core fraction in two redshift bins above and below $z=0.6$.
Around 30 to 40 per cent of the clusters have central cooling times below 2~Gyr.

\end{enumerate}

\section*{Source code}
The source code for \textsc{mbproj2} is publicly available and can be found at \url{https://github.com/jeremysanders/mbproj2}.

\section*{Acknowledgements}
ACF and HRR acknowledge support from the ERC Advanced Grant FEEDBACK.
SAW was supported by an appointment to the NASA Postdoctoral Program at the Goddard Space Flight Center, administered by the Universities Space Research Association through a contract with NASA.
The scientific results reported in this article are based on data obtained from the \emph{Chandra} Data Archive.
We thank M.~McDonald for providing the updated density values and discussions.

\bibliographystyle{mnras}
\bibliography{refs}

\appendix
\section{Data tables}
\subsection{Sample}
\label{append:sample}

Listed in Table \ref{tab:sample} for each cluster is the SPT identifier, any other main identifier, its redshift \citep[from][]{Bleem15}, the maximum radius used in the hydrostatic analysis (radii beyond this are used for background scaling), the uncertainty allowed on the binned emissivities, the list of \emph{Chandra} datasets examined and a numerical index which is used elsewhere in the paper.

\begin{table*}
\caption{
Sample of objects analysed here. Listed is the cluster index, SPT identifier, main identifier (if any), redshift from \protect\cite{Bleem15}, Galactic equivalent hydrogen column density ($N_\mathrm{H}$; $10^{20}\pcmsq$), maximum radius examined ($r_\mathrm{max}$; arcmin), percentage uncertainty on the emissivities in each bin after binning (Rebin), total-cleaned exposure (ks), background-subtracted counts (and background counts) within the maximum radius (k) and list of \emph{Chandra} observation identifiers.
}
  \input{taba1_1.tex}
\label{tab:sample}
\end{table*}

\begin{table*}
\contcaption{sample.}
  \input{taba1_2.tex}
\end{table*}

\subsection{Positions}
\label{append:posns}
Table \ref{tab:posns} lists the positions of the cluster peaks used as the centres of our annuli in the main part of the analysis.
We also list our positions computed using a centroid with a 250 to 500 kpc annulus (the same technique as used by MD13).
The difference between these two positions are given in arcsec and kpc.

\begin{table*}
\caption{Positions of the clusters.
Shown are the J2000 coordinates of the peak in degrees used as the cluster centres for our analysis.
Also shown are our positions of the centroid using a 250 to 500 kpc annulus.
The offset between the two is shown in arcsec and kpc.
}
\centering
\label{tab:posns}
\input{taba2_1.tex}
\end{table*}

\begin{table*}
\contcaption{positions.}
\centering
\input{taba2_2.tex}
\end{table*}

\subsection{Goodness of fits}
\label{append:goodness}
Although the Bayesian analysis does not directly produce a goodness of fit, we calculate these to highlight those clusters for which a model poorly reproduces the data (Table \ref{tab:fitstats}).
We have fitted the BIN-NFW, BIN-GNFW, BIN-NONHYDRO, INT-NFW, MBETA-NFW, KPLAW-NFW and GRAD-NFW models to each cluster to minimise the C-statistic (a modified Cash statistic; see \citealt{Kaastra17}), here ignoring the model priors.
Using the technique of \cite{Kaastra17} we compute the absolute deviation of the fit statistic from the expected value, dividing by the expected standard deviation of the statistic.
68 per cent of the time a perfectly-fitting model should lie in the range $-1$ to $1$.
For many of the clusters the models appear reasonable fits.
Often the model fits to the binned profiles are statistically better than the finely-binned profiles, as the binning process destroys signal in the data.
In the Bullet cluster, the models are all very poor fits to the data, likely due to the very high data quality and strong substructure in the system.
In some systems, the parametric MBETA-NFW and KPLAW-NFW models appear poor fits, including SPT-CLJ0102-4915, SPT-CLJ0411-4819 and SPT-CLJ2344-4243.

\begin{table*}
\centering
\caption{
Goodness of fits.
Values show the difference in fit statistic from that expected on average for the model divided by the expected standard deviation.
}
\input{taba3_1.tex}
\label{tab:fitstats}
\end{table*}

\begin{table*}
\contcaption{goodness of fits.}
\centering
\input{taba3_2.tex}
\end{table*}

\section{\textsc{mbproj2} modelling}
\label{append:mbproj}
The surface-brightness profiles were fit with a new \textsc{mbproj2} multiband projection code, based around the techniques used in \textsc{mbproj} \citep{Sanders14PKS0745}.
The advantages of this code over the original include more flexible modelling, increased speed, decreased complexity and a more modular codebase.
In addition, the bin pressures are now computed at the centres of the radial bins which improves the consistency between different radial binning schemes.
The software is made up of a number of Python objects representing the data, model profile components, model parameters, fit and MCMC state.
A user can define new model components or extend existing ones.

The code is capable of fitting surface-brightness profiles with or without assuming hydrostatic equilibrium.
If assuming hydrostatic equilibrium, the data are fit by a model parametrizing the gas electron-density profile, dark-matter mass profile, outer pressure and a metallicity profile.
When fitting without assuming hydrostatic equilibrium, the model parametrizes the temperature profile instead of the outer pressure and dark-matter profile.

When computing the profile assuming hydrostatic equilibrium, the code works by computing the pressure profile by summing inwards the contributions to the pressure inwards using $\delta P = \rho g \: \delta r$ and the outer pressure $P_\mathrm{out}$.
The gas density, $\rho$, as a function of radius, $r$, is calculated from either a binned or a parametric electron density model, $n_\mathrm{e}(r)$.
The gravitation acceleration, $g(r)$, is computed from the parametrized dark-matter profile and $\rho(r)$.
Given $P(r)$ and $\rho(r)$ the code calculates the temperature, $T(r)$.
The X-ray emissivity in several bands is then computed from $n_\mathrm{e}(r)$, $T(r)$, the metallicity model, $Z(r)$ and the absorbing column density $N_\mathrm{H}$.
These three-dimensional profiles are then projected to produce the observed model count-rate profiles.

We detail the steps used to compute the model surface-brightness profiles below.
\begin{enumerate}
\item The model dark-matter-mass and gas-density profiles are used to calculate the total gravitational acceleration for each shell.
For the dark-matter profile the code computes the acceleration at each mass-weighted shell radius assuming constant density in a shell,
\begin{equation}
r_\mathrm{mass} = \frac{3}{4} \frac{r_\mathrm{out}^4 - r_\mathrm{in}^4} {r_\mathrm{out}^3 - r_\mathrm{in}^3},
\label{eqn:massweighted}
\end{equation}
where $r_\mathrm{in}$ and $r_\mathrm{out}$ are the shell inner and outer radii.
The gas acceleration component is calculated from the average acceleration on a shell, assuming it has constant density,
\begin{equation}
  g_\mathrm{gas} = G \frac {3 \: M_\mathrm{gas}(r<r_\mathrm{in}) +
    \rho ( r_\mathrm{out} - r_\mathrm{in}) \left[ (r_\mathrm{out}+r_\mathrm{in})^2 + 2 r_\mathrm{in}^2 \right]}
{ r_\mathrm{in}^2 + r_\mathrm{in} r_\mathrm{out} + r_\mathrm{out}^2 },
\end{equation}
where $\rho$ is the gas-mass density in the shell, calculated from the electron number density, $n_\mathrm{e}$, assuming the plasma is fully ionized and using solar helium to hydrogen ratios.
$M_\mathrm{gas}(r<r_\mathrm{in})$ is the total gas mass in interior shells.
The total gravitational acceleration is therefore
\begin{equation}
g_i = \frac{ G \: M_{DM} (r < r_{\mathrm{mass},i})  } {r_{\mathrm{mass},i}} + g_{\mathrm{gas},i}
\end{equation}

\item The pressure at the mass-averaged centre of each shell is calculated by summing up the contribution to pressure from gas at larger radius,
\begin{equation}
P_i = (r_{\mathrm{out},i}-r_{\mathrm{mass},i}) \rho_{i} g_{i} +
\sum_{j=i+1}^{N} (r_{\mathrm{out},j}-r_{\mathrm{in},j}) \rho_{j} g_{j} +
P_\mathrm{out}
\end{equation}
where $P_\mathrm{out}$ is the outer pressure and there are $N$ shells.

\item \label{itm:calct} The temperature in each shell is computed from the total pressure using the ideal gas law, $k_B T_i = P_i / (n_{\mathrm{e}, i} \: X)$, where $X$ is the average number of particles per electron ($\sim 1.83$).

\item For each shell, given $T_i$, $n_{\mathrm{e},i}$, the model metallicity ($Z_i$) and Galactic absorbing column density, the emissivity in each of the X-ray energy bands is computed.
The computation is done by interpolating within a table of emissivity values tabulated for a range of temperature values.
The computation is done for unit density and for metallicity values of $0$ and $1\Zsun$, allowing emissivities at other densities and metallicities to be calculated by scaling.
In this paper, we used \textsc{xspec} 12.9.0o \citep{ArnaudXspec}, the \textsc{apec} 2.0.2 plasma emission model \citep{SmithApec01} and the \textsc{phabs} photoelectric absorption model \citep{BalucinskaChurchPhabs92} to do the conversions.
A single central response and ancillary response is used for each cluster, but we correct for vignetting below.

\item Given the emissivity profiles in each of the shells, the code computes the projected profile in a band by multiplying its emissivity profile with a matrix containing the volume of each shell ($V_i$) visible in an annulus on the sky $j$.
An area-scaling factor accounts for the difference between the purely geometric area on the sky in the projection code and the pixelized area of the extracted surface-brightness profiles.
In addition, exposure maps are used to scale the exposure time in each radial bin and band relative to the central exposure to account for vignetting, bad pixels and the edge of the detector.

\item Optionally, the instrument point spread function (PSF) can be accounted for by multiplying the model surface-brightness profiles with mixing matrices which account for the fraction of flux spread from one annulus to every other annulus for each energy band.
These matrices are precalculated assuming constant densities within each shell.
Mixing between shells can give difficulties in convergence if the effect of the PSF is similar to projection.
These effects can be alleviated by assuming a functional form for the density profile rather than using binning.
In our analysis here we do no account for the \emph{Chandra} PSF as it is usually small in the cluster centres.

\item Background profiles are added to the projected model profiles in each band, optionally scaling the background by an additional variable model parameter.
In this project, we use \emph{Chandra} blank sky backgrounds (see Section \ref{sect:background}).

\end{enumerate}

The description above parametrizes the gas density as a function of radius when assuming hydrostatic equilibrium.
It is sometimes useful to instead parametrize a different thermodynamic quantity if its model parameters are of interest (e.g. entropy floors).
With a different parametrization, instead of step \ref{itm:calct}, the temperature and density can be calculated from the pressure and the alternative parameter.
However, the gravitational acceleration used in the calculation of hydrostatic equilibrium depends partially on the gas density, which is not known until the end of the procedure.
To work around this problem, we iterate the computation of the profiles a number of times, using the density profile from the previous computation in the calculation of the gravitation acceleration.

\section{Individual profiles}
\label{append:individual}
In Fig.~\ref{fig:individual} are shown the individual profiles for each cluster.
The profiles are similar to Fig.~\ref{fig:example}, plotting the electron density, temperature, pressure, entropy, cooling time and cumulative-gas and -total masses.
The data in the profiles are provided in the electronic-only Table \ref{tab:individual}.

\begin{figure*}
  \includegraphics[width=0.3\textwidth]{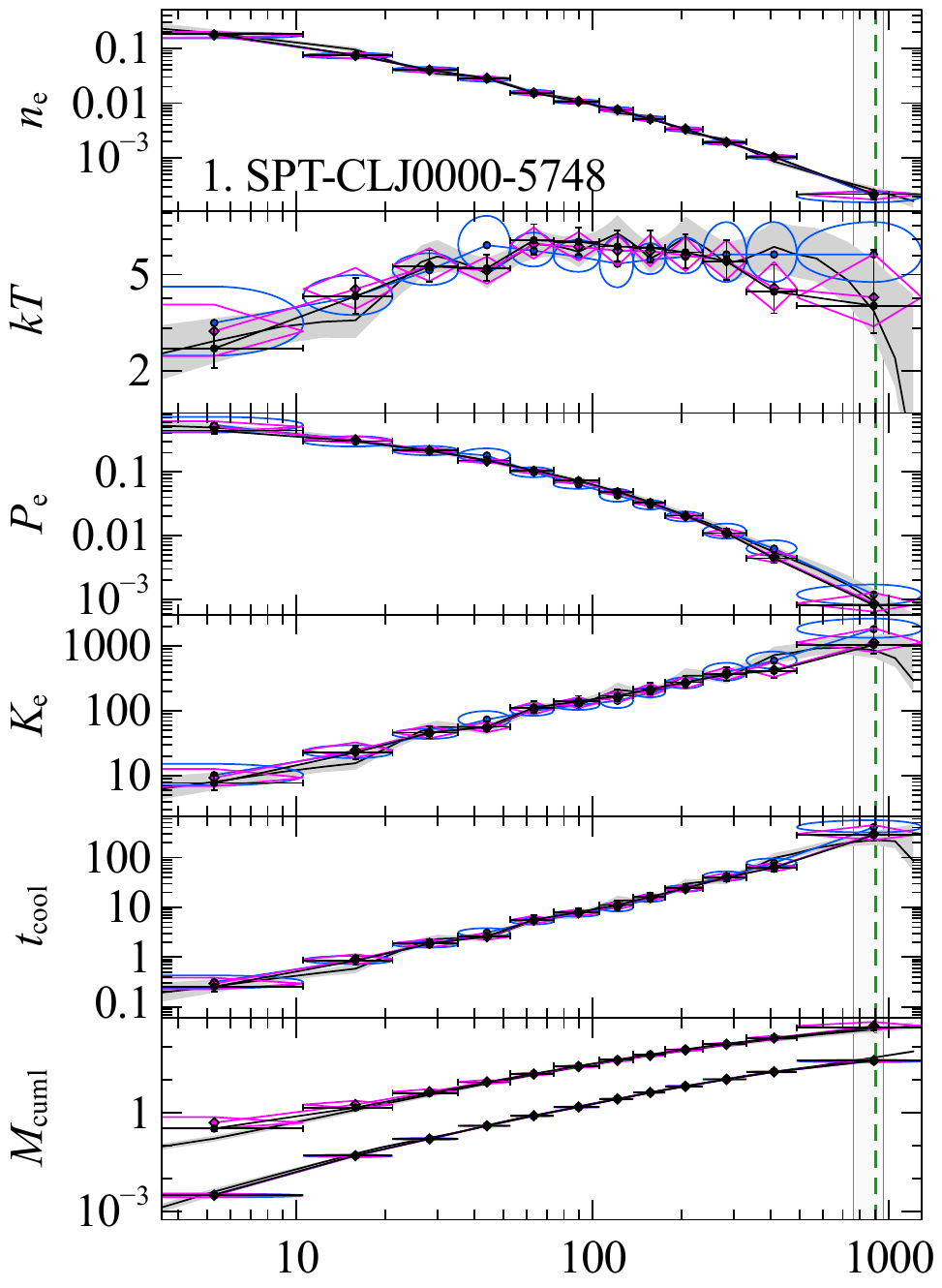}
  \includegraphics[width=0.3\textwidth]{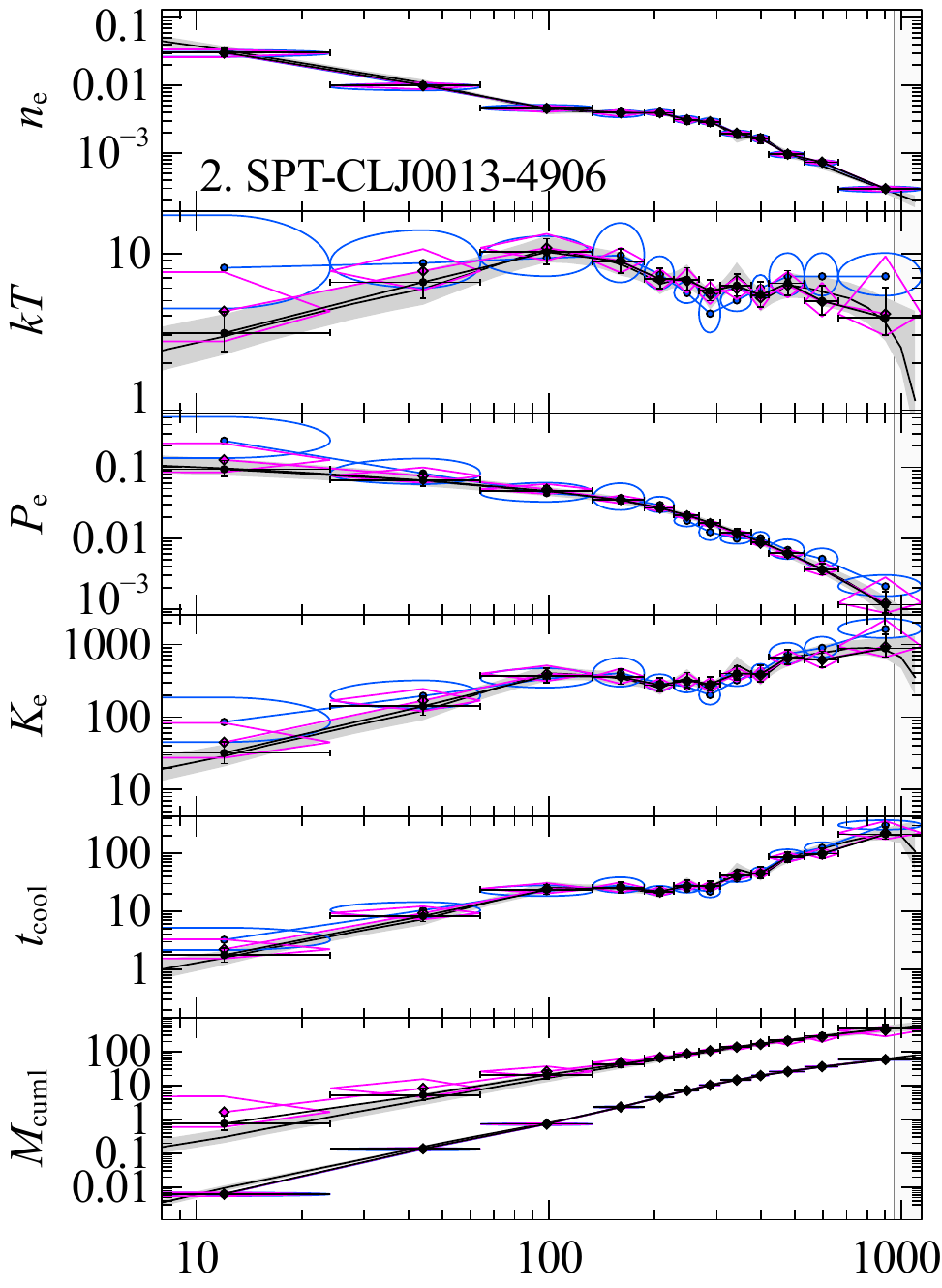}
  \includegraphics[width=0.3\textwidth]{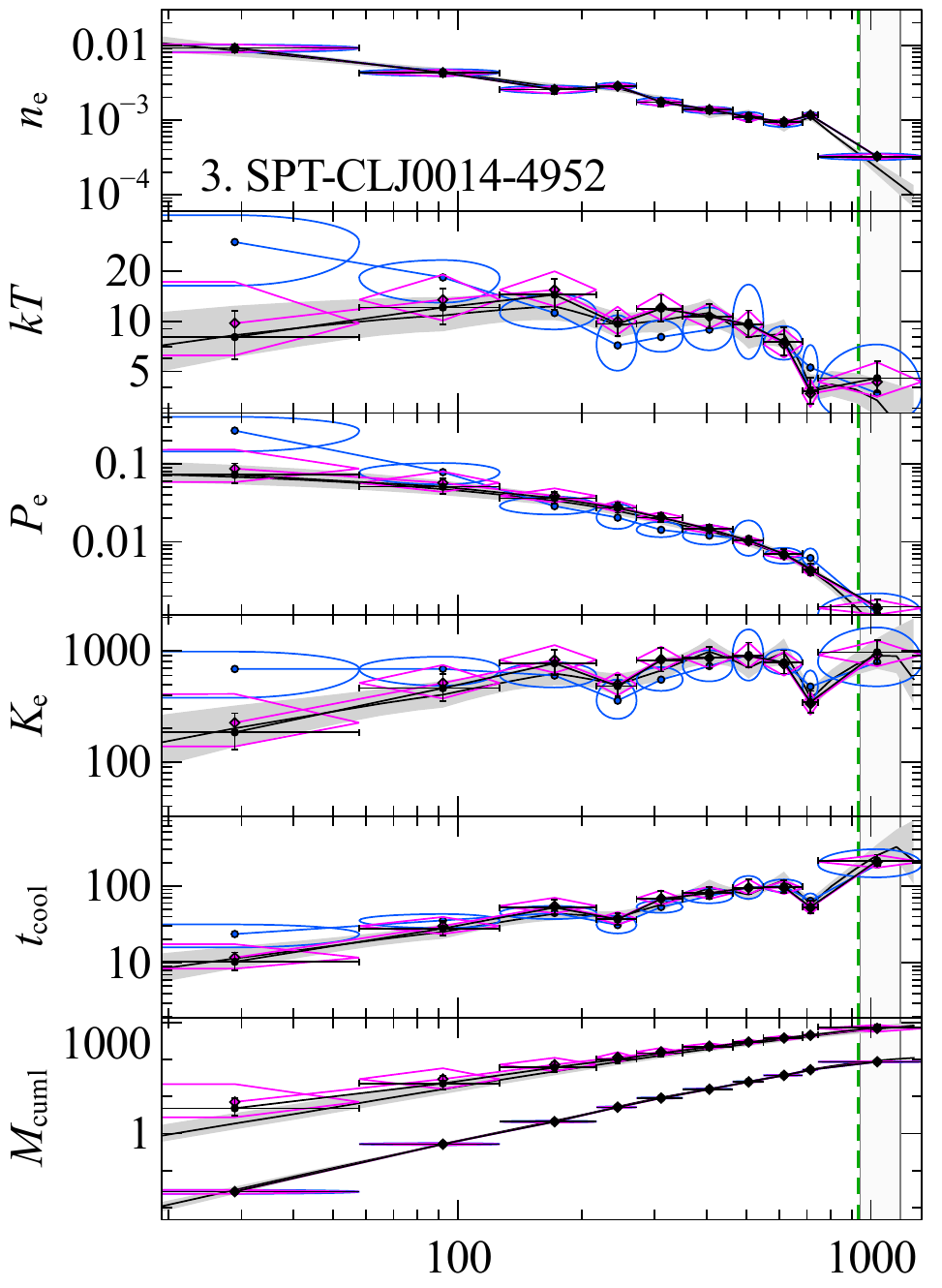}\\
  \includegraphics[width=0.3\textwidth]{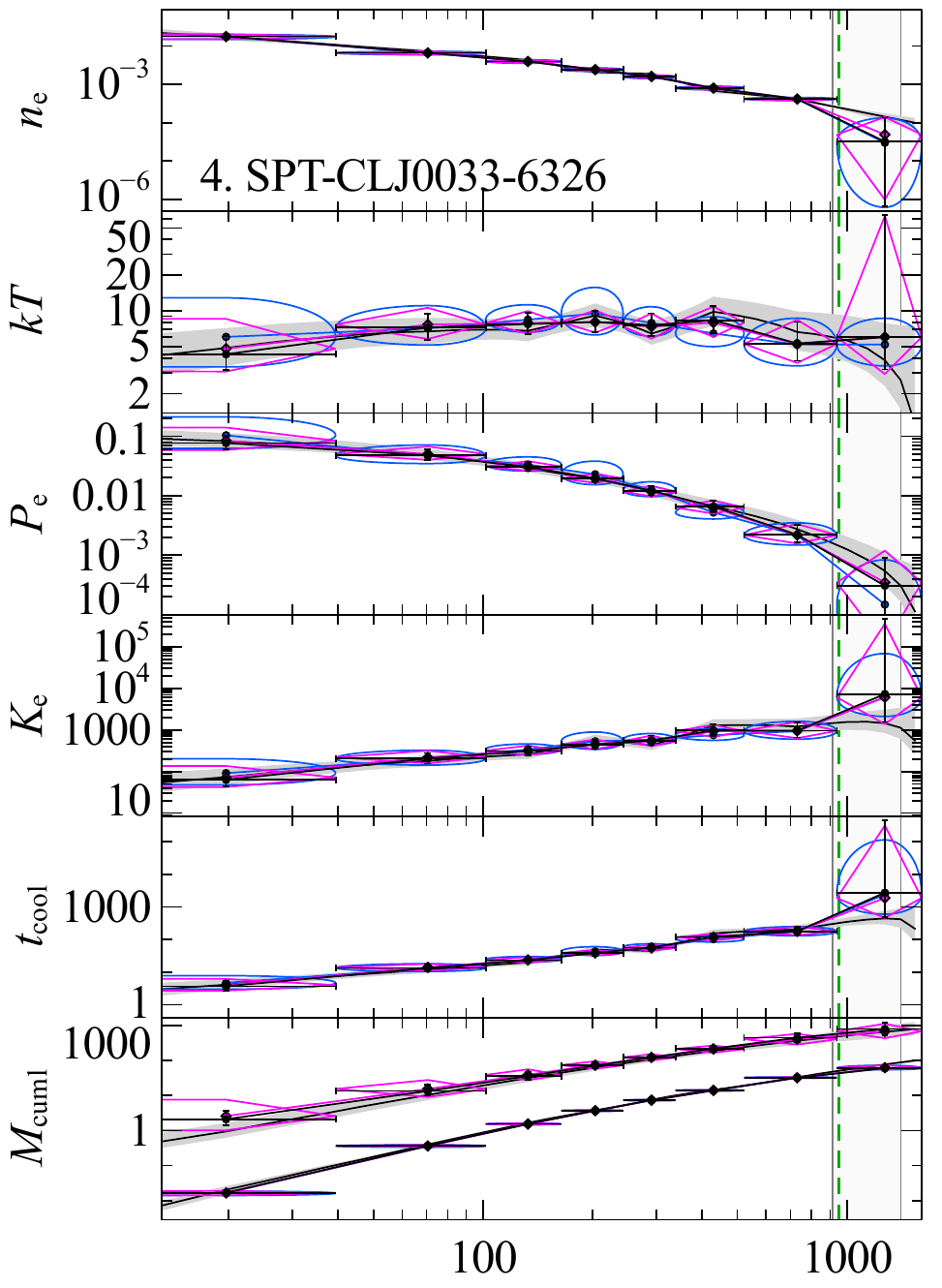}
  \includegraphics[width=0.3\textwidth]{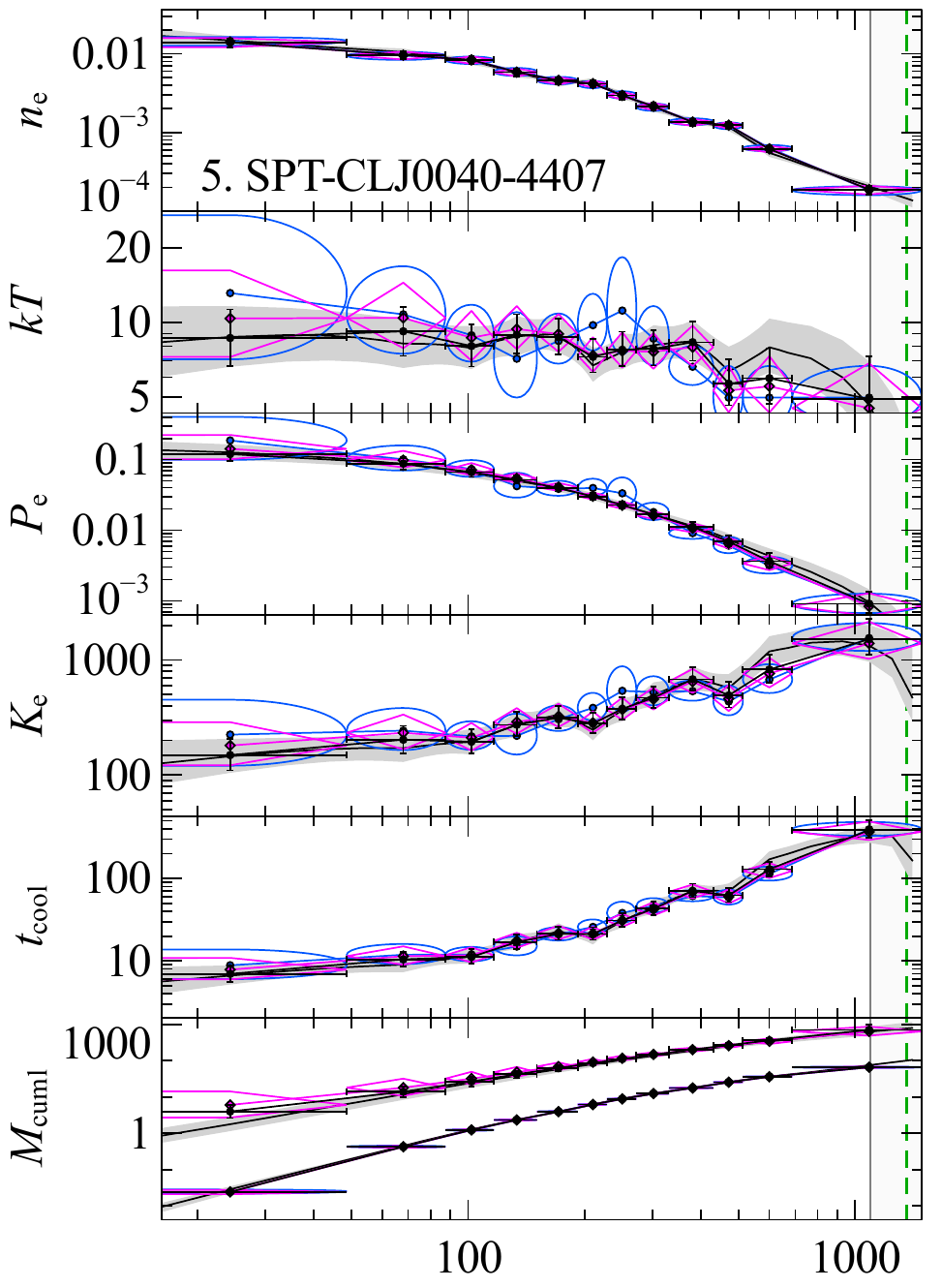}
  \includegraphics[width=0.3\textwidth]{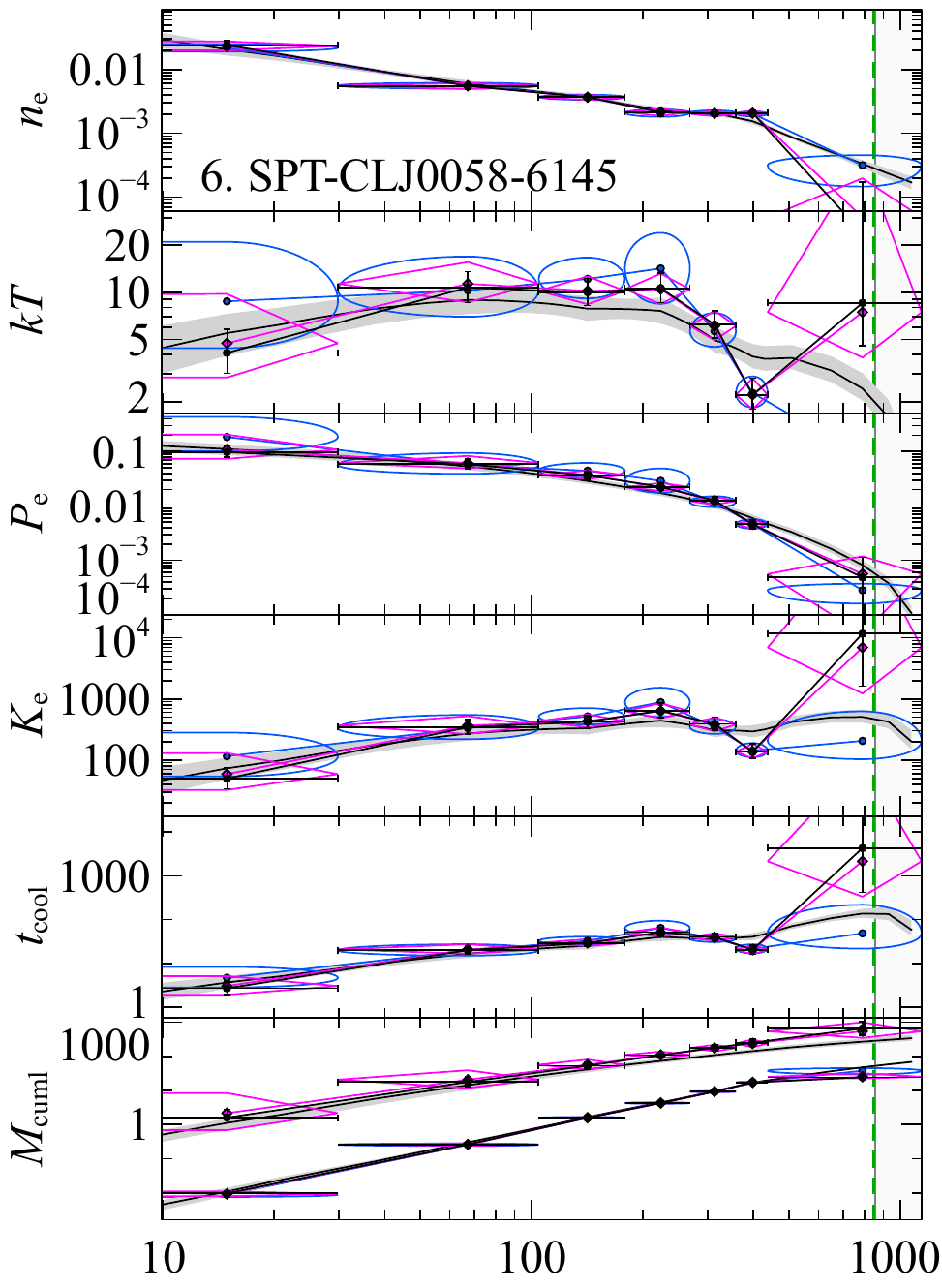}\\
  \includegraphics[width=0.3\textwidth]{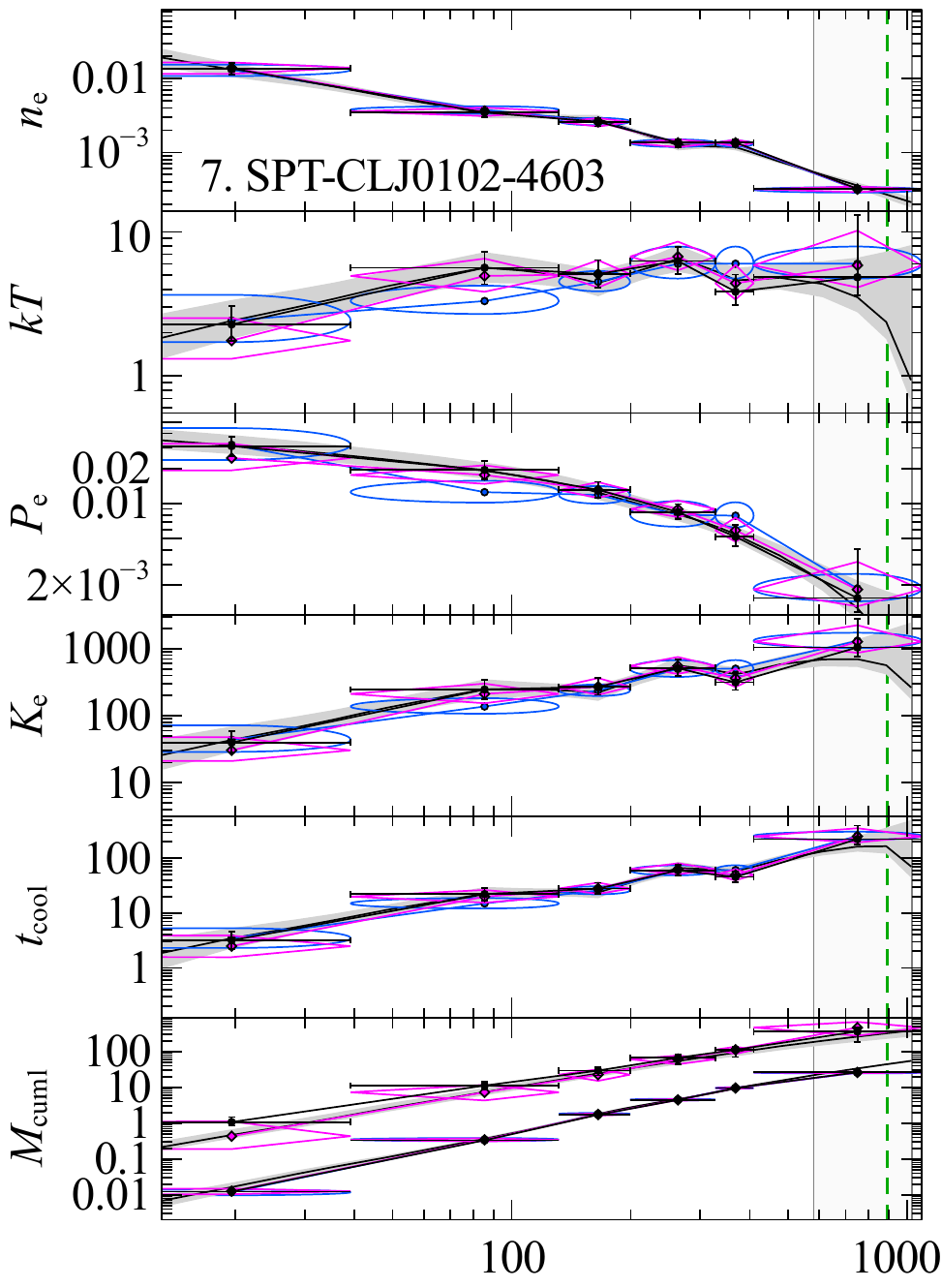}
  \includegraphics[width=0.3\textwidth]{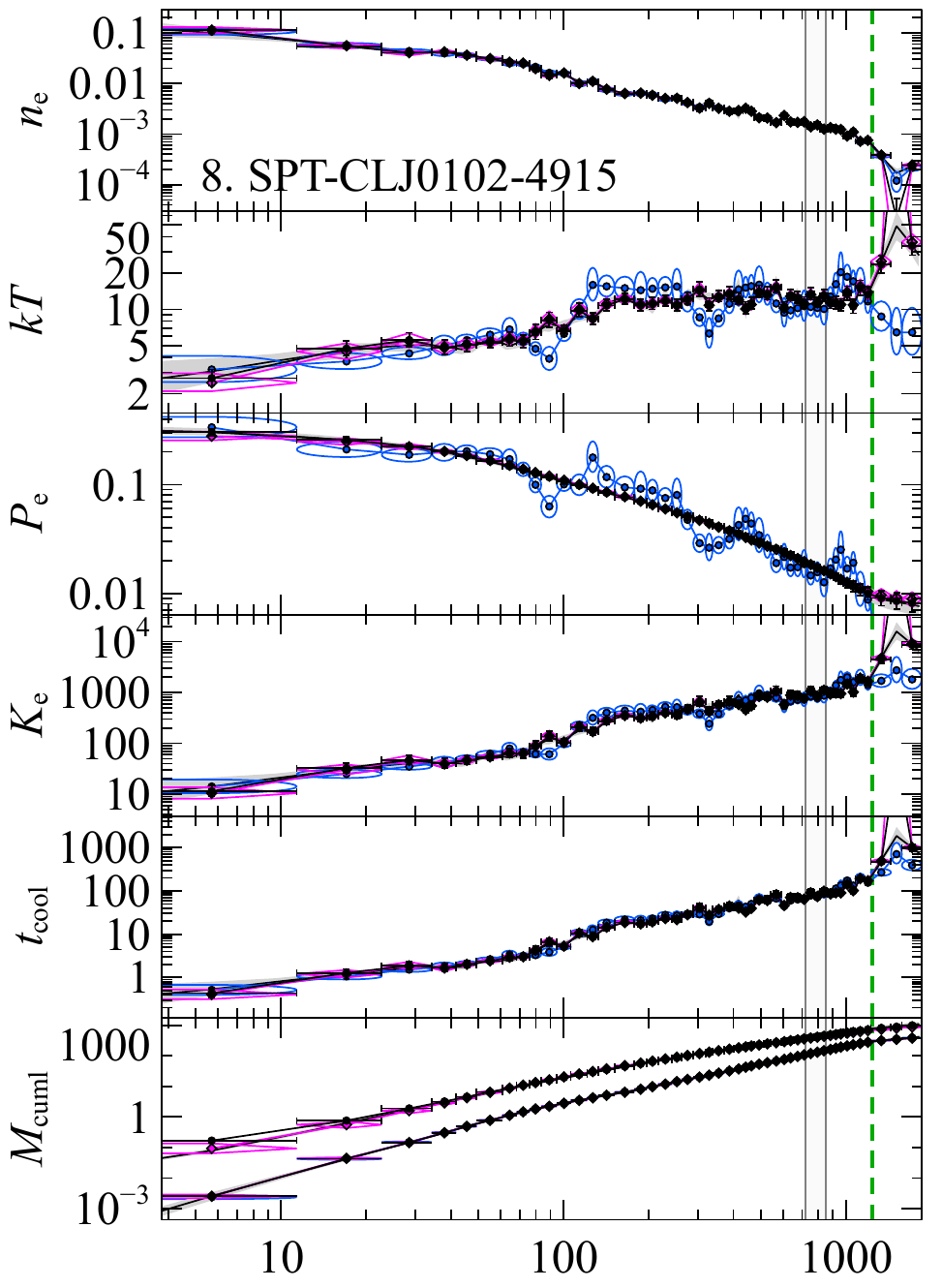}
  \includegraphics[width=0.3\textwidth]{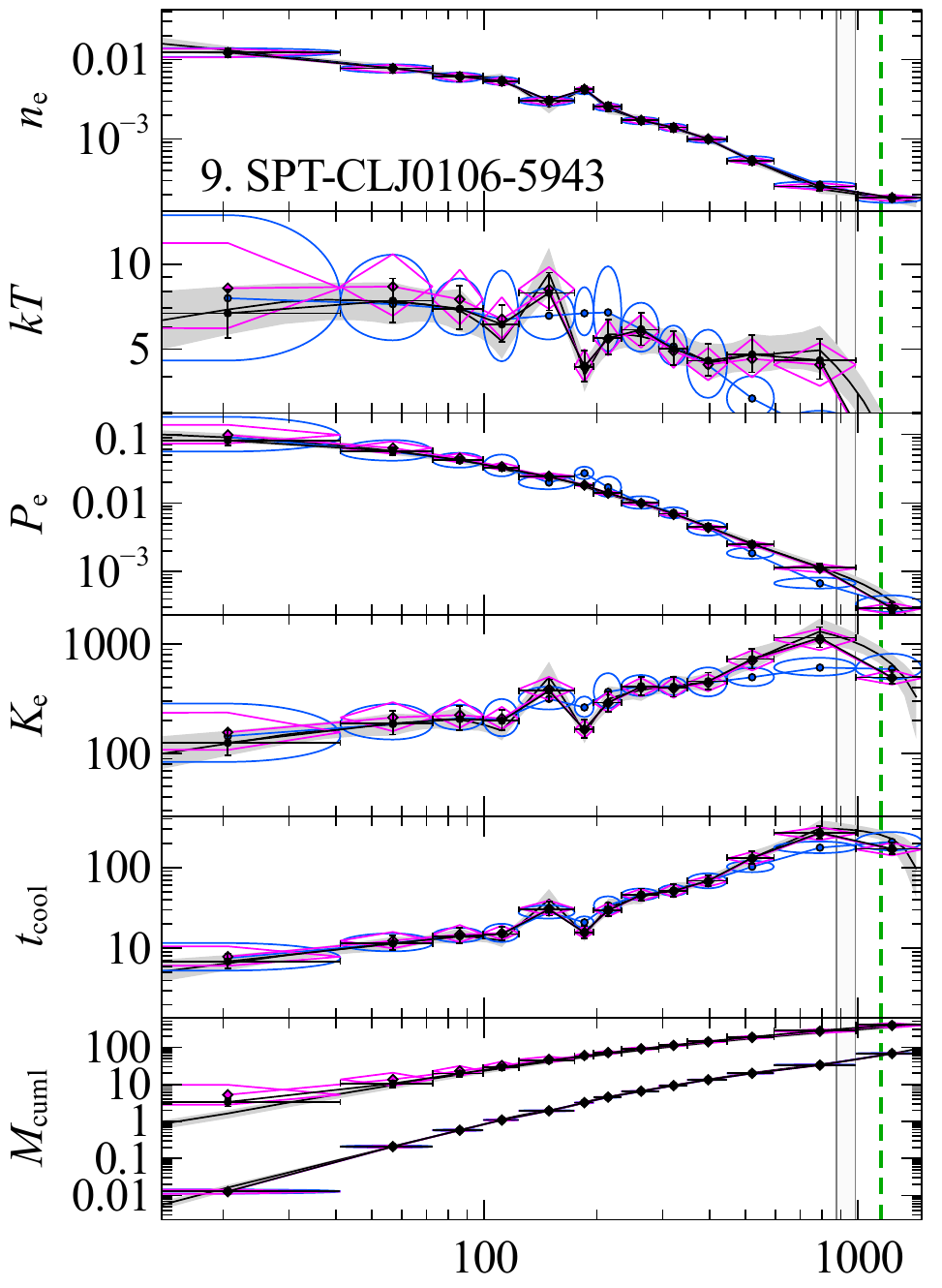}\\
\caption{
Profiles for individual clusters. Plotted for each system are the electron density (cm$^{-3}$), temperature (keV), electron pressure (keV cm$^{-3}$), entropy (keV cm$^{2}$), cooling time (Gyr) and cumulative gas and total masses ($10^{12} \Msun$), plotted against radius in kpc.
Similarly to Fig.~\ref{fig:example} are shown the results for the NFW and GNFW mass models in bins, the results not assuming hydrostatic equilibrium and an interpolated density profile assuming an NFW model.
The vertical dashed line is the SPT value of $R_{500}$, while the bounded radial region is the binned-hydrostatic range of $R_{500}$.
}
\label{fig:individual}
\end{figure*}
\begin{figure*}
  \centering
  \includegraphics[width=0.3\textwidth]{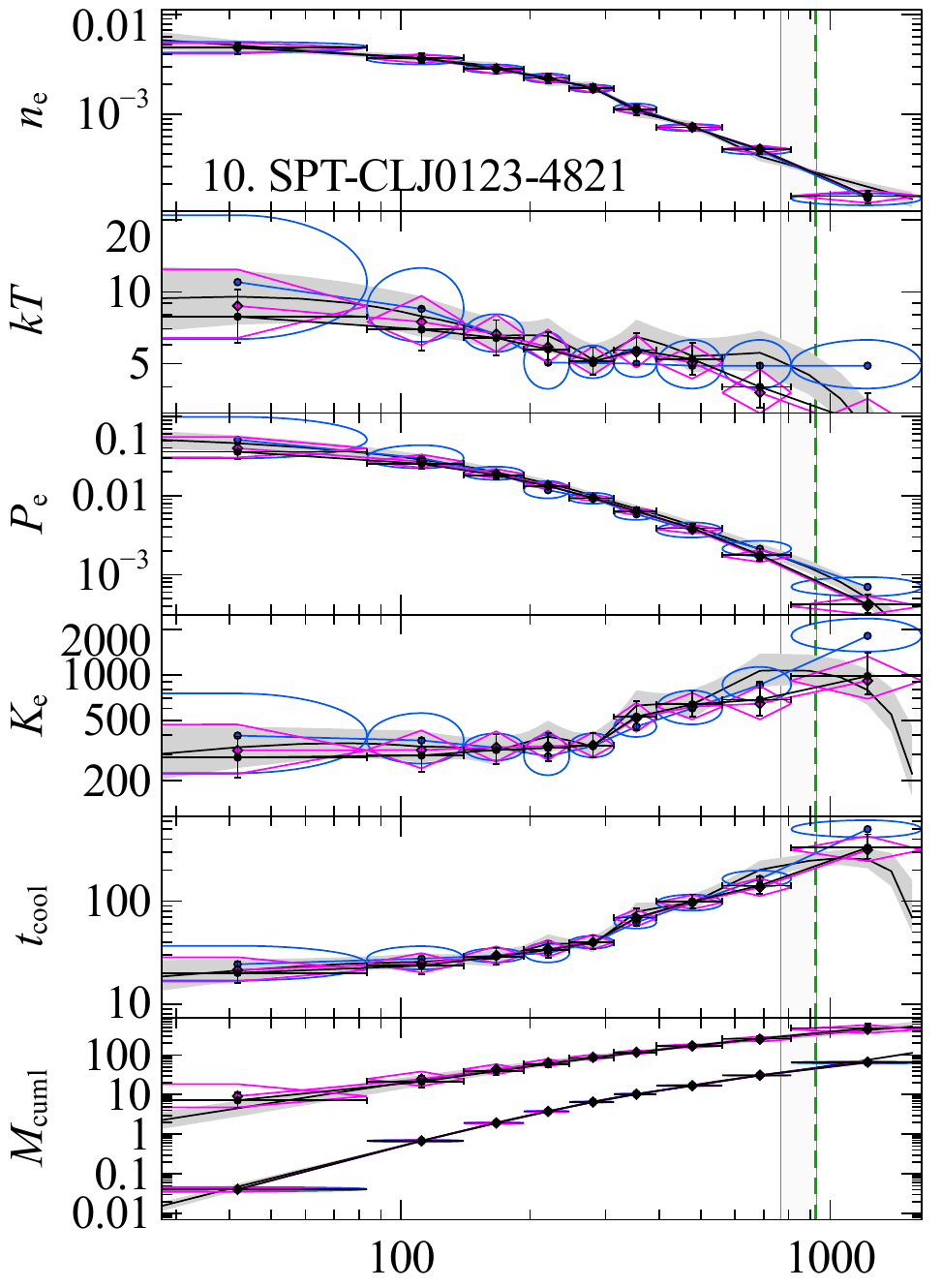}
  \includegraphics[width=0.3\textwidth]{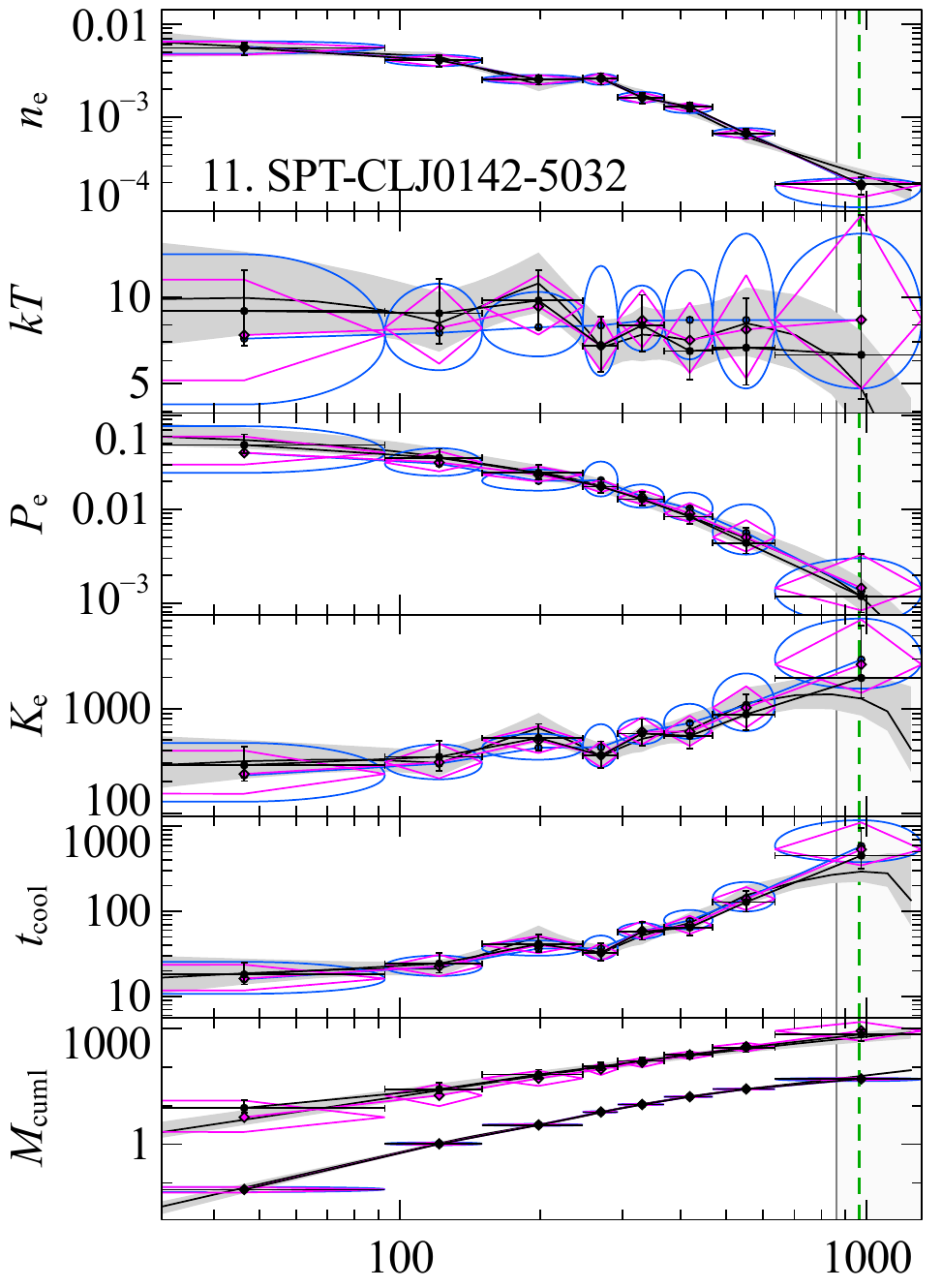}
  \includegraphics[width=0.3\textwidth]{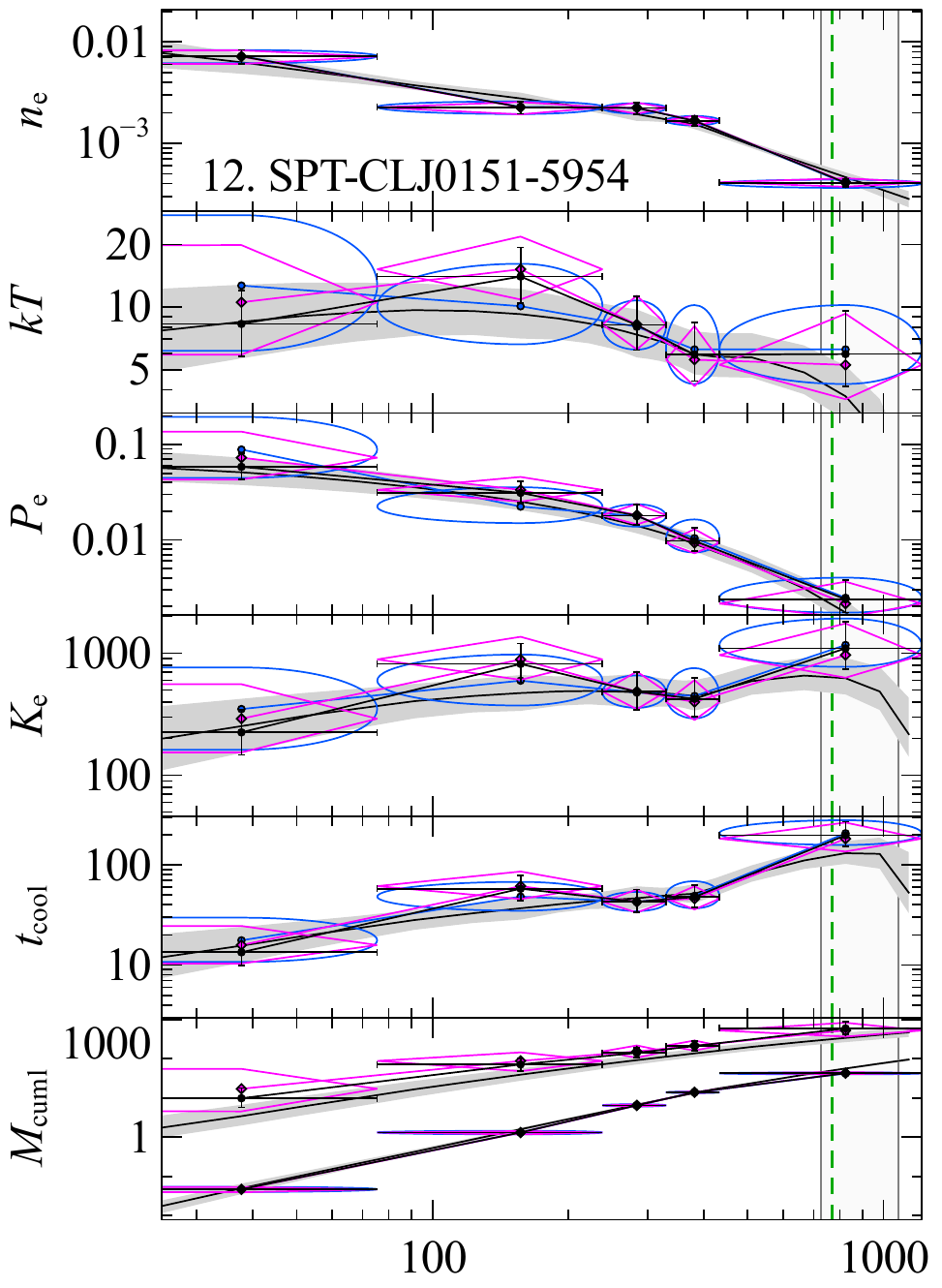}\\
  \includegraphics[width=0.3\textwidth]{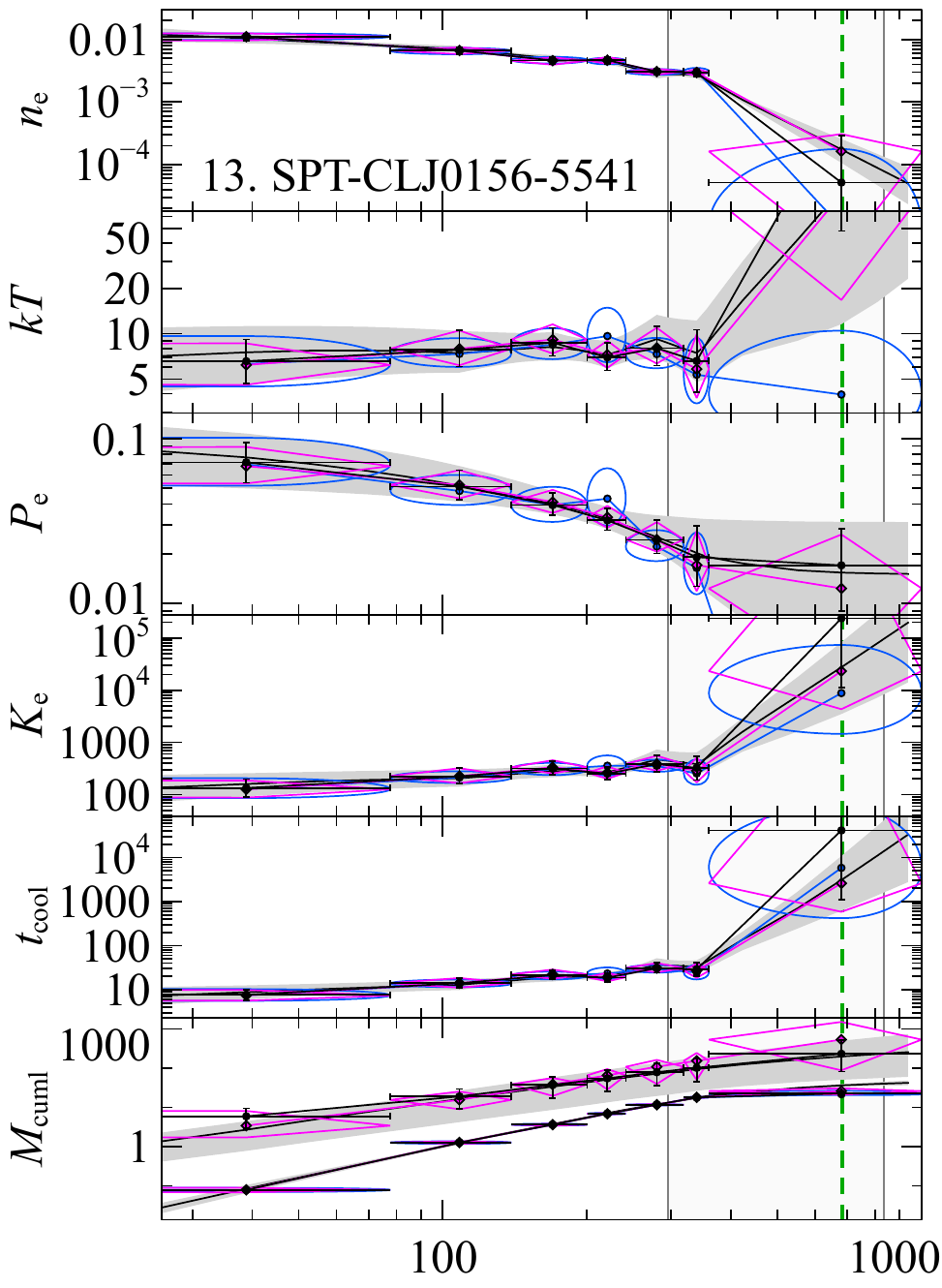}
  \includegraphics[width=0.3\textwidth]{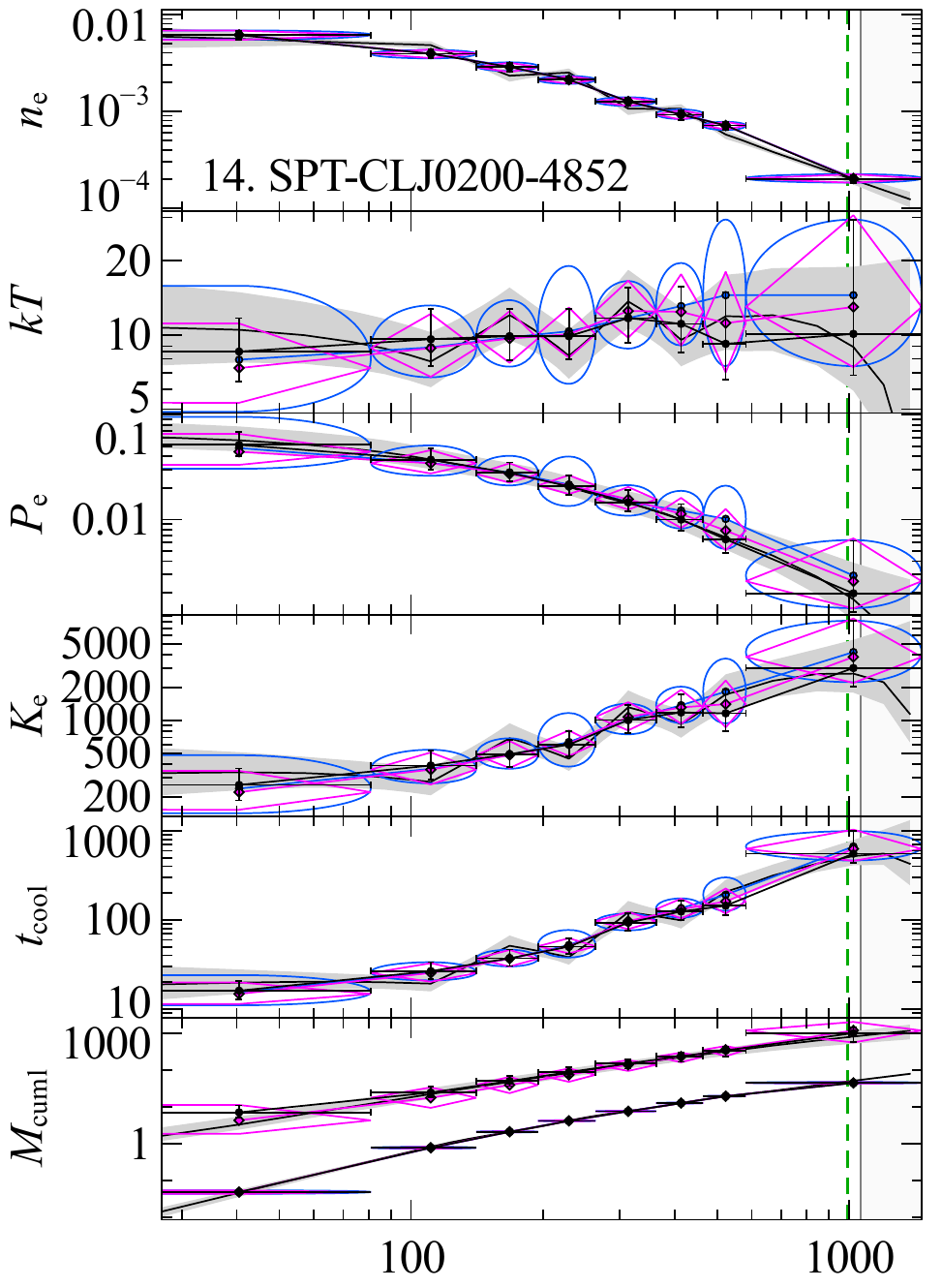}
  \includegraphics[width=0.3\textwidth]{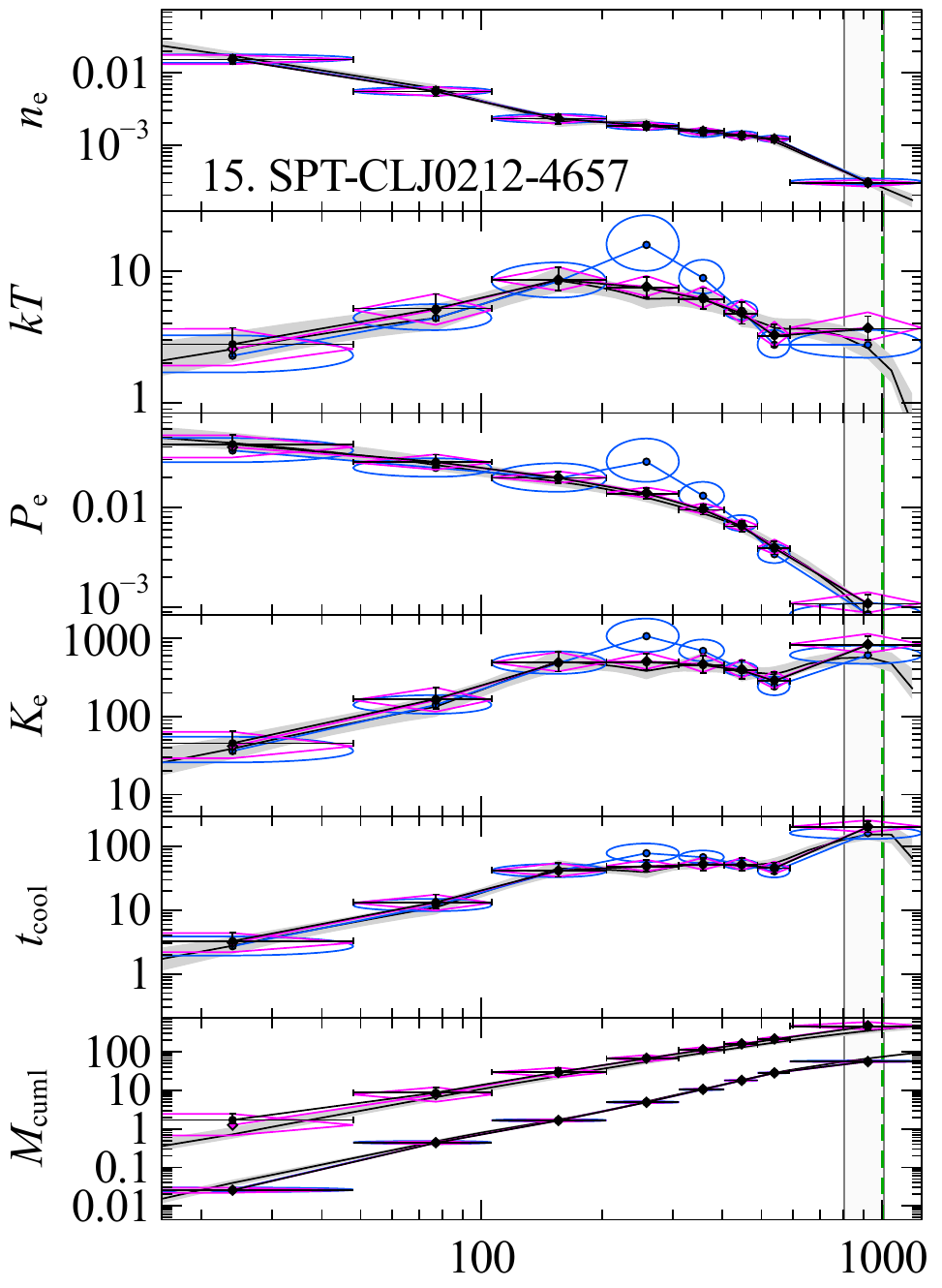}\\
  \includegraphics[width=0.3\textwidth]{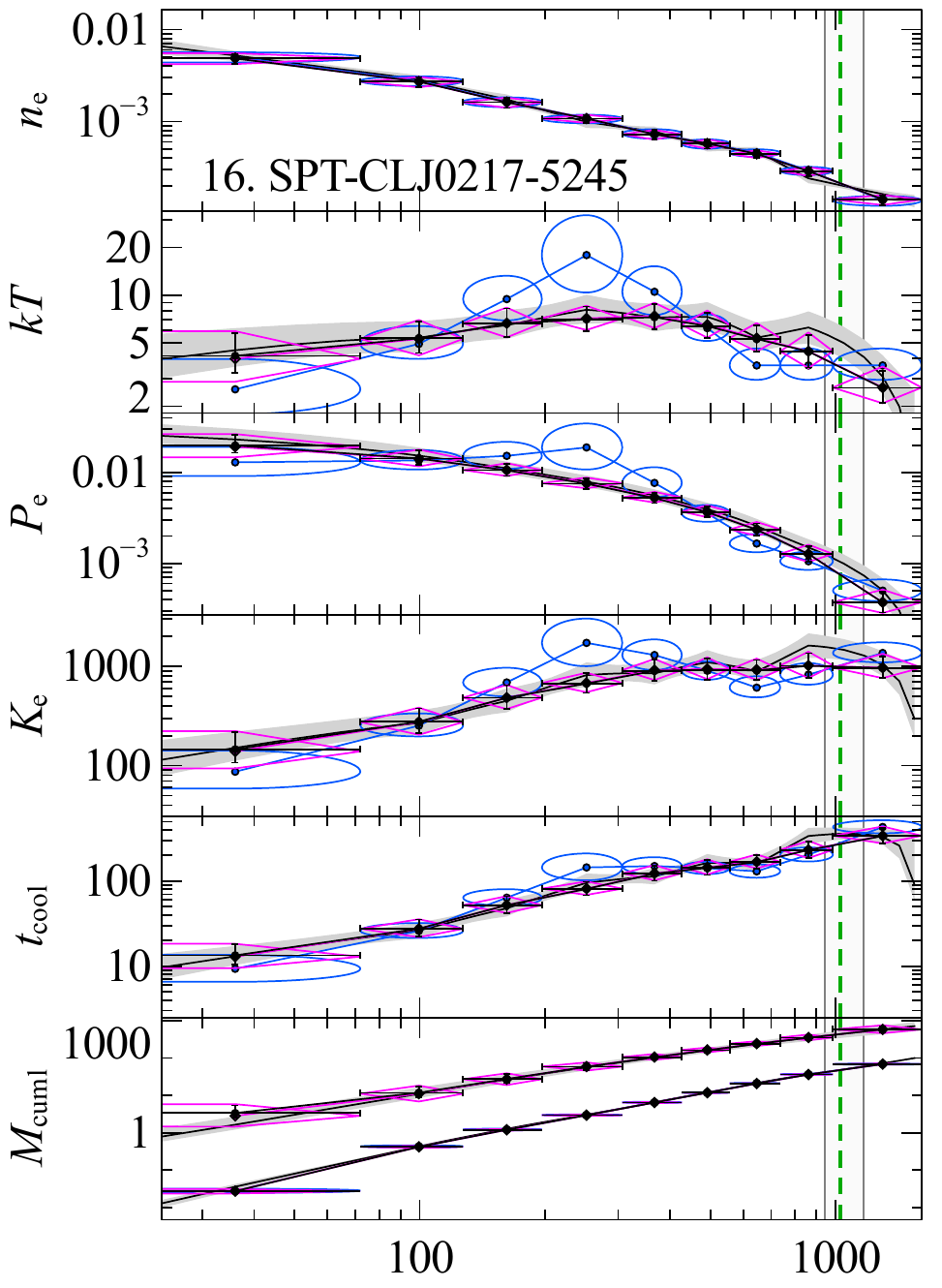}
  \includegraphics[width=0.3\textwidth]{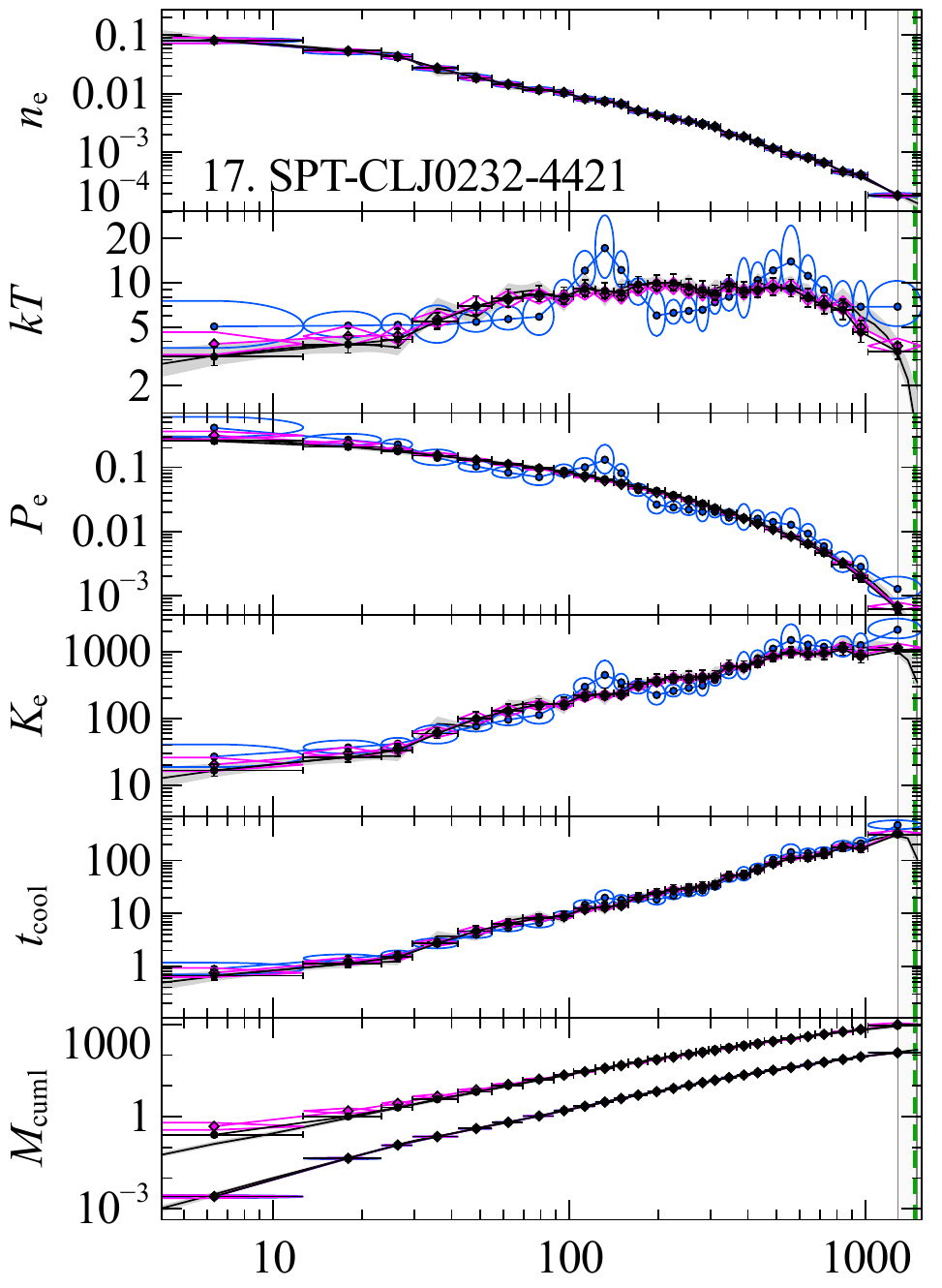}
  \includegraphics[width=0.3\textwidth]{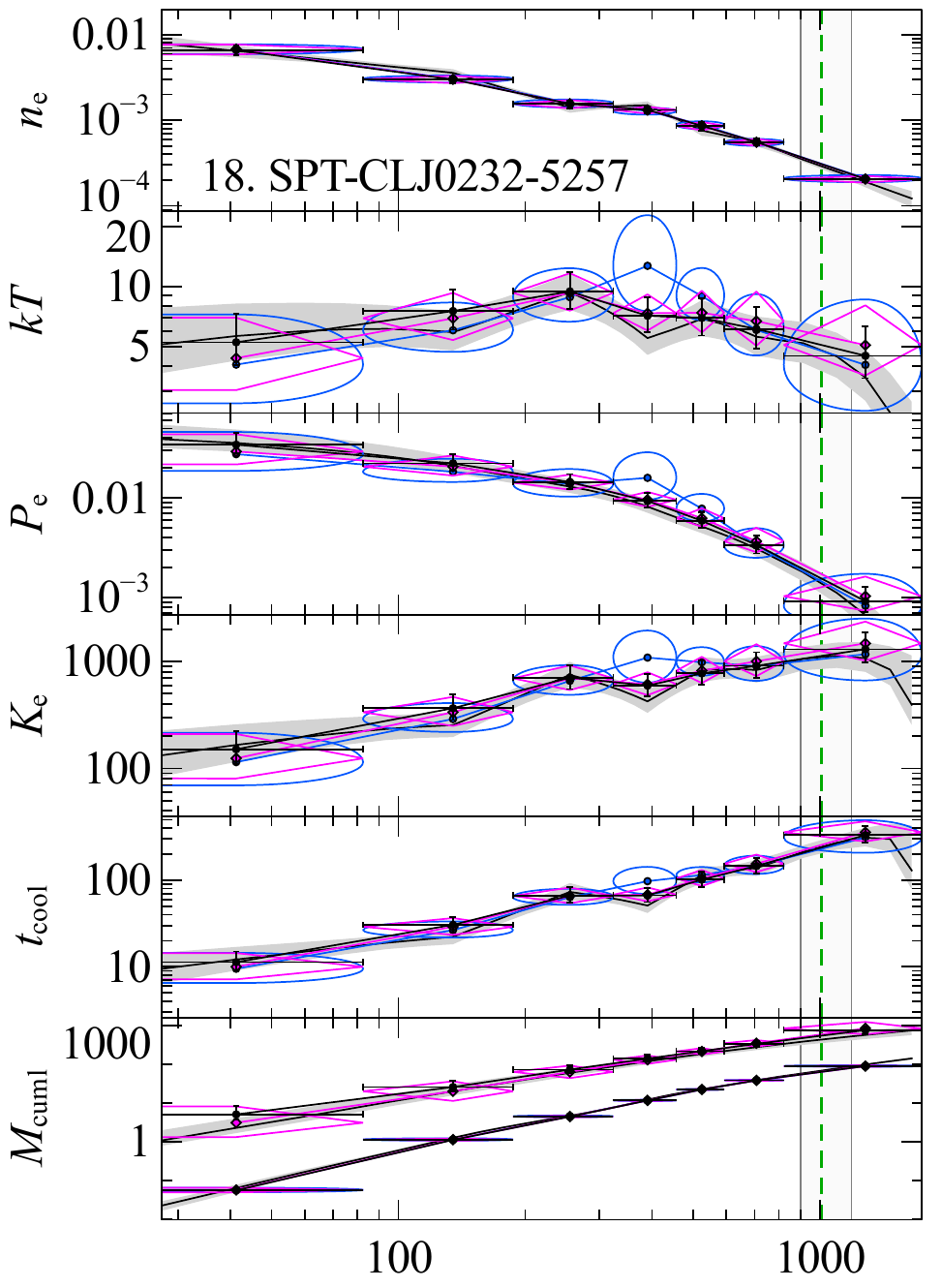}\\
  \contcaption{individual cluster profiles.}
\end{figure*}
\begin{figure*}
  \centering
  \includegraphics[width=0.3\textwidth]{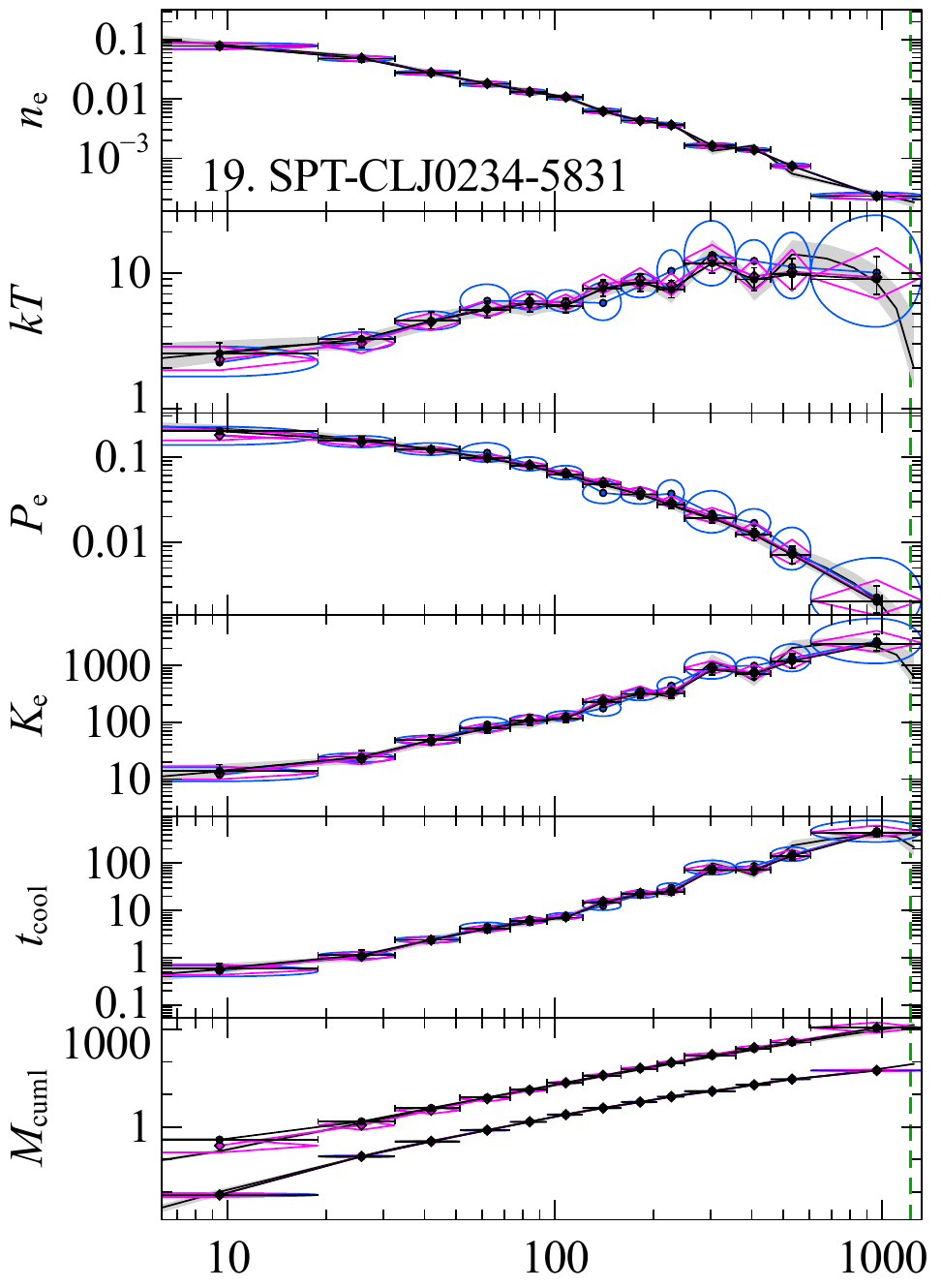}
  \includegraphics[width=0.3\textwidth]{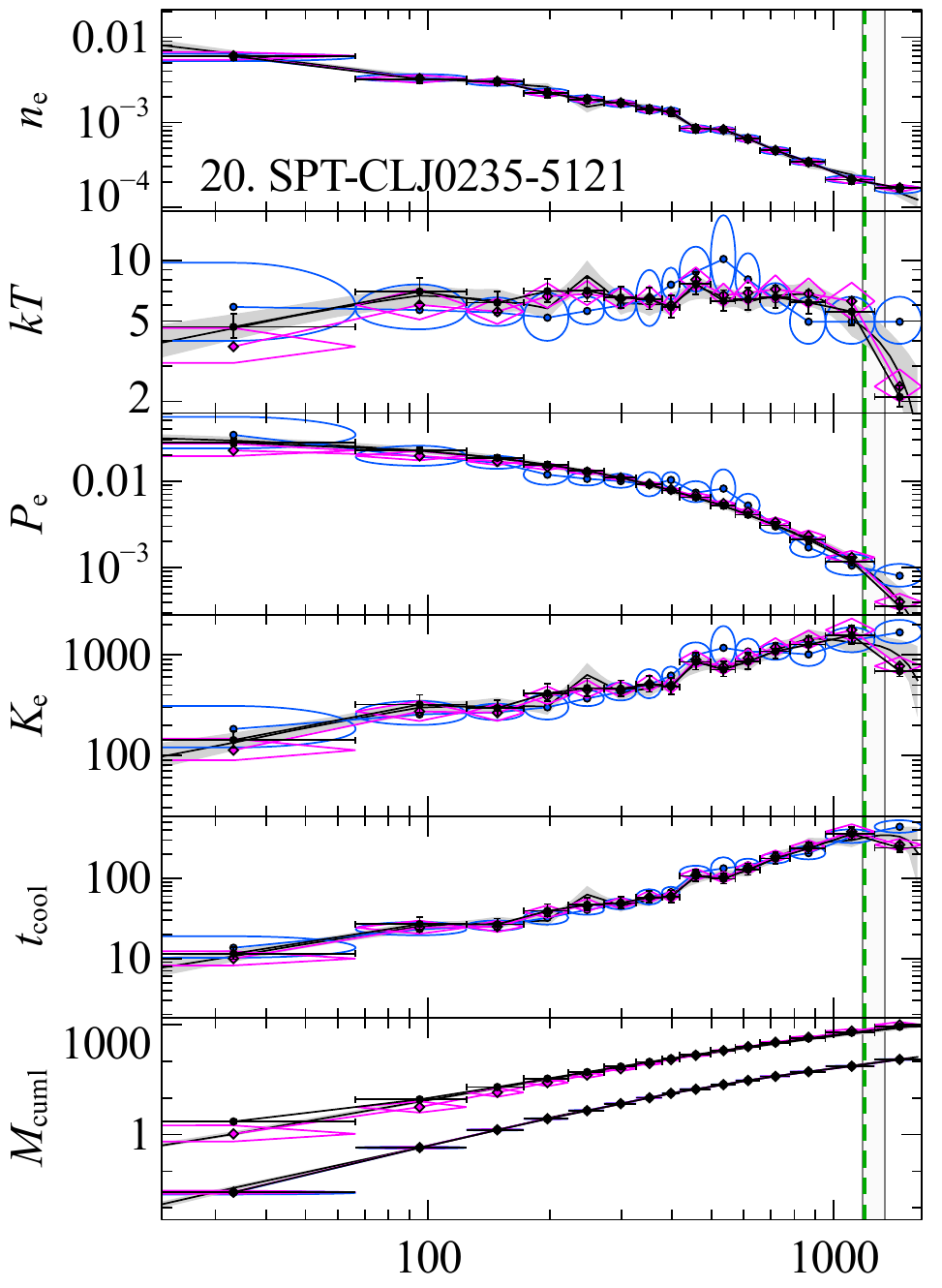}
  \includegraphics[width=0.3\textwidth]{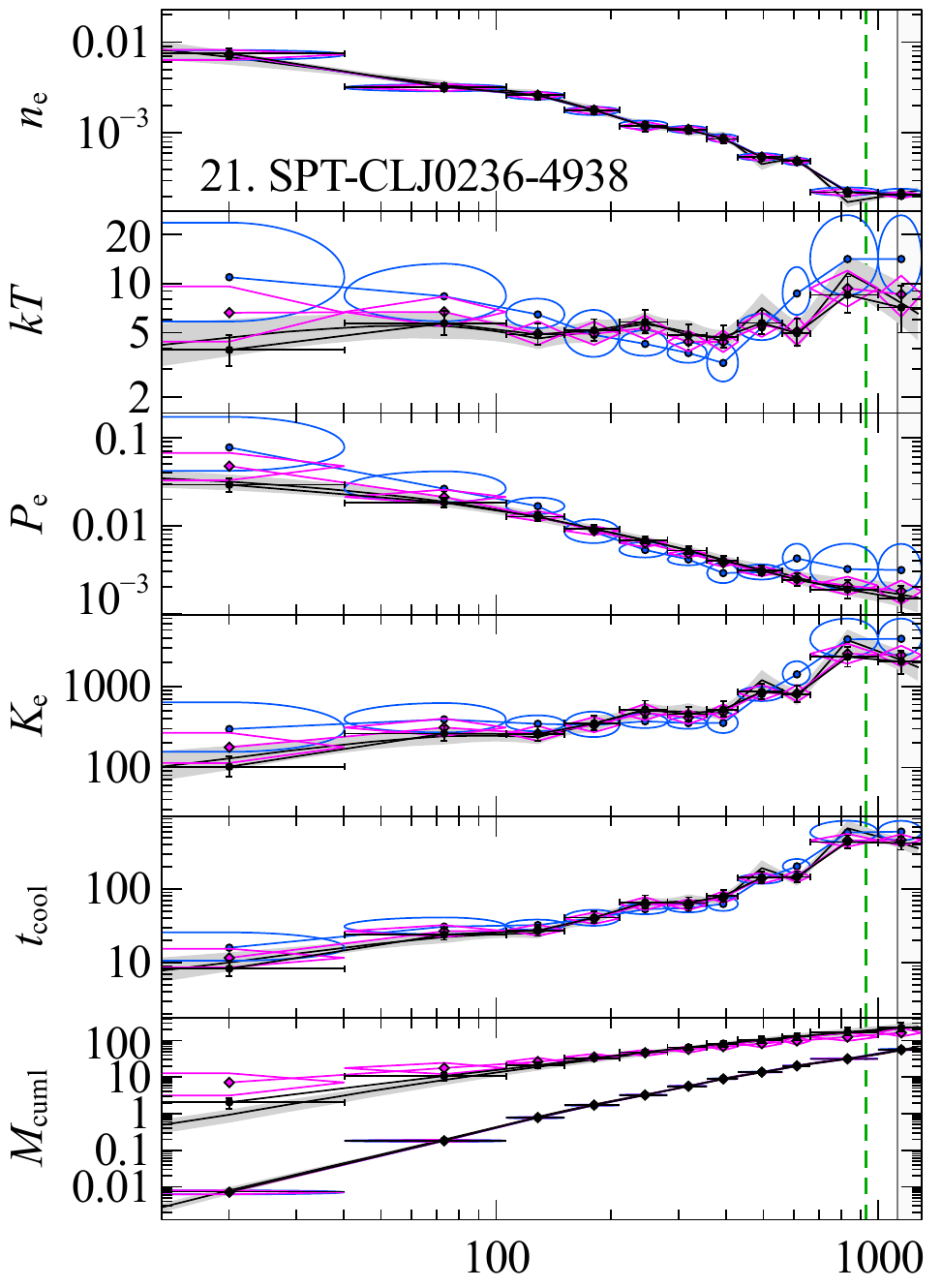}\\
  \includegraphics[width=0.3\textwidth]{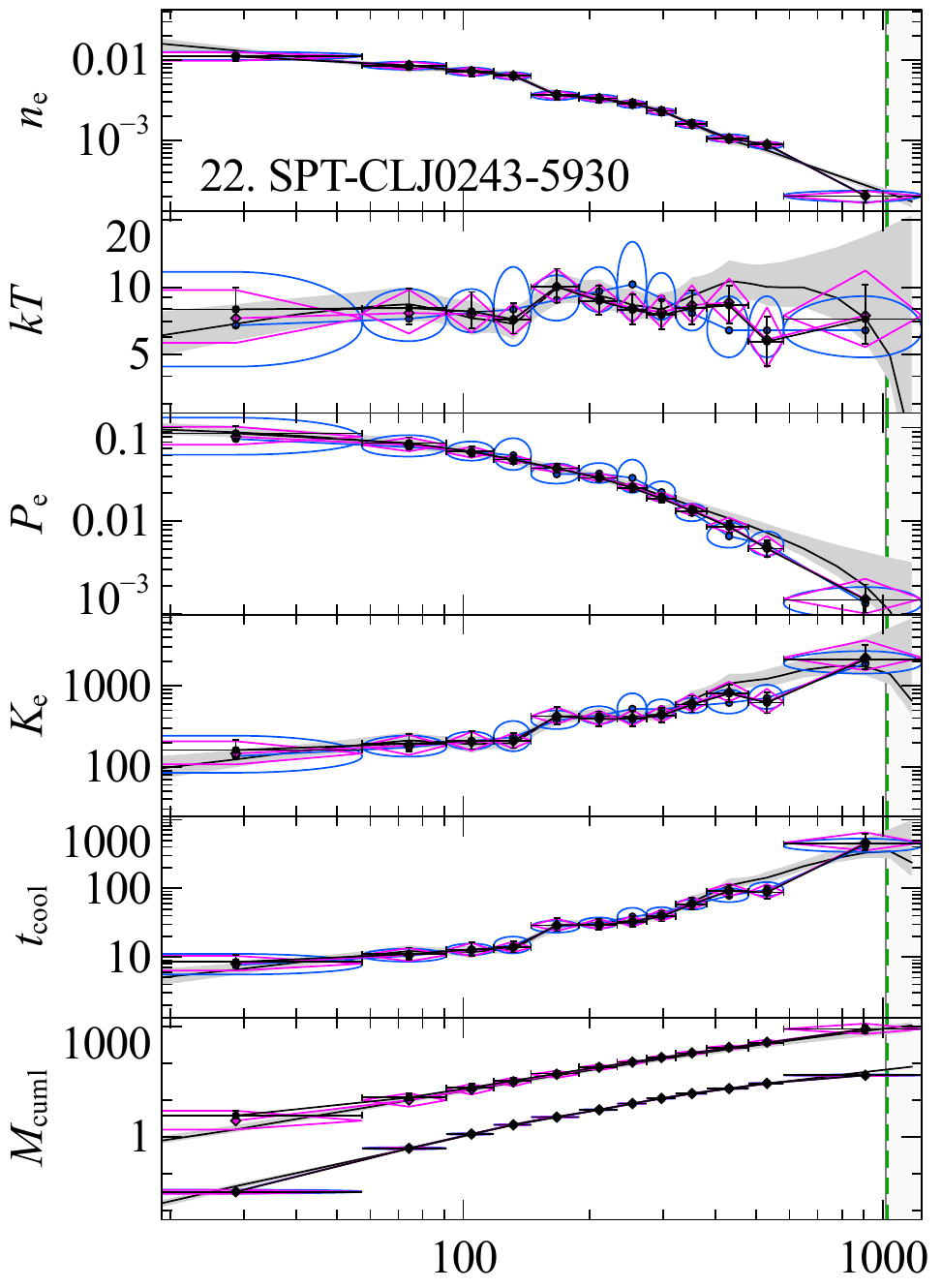}
  \includegraphics[width=0.3\textwidth]{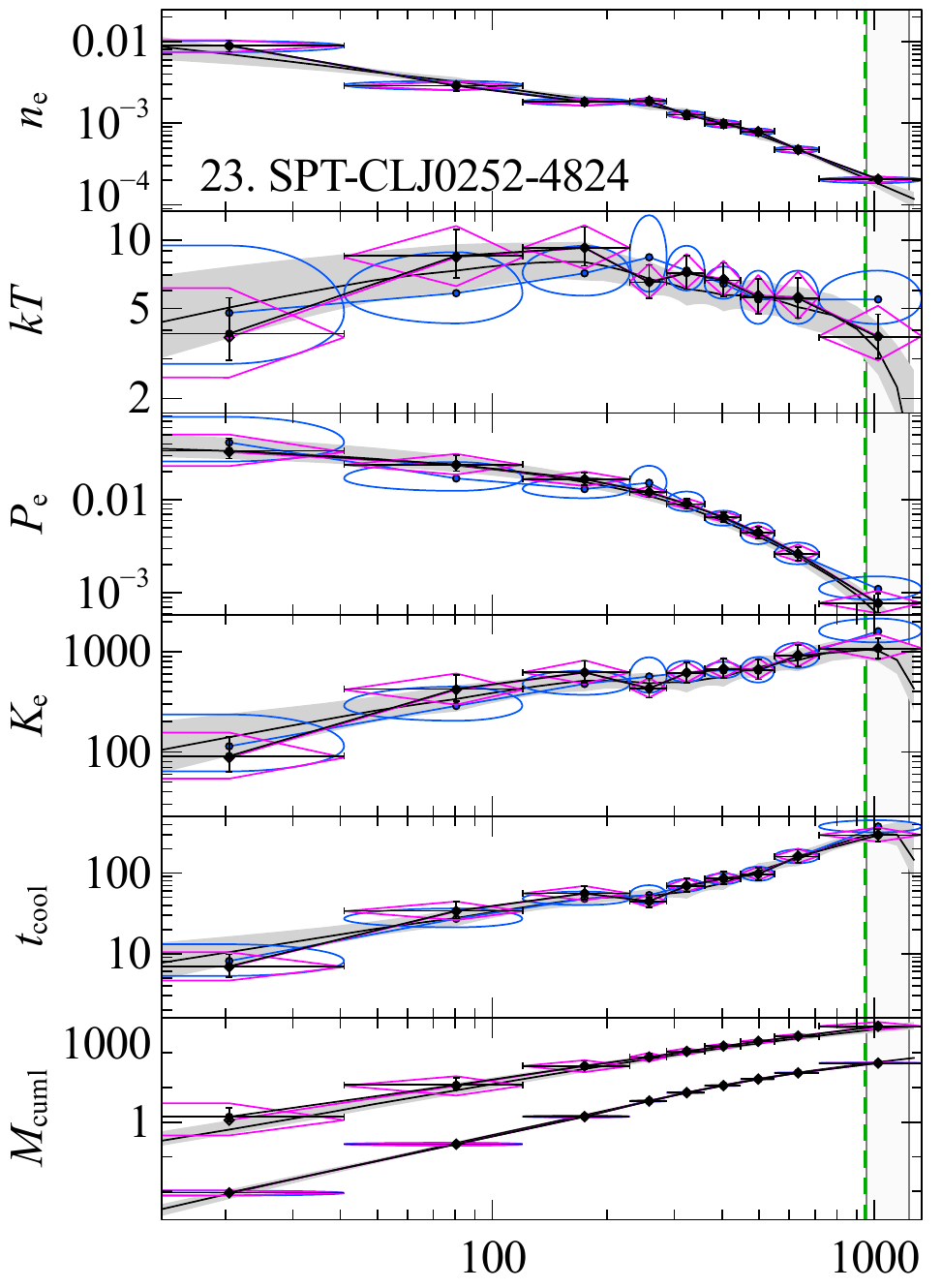}
  \includegraphics[width=0.3\textwidth]{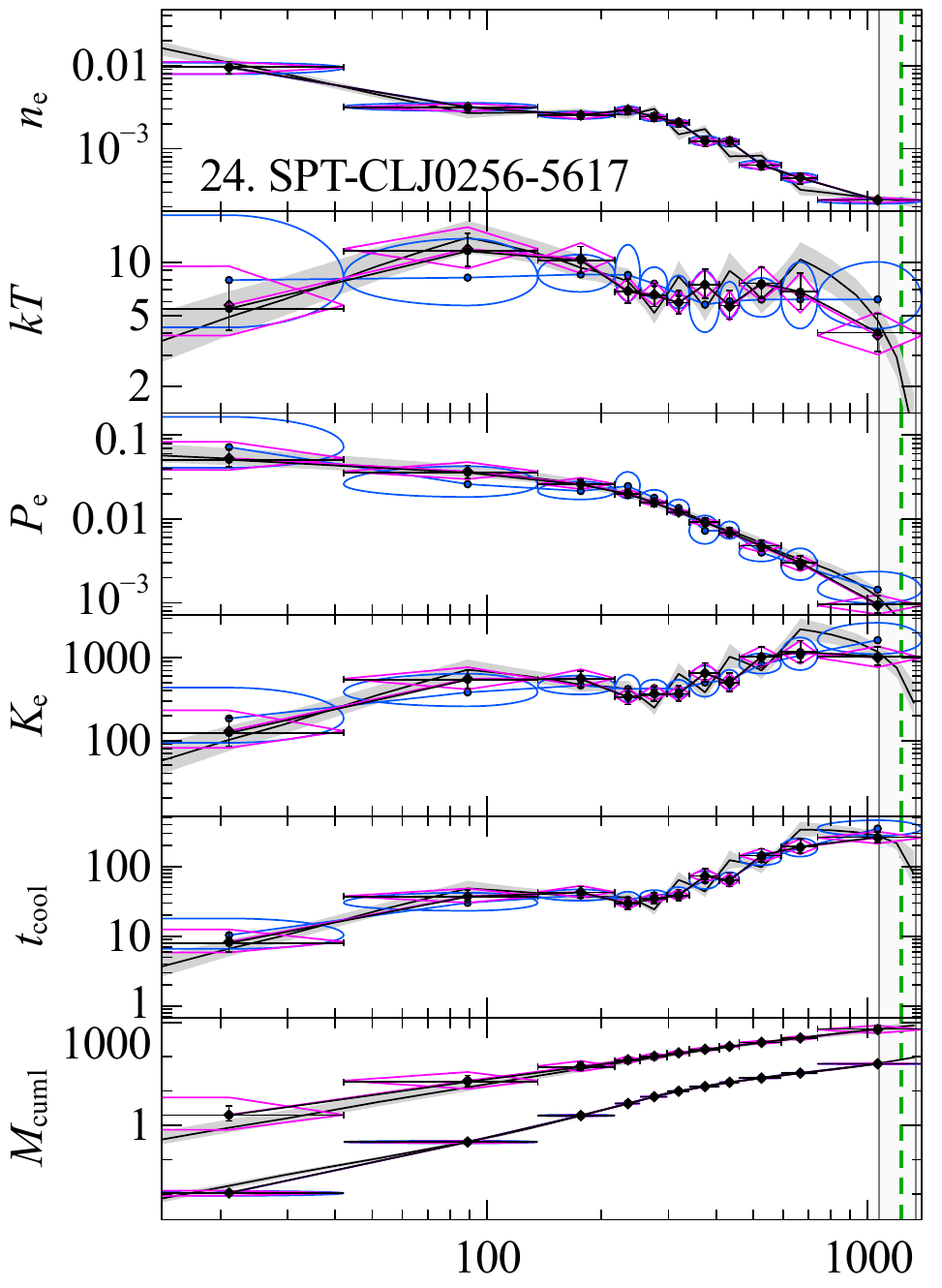}\\
  \includegraphics[width=0.3\textwidth]{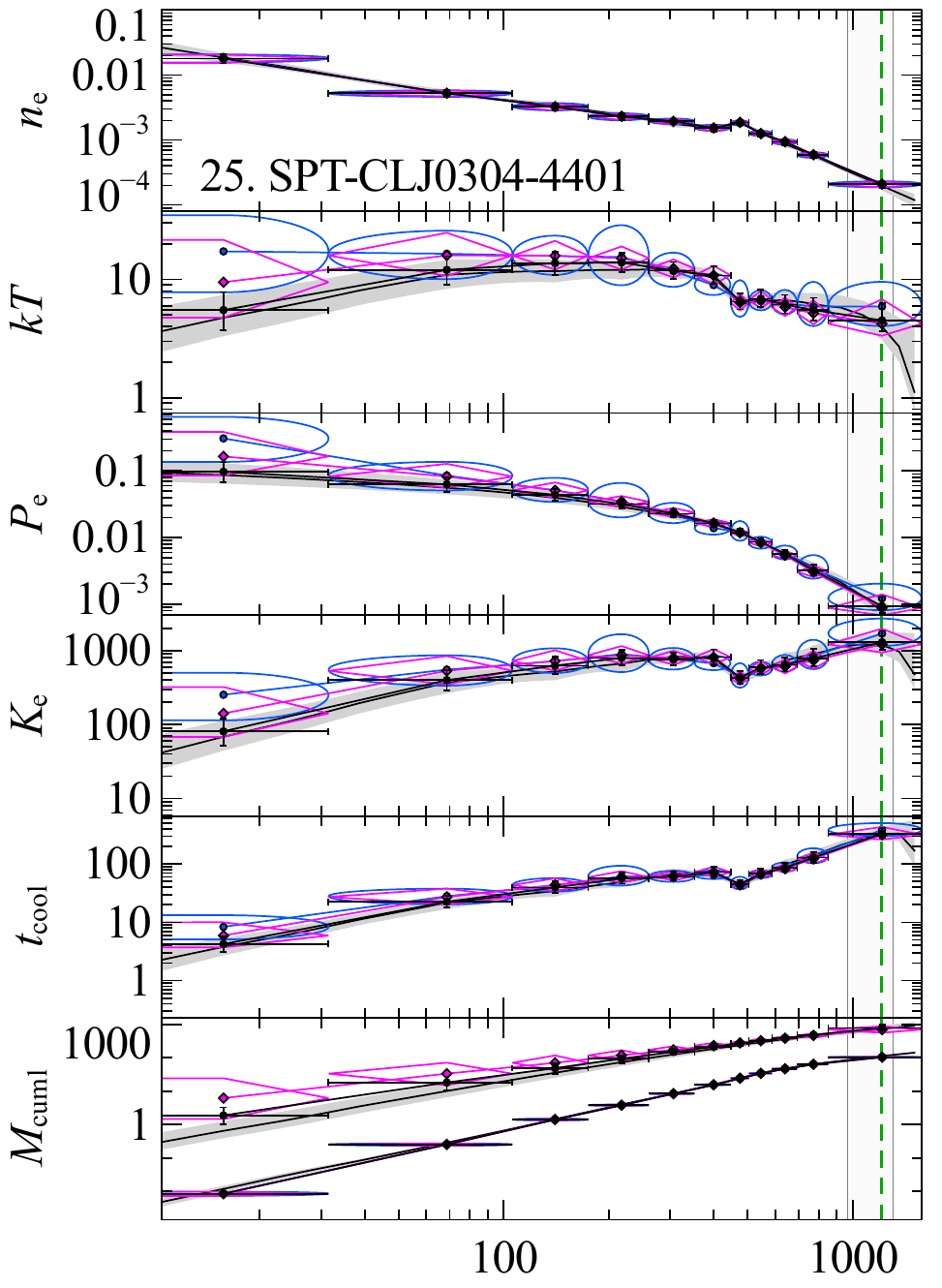}
  \includegraphics[width=0.3\textwidth]{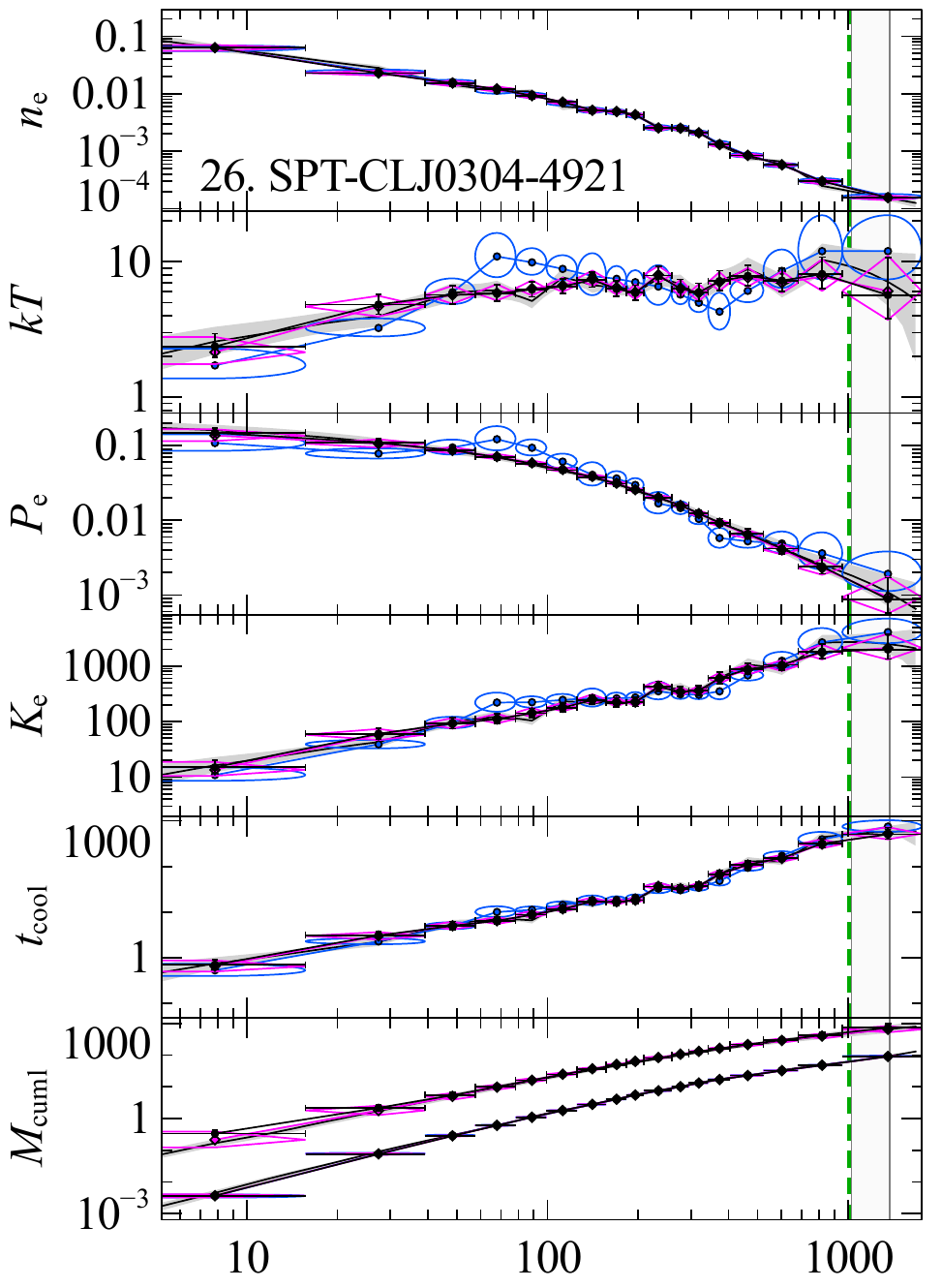}
  \includegraphics[width=0.3\textwidth]{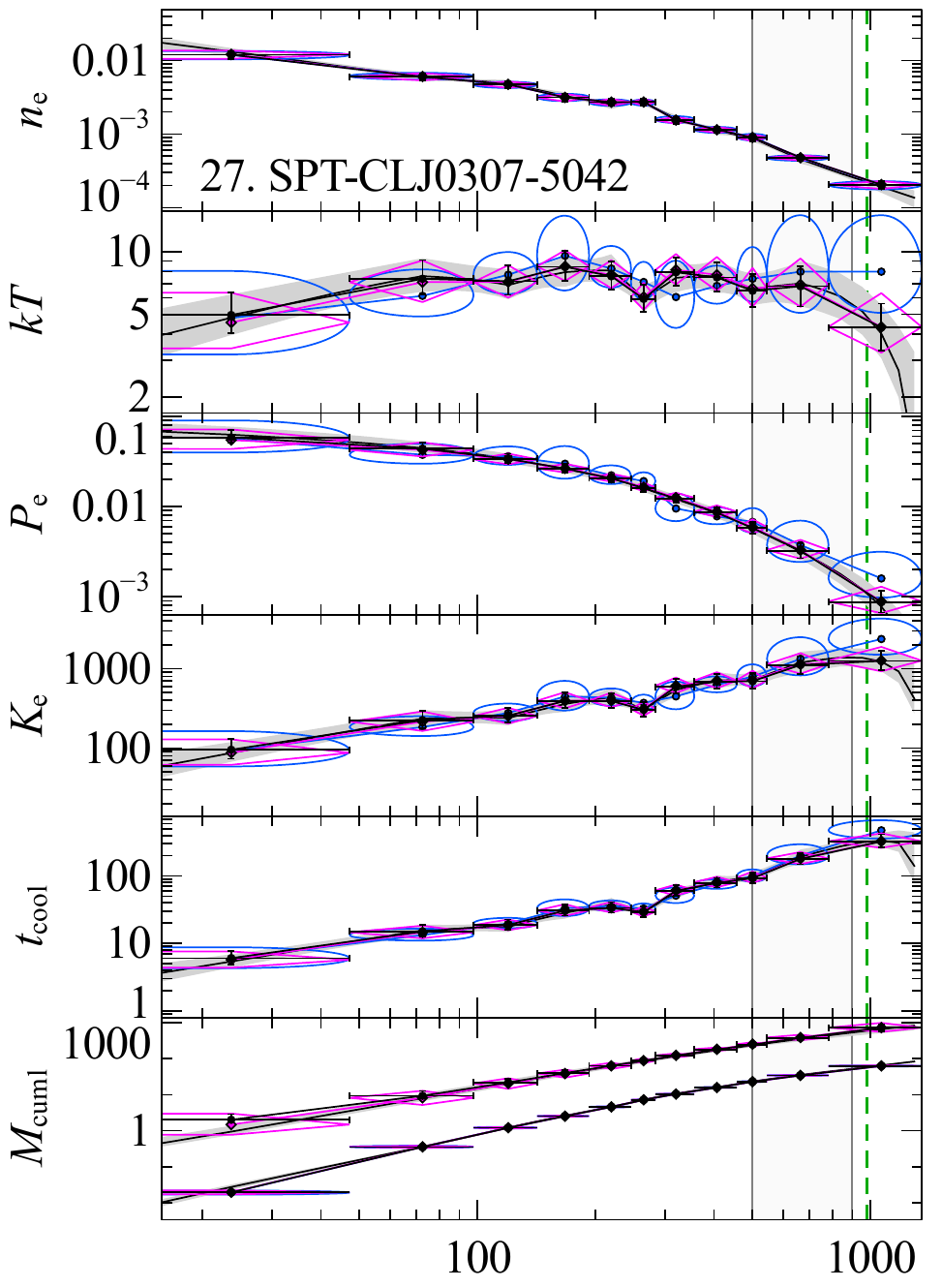}\\
  \contcaption{individual cluster profiles.}
\end{figure*}
\begin{figure*}
  \centering
  \includegraphics[width=0.3\textwidth]{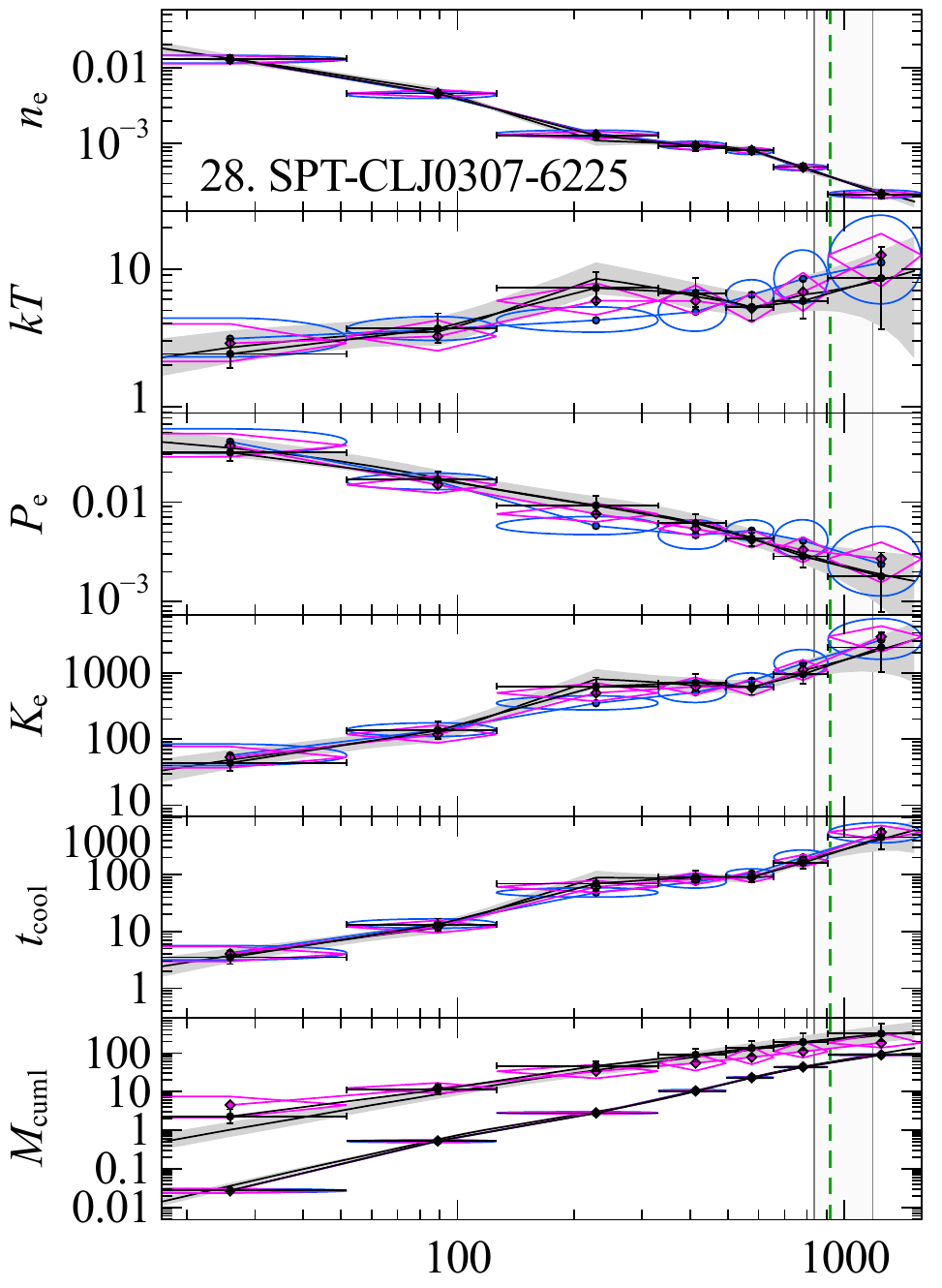}
  \includegraphics[width=0.3\textwidth]{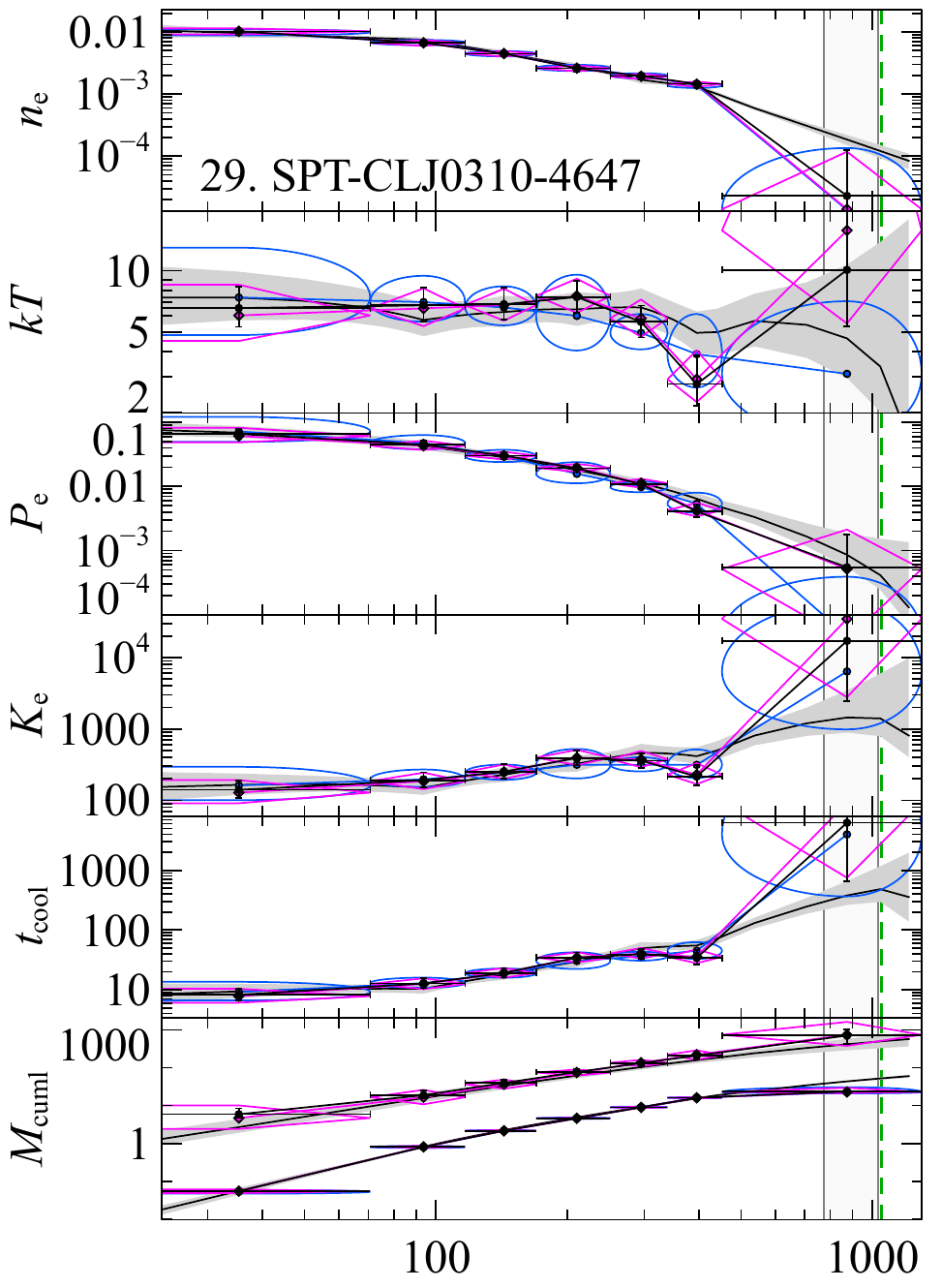}
  \includegraphics[width=0.3\textwidth]{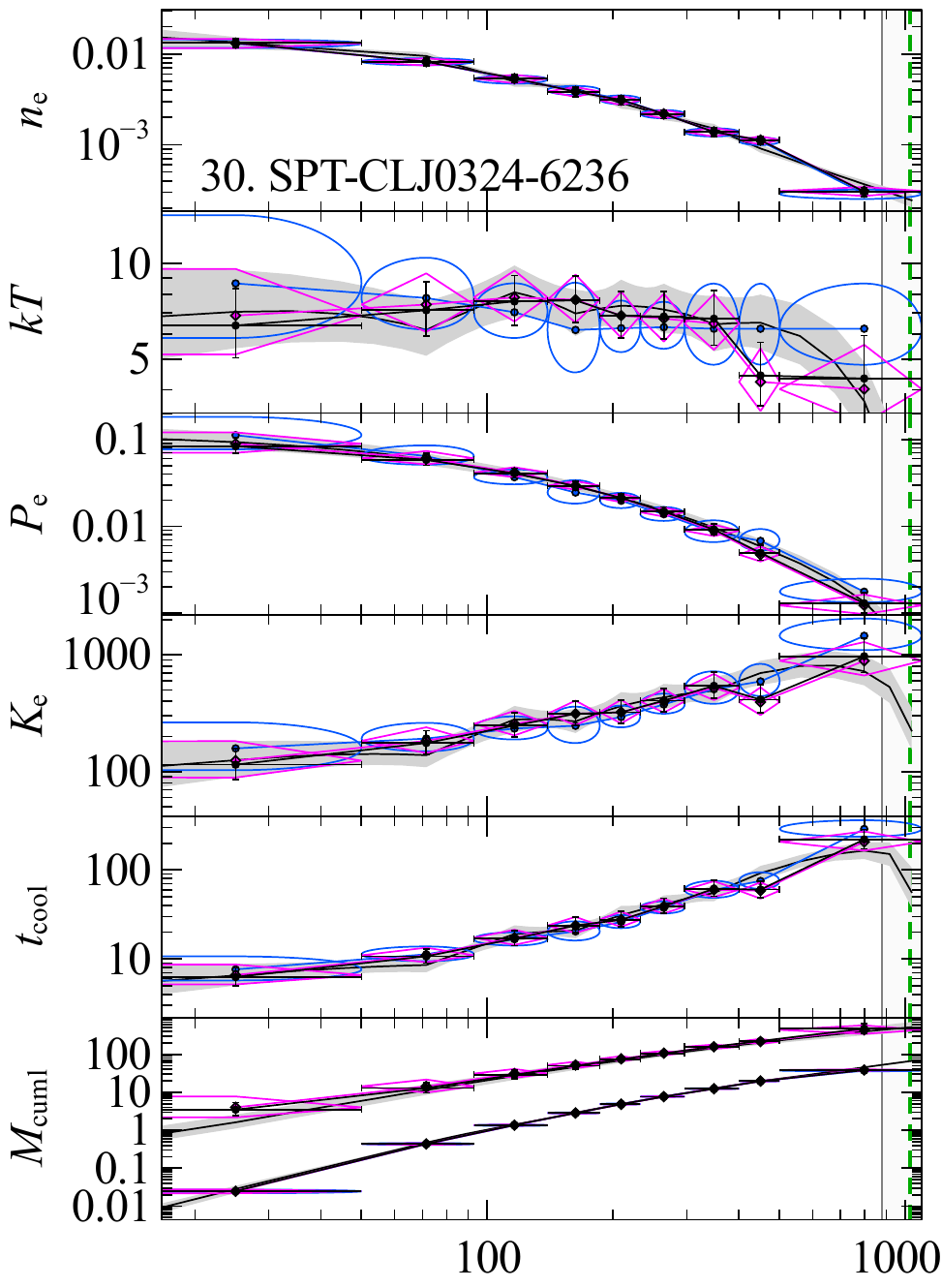}\\
  \includegraphics[width=0.3\textwidth]{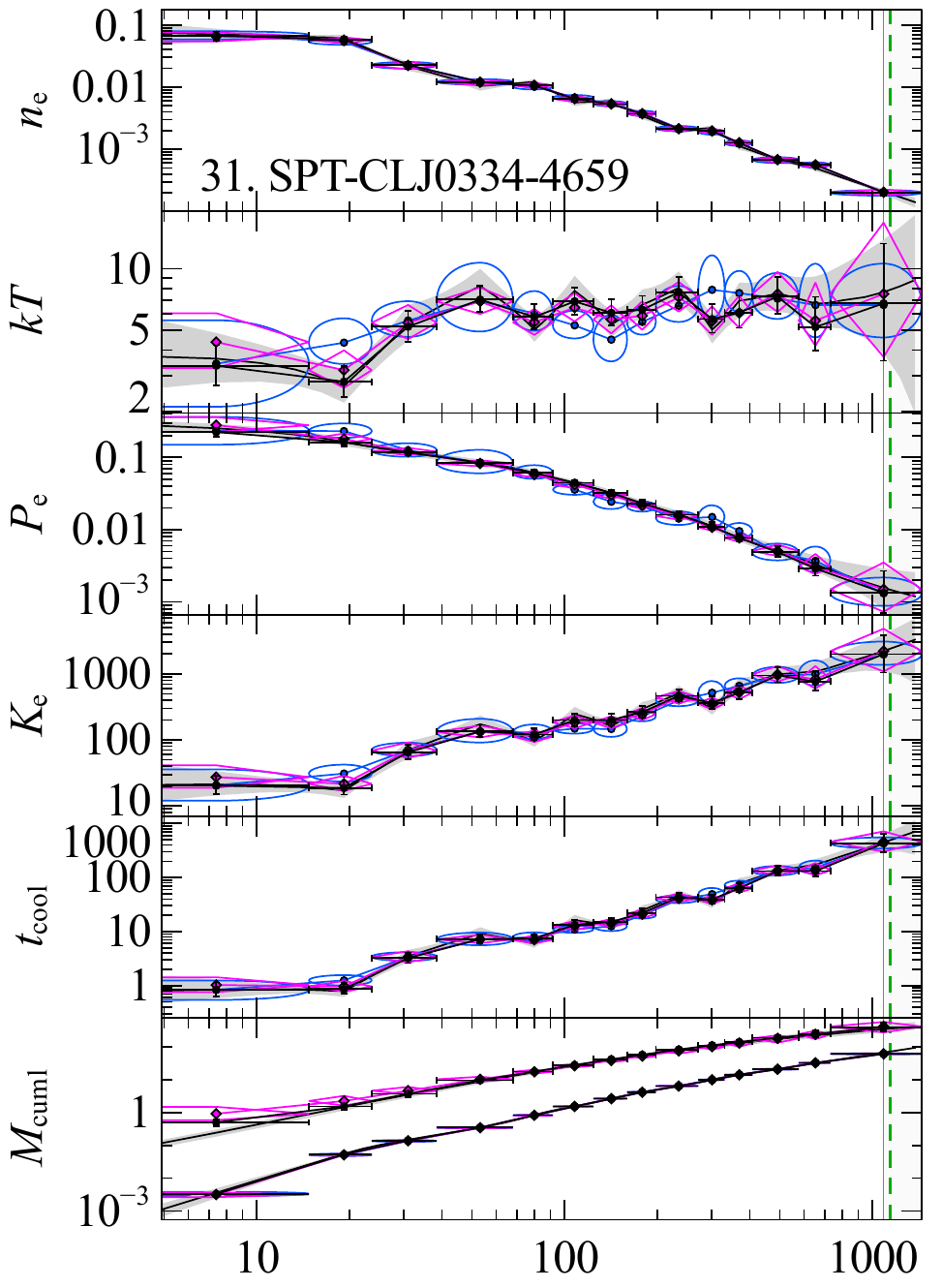}
  \includegraphics[width=0.3\textwidth]{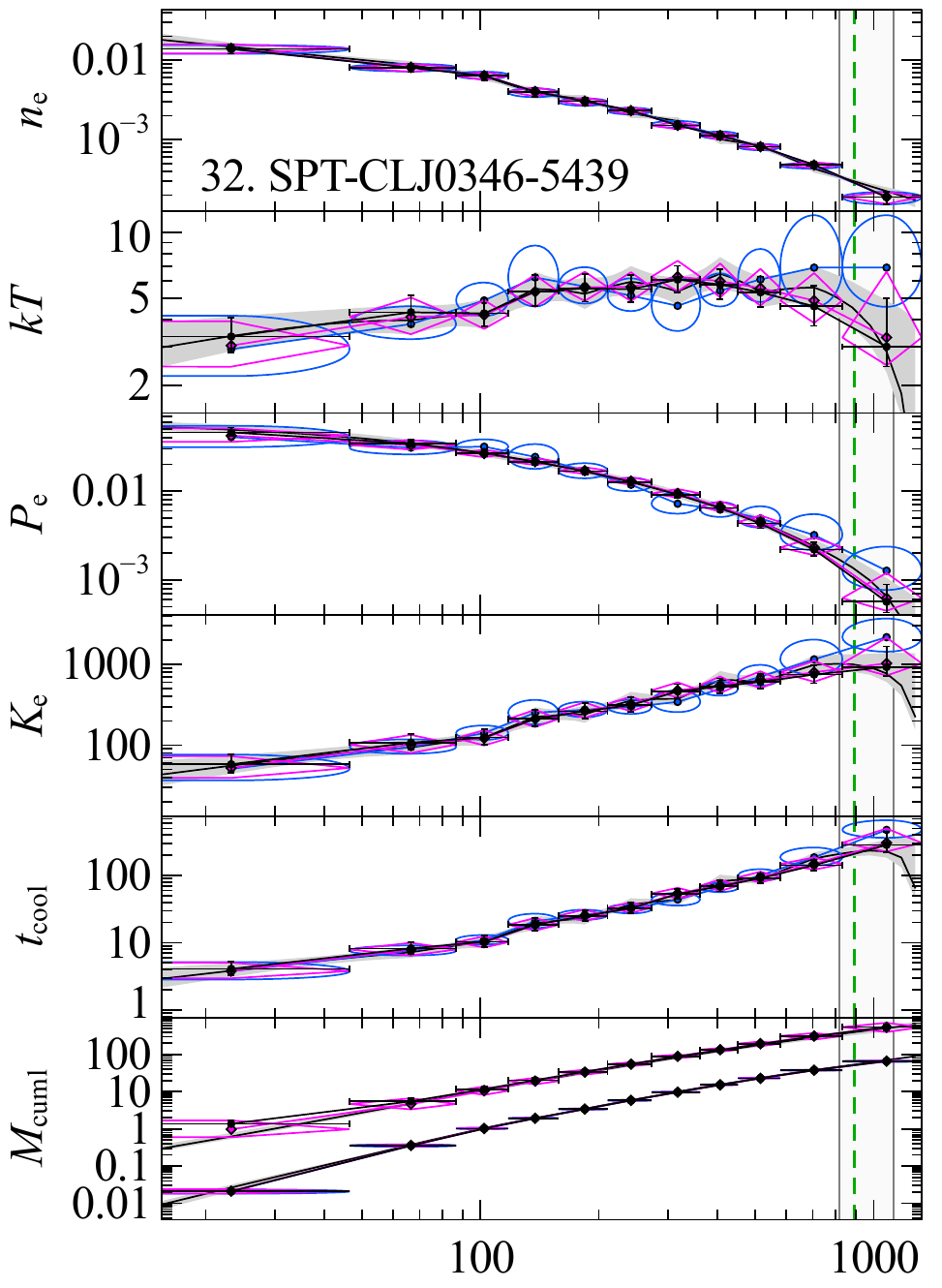}
  \includegraphics[width=0.3\textwidth]{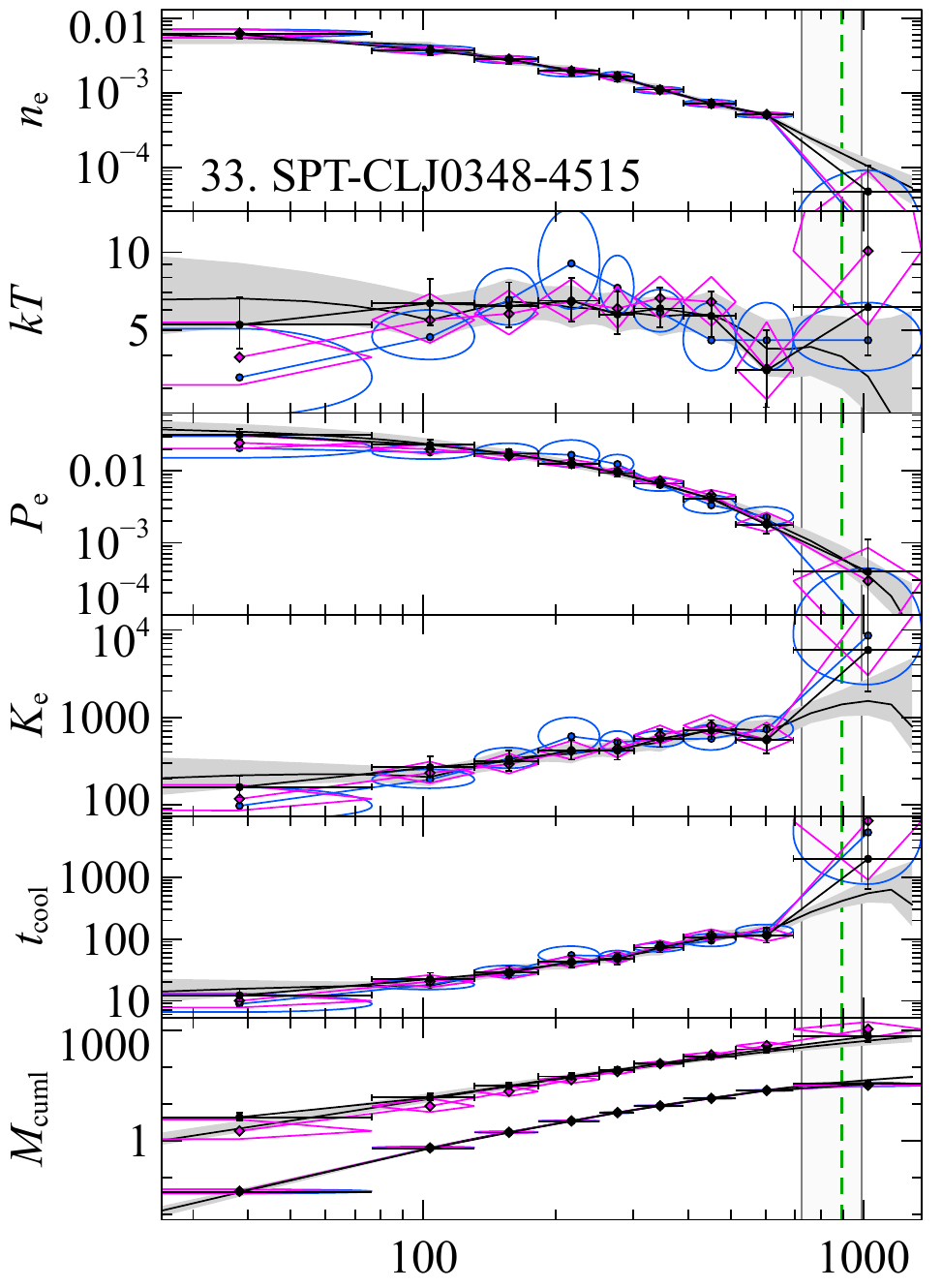}\\
  \includegraphics[width=0.3\textwidth]{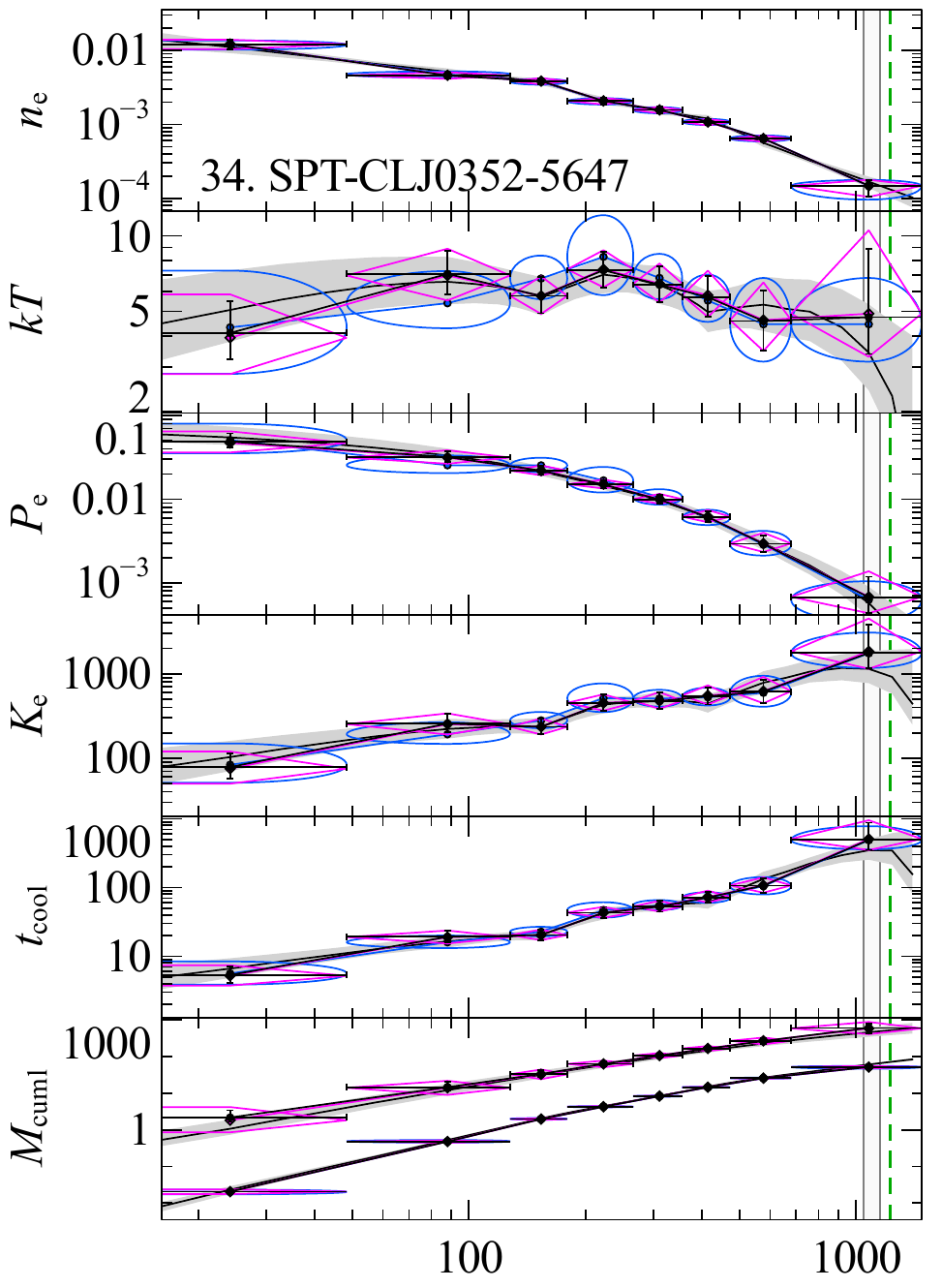}
  \includegraphics[width=0.3\textwidth]{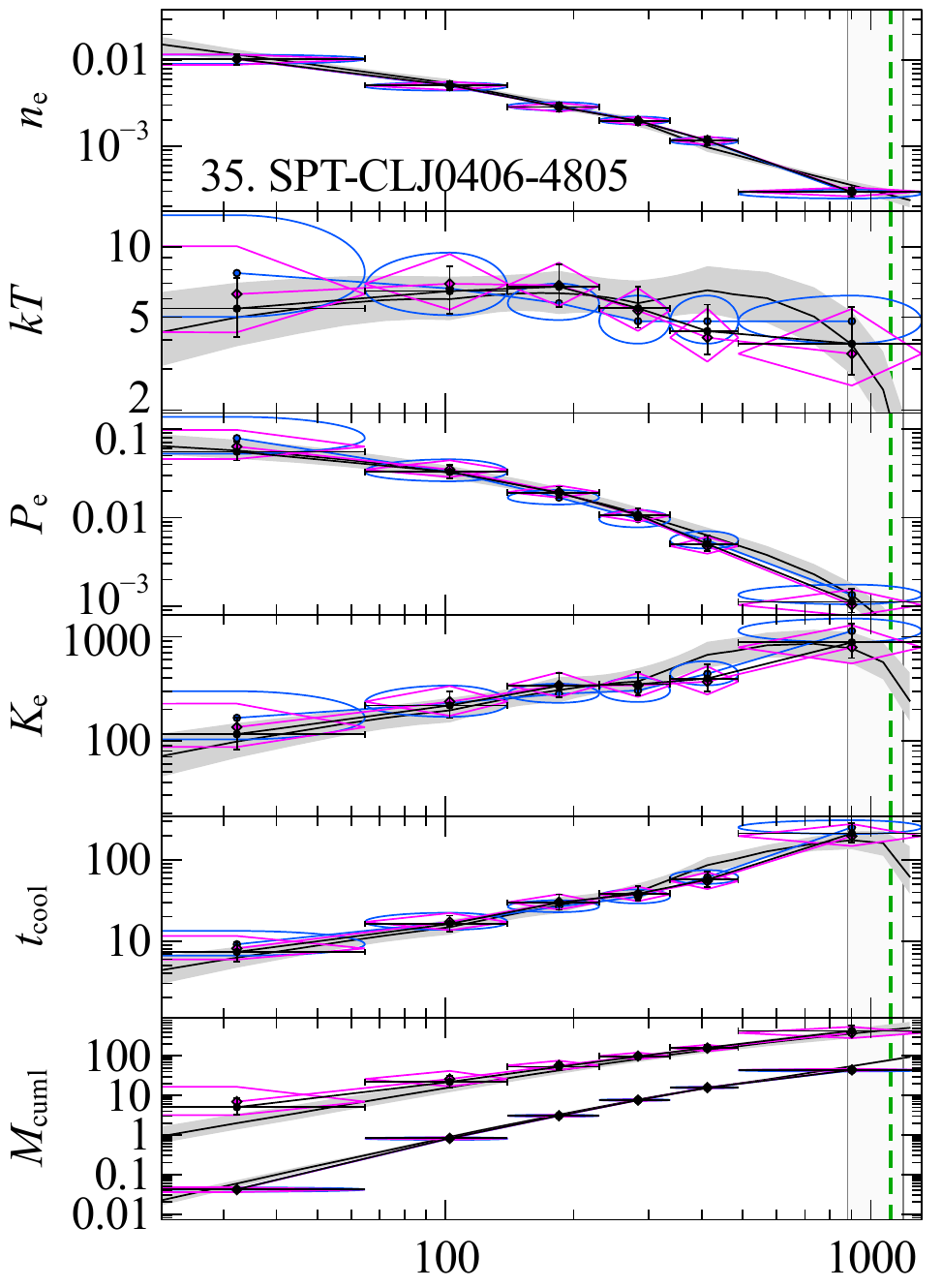}
  \includegraphics[width=0.3\textwidth]{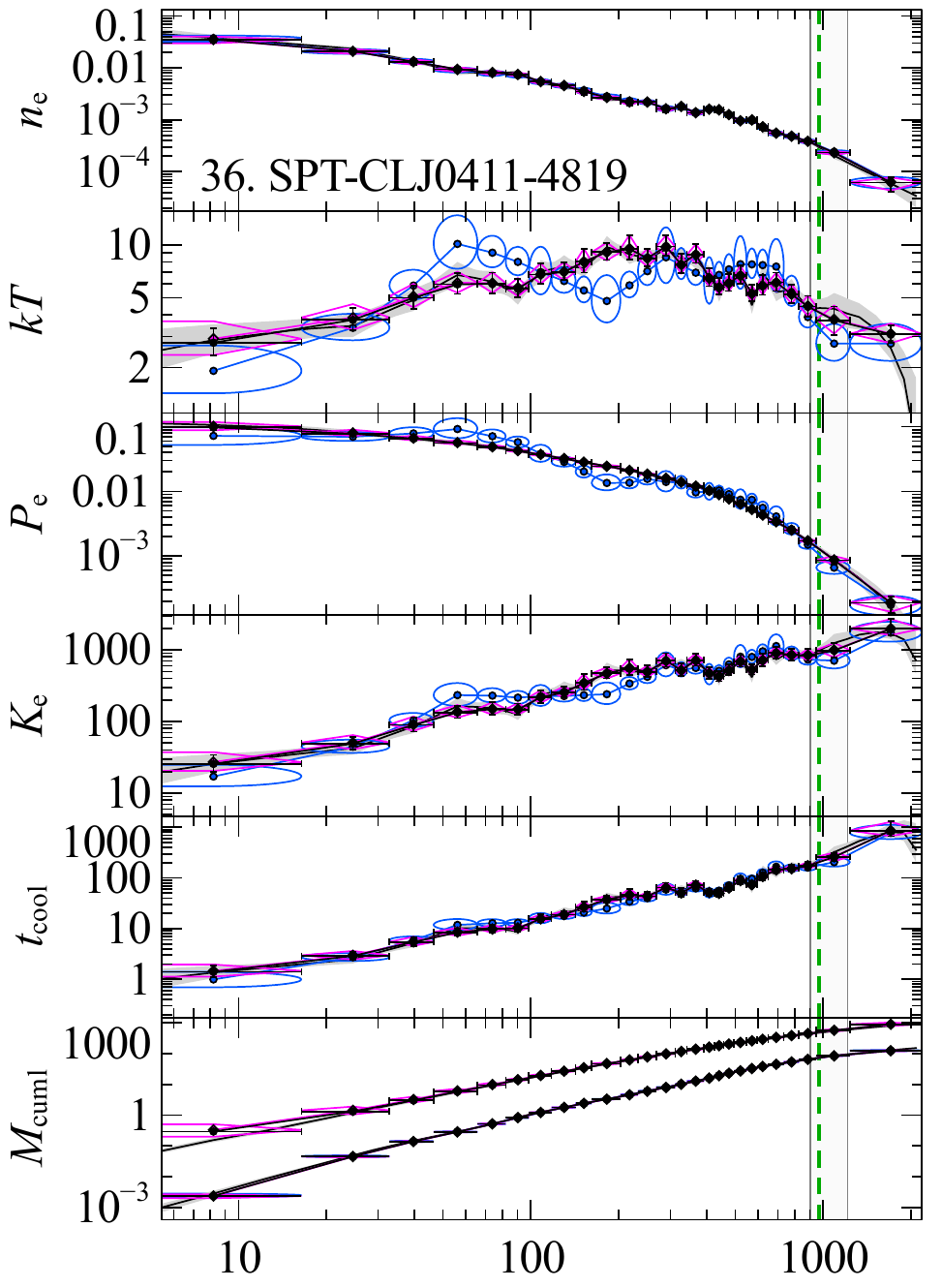}\\
  \contcaption{individual cluster profiles.}
\end{figure*}
\begin{figure*}
  \centering
  \includegraphics[width=0.3\textwidth]{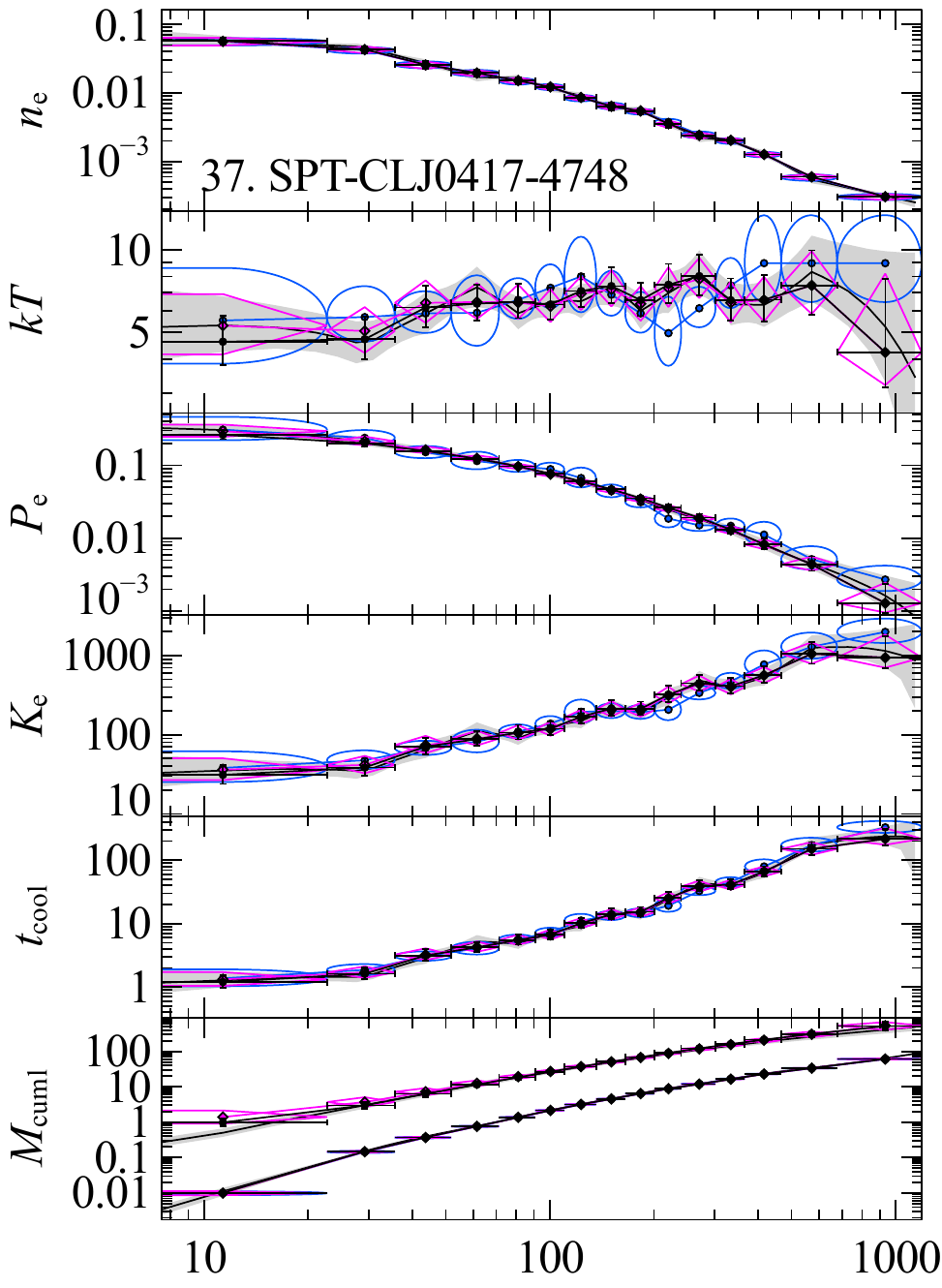}
  \includegraphics[width=0.3\textwidth]{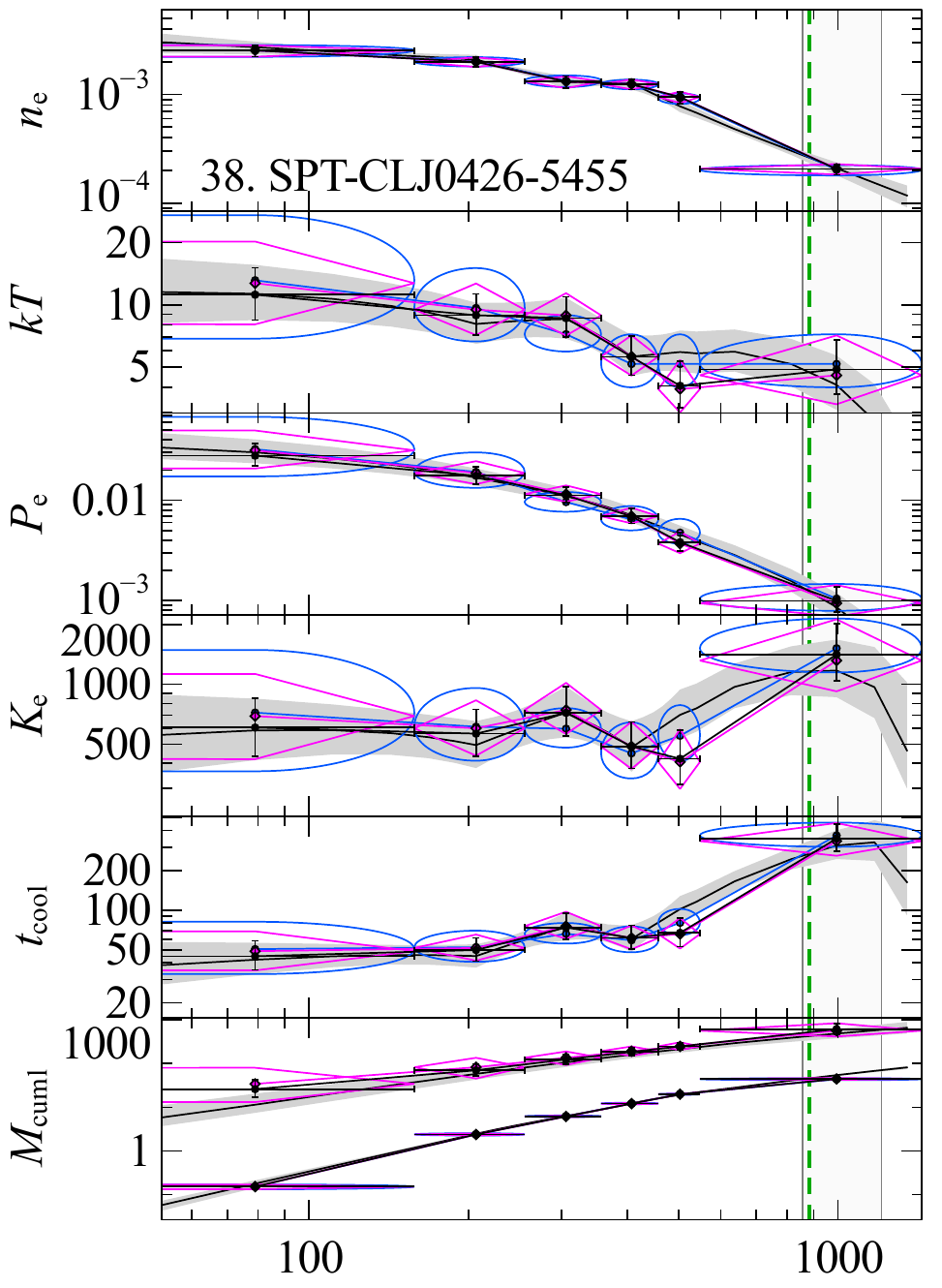}
  \includegraphics[width=0.3\textwidth]{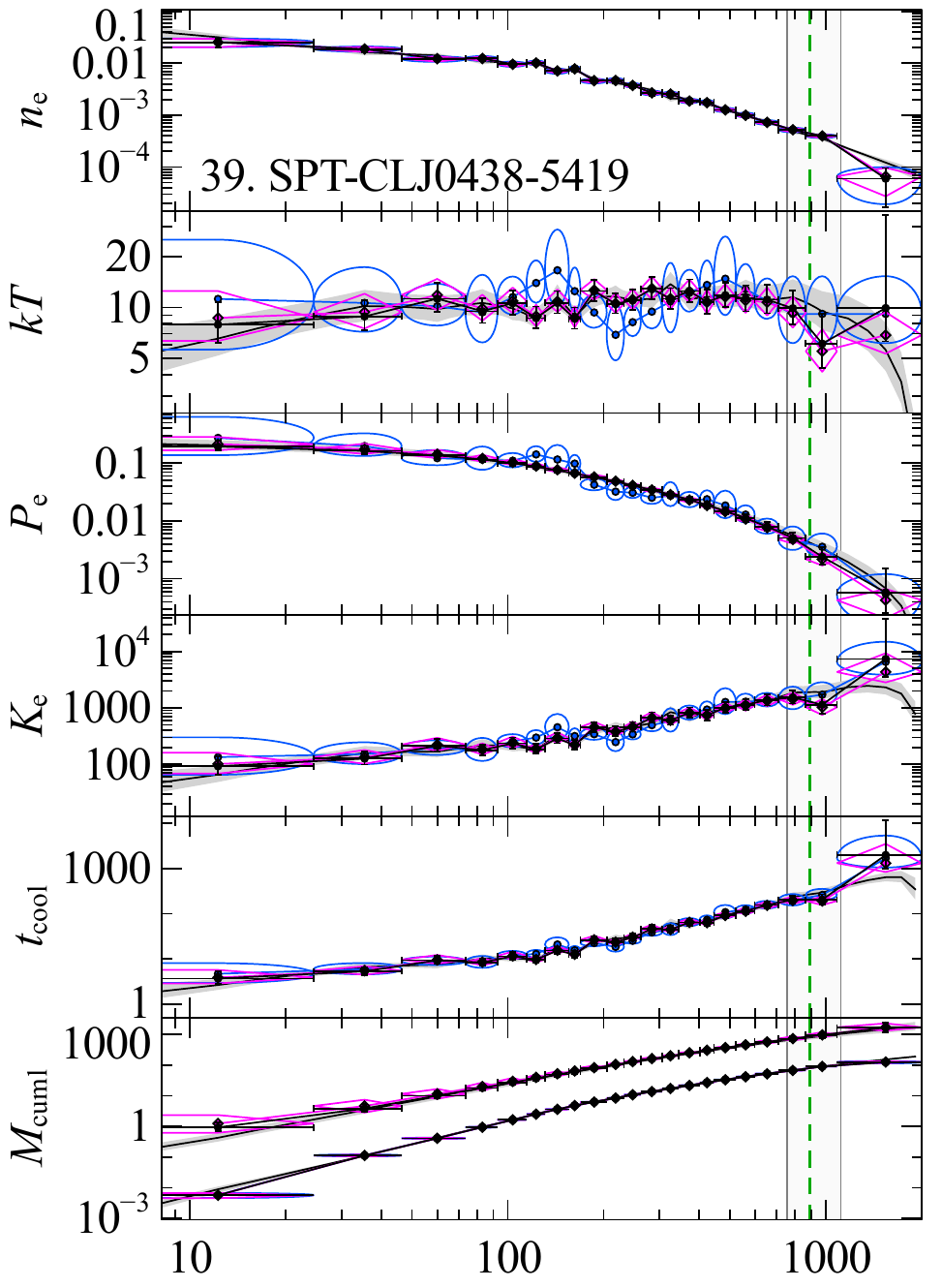}\\
  \includegraphics[width=0.3\textwidth]{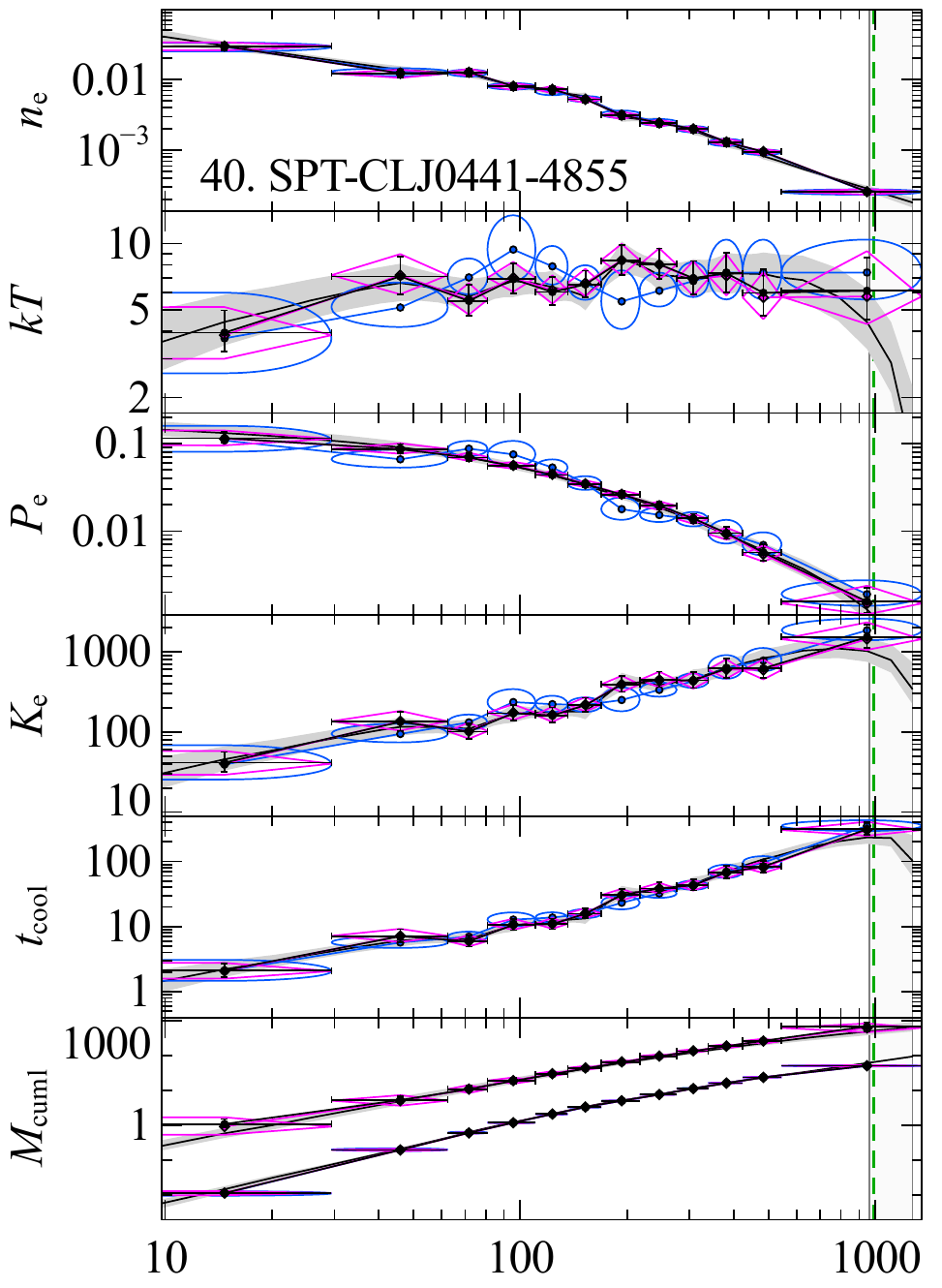}
  \includegraphics[width=0.3\textwidth]{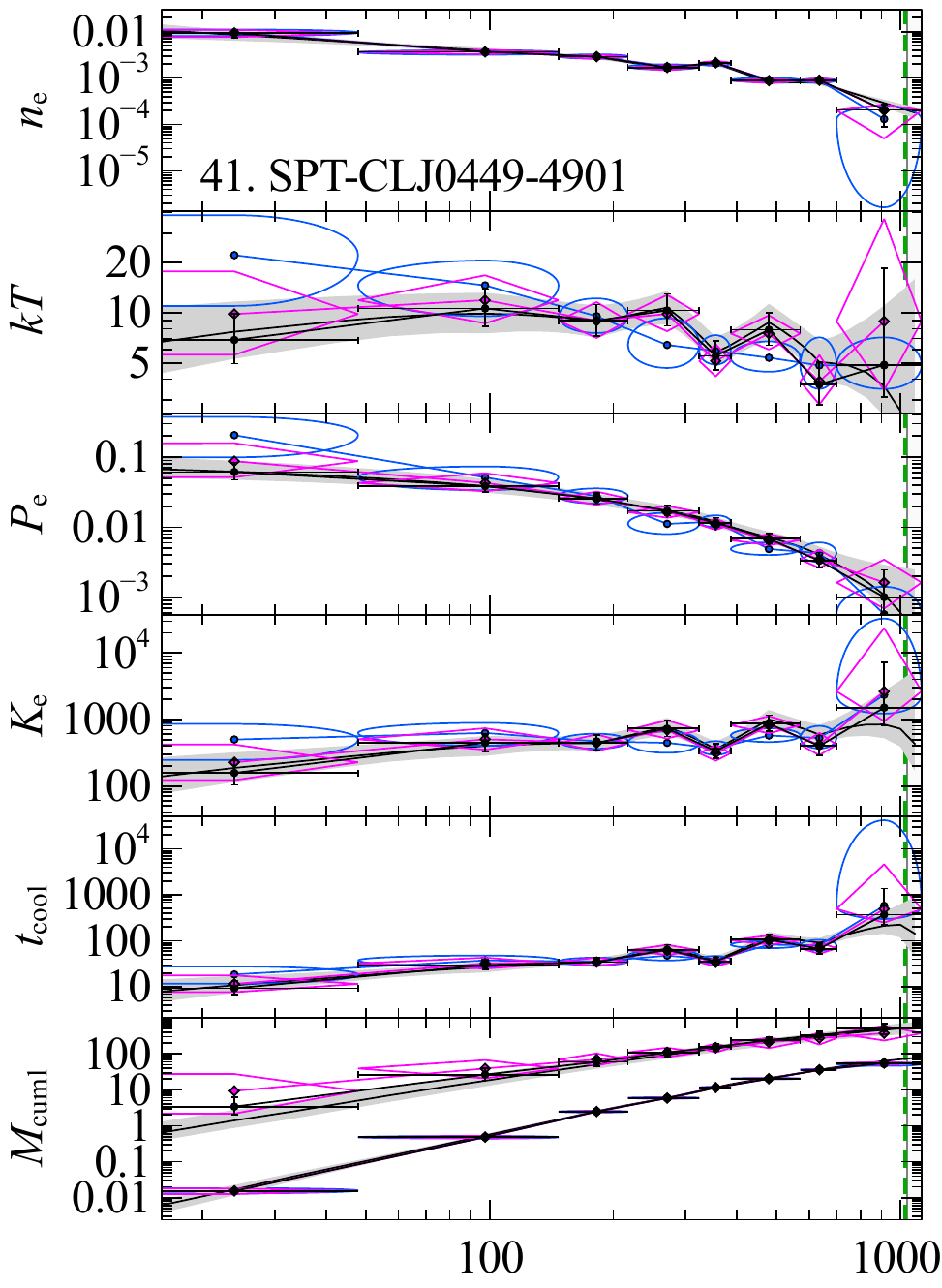}
  \includegraphics[width=0.3\textwidth]{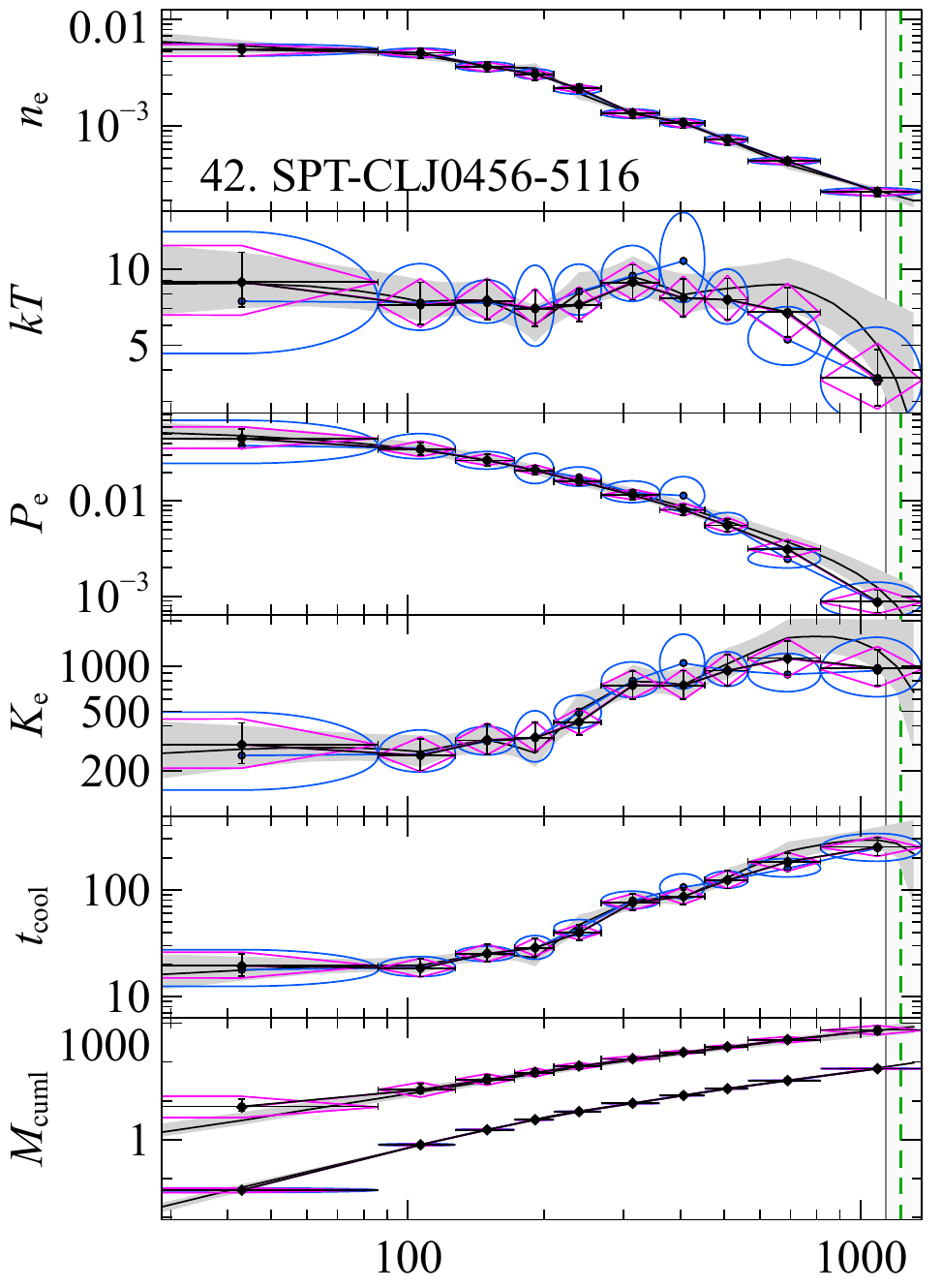}\\
  \includegraphics[width=0.3\textwidth]{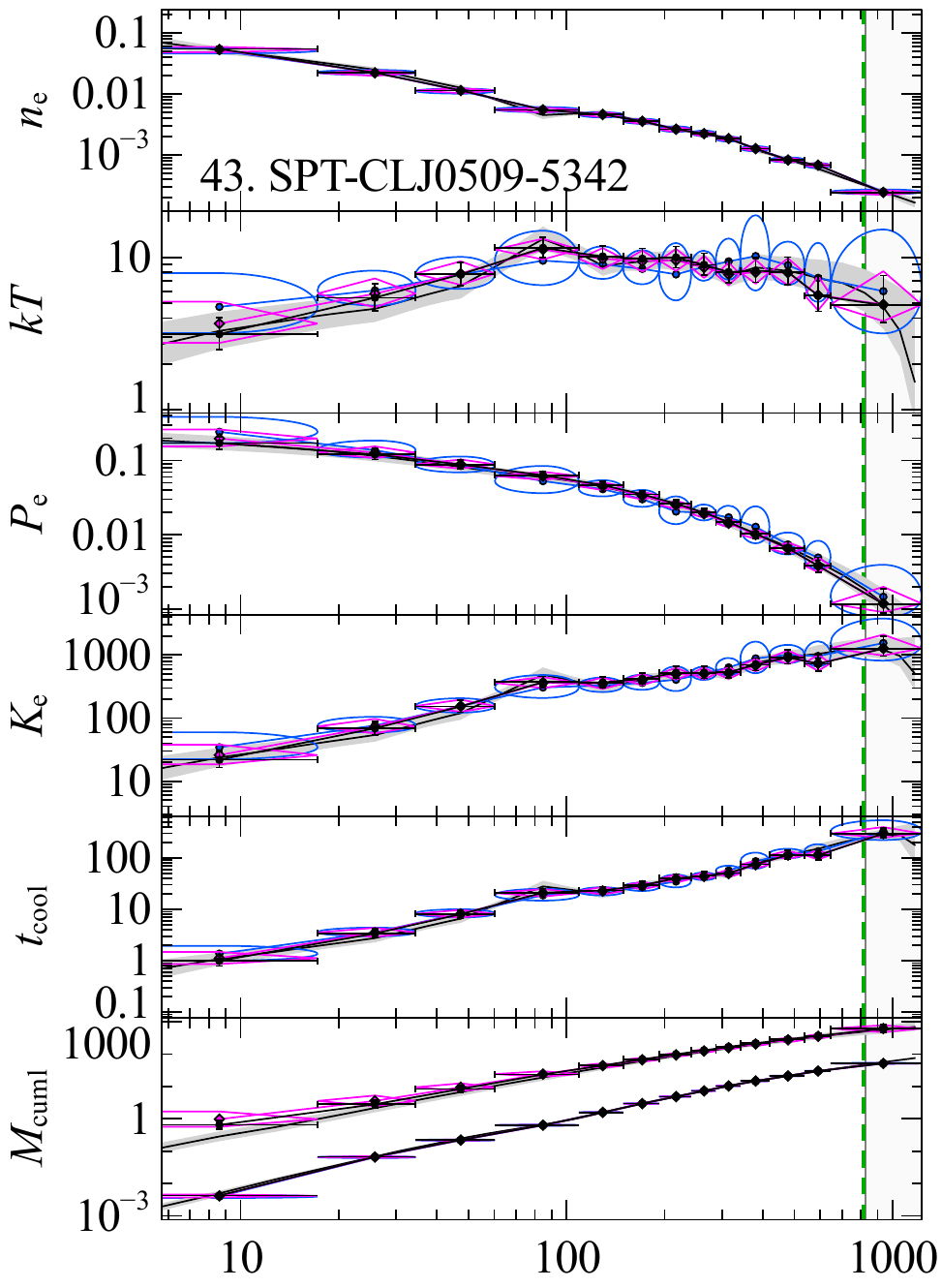}
  \includegraphics[width=0.3\textwidth]{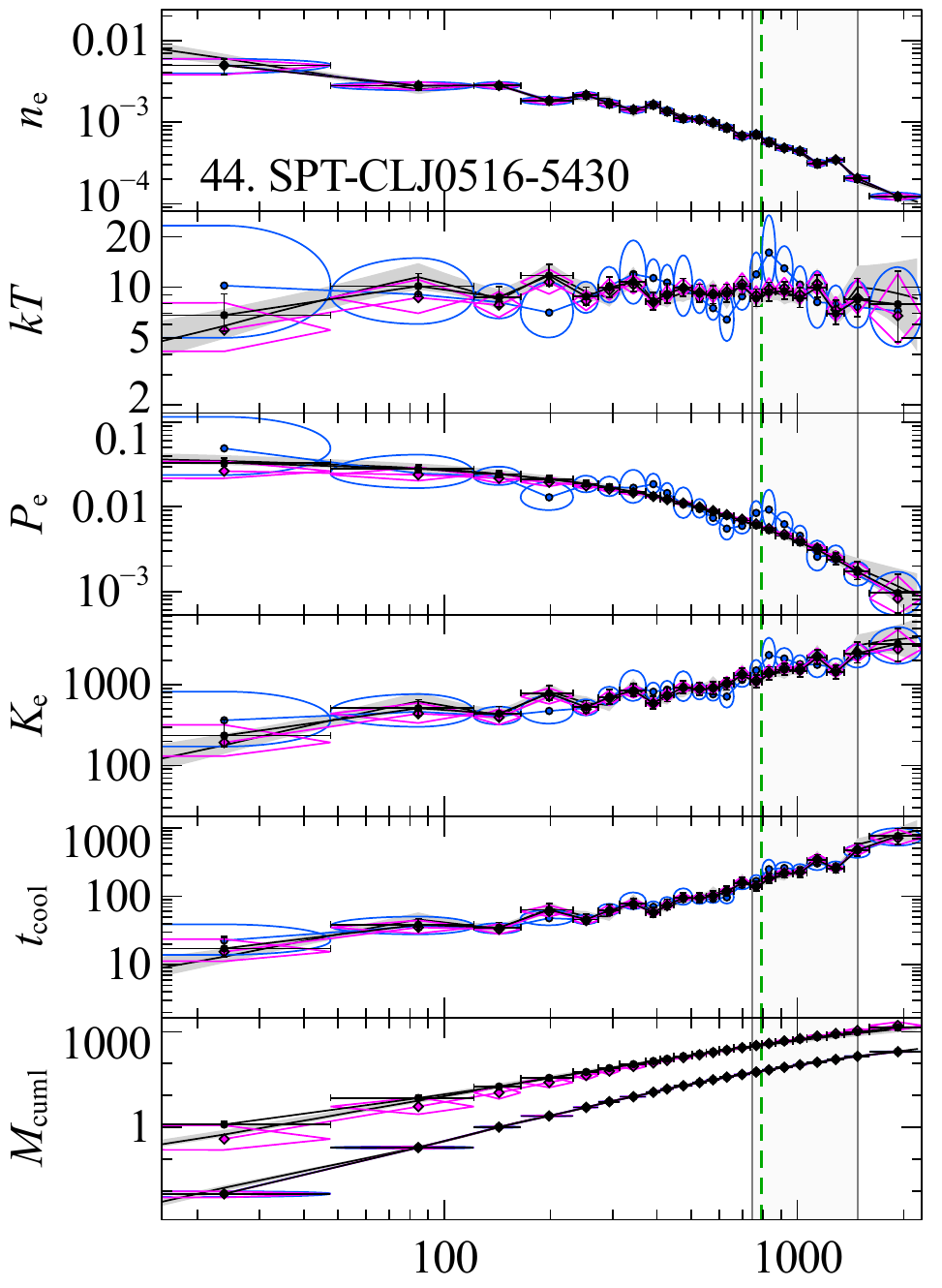}
  \includegraphics[width=0.3\textwidth]{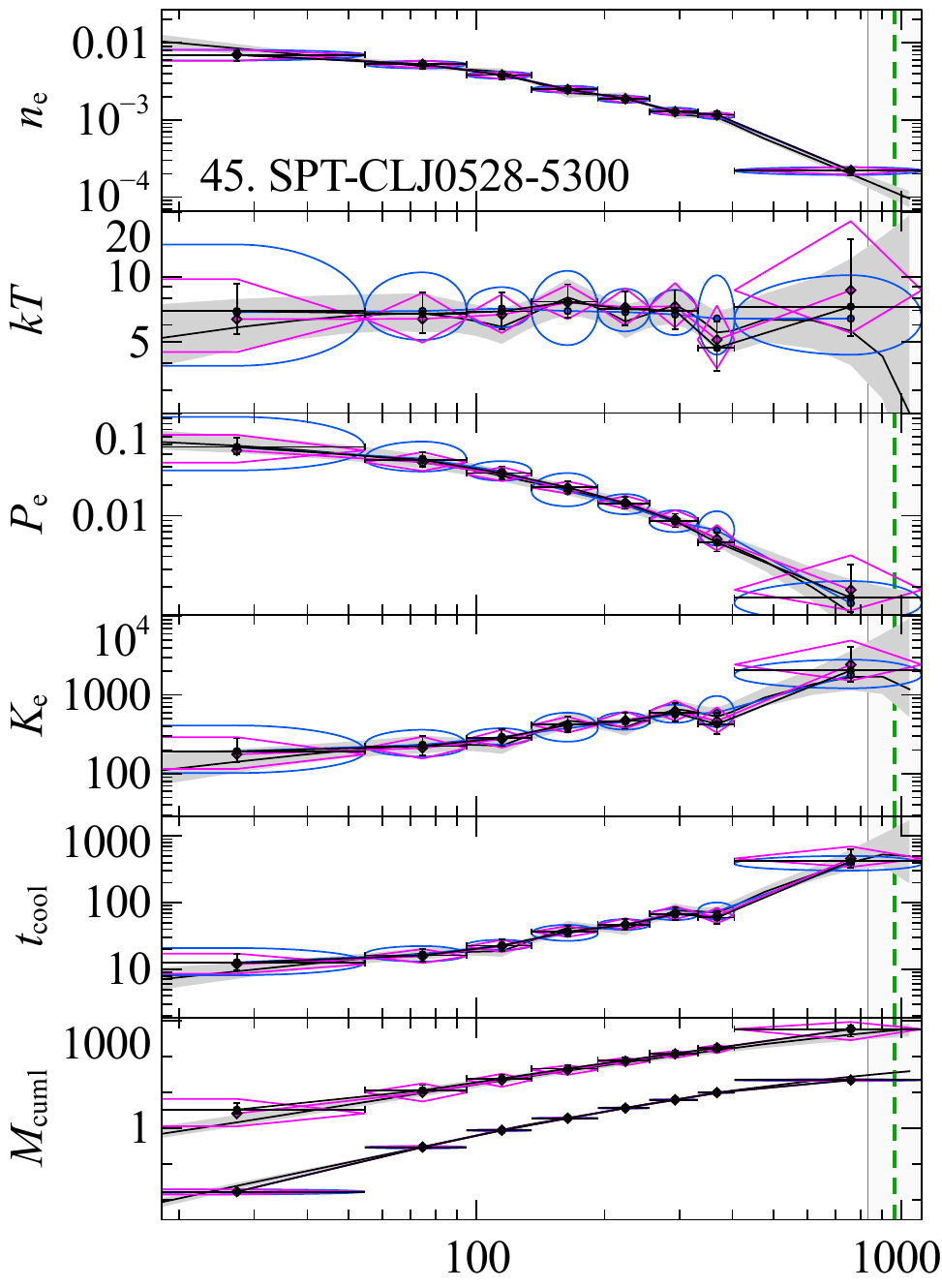}\\
  \contcaption{individual cluster profiles.}
\end{figure*}
\begin{figure*}
  \centering
  \includegraphics[width=0.3\textwidth]{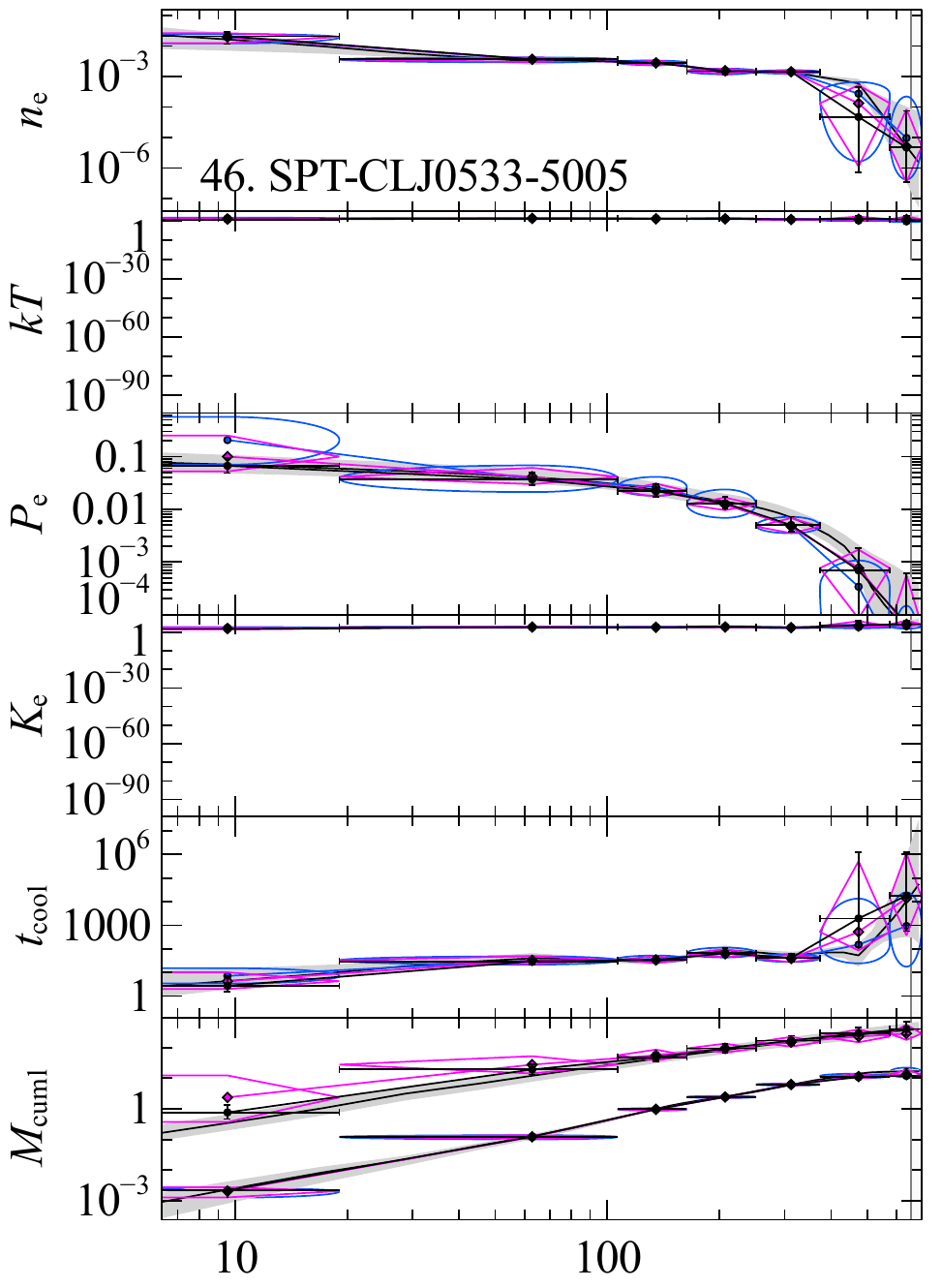}
  \includegraphics[width=0.3\textwidth]{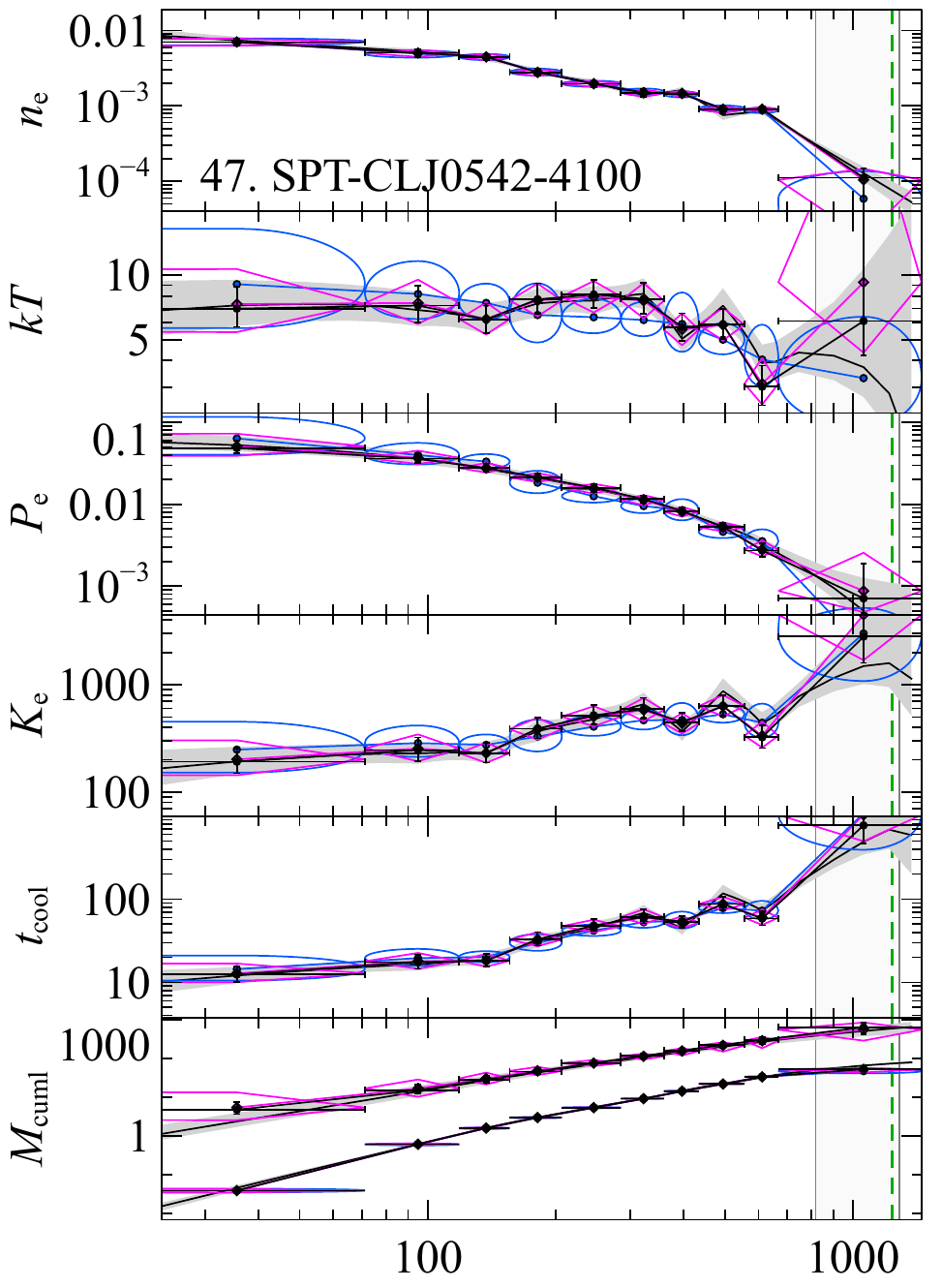}
  \includegraphics[width=0.3\textwidth]{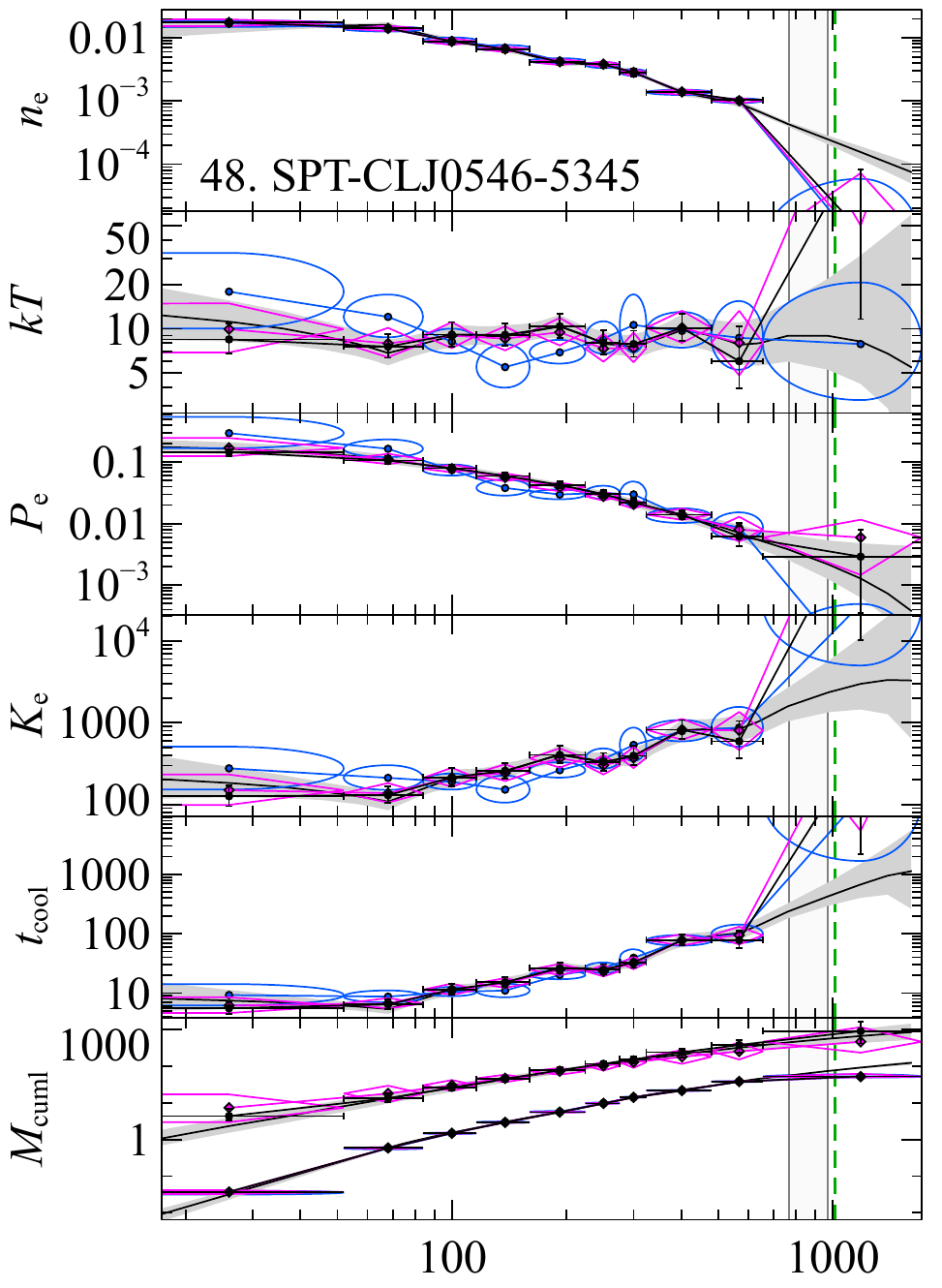}\\
  \includegraphics[width=0.3\textwidth]{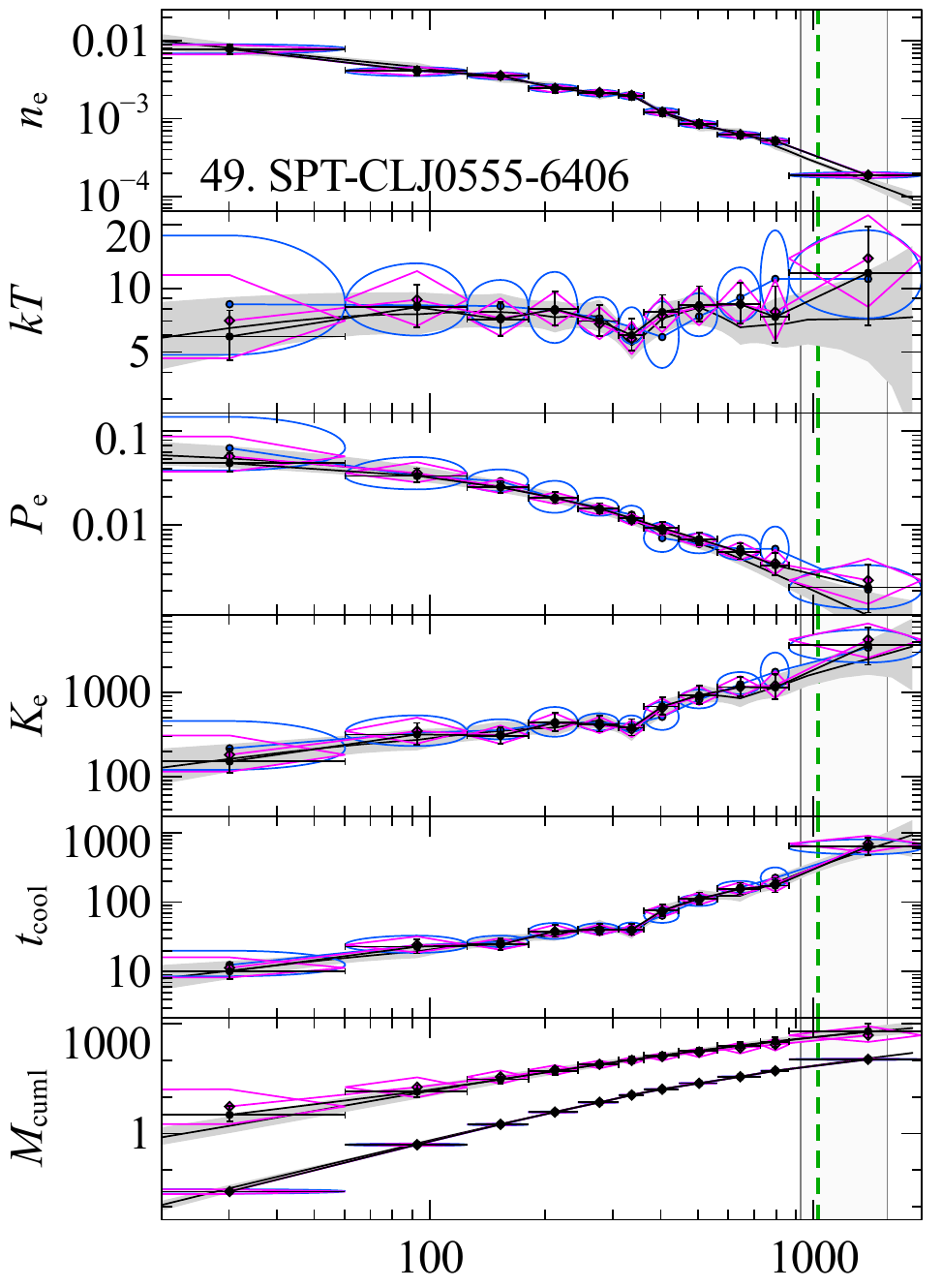}
  \includegraphics[width=0.3\textwidth]{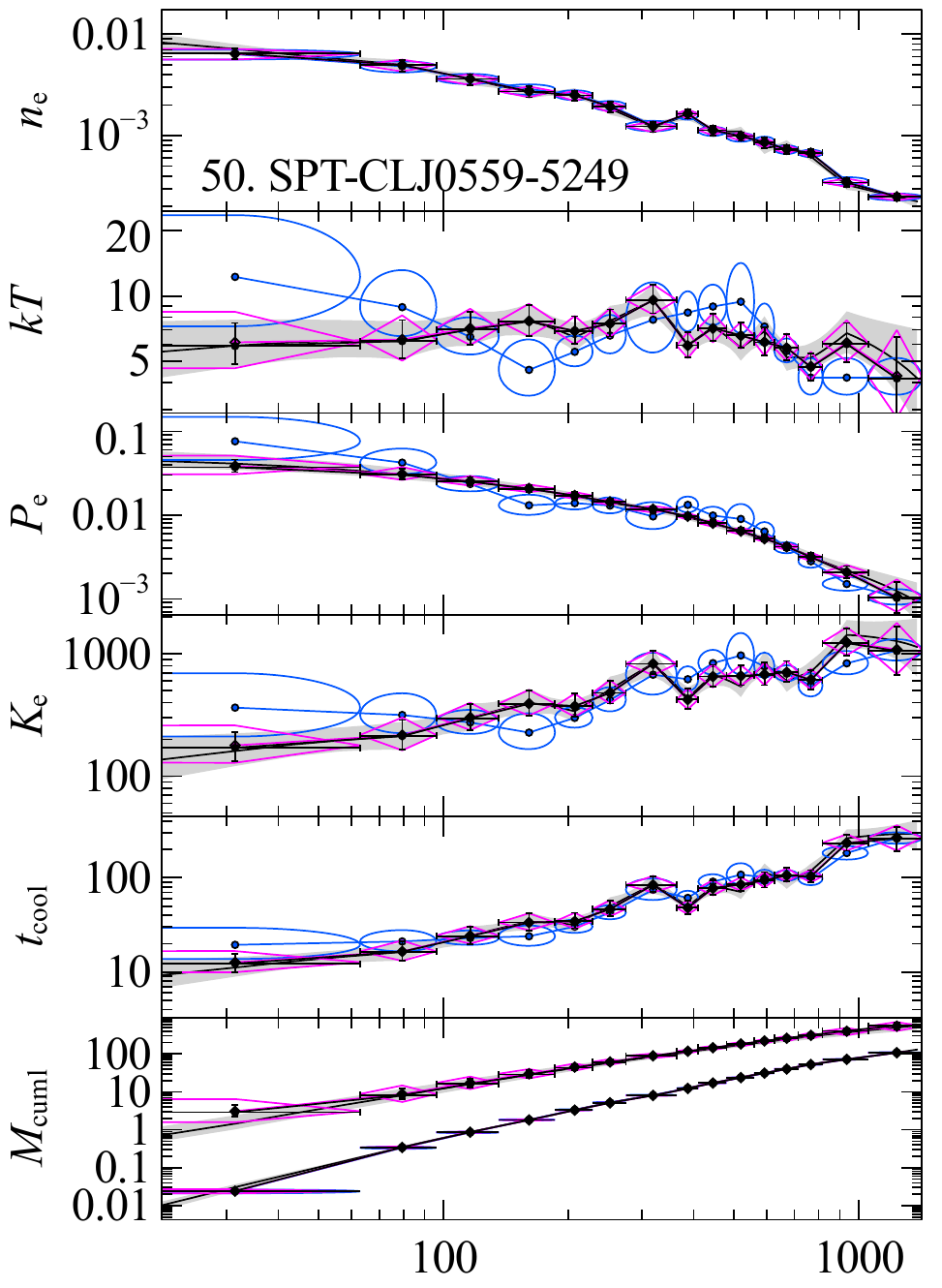}
  \includegraphics[width=0.3\textwidth]{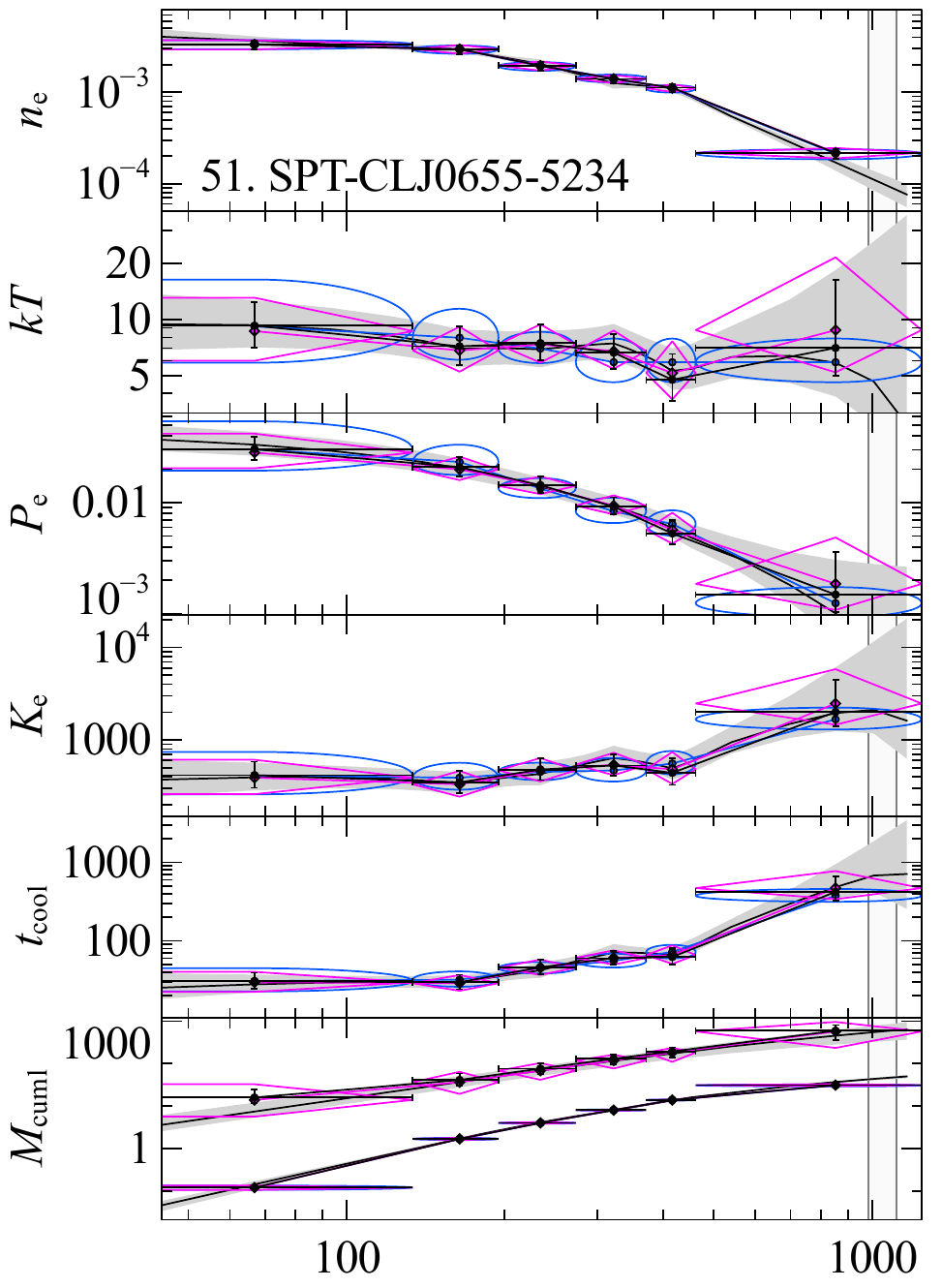}\\
  \includegraphics[width=0.3\textwidth]{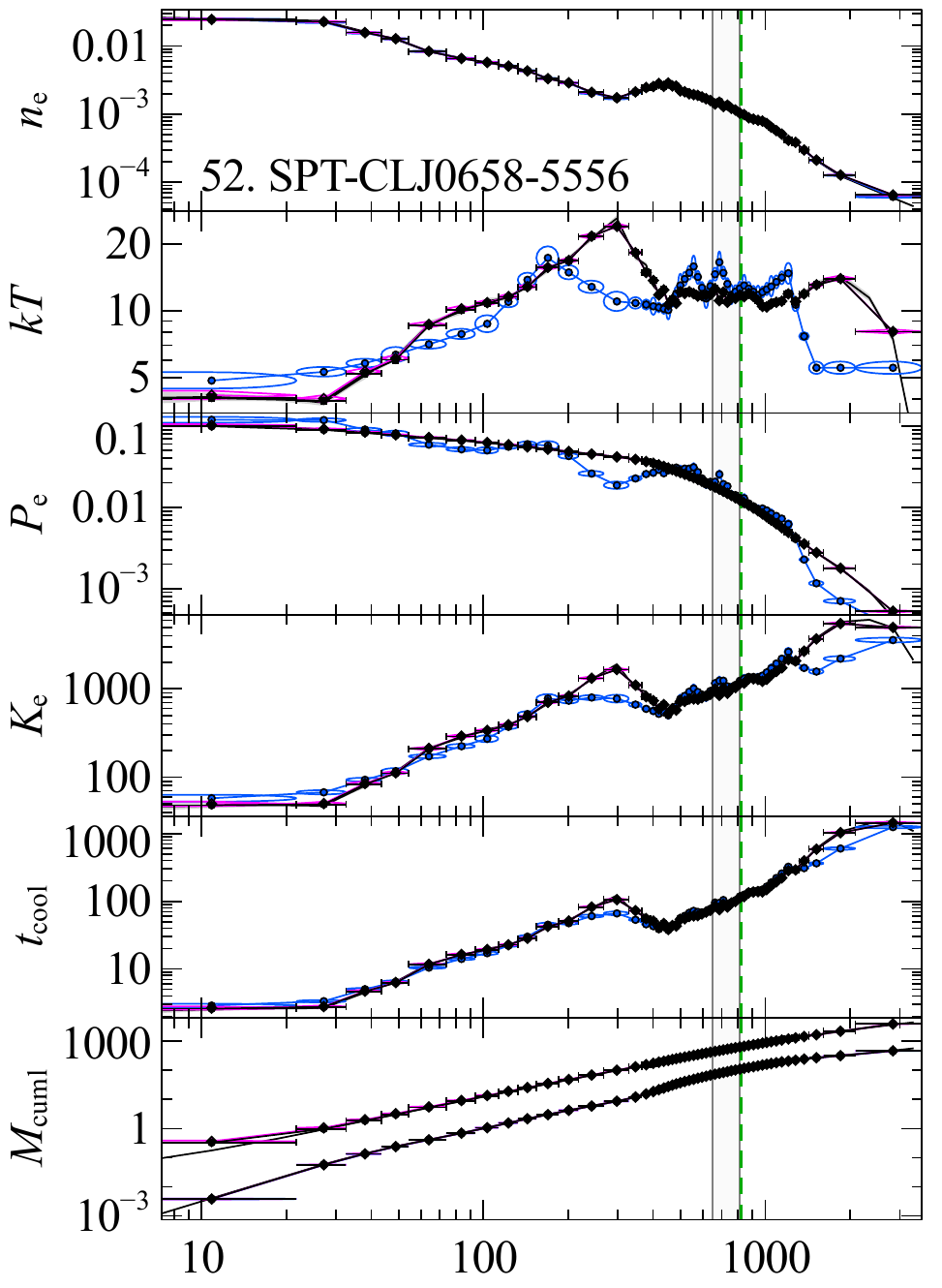}
  \includegraphics[width=0.3\textwidth]{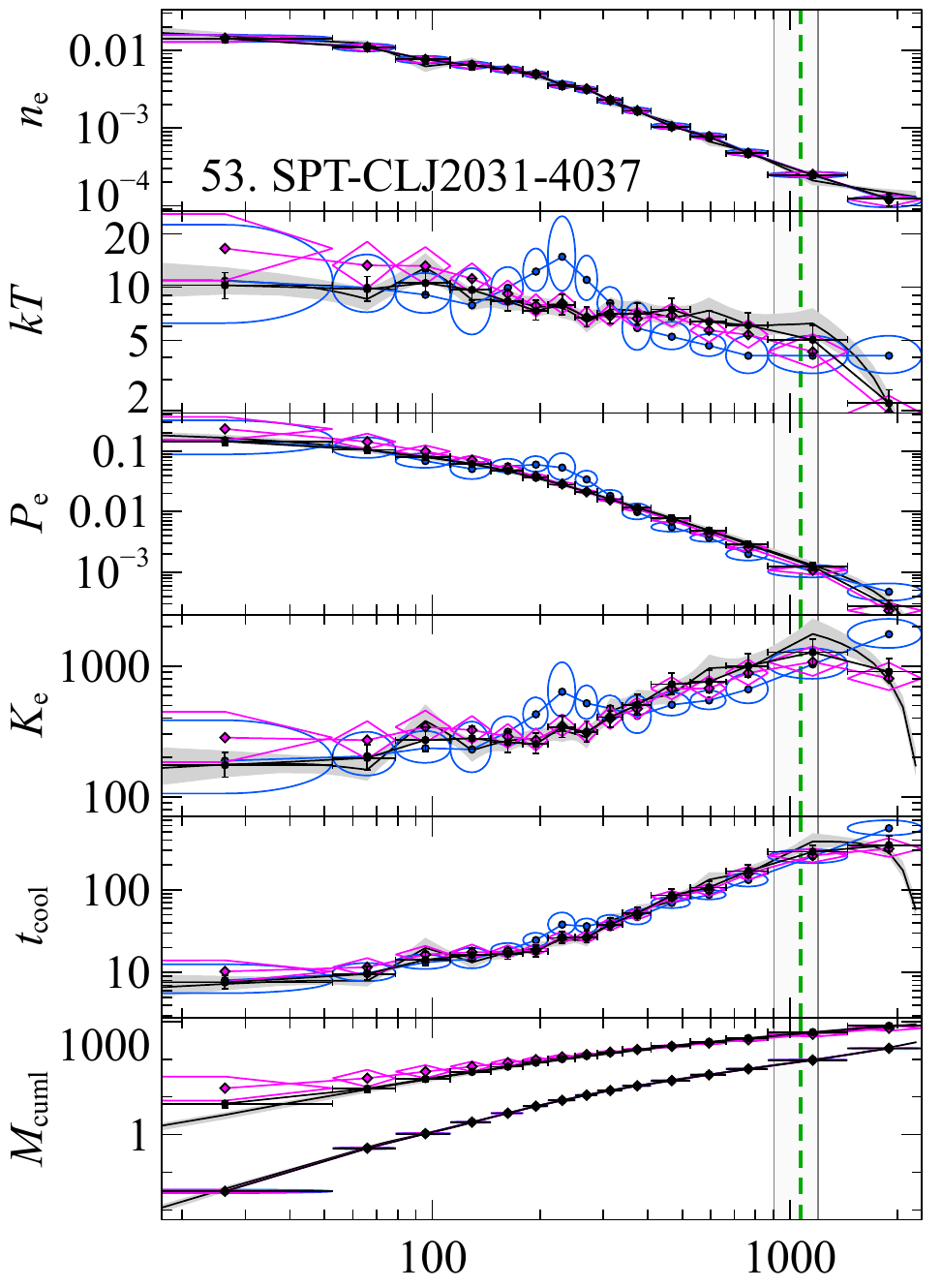}
  \includegraphics[width=0.3\textwidth]{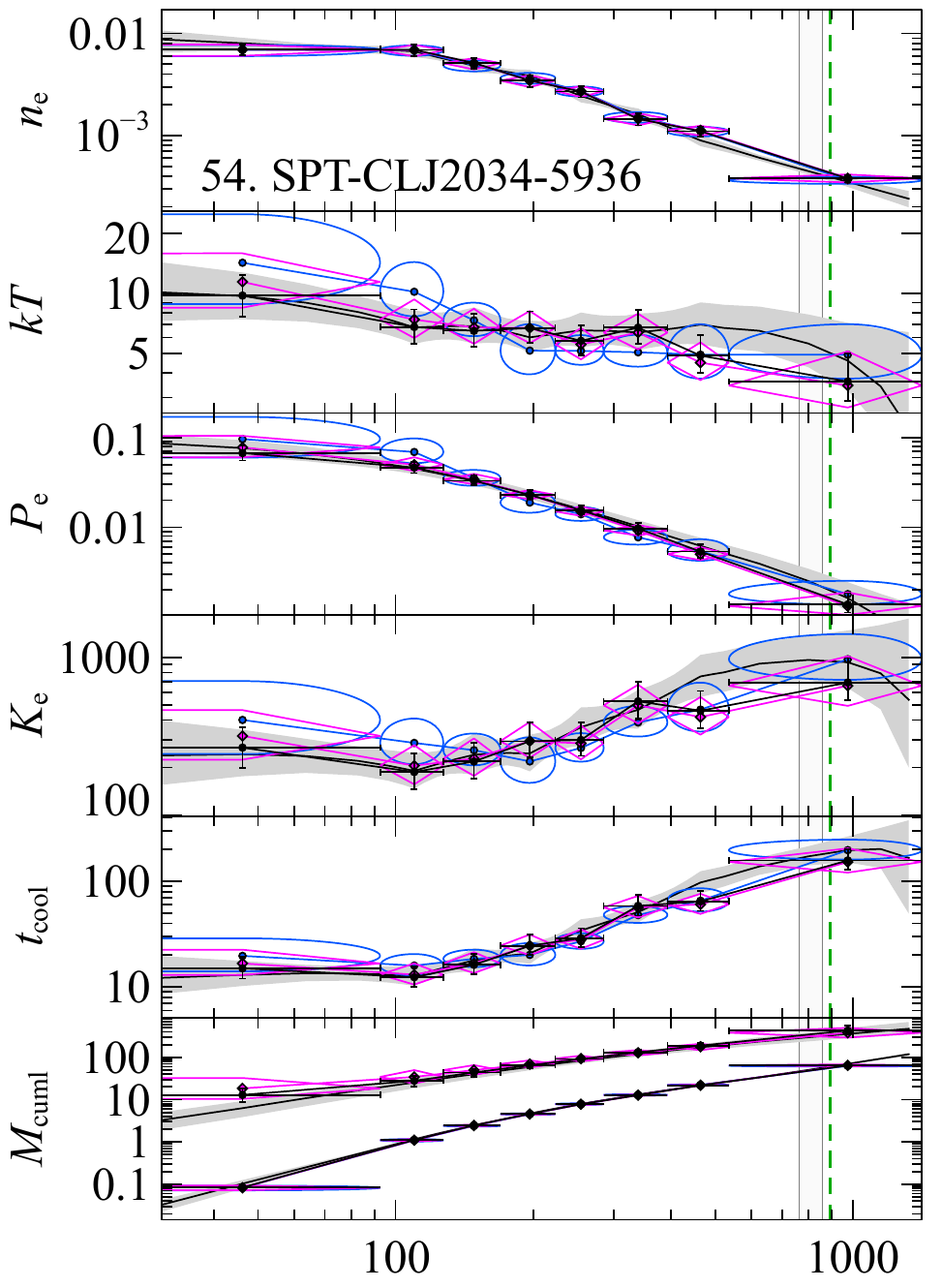}\\
  \contcaption{individual cluster profiles.}
\end{figure*}
\begin{figure*}
  \centering
  \includegraphics[width=0.3\textwidth]{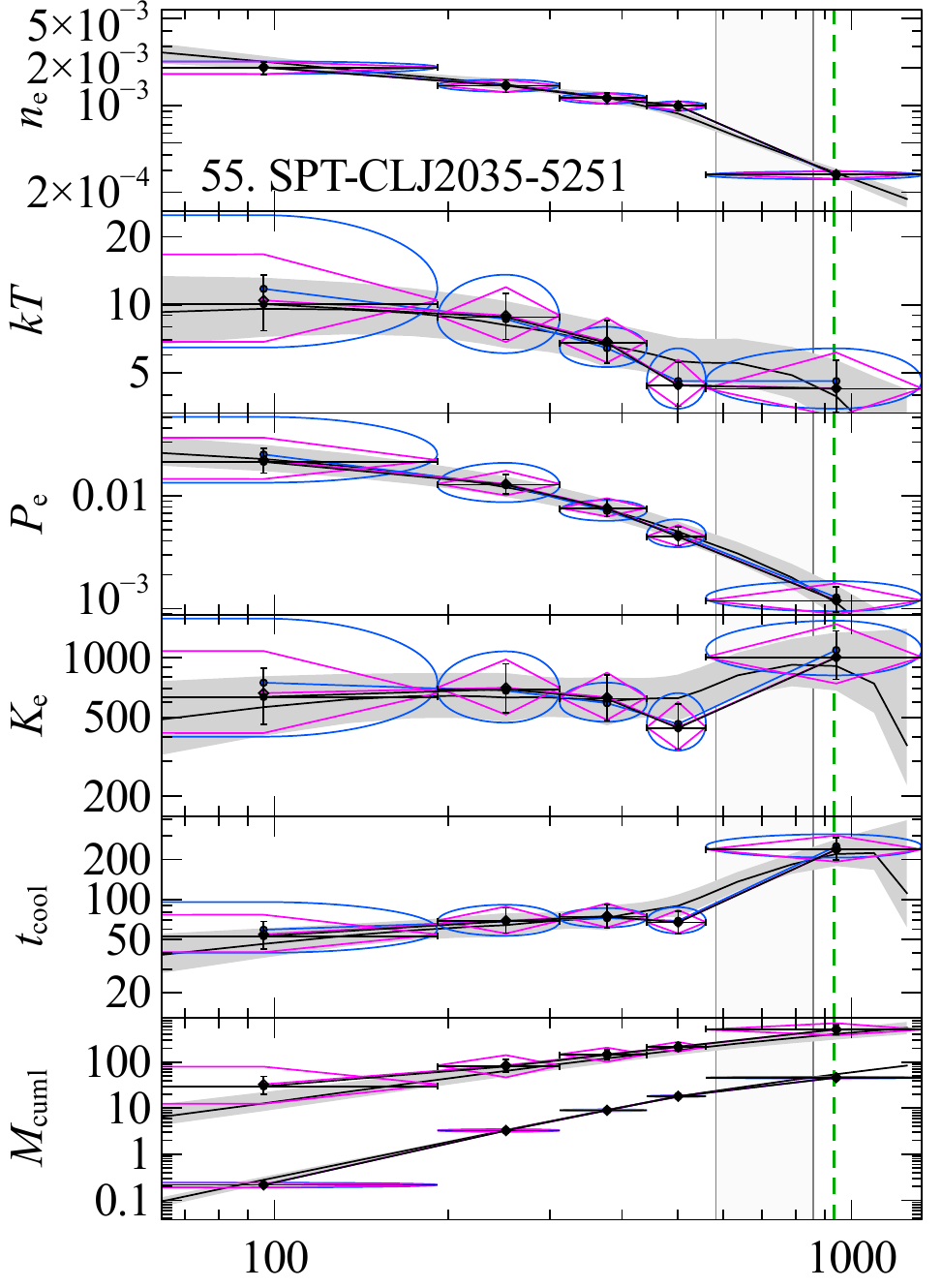}
  \includegraphics[width=0.3\textwidth]{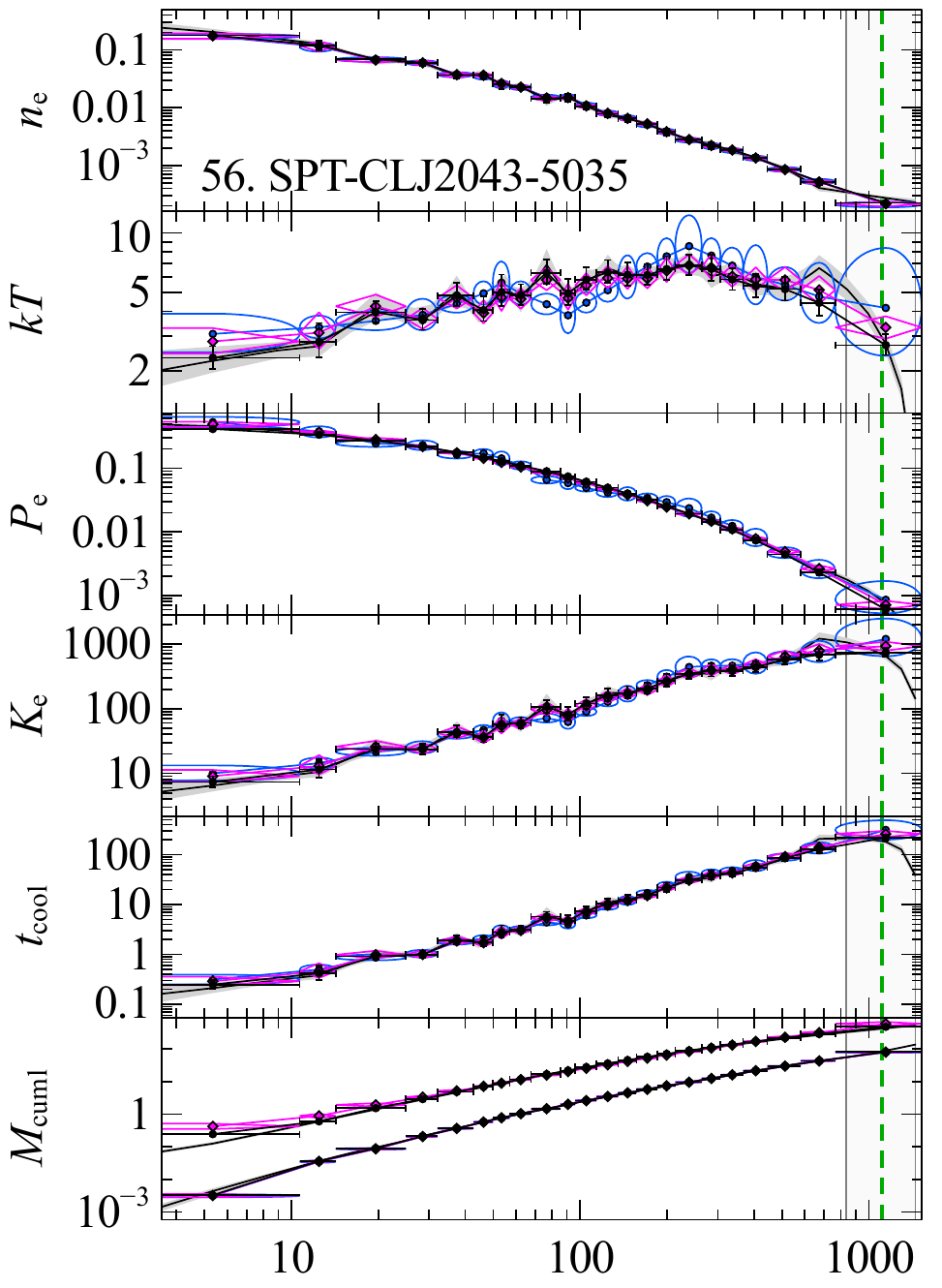}
  \includegraphics[width=0.3\textwidth]{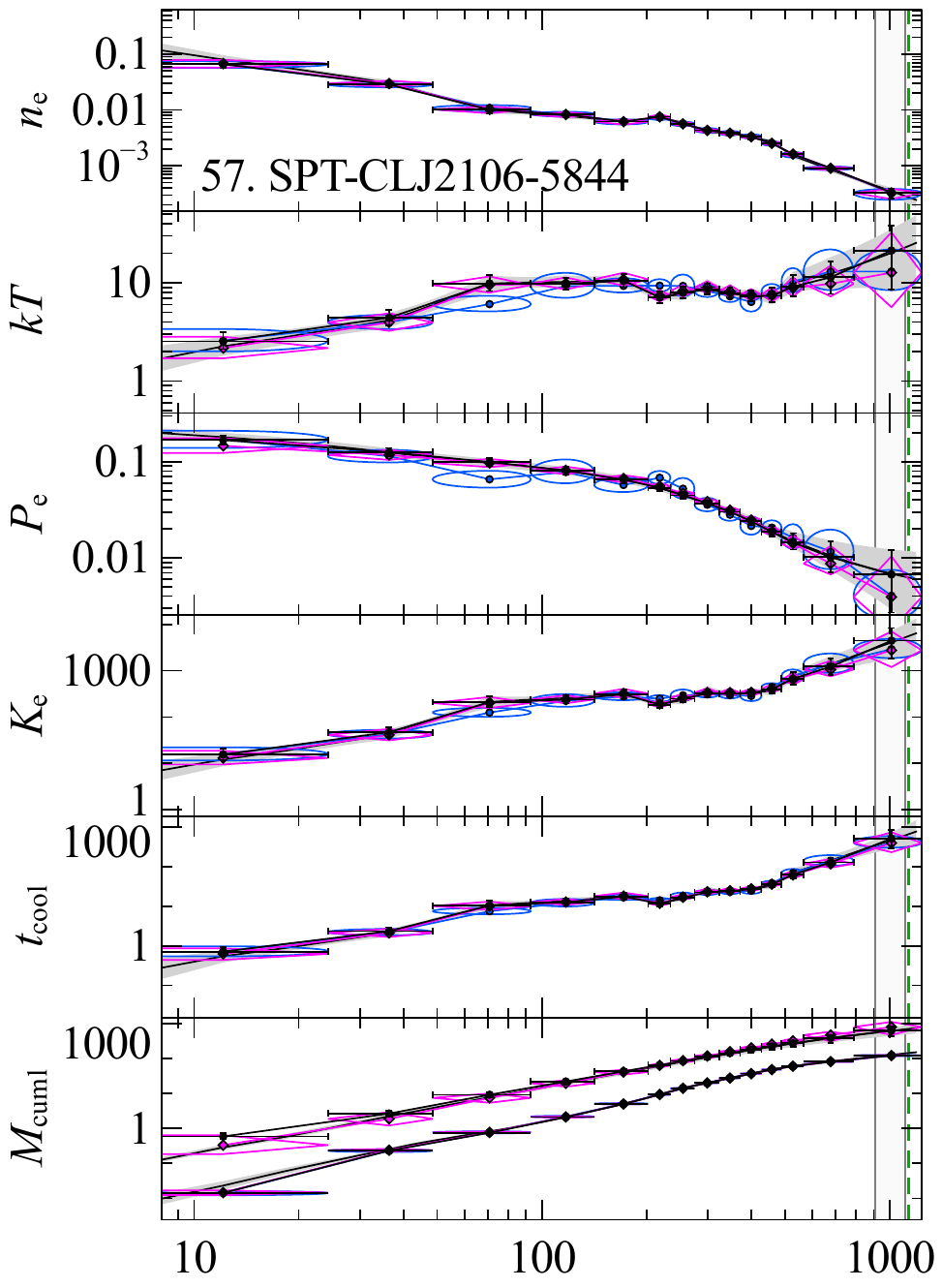}\\
  \includegraphics[width=0.3\textwidth]{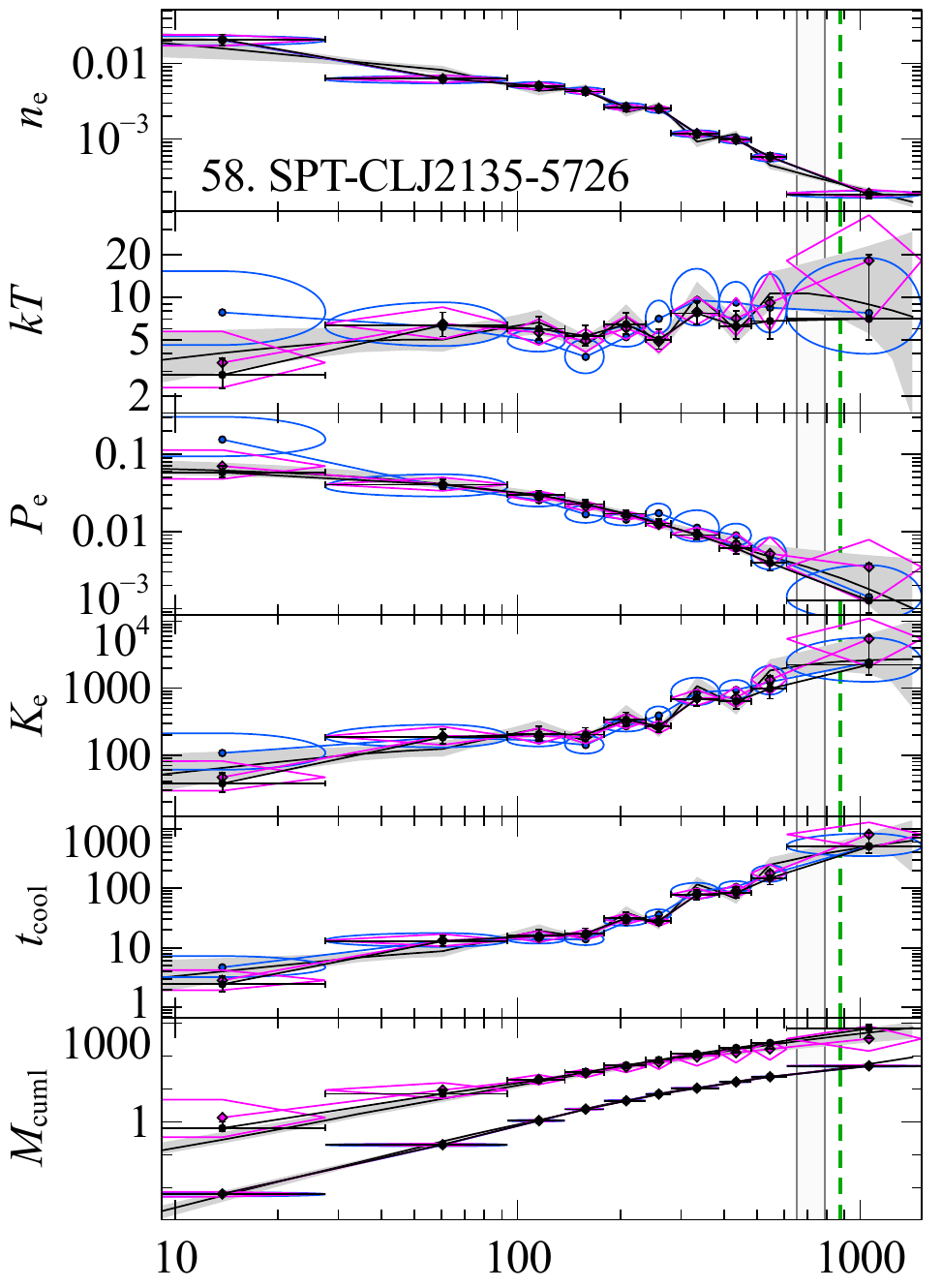}
  \includegraphics[width=0.3\textwidth]{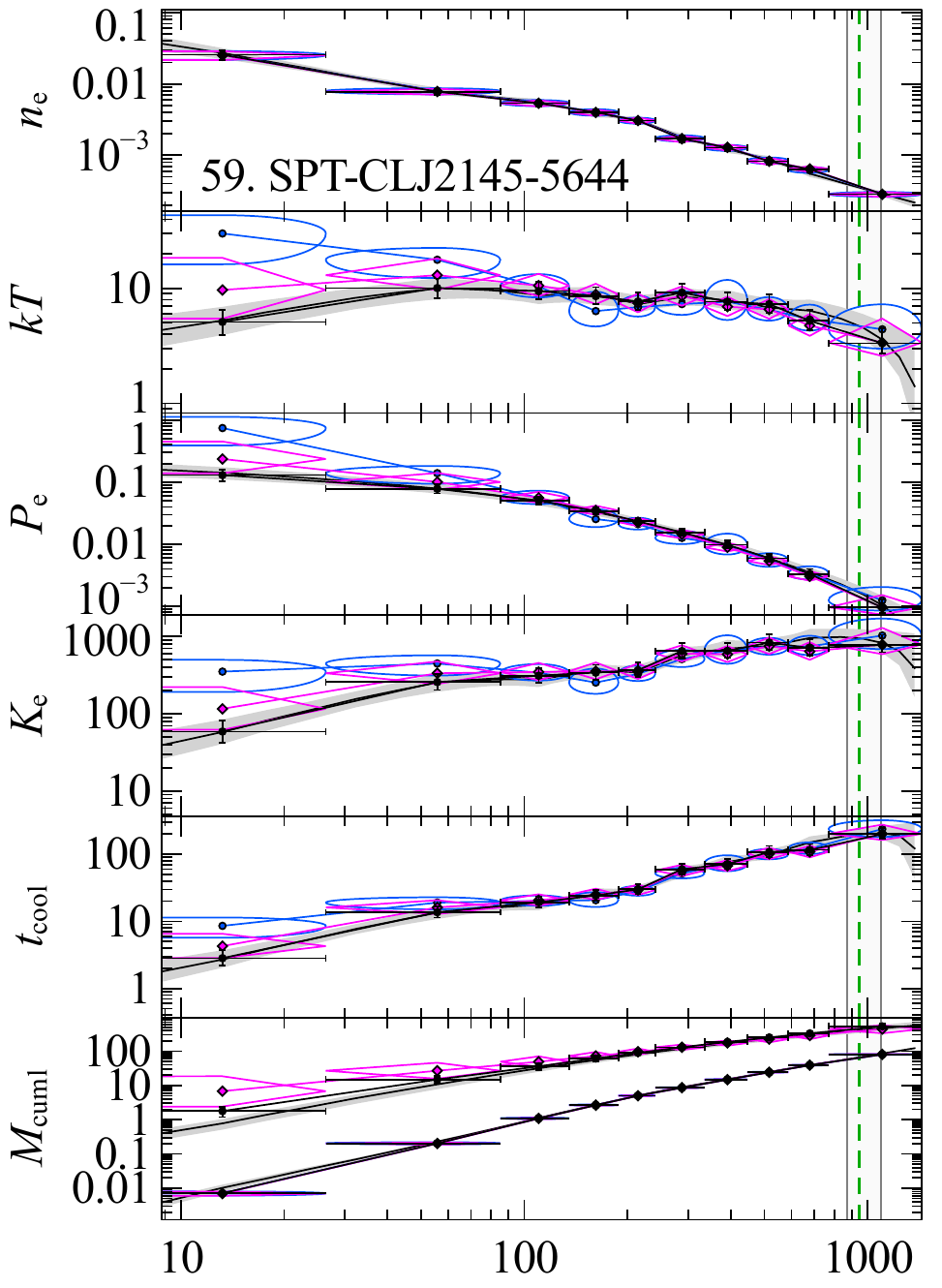}
  \includegraphics[width=0.3\textwidth]{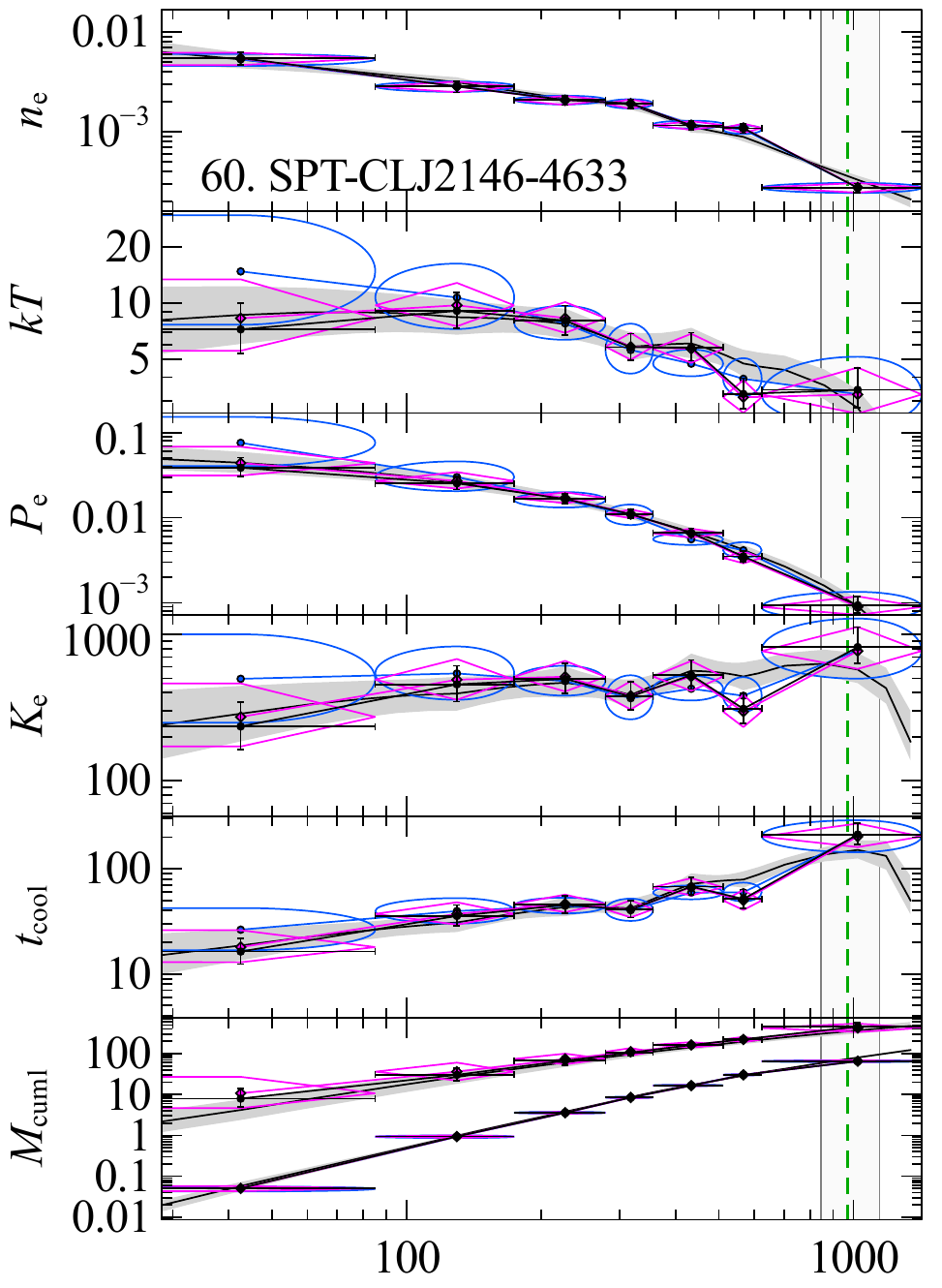}\\
  \includegraphics[width=0.3\textwidth]{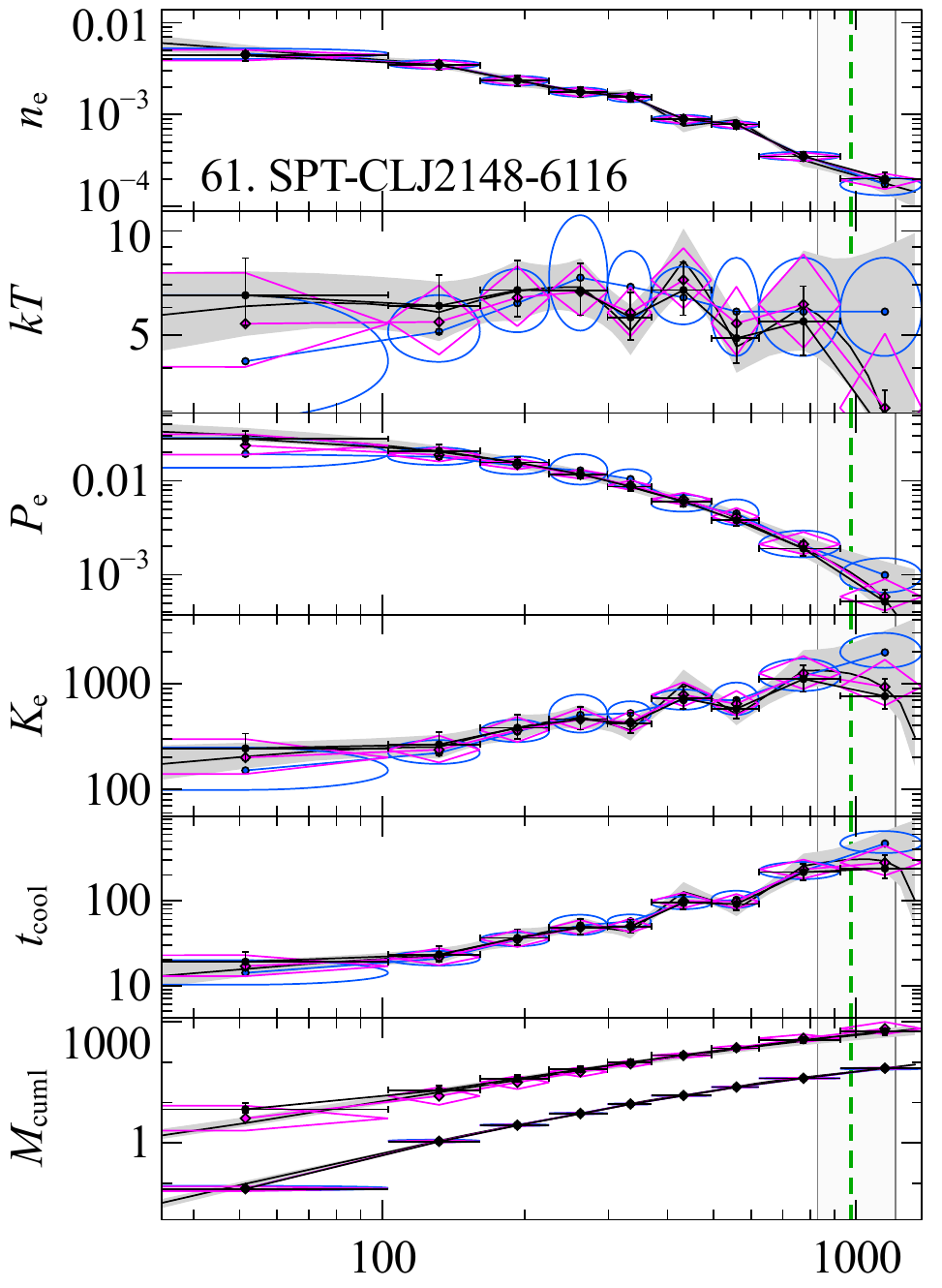}
  \includegraphics[width=0.3\textwidth]{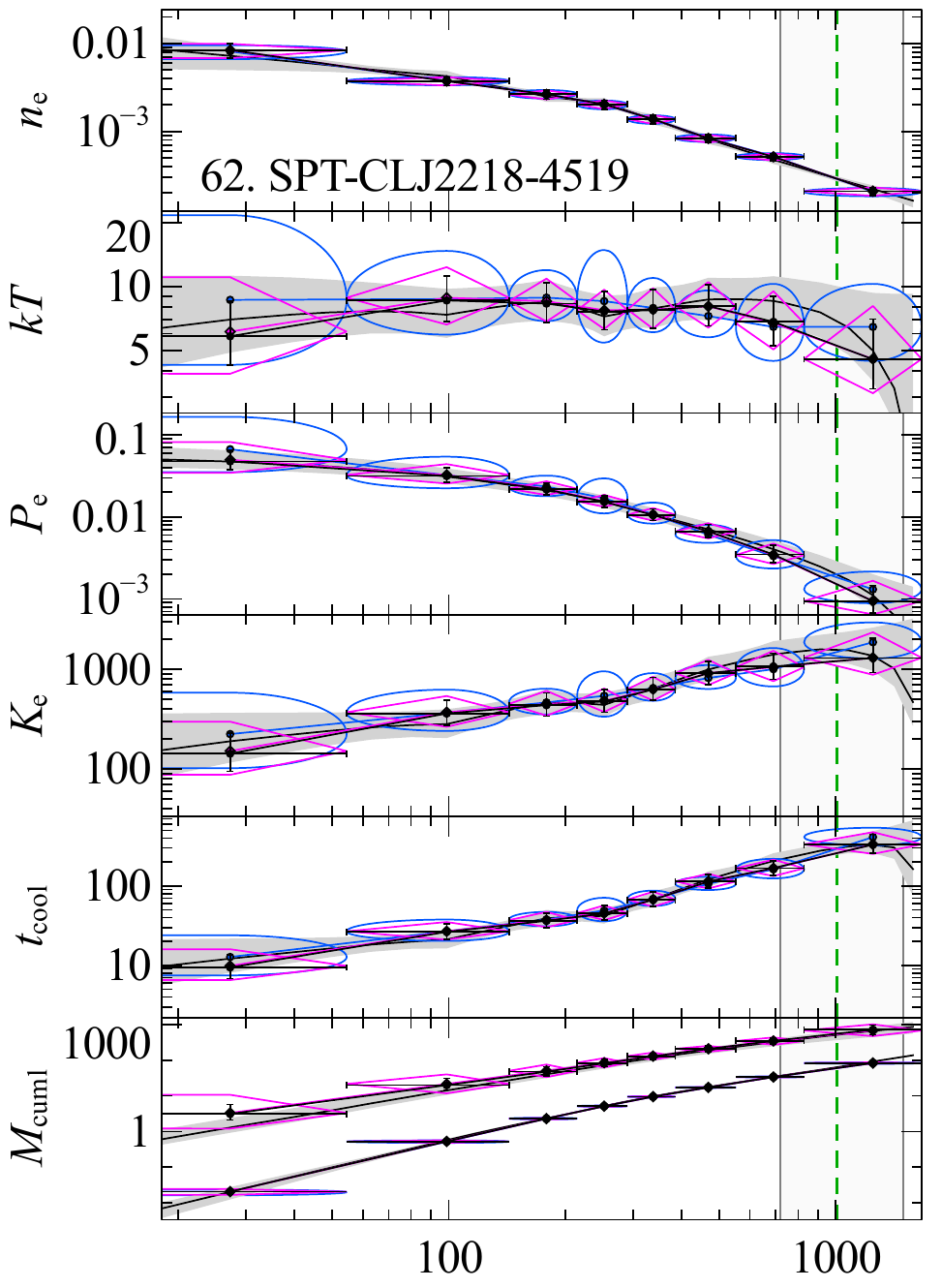}
  \includegraphics[width=0.3\textwidth]{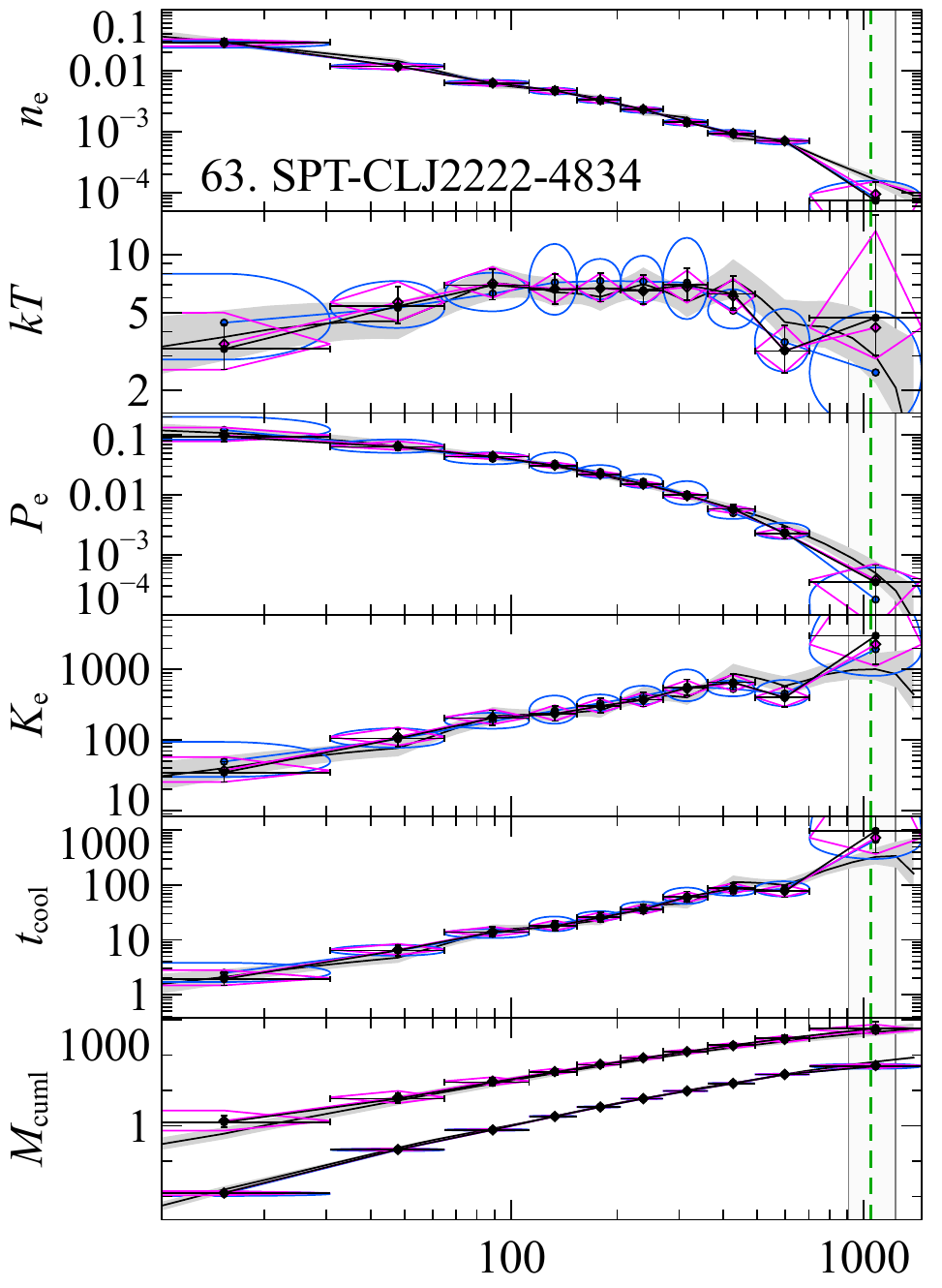}\\
  \contcaption{individual cluster profiles.}
\end{figure*}
\begin{figure*}
  \centering
  \includegraphics[width=0.3\textwidth]{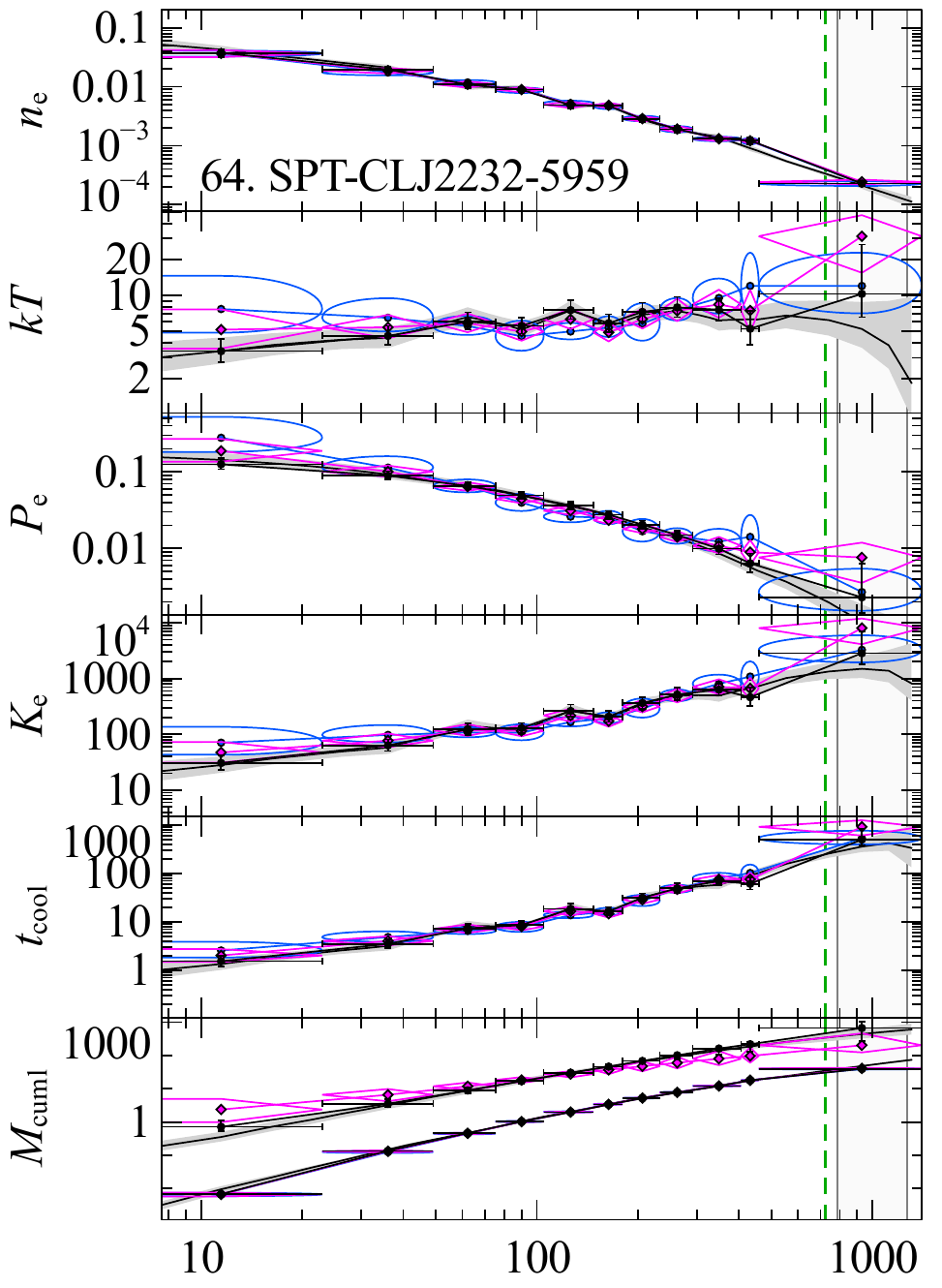}
  \includegraphics[width=0.3\textwidth]{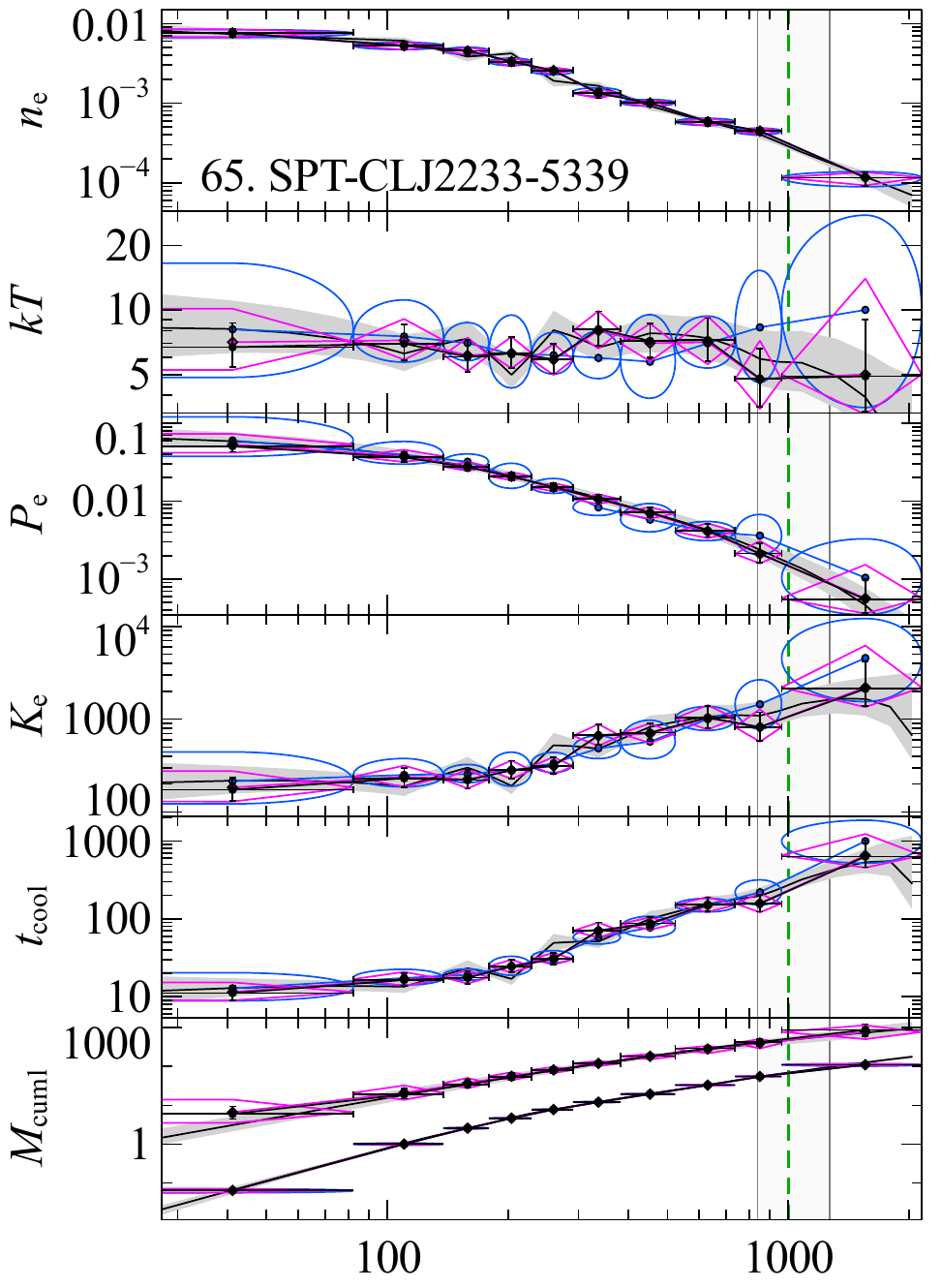}
  \includegraphics[width=0.3\textwidth]{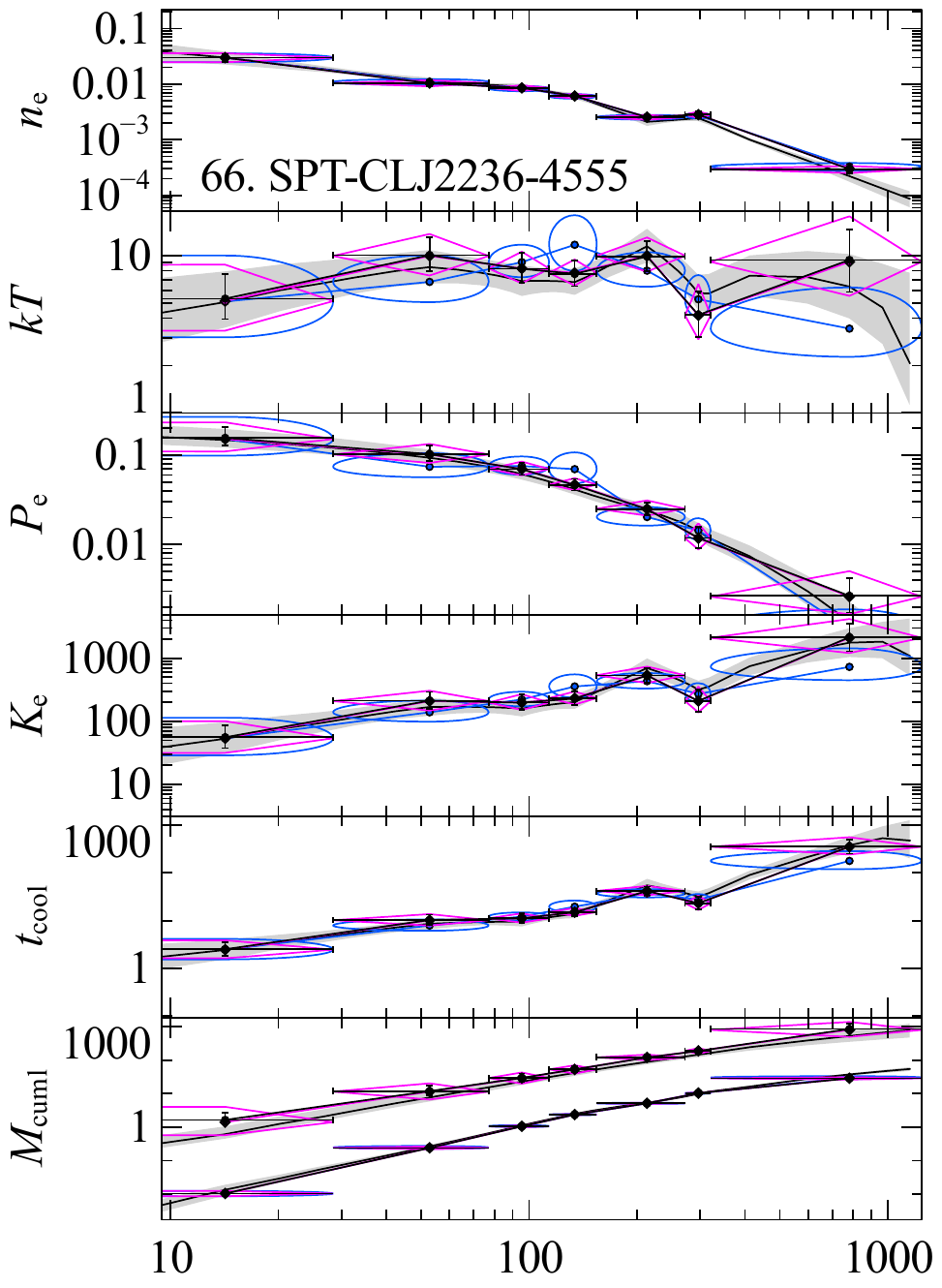}\\
  \includegraphics[width=0.3\textwidth]{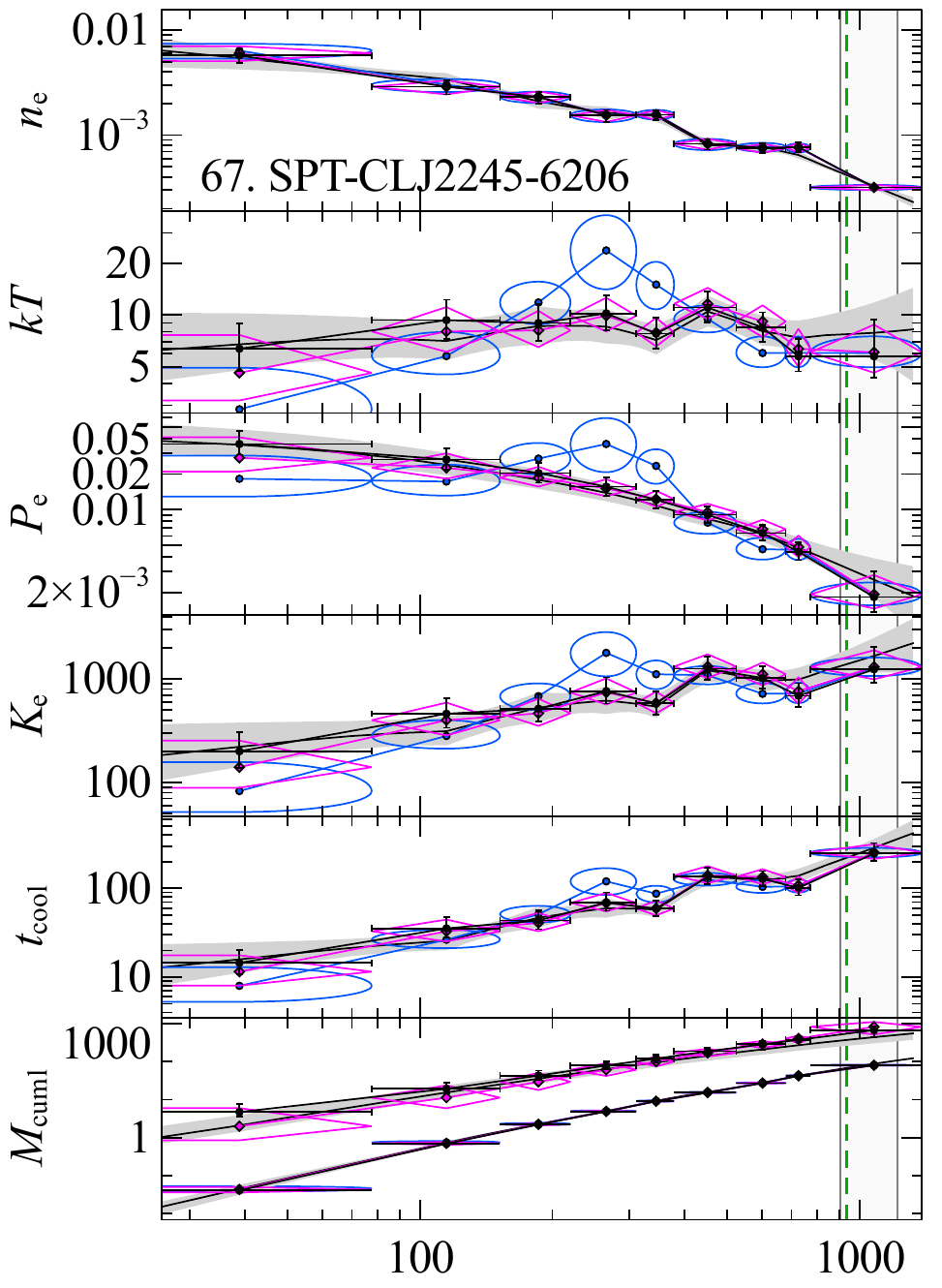}
  \includegraphics[width=0.3\textwidth]{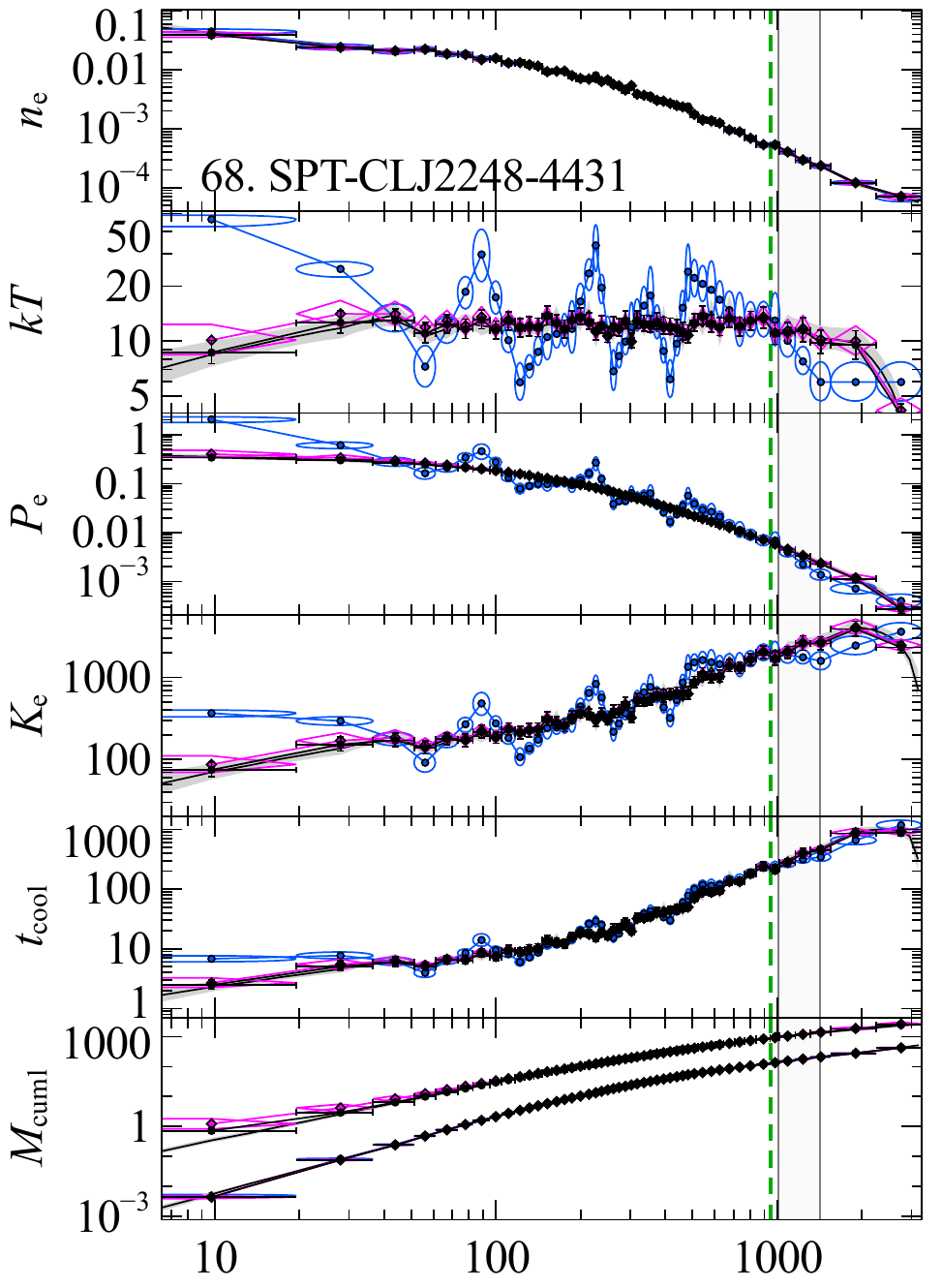}
  \includegraphics[width=0.3\textwidth]{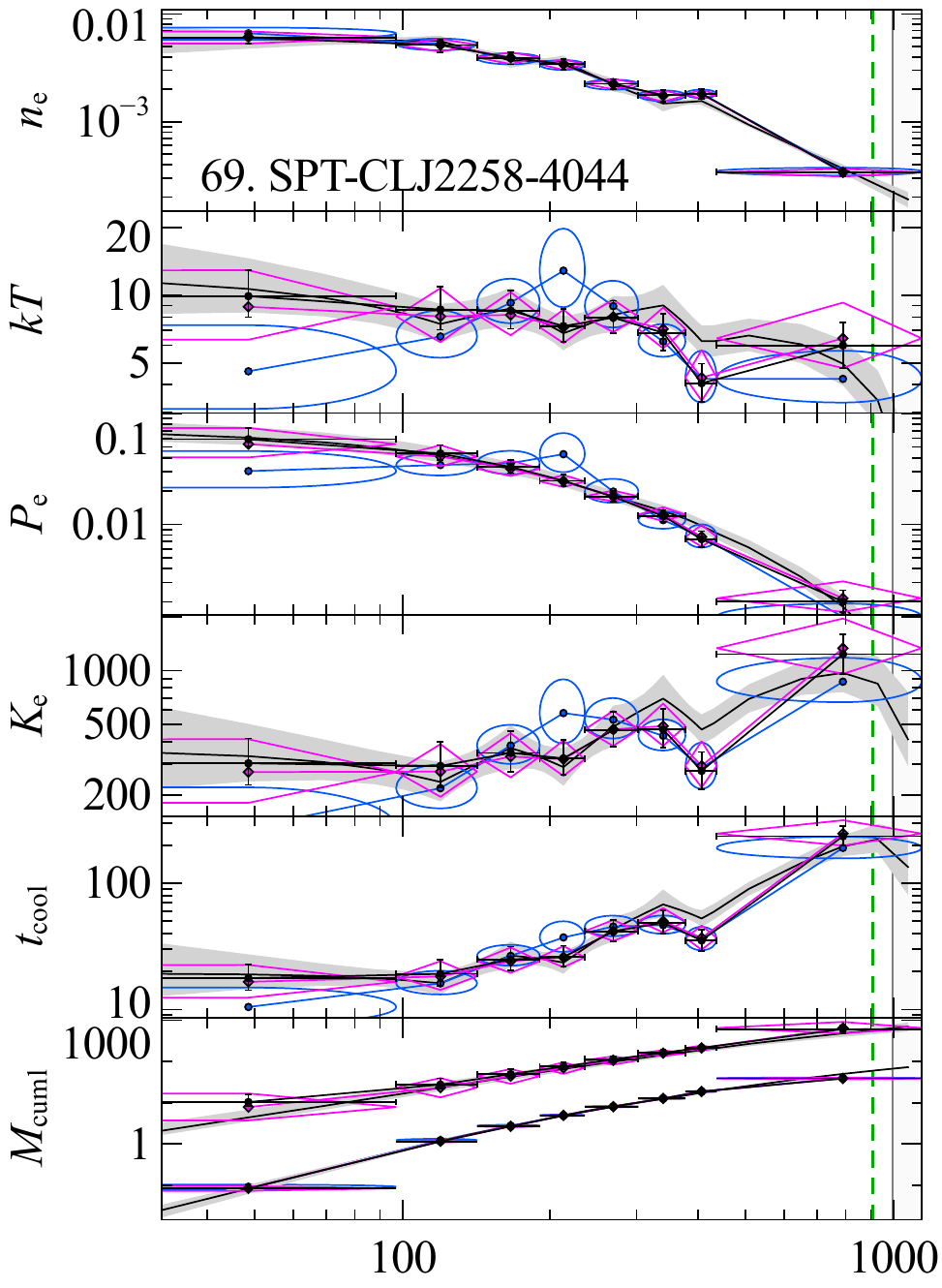}\\
  \includegraphics[width=0.3\textwidth]{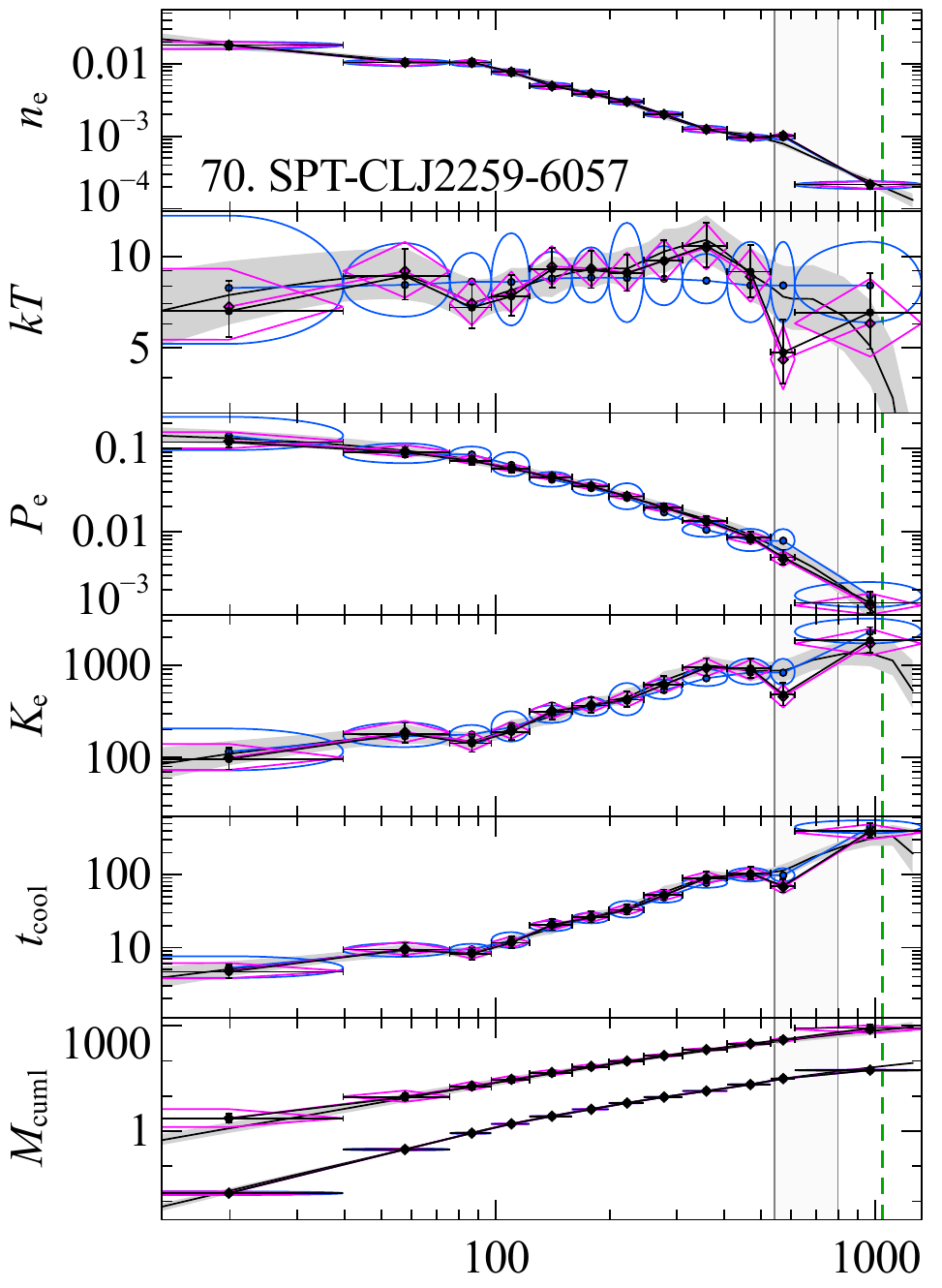}
  \includegraphics[width=0.3\textwidth]{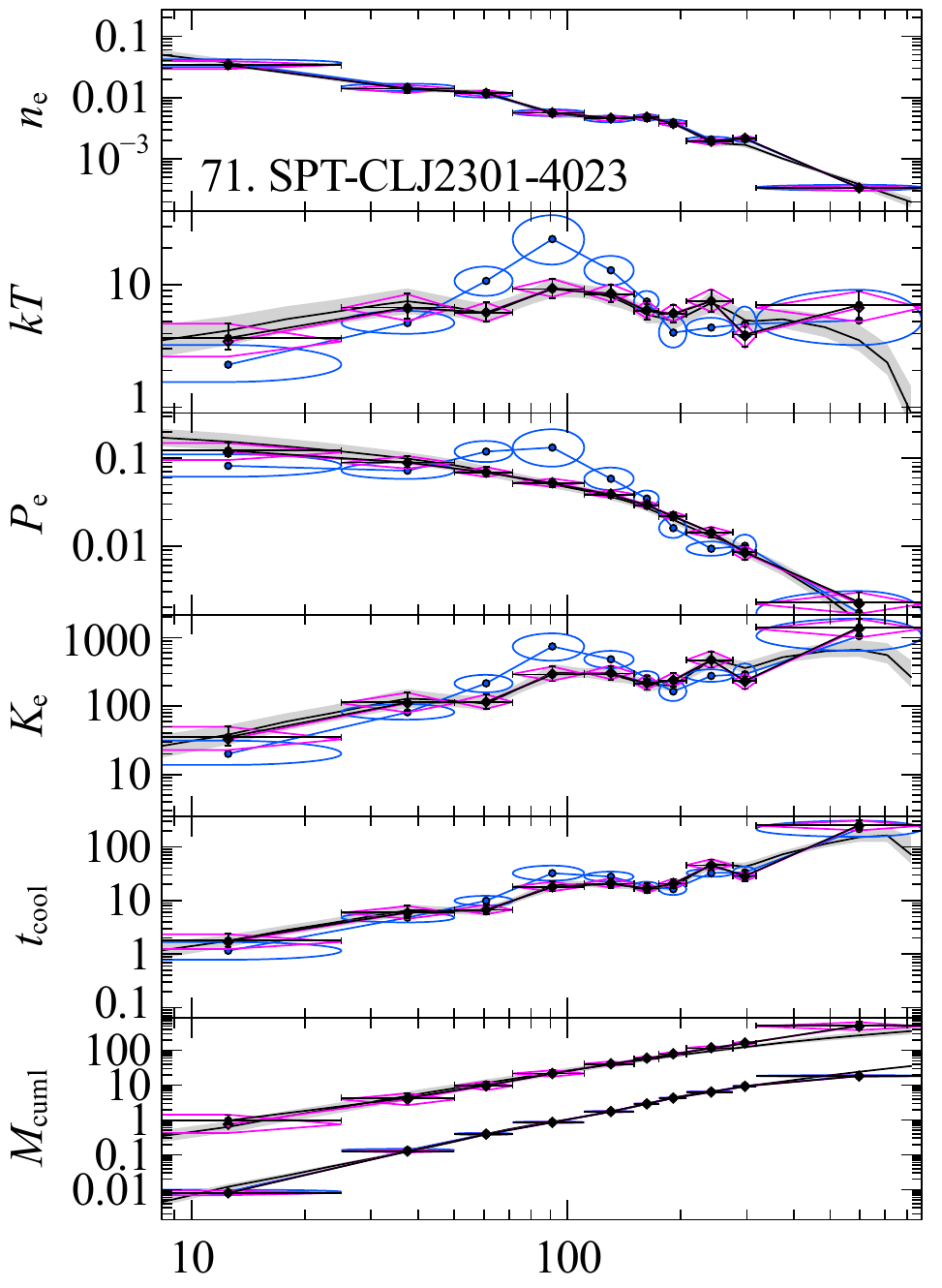}
  \includegraphics[width=0.3\textwidth]{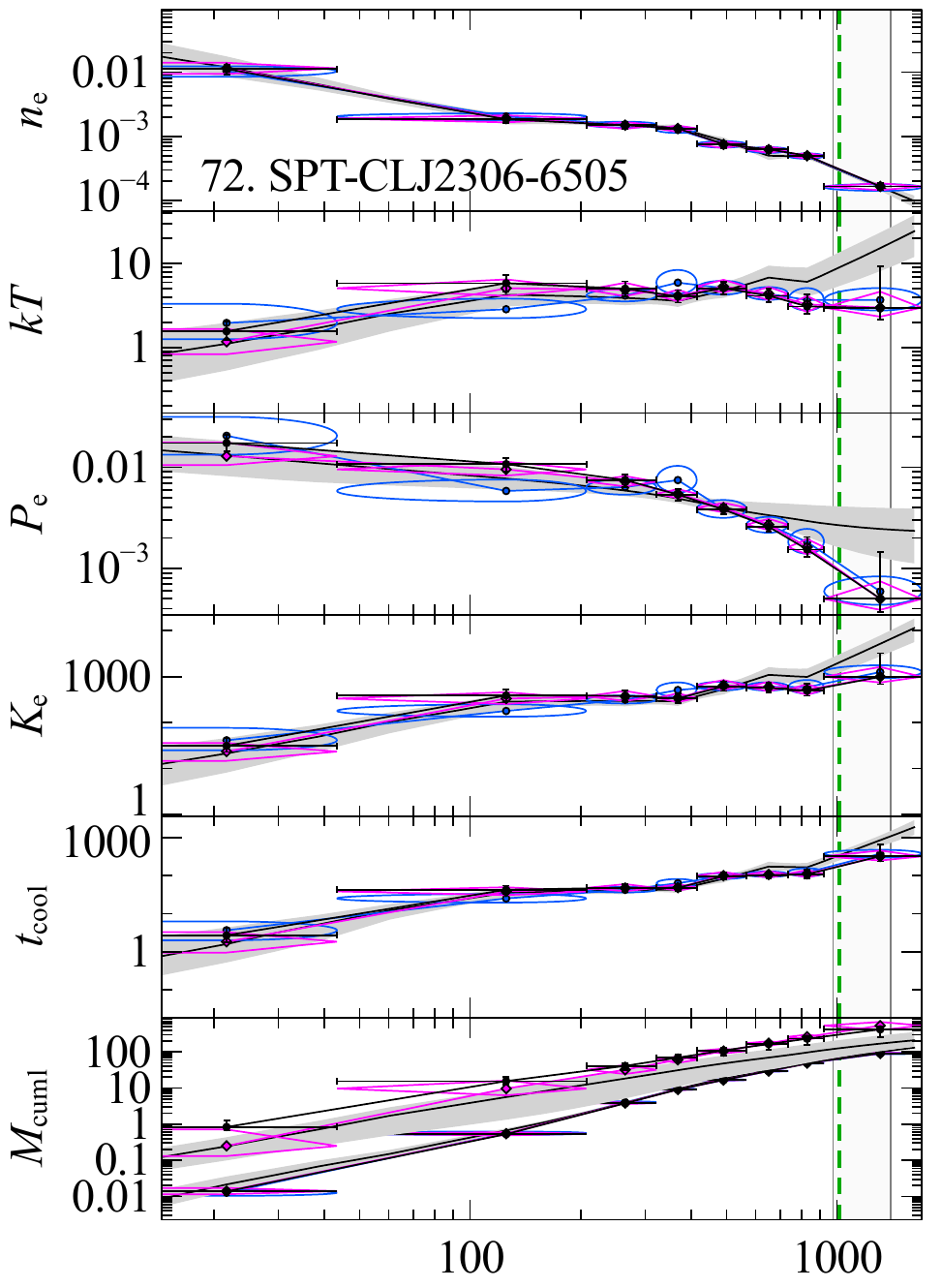}\\
  \contcaption{individual cluster profiles.}
\end{figure*}
\begin{figure*}
  \centering
  \includegraphics[width=0.3\textwidth]{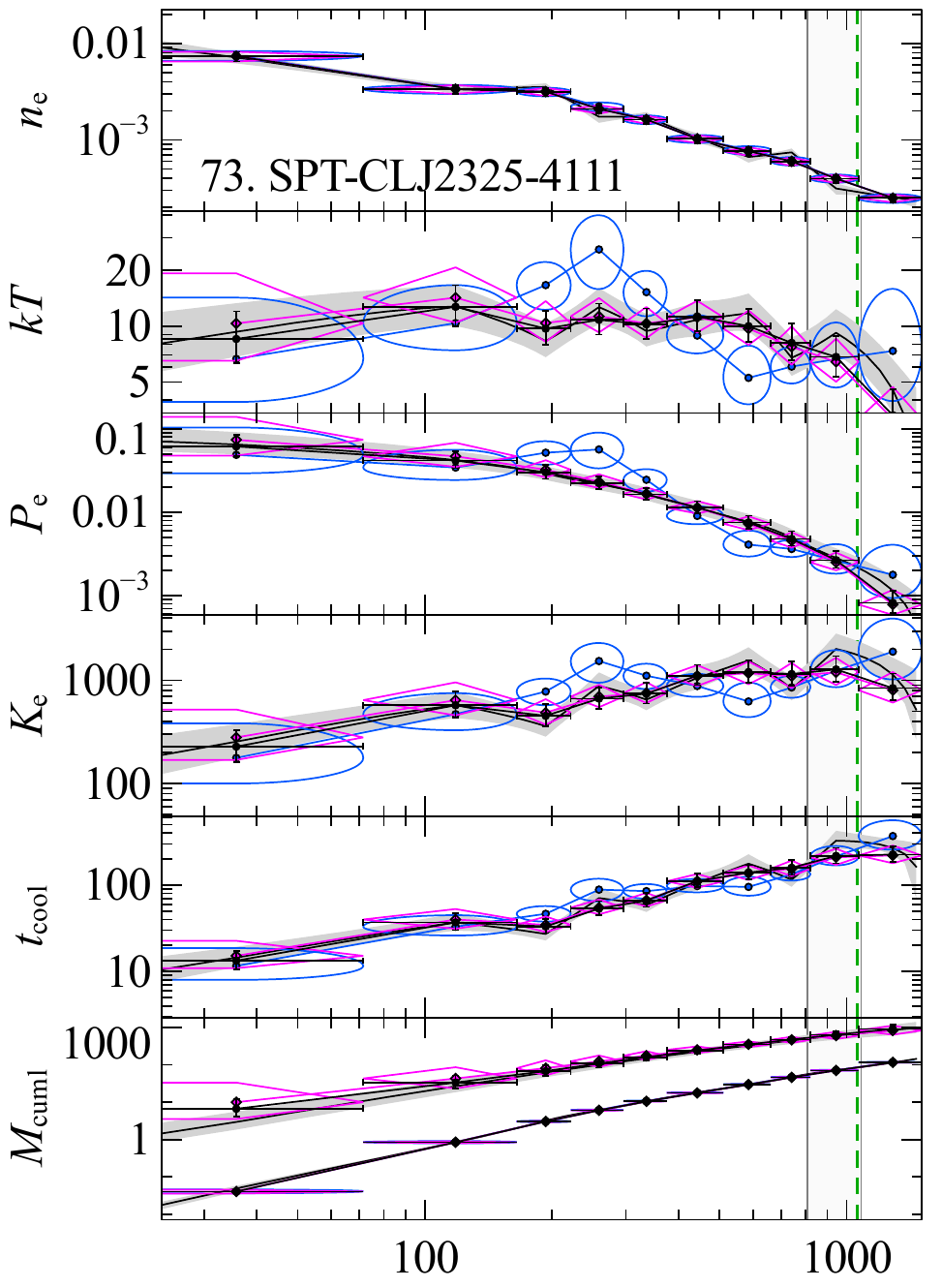}
  \includegraphics[width=0.3\textwidth]{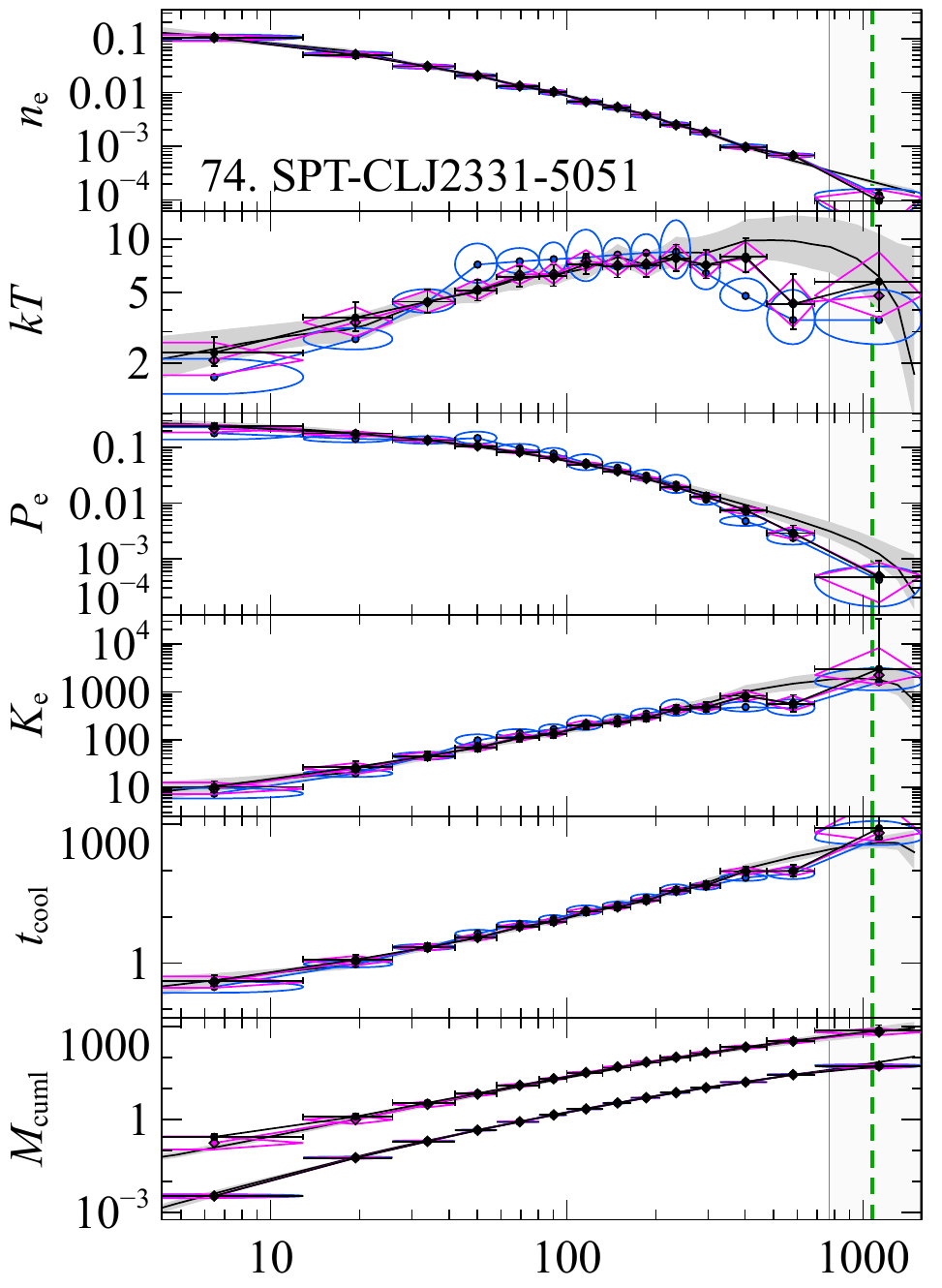}
  \includegraphics[width=0.3\textwidth]{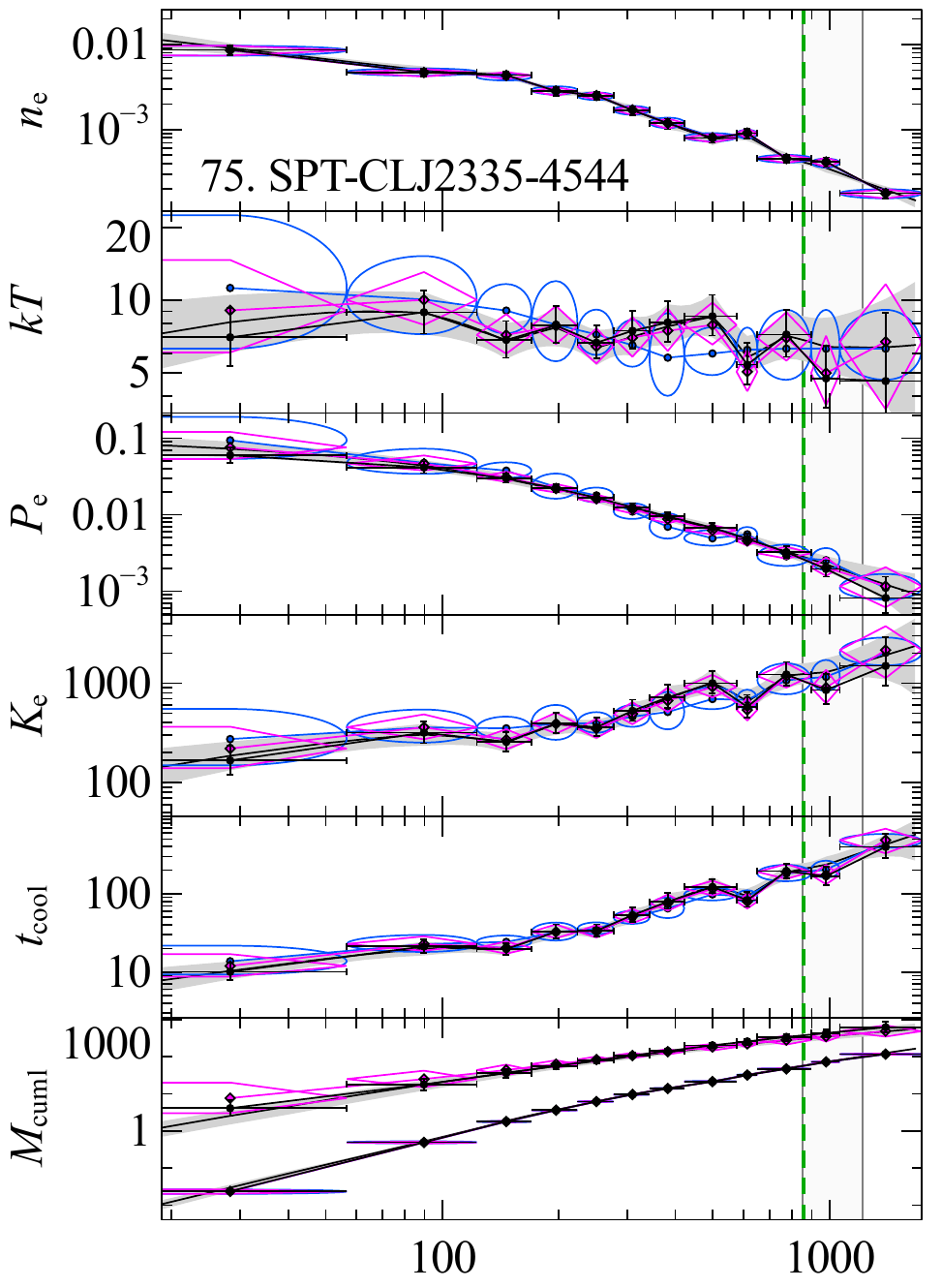}\\
  \includegraphics[width=0.3\textwidth]{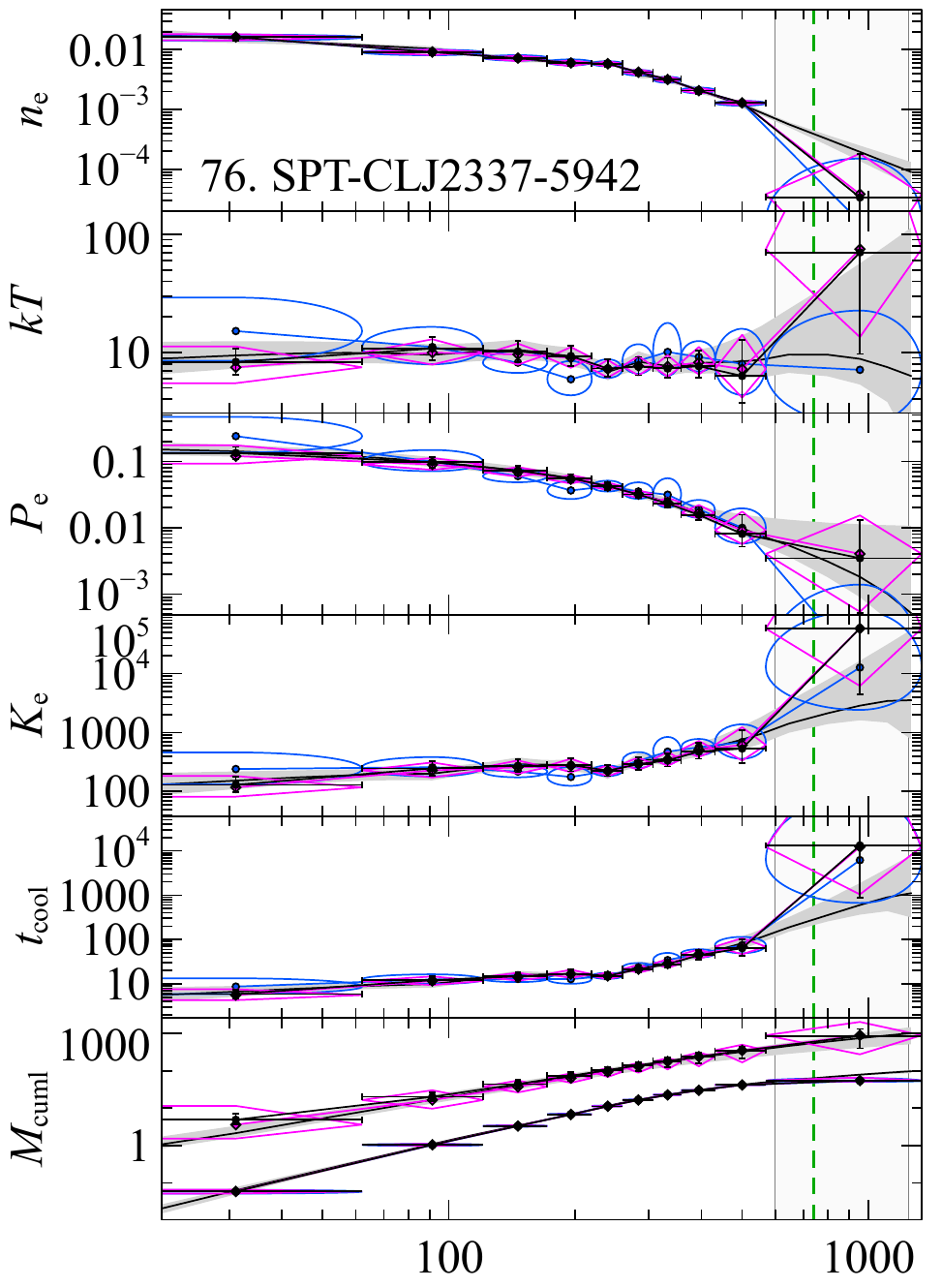}
  \includegraphics[width=0.3\textwidth]{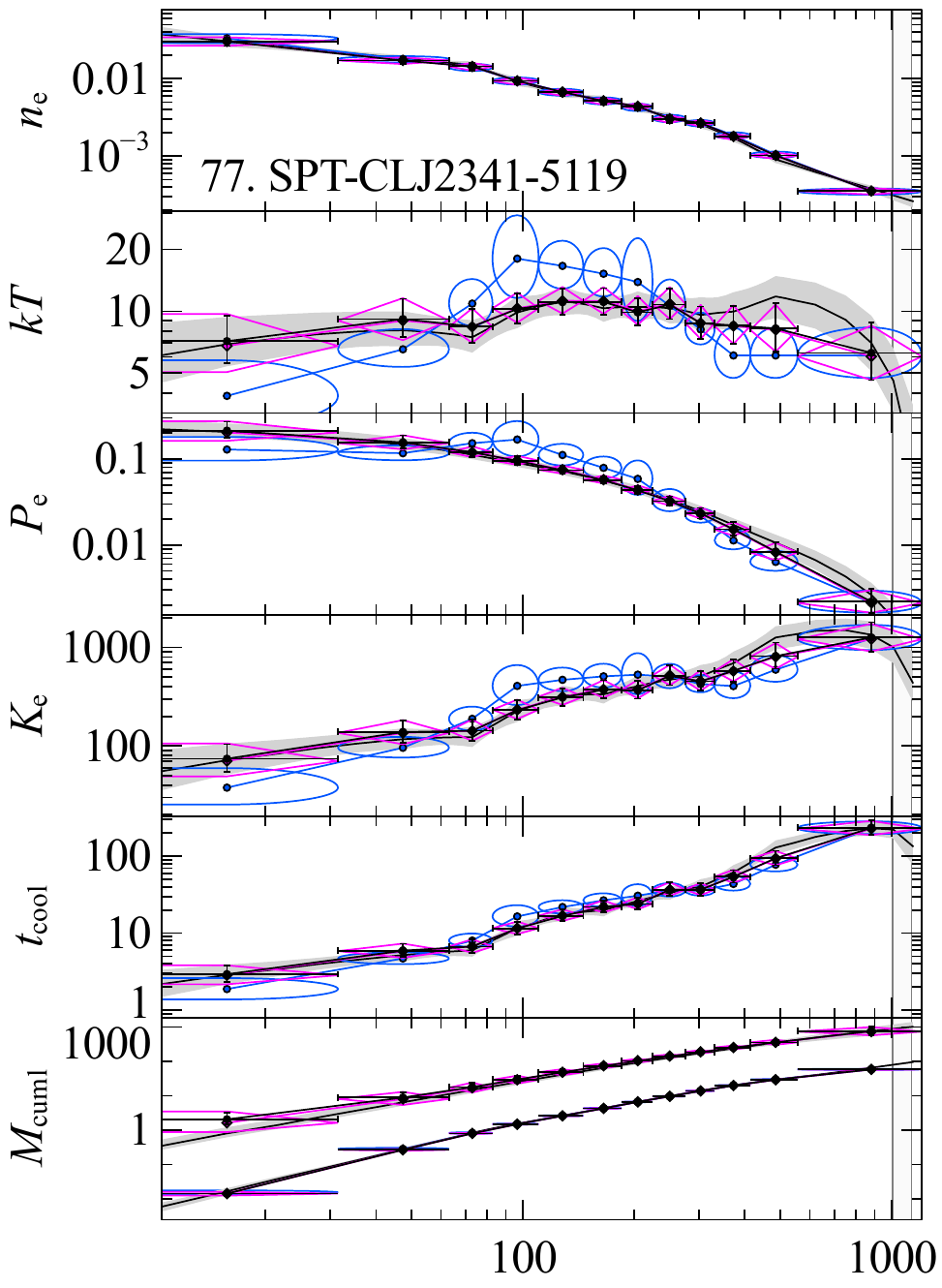}
  \includegraphics[width=0.3\textwidth]{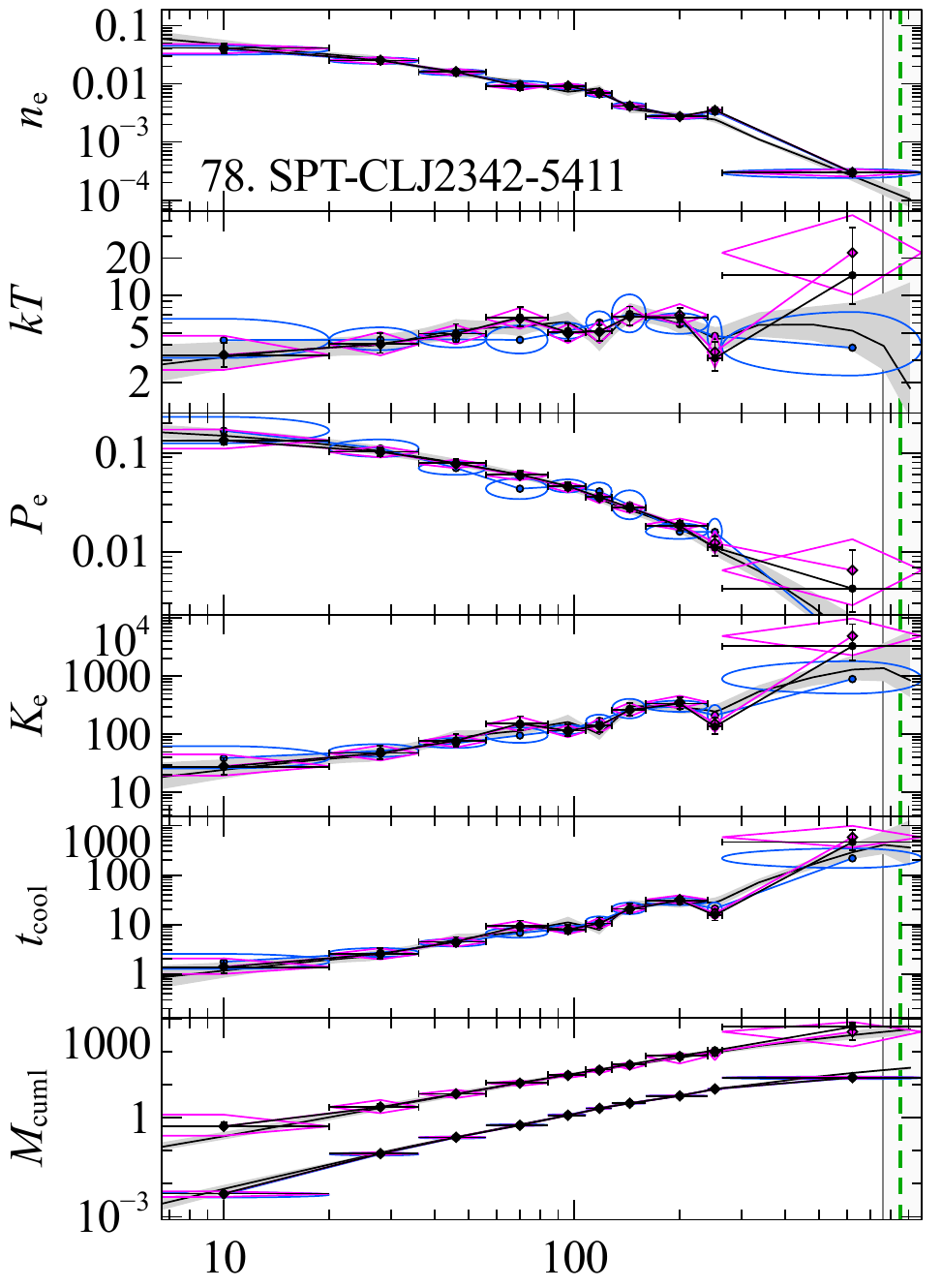}\\
  \includegraphics[width=0.3\textwidth]{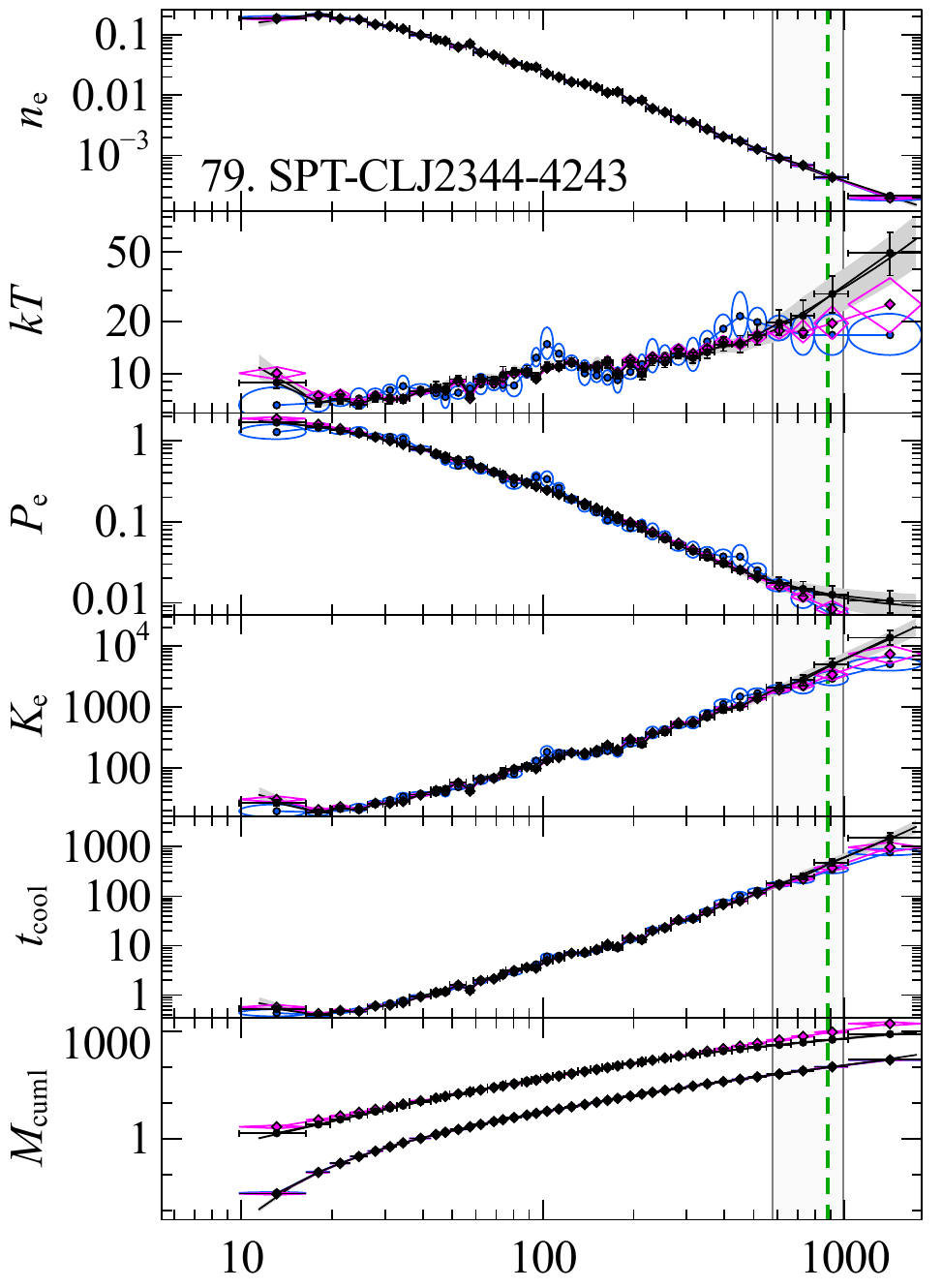}
  \includegraphics[width=0.3\textwidth]{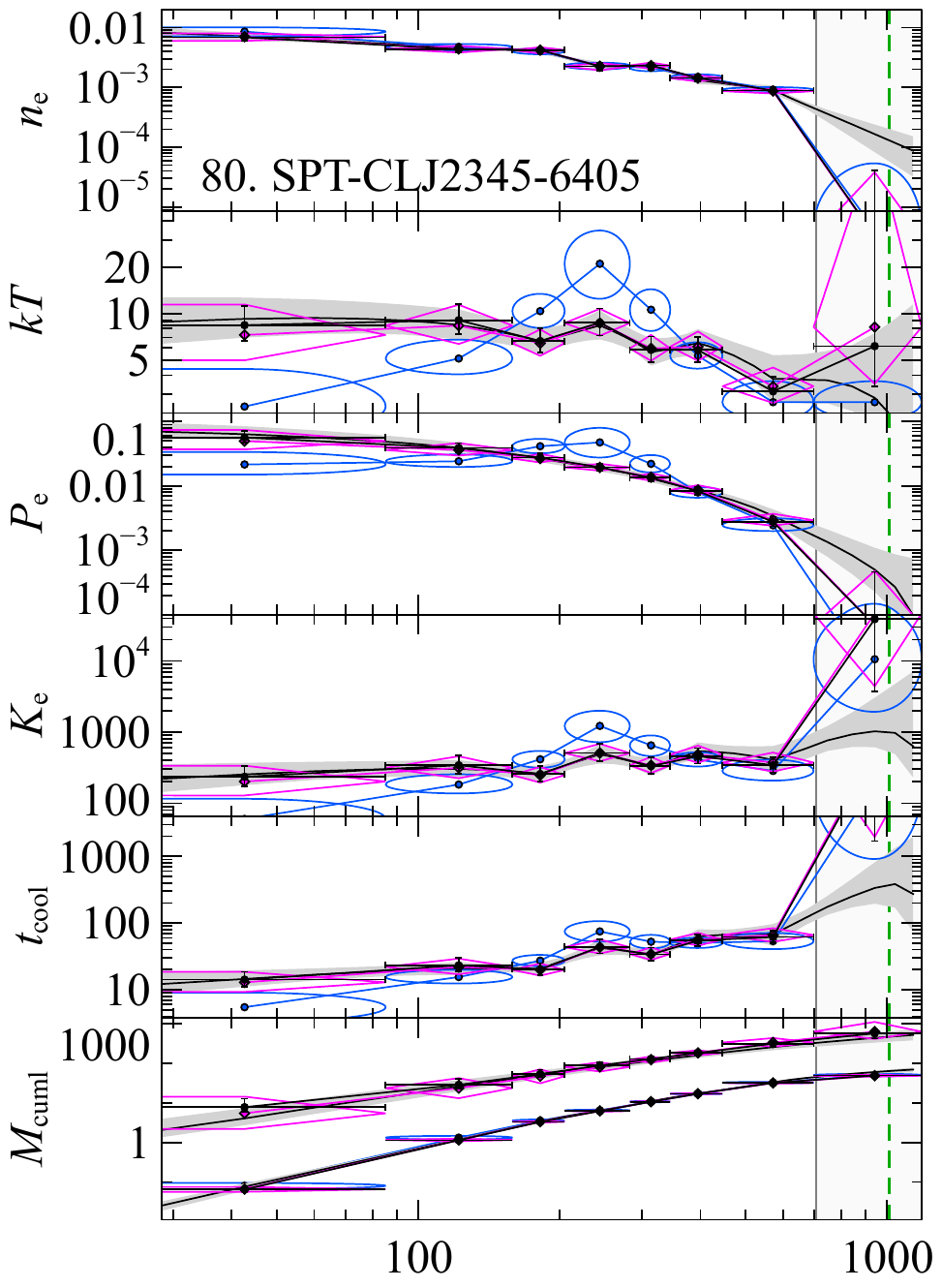}
  \includegraphics[width=0.3\textwidth]{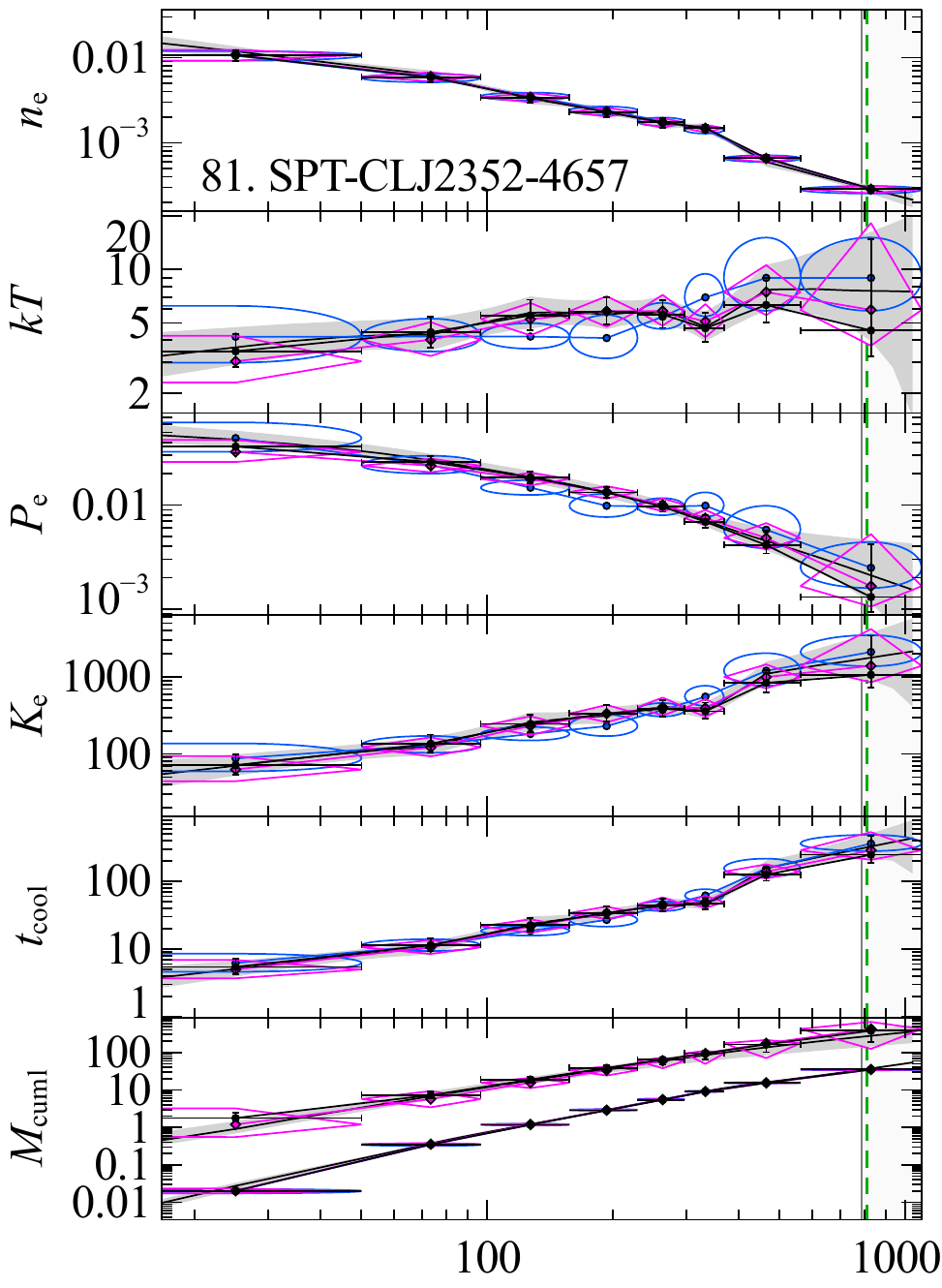}\\
  \contcaption{individual cluster profiles.}
\end{figure*}
\begin{figure*}
  \centering
  \includegraphics[width=0.3\textwidth]{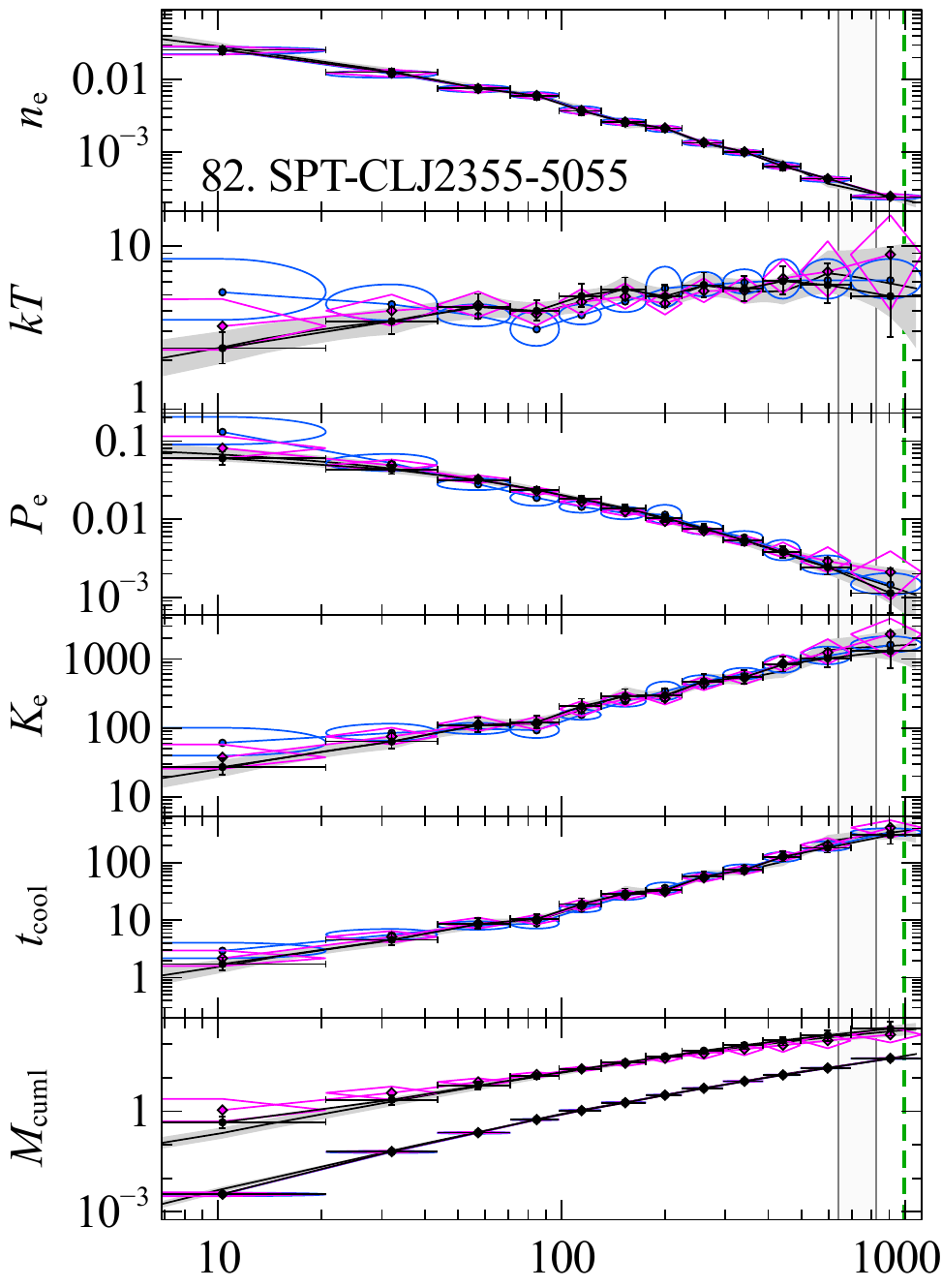}
  \includegraphics[width=0.3\textwidth]{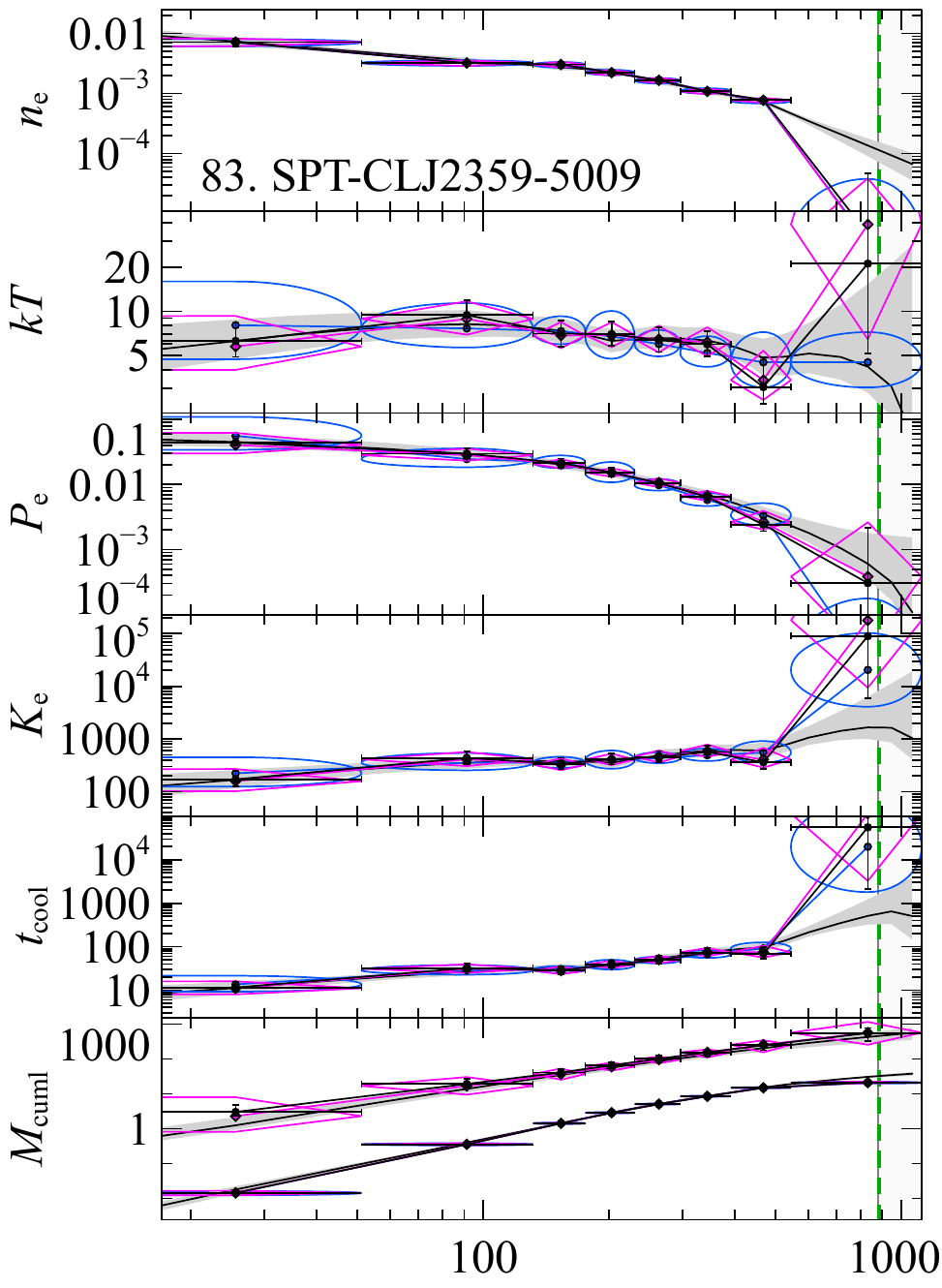}\\
  \contcaption{individual cluster profiles.}
\end{figure*}

\begin{table*}
\caption{
Table detailing individual profiles, provided electronically. The profiles for each cluster are repeated for each model (BIN-NFW, BIN-GNFW, BIN-NONHYDRO, INT-NFW, MBETA-NFW and KPLAW-NFW).
$r_\mathrm{in}$ and $r_\mathrm{out}$ are the inner and outer radii (kpc) of the annulus on the sky.
The median, 84.1 percentile (suffix hi) and 15.9 percentile (suffix lo) values for each physical quantity are given in the table.
The quantities and their units are, in order, temperature ($kT$; keV), electron density ($n_\mathrm{e}$; $\pcmcu$), electron entropy ($K_\mathrm{e}$; $\keV\cm^2$), electron pressure ($P_\mathrm{e}$; $\keV\pcmcu$), gravitational acceleration ($g$; $\cm\s^{-2}$; not valid for BIN-NONHYDRO), mean radiative cooling time ($t_\mathrm{cool}$; yr), mass deposition rate ($\dot{M}$; $\Msunpyr$), total cumulative mass ($M_\mathrm{tot}$; $\Msun$; not valid for BIN-NONHYDRO) and cumulative gas mass ($M_\mathrm{gas}$; $\Msun$; not valid for BIN-NONHYDRO).
}
\begin{tabular}{cccccccccccccc}
Name & Model & $r_\mathrm{in}$ & $r_\mathrm{out}$ & $kT$ & $kT_\mathrm{hi}$ & $kT_\mathrm{lo}$ & $n_\mathrm{e}$ & $n_\mathrm{e,hi}$ & $n_\mathrm{e,lo}$ & $K_\mathrm{e}$ & $K_\mathrm{e,hi}$ & $K_\mathrm{e,lo}$ & ... \\ \hline
SPT-CLJ0000-5748 & BIN-NFW & 0.0 & 10.6 & 2.47 & 2.99 & 2.06 &  0.181 & 0.203 & 0.158  &   8 &  10 &   6 & ... \\
SPT-CLJ0000-5748 & BIN-NFW & 10.6& 21.1 & 4.07 & 4.86 & 3.43 &  0.075 & 0.084 & 0.066  &  23 &  29 &  18 & ... \\
SPT-CLJ0000-5748 & BIN-NFW & 21.1& 35.2 & 5.40 & 6.31 & 4.67 &  0.040 & 0.045 & 0.036  &  46 &  58 &  38 & ... \\
SPT-CLJ0000-5748 & BIN-NFW & 35.2& 52.8 & 5.31 & 6.07 & 4.70 &  0.028 & 0.031 & 0.026  &  57 &  69 &  48 & ... \\
SPT-CLJ0000-5748 & BIN-NFW & 52.8& 73.9 & 6.93 & 8.08 & 6.06 &  0.015 & 0.017 & 0.013  & 112 & 142 &  92 & ... \\
...              & ...     & ... & .. . & ...  & ...  & ...  & ...    & ...   & ...    & ... & ... & ... & ... \\
\end{tabular}
\label{tab:individual}
\end{table*}

\end{document}

%% file: defn.tex
\newcommand{\Mpc}{\rm\thinspace Mpc}
\newcommand{\kpc}{\rm\thinspace kpc}

\newcommand{\km}{\rm\thinspace km}

\newcommand{\cm}{\rm\thinspace cm}

\newcommand{\cmsq}{\hbox{$\cm^2\,$}}

\newcommand{\pcmcu}{\hbox{$\cm^{-3}\,$}}

\newcommand{\yr}{\rm\thinspace yr}
\newcommand{\Gyr}{\rm\thinspace Gyr}
\newcommand{\s}{\rm\thinspace s}

\newcommand{\Msun}{\hbox{$\rm\thinspace M_{\odot}$}}

\newcommand{\Msunpyr}{\hbox{$\Msun\yr^{-1}\,$}}
\newcommand{\keV}{\rm\thinspace keV}

\newcommand{\kmps}{\hbox{$\km\s^{-1}\,$}}

\newcommand{\kmpspMpc}{\hbox{$\kmps\Mpc^{-1}$}}
\newcommand{\Zsun}{\hbox{$\thinspace \mathrm{Z}_{\odot}$}}

\newcommand{\pcmsq}{\hbox{$\cm^{-2}\,$}}

%% file: tab2.tex
10 kpc
&       $n_\mathrm{e}$ & $-1.61 \pm 0.04$ & $ 0.14 \pm 0.22$ & $ 0.31 \pm 0.03$ \\
&       $K_\mathrm{e}$ & $ 1.66 \pm 0.04$ & $-0.03 \pm 0.23$ & $ 0.29 \pm 0.04$ \\
&    $t_\mathrm{cool}$ & $ 9.39 \pm 0.05$ & $-0.07 \pm 0.26$ & $ 0.35 \pm 0.04$ \\
&       $P_\mathrm{e}$ & $-1.01 \pm 0.04$ & $ 0.27 \pm 0.22$ & $ 0.31 \pm 0.03$ \\
\hline
$0.015 R_{500}$
&       $n_\mathrm{e}$ & $ 1.36 \pm 0.04$ & $-0.31 \pm 0.22$ & $ 0.31 \pm 0.03$ \\
&       $K_\mathrm{e}$ & $-1.14 \pm 0.05$ & $ 0.26 \pm 0.25$ & $ 0.33 \pm 0.04$ \\
&    $t_\mathrm{cool}$ & $-1.50 \pm 0.05$ & $ 0.34 \pm 0.27$ & $ 0.38 \pm 0.04$ \\
&       $P_\mathrm{e}$ & $ 1.15 \pm 0.03$ & $-0.24 \pm 0.18$ & $ 0.25 \pm 0.02$ \\
\hline

%% file: taba1_1.tex
\begin{tabular}{lllllllrlp{3cm}}
Index & SPT ID & Identifier & $z$ & $N_\mathrm{H}$ & $r_\mathrm{max}$ & Rebin & Exp. & Counts & OBSIDs  \\ \hline
1 & SPT-CLJ0000-5748 &  & 0.702 & 1.4 & 3.0 & 20 & 29.7 & $1.8$ ($2.6$) & 9335 \\
2 & SPT-CLJ0013-4906 &  & 0.406 & 1.5 & 3.5 & 20 & 13.7 & $2.3$ ($1.4$) & 13462 \\
3 & SPT-CLJ0014-4952 &  & 0.752 & 1.5 & 3.0 & 20 & 54.5 & $2.4$ ($3.3$) & 13471 \\
4 & SPT-CLJ0033-6326 &  & 0.597 & 1.8 & 4.0 & 20 & 20.7 & $1.1$ ($2.1$) & 13483 \\
5 & SPT-CLJ0040-4407 &  & 0.35 & 3.5 & 5.0 & 20 & 7.9 & $2.1$ ($1.6$) & 13395 \\
6 & SPT-CLJ0058-6145 &  & 0.826 & 1.6 & 2.5 & 20 & 50.0 & $1.1$ ($2.4$) & 13479 \\
7 & SPT-CLJ0102-4603 & El Gordo & 0.722 & 1.7 & 2.5 & 20 & 59.0 & $0.9$ ($3.0$) & 13485 \\
8 & SPT-CLJ0102-4915 &  & 0.870 & 1.7 & 4.0 & 20 & 171.0 & $36.0$ ($20.9$) & 12258, 14023 \\
9 & SPT-CLJ0106-5943 &  & 0.348 & 1.8 & 5.0 & 20 & 17.3 & $2.6$ ($2.8$) & 13468 \\
10 & SPT-CLJ0123-4821 &  & 0.62 & 1.8 & 4.0 & 20 & 70.1 & $2.3$ ($6.9$) & 13491 \\
11 & SPT-CLJ0142-5032 &  & 0.73 & 2.2 & 3.0 & 20 & 28.6 & $1.0$ ($1.8$) & 13467 \\
12 & SPT-CLJ0151-5954 &  & 1.035 & 2.2 & 2.5 & 20 & 48.0 & $0.6$ ($2.4$) & 13480 \\
13 & SPT-CLJ0156-5541 &  & 1.221 & 3.1 & 2.0 & 20 & 76.6 & $1.0$ ($2.3$) & 13489 \\
14 & SPT-CLJ0200-4852 &  & 0.498 & 1.8 & 4.0 & 20 & 23.3 & $1.3$ ($2.5$) & 13487 \\
15 & SPT-CLJ0212-4657 &  & 0.655 & 1.7 & 3.0 & 20 & 27.8 & $1.1$ ($1.9$) & 13464 \\
16 & SPT-CLJ0217-5245 & MCXC J0217.2-5244 & 0.343 & 2.7 & 5.5 & 20 & 19.4 & $1.9$ ($5.0$) & 12269 \\
17 & SPT-CLJ0232-4421 & 2MAXI J0231-440 & 0.284 & 1.7 & 6.0 & 20 & 11.3 & $8.8$ ($3.0$) & 4993 \\
18 & SPT-CLJ0232-5257 &  & 0.556 & 2.8 & 4.5 & 20 & 19.4 & $1.1$ ($2.9$) & 12263 \\
19 & SPT-CLJ0234-5831 & 1RXS J023443.1-583114 & 0.415 & 2.7 & 4.0 & 20 & 9.3 & $2.0$ ($1.2$) & 13403 \\
20 & SPT-CLJ0235-5121 & PSZ1 G270.90-58.78 & 0.278 & 3.0 & 6.5 & 20 & 19.6 & $5.4$ ($5.1$) & 12262 \\
21 & SPT-CLJ0236-4938 & ACT-CLJ0237-4939 & 0.334 & 2.5 & 4.5 & 20 & 38.6 & $3.0$ ($6.2$) & 12266 \\
22 & SPT-CLJ0243-5930 &  & 0.635 & 2.4 & 3.0 & 20 & 45.6 & $2.5$ ($2.9$) & 13484, 15573 \\
23 & SPT-CLJ0252-4824 &  & 0.421 & 2.5 & 4.0 & 20 & 29.6 & $1.8$ ($3.3$) & 13494 \\
24 & SPT-CLJ0256-5617 &  & 0.58 & 1.4 & 3.5 & 20 & 46.8 & $2.4$ ($4.3$) & 13481, 14448 \\
25 & SPT-CLJ0304-4401 &  & 0.458 & 1.3 & 4.5 & 20 & 14.7 & $2.2$ ($2.1$) & 13402 \\
26 & SPT-CLJ0304-4921 &  & 0.392 & 1.8 & 5.5 & 20 & 20.8 & $4.1$ ($4.4$) & 12265 \\
27 & SPT-CLJ0307-5042 &  & 0.55 & 1.8 & 3.5 & 20 & 38.2 & $2.4$ ($3.4$) & 13476 \\
28 & SPT-CLJ0307-6225 &  & 0.579 & 2.1 & 4.0 & 20 & 24.2 & $1.2$ ($3.4$) & 12191 \\
29 & SPT-CLJ0310-4647 &  & 0.709 & 1.8 & 3.0 & 20 & 36.1 & $1.1$ ($2.4$) & 13492 \\
30 & SPT-CLJ0324-6236 &  & 0.73 & 2.8 & 2.5 & 20 & 53.3 & $1.7$ ($3.0$) & 12181, 13137, 13213 \\
31 & SPT-CLJ0334-4659 &  & 0.485 & 1.2 & 4.0 & 20 & 25.3 & $2.4$ ($2.6$) & 13470 \\
32 & SPT-CLJ0346-5439 & ACT-CLJ0346-5438 & 0.530 & 1.4 & 3.5 & 20 & 33.6 & $2.4$ ($3.5$) & 12270, 13155 \\
33 & SPT-CLJ0348-4515 &  & 0.358 & 0.9 & 4.5 & 20 & 12.4 & $1.3$ ($1.7$) & 13465 \\
34 & SPT-CLJ0352-5647 &  & 0.67 & 1.4 & 3.5 & 20 & 41.9 & $1.5$ ($3.4$) & 13490, 15571 \\
35 & SPT-CLJ0406-4805 &  & 0.737 & 1.3 & 3.0 & 20 & 25.8 & $0.7$ ($1.8$) & 13477 \\
36 & SPT-CLJ0411-4819 & PLCKESZ G255.62-46.16 & 0.424 & 1.5 & 6.5 & 20 & 65.1 & $11.9$ ($15.7$) & 13396, 16355, 17536 \\
37 & SPT-CLJ0417-4748 &  & 0.581 & 1.3 & 3.0 & 20 & 21.2 & $2.7$ ($1.5$) & 13397 \\
38 & SPT-CLJ0426-5455 &  & 0.63 & 0.8 & 3.5 & 20 & 31.8 & $1.0$ ($2.5$) & 13472 \\
39 & SPT-CLJ0438-5419 & PLCKESZ G262.71-40.91 & 0.421 & 1.0 & 6.0 & 20 & 19.6 & $6.4$ ($5.0$) & 12259 \\
40 & SPT-CLJ0441-4855 &  & 0.79 & 1.5 & 3.0 & 20 & 67.5 & $2.3$ ($4.8$) & 13475, 14371, 14372 \\
41 & SPT-CLJ0449-4901 &  & 0.792 & 1.2 & 2.5 & 20 & 49.8 & $1.4$ ($2.4$) & 13473 \\
42 & SPT-CLJ0456-5116 &  & 0.562 & 1.1 & 3.5 & 20 & 49.8 & $2.5$ ($4.6$) & 13474 \\
43 & SPT-CLJ0509-5342 & ACT-CLJ0509-5341 & 0.461 & 1.5 & 3.5 & 20 & 28.2 & $2.8$ ($3.2$) & 9432 \\
44 & SPT-CLJ0516-5430 & Abell S520 & 0.295 & 2.1 & 8.5 & 20 & 30.4 & $13.5$ ($11.4$) & 9331, 15099 \\
45 & SPT-CLJ0528-5300 &  & 0.768 & 3.2 & 2.5 & 20 & 122.1 & $1.6$ ($6.9$) & 11747, 11874, 12092, 13126, 9341, 10862, 11996 \\
46 & SPT-CLJ0533-5005 &  & 0.881 & 2.9 & 1.5 & 40 & 71.7 & $0.4$ ($1.6$) & 11748, 12001, 12002 \\
47 & SPT-CLJ0542-4100 &  & 0.642 & 3.2 & 3.5 & 20 & 48.7 & $2.5$ ($3.7$) & 914 \\
48 & SPT-CLJ0546-5345 & 1RXS J054638.7-534434 & 1.066 & 6.8 & 3.5 & 20 & 69.0 & $2.1$ ($7.7$) & 9332, 9336, 10851, 10864, 11739 \\
49 & SPT-CLJ0555-6406 &  & 0.345 & 4.0 & 6.5 & 20 & 10.9 & $2.3$ ($2.9$) & 13404 \\
50 & SPT-CLJ0559-5249 & 1RXS J055942.1-524950 & 0.609 & 5.1 & 3.5 & 20 & 106.9 & $6.1$ ($11.5$) & 12264, 13116, 13117 \\
51 & SPT-CLJ0655-5234 &  & 0.470 & 4.6 & 3.5 & 20 & 20.0 & $0.9$ ($2.1$) & 13486 \\
52 & SPT-CLJ0658-5556 & Bullet cluster & 0.296 & 4.9 & - & 5 & 531.2 & $729.6$ ($282.5$) & 5355, 5356, 5357, 5358, 3184, 5361, 4984, 4985, 4986 \\
53 & SPT-CLJ2031-4037 & MCXC J2031.8-4037 & 0.342 & 3.4 & 8.0 & 20 & 9.9 & $3.9$ ($3.2$) & 13517 \\
54 & SPT-CLJ2034-5936 &  & 0.919 & 5.7 & 3.0 & 20 & 57.7 & $1.4$ ($3.0$) & 12182 \\
55 & SPT-CLJ2035-5251 &  & 0.528 & 2.9 & 3.5 & 20 & 18.0 & $0.9$ ($1.8$) & 13466 \\
56 & SPT-CLJ2043-5035 &  & 0.723 & 2.4 & 3.5 & 20 & 77.8 & $5.7$ ($7.6$) & 13478 \\
\hline
\end{tabular}

%% file: taba1_2.tex
\begin{tabular}{lllllllrlp{3cm}}
Index & SPT ID & Identifier & $z$ & $N_\mathrm{H}$ & $r_\mathrm{max}$ & Rebin & Exp. & Counts & OBSIDs  \\ \hline
57 & SPT-CLJ2106-5844 &  & 1.132 & 4.3 & 2.5 & 20 & 72.1 & $3.7$ ($3.8$) & 12180, 12189 \\
58 & SPT-CLJ2135-5726 &  & 0.427 & 2.8 & 4.5 & 20 & 16.5 & $1.6$ ($2.3$) & 13463 \\
59 & SPT-CLJ2145-5644 & 1RXS J214559.3-564455 & 0.48 & 2.6 & 4.0 & 20 & 14.1 & $1.7$ ($1.3$) & 13398 \\
60 & SPT-CLJ2146-4633 &  & 0.933 & 1.6 & 3.0 & 20 & 79.6 & $1.6$ ($5.6$) & 13469 \\
61 & SPT-CLJ2148-6116 &  & 0.571 & 3.3 & 3.5 & 20 & 36.3 & $2.0$ ($2.8$) & 13488 \\
62 & SPT-CLJ2218-4519 &  & 0.65 & 1.3 & 4.0 & 20 & 33.8 & $1.4$ ($3.4$) & 13501 \\
63 & SPT-CLJ2222-4834 &  & 0.652 & 1.2 & 3.5 & 20 & 32.0 & $1.5$ ($2.7$) & 13497 \\
64 & SPT-CLJ2232-5959 &  & 0.594 & 1.9 & 3.5 & 20 & 31.5 & $1.8$ ($3.0$) & 13502 \\
65 & SPT-CLJ2233-5339 &  & 0.48 & 1.8 & 6.0 & 20 & 16.8 & $2.0$ ($4.3$) & 13504 \\
66 & SPT-CLJ2236-4555 &  & 1.162 & 1.1 & 2.5 & 20 & 81.7 & $1.0$ ($3.5$) & 13507, 15266 \\
67 & SPT-CLJ2245-6206 &  & 0.58 & 2.1 & 3.5 & 20 & 28.7 & $1.7$ ($2.3$) & 13499 \\
68 & SPT-CLJ2248-4431 & Abell S1063 & 0.351 & 1.2 & - & 20 & 26.1 & $34.4$ ($14.2$) & 4966 \\
69 & SPT-CLJ2258-4044 &  & 0.826 & 1.1 & 2.5 & 20 & 53.4 & $1.4$ ($2.4$) & 13495 \\
70 & SPT-CLJ2259-6057 &  & 0.75 & 1.9 & 3.0 & 20 & 62.7 & $2.7$ ($4.4$) & 13498 \\
71 & SPT-CLJ2301-4023 &  & 0.73 & 1.1 & 2.0 & 20 & 56.4 & $1.3$ ($1.6$) & 13505 \\
72 & SPT-CLJ2306-6505 &  & 0.530 & 2.3 & 4.5 & 20 & 25.1 & $1.7$ ($3.2$) & 13503 \\
73 & SPT-CLJ2325-4111 & Abell S1121 & 0.358 & 1.6 & 5.0 & 20 & 8.7 & $2.0$ ($1.6$) & 13405 \\
74 & SPT-CLJ2331-5051 &  & 0.576 & 1.1 & 4.0 & 20 & 28.5 & $2.5$ ($3.9$) & 9333 \\
75 & SPT-CLJ2335-4544 &  & 0.547 & 1.3 & 4.5 & 20 & 34.9 & $2.8$ ($4.6$) & 13496, 17477 \\
76 & SPT-CLJ2337-5942 &  & 0.775 & 1.5 & 3.0 & 20 & 19.2 & $1.7$ ($1.7$) & 11859 \\
77 & SPT-CLJ2341-5119 &  & 1.003 & 1.2 & 2.5 & 20 & 79.4 & $2.6$ ($4.9$) & 11799, 9345 \\
78 & SPT-CLJ2342-5411 &  & 1.075 & 1.5 & 2.0 & 20 & 173.2 & $1.7$ ($6.7$) & 11741, 11870, 12014, 12091 \\
79 & SPT-CLJ2344-4243 & Phoenix cluster & 0.596 & 1.5 & 4.5 & 10 & 128.3 & $65.3$ ($15.4$) & 13401, 16135, 16545 \\
80 & SPT-CLJ2345-6405 &  & 0.937 & 2.2 & 2.5 & 20 & 64.4 & $1.5$ ($2.7$) & 13500 \\
81 & SPT-CLJ2352-4657 &  & 0.734 & 1.3 & 2.5 & 20 & 78.4 & $1.5$ ($3.5$) & 13506 \\
82 & SPT-CLJ2355-5055 &  & 0.320 & 1.3 & 4.0 & 20 & 21.3 & $2.5$ ($3.3$) & 11746, 11998 \\
83 & SPT-CLJ2359-5009 &  & 0.775 & 1.3 & 2.5 & 20 & 128.2 & $2.1$ ($7.5$) & 9334, 11742, 11864, 11997 \\
\hline
\end{tabular}

%% file: taba2_1.tex
\begin{tabular}{lrrrrrr}
SPT ID & RA (peak) & Dec (peak) & RA (annulus) & Dec (annulus) & Offset (arcsec) & Offset (kpc) \\ \hline
SPT-CLJ0000-5748 & $0.2499$ & $-57.8093$ & $0.2490$ & $-57.8100$ & $3$ & $21$ \\
SPT-CLJ0013-4906 & $3.3306$ & $-49.1103$ & $3.3304$ & $-49.1159$ & $20$ & $110$ \\
SPT-CLJ0014-4952 & $3.7044$ & $-49.8837$ & $3.6905$ & $-49.8806$ & $34$ & $251$ \\
SPT-CLJ0033-6326 & $8.4691$ & $-63.4444$ & $8.4695$ & $-63.4423$ & $8$ & $51$ \\
SPT-CLJ0040-4407 & $10.2102$ & $-44.1317$ & $10.2088$ & $-44.1328$ & $5$ & $26$ \\
SPT-CLJ0058-6145 & $14.5886$ & $-61.7678$ & $14.5842$ & $-61.7694$ & $9$ & $72$ \\
SPT-CLJ0102-4603 & $15.6774$ & $-46.0716$ & $15.6739$ & $-46.0658$ & $22$ & $162$ \\
SPT-CLJ0102-4915 & $15.7423$ & $-49.2743$ & $15.7347$ & $-49.2664$ & $33$ & $258$ \\
SPT-CLJ0106-5943 & $16.6166$ & $-59.7210$ & $16.6142$ & $-59.7200$ & $6$ & $28$ \\
SPT-CLJ0123-4821 & $20.7980$ & $-48.3559$ & $20.7936$ & $-48.3573$ & $12$ & $79$ \\
SPT-CLJ0142-5032 & $25.5422$ & $-50.5400$ & $25.5452$ & $-50.5401$ & $7$ & $50$ \\
SPT-CLJ0151-5954 & $27.8457$ & $-59.9079$ & $27.8583$ & $-59.9073$ & $23$ & $184$ \\
SPT-CLJ0156-5541 & $29.0437$ & $-55.6984$ & $29.0407$ & $-55.6988$ & $6$ & $53$ \\
SPT-CLJ0200-4852 & $30.1455$ & $-48.8711$ & $30.1391$ & $-48.8739$ & $18$ & $111$ \\
SPT-CLJ0212-4657 & $33.0998$ & $-46.9540$ & $33.1087$ & $-46.9496$ & $27$ & $187$ \\
SPT-CLJ0217-5245 & $34.3047$ & $-52.7632$ & $34.2949$ & $-52.7512$ & $48$ & $236$ \\
SPT-CLJ0232-4421 & $38.0778$ & $-44.3466$ & $38.0710$ & $-44.3513$ & $24$ & $104$ \\
SPT-CLJ0232-5257 & $38.2039$ & $-52.9529$ & $38.1977$ & $-52.9554$ & $16$ & $103$ \\
SPT-CLJ0234-5831 & $38.6745$ & $-58.5235$ & $38.6791$ & $-58.5241$ & $9$ & $49$ \\
SPT-CLJ0235-5121 & $38.9356$ & $-51.3519$ & $38.9351$ & $-51.3576$ & $21$ & $87$ \\
SPT-CLJ0236-4938 & $39.2595$ & $-49.6365$ & $39.2509$ & $-49.6345$ & $21$ & $102$ \\
SPT-CLJ0243-5930 & $40.8620$ & $-59.5172$ & $40.8646$ & $-59.5171$ & $5$ & $32$ \\
SPT-CLJ0252-4824 & $43.2074$ & $-48.4163$ & $43.1949$ & $-48.4139$ & $31$ & $172$ \\
SPT-CLJ0256-5617 & $44.1054$ & $-56.2983$ & $44.1046$ & $-56.2980$ & $2$ & $12$ \\
SPT-CLJ0304-4401 & $46.0700$ & $-44.0253$ & $46.0669$ & $-44.0323$ & $26$ & $153$ \\
SPT-CLJ0304-4921 & $46.0675$ & $-49.3569$ & $46.0665$ & $-49.3573$ & $3$ & $14$ \\
SPT-CLJ0307-5042 & $46.9604$ & $-50.7021$ & $46.9598$ & $-50.7044$ & $9$ & $55$ \\
SPT-CLJ0307-6225 & $46.8160$ & $-62.4474$ & $46.8273$ & $-62.4352$ & $48$ & $316$ \\
SPT-CLJ0310-4647 & $47.6352$ & $-46.7855$ & $47.6357$ & $-46.7831$ & $9$ & $61$ \\
SPT-CLJ0324-6236 & $51.0529$ & $-62.5982$ & $51.0515$ & $-62.5987$ & $3$ & $21$ \\
SPT-CLJ0334-4659 & $53.5460$ & $-46.9958$ & $53.5496$ & $-46.9960$ & $9$ & $53$ \\
SPT-CLJ0346-5439 & $56.7328$ & $-54.6484$ & $56.7315$ & $-54.6472$ & $5$ & $34$ \\
SPT-CLJ0348-4515 & $57.0739$ & $-45.2477$ & $57.0703$ & $-45.2501$ & $13$ & $63$ \\
SPT-CLJ0352-5647 & $58.2406$ & $-56.7959$ & $58.2394$ & $-56.7985$ & $10$ & $67$ \\
SPT-CLJ0406-4805 & $61.7311$ & $-48.0819$ & $61.7271$ & $-48.0850$ & $14$ & $105$ \\
SPT-CLJ0411-4819 & $62.8183$ & $-48.3153$ & $62.8093$ & $-48.3217$ & $32$ & $176$ \\
SPT-CLJ0417-4748 & $64.3463$ & $-47.8135$ & $64.3456$ & $-47.8146$ & $4$ & $29$ \\
SPT-CLJ0426-5455 & $66.5226$ & $-54.9217$ & $66.5201$ & $-54.9169$ & $18$ & $122$ \\
SPT-CLJ0438-5419 & $69.5725$ & $-54.3226$ & $69.5778$ & $-54.3201$ & $14$ & $78$ \\
SPT-CLJ0441-4855 & $70.4489$ & $-48.9236$ & $70.4498$ & $-48.9226$ & $4$ & $31$ \\
SPT-CLJ0449-4901 & $72.2765$ & $-49.0267$ & $72.2741$ & $-49.0248$ & $9$ & $66$ \\
SPT-CLJ0456-5116 & $74.1147$ & $-51.2789$ & $74.1207$ & $-51.2779$ & $14$ & $89$ \\
SPT-CLJ0509-5342 & $77.3388$ & $-53.7037$ & $77.3375$ & $-53.7036$ & $3$ & $16$ \\
SPT-CLJ0516-5430 & $79.1572$ & $-54.5134$ & $79.1490$ & $-54.5126$ & $17$ & $76$ \\
SPT-CLJ0528-5300 & $82.0217$ & $-52.9969$ & $82.0219$ & $-52.9962$ & $3$ & $20$ \\
SPT-CLJ0533-5005 & $83.4068$ & $-50.0971$ & $83.4048$ & $-50.0972$ & $5$ & $35$ \\
SPT-CLJ0542-4100 & $85.7090$ & $-40.9987$ & $85.7118$ & $-41.0021$ & $14$ & $100$ \\
SPT-CLJ0546-5345 & $86.6552$ & $-53.7593$ & $86.6529$ & $-53.7613$ & $9$ & $69$ \\
SPT-CLJ0555-6406 & $88.8578$ & $-64.1070$ & $88.8667$ & $-64.1056$ & $15$ & $73$ \\
SPT-CLJ0559-5249 & $89.9282$ & $-52.8317$ & $89.9354$ & $-52.8249$ & $29$ & $196$ \\
SPT-CLJ0655-5234 & $103.9714$ & $-52.5695$ & $103.9731$ & $-52.5701$ & $4$ & $25$ \\
SPT-CLJ0658-5556 & $104.5829$ & $-55.9418$ & $104.6188$ & $-55.9453$ & $73$ & $324$ \\
SPT-CLJ2031-4037 & $307.9696$ & $-40.6227$ & $307.9646$ & $-40.6219$ & $14$ & $67$ \\
SPT-CLJ2034-5936 & $308.5378$ & $-59.6052$ & $308.5369$ & $-59.6038$ & $5$ & $41$ \\
SPT-CLJ2035-5251 & $308.7974$ & $-52.8556$ & $308.7923$ & $-52.8546$ & $12$ & $74$ \\
SPT-CLJ2043-5035 & $310.8238$ & $-50.5923$ & $310.8246$ & $-50.5933$ & $4$ & $29$ \\
SPT-CLJ2106-5844 & $316.5224$ & $-58.7422$ & $316.5185$ & $-58.7427$ & $8$ & $62$ \\
SPT-CLJ2135-5726 & $323.9093$ & $-57.4411$ & $323.9132$ & $-57.4392$ & $10$ & $56$ \\
\hline
\end{tabular}

%% file: taba2_2.tex
\begin{tabular}{lrrrrrr}
SPT ID & RA (peak) & Dec (peak) & RA (annulus) & Dec (annulus) & Offset (arcsec) & Offset (kpc) \\ \hline
SPT-CLJ2145-5644 & $326.4676$ & $-56.7470$ & $326.4687$ & $-56.7490$ & $7$ & $44$ \\
SPT-CLJ2146-4633 & $326.6453$ & $-46.5475$ & $326.6441$ & $-46.5493$ & $7$ & $57$ \\
SPT-CLJ2148-6116 & $327.1770$ & $-61.2807$ & $327.1811$ & $-61.2787$ & $10$ & $66$ \\
SPT-CLJ2218-4519 & $334.7475$ & $-45.3161$ & $334.7458$ & $-45.3149$ & $6$ & $42$ \\
SPT-CLJ2222-4834 & $335.7119$ & $-48.5764$ & $335.7135$ & $-48.5769$ & $4$ & $28$ \\
SPT-CLJ2232-5959 & $338.1410$ & $-59.9986$ & $338.1423$ & $-59.9986$ & $2$ & $15$ \\
SPT-CLJ2233-5339 & $338.3177$ & $-53.6564$ & $338.3226$ & $-53.6531$ & $16$ & $94$ \\
SPT-CLJ2236-4555 & $339.2181$ & $-45.9309$ & $339.2194$ & $-45.9277$ & $12$ & $98$ \\
SPT-CLJ2245-6206 & $341.2578$ & $-62.1268$ & $341.2576$ & $-62.1196$ & $26$ & $172$ \\
SPT-CLJ2248-4431 & $342.1827$ & $-44.5298$ & $342.1875$ & $-44.5289$ & $13$ & $63$ \\
SPT-CLJ2258-4044 & $344.7024$ & $-40.7390$ & $344.7067$ & $-40.7398$ & $12$ & $92$ \\
SPT-CLJ2259-6057 & $344.7535$ & $-60.9606$ & $344.7509$ & $-60.9590$ & $7$ & $54$ \\
SPT-CLJ2301-4023 & $345.4715$ & $-40.3851$ & $345.4707$ & $-40.3893$ & $15$ & $111$ \\
SPT-CLJ2306-6505 & $346.7276$ & $-65.0926$ & $346.7277$ & $-65.0898$ & $10$ & $65$ \\
SPT-CLJ2325-4111 & $351.2986$ & $-41.2015$ & $351.3015$ & $-41.1959$ & $21$ & $107$ \\
SPT-CLJ2331-5051 & $352.9636$ & $-50.8649$ & $352.9606$ & $-50.8635$ & $8$ & $56$ \\
SPT-CLJ2335-4544 & $353.7836$ & $-45.7399$ & $353.7862$ & $-45.7390$ & $7$ & $46$ \\
SPT-CLJ2337-5942 & $354.3541$ & $-59.7064$ & $354.3525$ & $-59.7062$ & $3$ & $23$ \\
SPT-CLJ2341-5119 & $355.3017$ & $-51.3287$ & $355.2989$ & $-51.3287$ & $6$ & $50$ \\
SPT-CLJ2342-5411 & $355.6921$ & $-54.1852$ & $355.6913$ & $-54.1827$ & $9$ & $73$ \\
SPT-CLJ2344-4243 & $356.1831$ & $-42.7202$ & $356.1839$ & $-42.7207$ & $3$ & $18$ \\
SPT-CLJ2345-6405 & $356.2405$ & $-64.0960$ & $356.2498$ & $-64.0998$ & $20$ & $158$ \\
SPT-CLJ2352-4657 & $358.0683$ & $-46.9594$ & $358.0687$ & $-46.9597$ & $1$ & $11$ \\
SPT-CLJ2355-5055 & $358.9479$ & $-50.9280$ & $358.9496$ & $-50.9290$ & $5$ & $24$ \\
SPT-CLJ2359-5009 & $359.9318$ & $-50.1718$ & $359.9318$ & $-50.1707$ & $4$ & $32$ \\
\hline
\end{tabular}

%% file: taba3_1.tex
\begin{tabular}{lccccccc}
Name & BIN-NFW & BIN-GNFW & BIN-NONHYDRO & INT-NFW & MBETA-NFW & KPLAW-NFW & GRAD-NFW \\
\hline
SPT-CLJ0000-5748 & $ -0.2$ & $ -0.2$ & $ -0.2$ & $ +0.4$ & $ +0.5$ & $ +0.6$ & $ +0.5$ \\
SPT-CLJ0013-4906 & $ -0.9$ & $ -1.0$ & $ -1.2$ & $ +0.0$ & $ +0.5$ & $ +1.4$ & $ +0.1$ \\
SPT-CLJ0014-4952 & $ +1.3$ & $ +1.2$ & $ +0.7$ & $ +1.1$ & $ +1.6$ & $ +1.6$ & $ +1.3$ \\
SPT-CLJ0033-6326 & $ -1.9$ & $ -1.9$ & $ -1.9$ & $ -0.6$ & $ -0.5$ & $ -0.5$ & $ -0.5$ \\
SPT-CLJ0040-4407 & $ -2.5$ & $ -2.5$ & $ -2.5$ & $ -1.8$ & $ -1.7$ & $ -1.7$ & $ -1.7$ \\
SPT-CLJ0058-6145 & $ +1.9$ & $ +1.9$ & $ +1.5$ & $ +0.5$ & $ +0.5$ & $ +1.2$ & $ +0.5$ \\
SPT-CLJ0102-4603 & $ -0.8$ & $ -0.9$ & $ -1.0$ & $ +1.2$ & $ +1.2$ & $ +1.2$ & $ +1.2$ \\
SPT-CLJ0102-4915 & $ +0.5$ & $ +0.5$ & $ -0.1$ & $ +1.1$ & $+12.9$ & $ +5.8$ & $ +5.9$ \\
SPT-CLJ0106-5943 & $ -1.3$ & $ -1.3$ & $ -1.3$ & $ -0.4$ & $ +0.3$ & $ +0.4$ & $ -0.1$ \\
SPT-CLJ0123-4821 & $ +0.8$ & $ +0.8$ & $ +1.0$ & $ +1.8$ & $ +1.9$ & $ +1.9$ & $ +1.8$ \\
SPT-CLJ0142-5032 & $ -1.0$ & $ -1.0$ & $ -1.0$ & $ -0.4$ & $ -0.2$ & $ -0.3$ & $ -0.3$ \\
SPT-CLJ0151-5954 & $ +1.5$ & $ +1.5$ & $ +1.5$ & $ +0.4$ & $ +0.4$ & $ +0.4$ & $ +0.3$ \\
SPT-CLJ0156-5541 & $ +1.0$ & $ +1.0$ & $ +1.0$ & $ +2.5$ & $ +2.6$ & $ +2.5$ & $ +2.6$ \\
SPT-CLJ0200-4852 & $ +1.3$ & $ +1.3$ & $ +1.3$ & $ +2.1$ & $ +2.3$ & $ +2.2$ & $ +2.2$ \\
SPT-CLJ0212-4657 & $ -1.6$ & $ -1.6$ & $ -1.8$ & $ -1.6$ & $ -1.5$ & $ -1.2$ & $ -1.6$ \\
SPT-CLJ0217-5245 & $ -0.4$ & $ -0.4$ & $ -0.5$ & $ -0.6$ & $ -0.5$ & $ -0.5$ & $ -0.5$ \\
SPT-CLJ0232-4421 & $ -1.4$ & $ -1.6$ & $ -1.9$ & $ +0.3$ & $ +0.8$ & $ +1.3$ & $ +1.1$ \\
SPT-CLJ0232-5257 & $ -0.4$ & $ -0.4$ & $ -0.4$ & $ -0.6$ & $ -0.4$ & $ -0.4$ & $ -0.6$ \\
SPT-CLJ0234-5831 & $ -1.0$ & $ -1.0$ & $ -1.1$ & $ -0.7$ & $ -0.2$ & $ -0.2$ & $ -0.3$ \\
SPT-CLJ0235-5121 & $ +1.3$ & $ +1.2$ & $ +1.4$ & $ +0.8$ & $ +1.0$ & $ +1.2$ & $ +1.0$ \\
SPT-CLJ0236-4938 & $ -0.4$ & $ -0.5$ & $ -0.6$ & $ -1.6$ & $ -0.8$ & $ -0.8$ & $ -1.2$ \\
SPT-CLJ0243-5930 & $ -1.1$ & $ -1.1$ & $ -1.1$ & $ +2.1$ & $ +2.3$ & $ +2.4$ & $ +2.4$ \\
SPT-CLJ0252-4824 & $ -1.3$ & $ -1.3$ & $ -1.3$ & $ +0.7$ & $ +0.8$ & $ +0.9$ & $ +0.7$ \\
SPT-CLJ0256-5617 & $ -2.1$ & $ -2.1$ & $ -2.0$ & $ -1.0$ & $ +0.3$ & $ +0.4$ & $ -0.7$ \\
SPT-CLJ0304-4401 & $ +0.1$ & $ +0.1$ & $ +0.1$ & $ +1.6$ & $ +2.1$ & $ +2.8$ & $ +2.3$ \\
SPT-CLJ0304-4921 & $ -0.2$ & $ -0.3$ & $ -0.4$ & $ +0.5$ & $ +1.0$ & $ +0.9$ & $ +0.8$ \\
SPT-CLJ0307-5042 & $ -1.0$ & $ -1.1$ & $ -1.0$ & $ -0.1$ & $ +0.1$ & $ +0.1$ & $ +0.1$ \\
SPT-CLJ0307-6225 & $ -0.5$ & $ -0.7$ & $ -0.5$ & $ -0.6$ & $ -0.0$ & $ -0.1$ & $ -0.6$ \\
SPT-CLJ0310-4647 & $ -0.4$ & $ -0.4$ & $ -0.5$ & $ +0.6$ & $ +0.5$ & $ +0.6$ & $ +0.5$ \\
SPT-CLJ0324-6236 & $ +1.2$ & $ +1.2$ & $ +1.2$ & $ +0.9$ & $ +1.0$ & $ +1.1$ & $ +0.9$ \\
SPT-CLJ0334-4659 & $ -1.6$ & $ -1.7$ & $ -1.6$ & $ -0.2$ & $ +0.6$ & $ +0.6$ & $ +0.5$ \\
SPT-CLJ0346-5439 & $ -0.8$ & $ -0.8$ & $ -0.8$ & $ -0.8$ & $ -0.7$ & $ -0.7$ & $ -0.7$ \\
SPT-CLJ0348-4515 & $ -0.1$ & $ -0.2$ & $ -0.2$ & $ -0.4$ & $ -0.4$ & $ -0.3$ & $ -0.4$ \\
\hline
\end{tabular}

%% file: taba3_2.tex
\begin{tabular}{lccccccc}
Name & BIN-NFW & BIN-GNFW & BIN-NONHYDRO & INT-NFW & MBETA-NFW & KPLAW-NFW & GRAD-NFW \\
\hline
SPT-CLJ0352-5647 & $ -0.3$ & $ -0.3$ & $ -0.3$ & $ -0.5$ & $ -0.3$ & $ -0.4$ & $ -0.4$ \\
SPT-CLJ0406-4805 & $ -0.9$ & $ -0.9$ & $ -0.9$ & $ -2.3$ & $ -2.2$ & $ -2.2$ & $ -2.3$ \\
SPT-CLJ0411-4819 & $ -0.6$ & $ -0.6$ & $ -1.2$ & $ +0.7$ & $ +2.6$ & $ +2.6$ & $ +1.5$ \\
SPT-CLJ0417-4748 & $ -1.8$ & $ -1.8$ & $ -1.9$ & $ -1.9$ & $ -1.6$ & $ -1.6$ & $ -1.6$ \\
SPT-CLJ0426-5455 & $ -0.6$ & $ -0.6$ & $ -0.5$ & $ +0.7$ & $ +0.8$ & $ +0.8$ & $ +0.6$ \\
SPT-CLJ0438-5419 & $ -1.8$ & $ -1.9$ & $ -2.0$ & $ +1.2$ & $ +1.5$ & $ +1.6$ & $ +1.4$ \\
SPT-CLJ0441-4855 & $ -1.8$ & $ -1.8$ & $ -1.8$ & $ -0.4$ & $ -0.2$ & $ -0.2$ & $ -0.3$ \\
SPT-CLJ0449-4901 & $ -0.5$ & $ -0.6$ & $ -0.4$ & $ +0.4$ & $ +0.6$ & $ +0.8$ & $ +0.5$ \\
SPT-CLJ0456-5116 & $ -1.6$ & $ -1.7$ & $ -1.8$ & $ -1.0$ & $ -0.8$ & $ -0.8$ & $ -1.0$ \\
SPT-CLJ0509-5342 & $ -0.4$ & $ -0.5$ & $ -0.4$ & $ -0.4$ & $ +0.2$ & $ +0.4$ & $ -0.1$ \\
SPT-CLJ0516-5430 & $ -0.7$ & $ -0.7$ & $ -0.8$ & $ +1.0$ & $ +1.4$ & $ +1.6$ & $ +1.4$ \\
SPT-CLJ0528-5300 & $ -1.4$ & $ -1.4$ & $ -1.4$ & $ -0.3$ & $ -0.2$ & $ -0.2$ & $ -0.4$ \\
SPT-CLJ0533-5005 & $ +0.1$ & $ +0.1$ & $ +0.0$ & $ +0.4$ & $ +0.7$ & $ +0.9$ & $ +0.5$ \\
SPT-CLJ0542-4100 & $ +1.2$ & $ +1.1$ & $ +1.1$ & $ -0.0$ & $ +0.3$ & $ +0.3$ & $ +0.2$ \\
SPT-CLJ0546-5345 & $ -1.3$ & $ -1.3$ & $ -1.5$ & $ +0.1$ & $ +0.3$ & $ +0.3$ & $ +0.2$ \\
SPT-CLJ0555-6406 & $ -0.2$ & $ -0.2$ & $ -0.1$ & $ +1.4$ & $ +1.6$ & $ +1.7$ & $ +1.7$ \\
SPT-CLJ0559-5249 & $ -1.7$ & $ -1.7$ & $ -1.9$ & $ -0.1$ & $ +0.4$ & $ +0.3$ & $ +0.3$ \\
SPT-CLJ0655-5234 & $ -2.1$ & $ -2.2$ & $ -2.1$ & $ +0.4$ & $ +0.4$ & $ +0.5$ & $ +0.3$ \\
SPT-CLJ0658-5556 & $+25.1$ & $+25.0$ & $+18.6$ & $+13.8$ & $+220.3$ & $+512.8$ & $+103.7$ \\
SPT-CLJ2031-4037 & $ -0.1$ & $ -0.1$ & $ -0.3$ & $ -1.3$ & $ -0.8$ & $ -0.7$ & $ -1.1$ \\
SPT-CLJ2034-5936 & $ -0.1$ & $ -0.1$ & $ -0.1$ & $ +1.7$ & $ +1.7$ & $ +1.9$ & $ +1.4$ \\
SPT-CLJ2035-5251 & $ -2.3$ & $ -2.3$ & $ -2.3$ & $ -0.8$ & $ -0.8$ & $ -0.8$ & $ -1.0$ \\
SPT-CLJ2043-5035 & $ -1.1$ & $ -1.3$ & $ -1.5$ & $ +1.0$ & $ +1.4$ & $ +1.5$ & $ +1.2$ \\
SPT-CLJ2106-5844 & $ -0.9$ & $ -0.9$ & $ -1.0$ & $ -0.1$ & $ +0.4$ & $ +2.7$ & $ +0.0$ \\
SPT-CLJ2135-5726 & $ -1.8$ & $ -1.9$ & $ -2.0$ & $ -1.0$ & $ -0.4$ & $ -0.5$ & $ -0.6$ \\
SPT-CLJ2145-5644 & $ -1.3$ & $ -1.4$ & $ -1.6$ & $ -1.6$ & $ -1.5$ & $ -1.3$ & $ -1.6$ \\
SPT-CLJ2146-4633 & $ -1.6$ & $ -1.6$ & $ -1.6$ & $ -1.1$ & $ -1.1$ & $ -0.9$ & $ -1.2$ \\
SPT-CLJ2148-6116 & $ -1.8$ & $ -1.8$ & $ -1.7$ & $ -0.2$ & $ +0.0$ & $ +0.1$ & $ -0.1$ \\
SPT-CLJ2218-4519 & $ -1.1$ & $ -1.1$ & $ -1.1$ & $ -1.0$ & $ -0.9$ & $ -0.9$ & $ -0.9$ \\
SPT-CLJ2222-4834 & $ -1.3$ & $ -1.3$ & $ -1.2$ & $ +1.7$ & $ +1.8$ & $ +1.8$ & $ +1.6$ \\
SPT-CLJ2232-5959 & $ +0.7$ & $ +0.6$ & $ +0.6$ & $ -0.1$ & $ +0.0$ & $ +0.0$ & $ -0.0$ \\
SPT-CLJ2233-5339 & $ -0.7$ & $ -0.7$ & $ -0.7$ & $ +1.1$ & $ +1.3$ & $ +1.2$ & $ +1.1$ \\
SPT-CLJ2236-4555 & $ -0.3$ & $ -0.3$ & $ -0.6$ & $ -0.0$ & $ +0.3$ & $ +0.3$ & $ -0.0$ \\
SPT-CLJ2245-6206 & $ -1.5$ & $ -1.5$ & $ -1.8$ & $ -0.2$ & $ -0.1$ & $ -0.1$ & $ -0.1$ \\
SPT-CLJ2248-4431 & $ +0.3$ & $ +0.2$ & $ -0.4$ & $ +0.5$ & $ +1.4$ & $ +1.4$ & $ +1.4$ \\
SPT-CLJ2258-4044 & $ -1.3$ & $ -1.3$ & $ -1.6$ & $ -1.2$ & $ -1.0$ & $ -1.0$ & $ -1.0$ \\
SPT-CLJ2259-6057 & $ +0.3$ & $ +0.2$ & $ +0.4$ & $ +1.3$ & $ +1.7$ & $ +1.5$ & $ +1.5$ \\
SPT-CLJ2301-4023 & $ -1.1$ & $ -1.2$ & $ -1.5$ & $ -0.3$ & $ +0.0$ & $ +0.3$ & $ -0.1$ \\
SPT-CLJ2306-6505 & $ -0.4$ & $ -0.4$ & $ -0.4$ & $ -0.6$ & $ -0.3$ & $ +0.1$ & $ -0.6$ \\
SPT-CLJ2325-4111 & $ -0.3$ & $ -0.3$ & $ -0.6$ & $ -1.1$ & $ -0.7$ & $ -0.6$ & $ -0.8$ \\
SPT-CLJ2331-5051 & $ +0.2$ & $ +0.0$ & $ -0.2$ & $ -0.2$ & $ -0.1$ & $ -0.1$ & $ -0.1$ \\
SPT-CLJ2335-4544 & $ -1.6$ & $ -1.6$ & $ -1.6$ & $ +0.5$ & $ +0.6$ & $ +0.8$ & $ +0.6$ \\
SPT-CLJ2337-5942 & $ -0.7$ & $ -0.7$ & $ -0.7$ & $ -1.6$ & $ -1.4$ & $ -1.4$ & $ -1.4$ \\
SPT-CLJ2341-5119 & $ -0.6$ & $ -0.7$ & $ -0.8$ & $ -0.7$ & $ -0.5$ & $ -0.5$ & $ -0.7$ \\
SPT-CLJ2342-5411 & $ -0.7$ & $ -0.7$ & $ -0.7$ & $ -1.2$ & $ -1.0$ & $ -1.0$ & $ -1.1$ \\
SPT-CLJ2344-4243 & $ +1.7$ & $ +1.6$ & $ +1.3$ & $ +1.6$ & $ +4.3$ & $ +4.9$ & $ +5.1$ \\
SPT-CLJ2345-6405 & $ +0.0$ & $ +0.0$ & $ -0.3$ & $ +1.5$ & $ +1.5$ & $ +1.6$ & $ +1.5$ \\
SPT-CLJ2352-4657 & $ -0.6$ & $ -0.7$ & $ -0.7$ & $ +0.9$ & $ +1.1$ & $ +1.0$ & $ +0.9$ \\
SPT-CLJ2355-5055 & $ +1.1$ & $ +1.0$ & $ +0.9$ & $ +2.5$ & $ +2.6$ & $ +2.6$ & $ +2.4$ \\
SPT-CLJ2359-5009 & $ +0.3$ & $ +0.3$ & $ +0.3$ & $ +1.5$ & $ +1.6$ & $ +1.6$ & $ +1.5$ \\
\hline
\end{tabular}